%% file: main.tex
\documentclass[a4paper,11pt]{article}
\usepackage{jheppub} 
\usepackage{lineno}

\usepackage[T1]{fontenc}
\usepackage{lmodern} 
\usepackage[utf8]{inputenc} 
\usepackage{mathrsfs}
\usepackage{amssymb,amsmath,mathtools}
\usepackage{amsthm}
\newtheorem{postulate}{Postulate}

\newtheorem{weakpostulate}{Assumption}

\newtheorem{conjecture}{Conjecture}[section]
\usepackage{graphicx}
\usepackage{xcolor}
\usepackage{placeins}
\usepackage[normalem]{ulem}
\usepackage{stmaryrd}
\usepackage{booktabs}
\usepackage{fancyhdr}
\usepackage{pdflscape}
\usepackage{enumitem}
\setlist[itemize]{noitemsep} 

\definecolor{lcolor}{rgb}{0.5,0,0}
\definecolor{citcolor}{rgb}{0,0.3,0.0}
\definecolor{coloryksi}{rgb}{0.5,0.0,0.0}
\definecolor{colorkaksi}{rgb}{0.0,0.0,0.5}
\definecolor{colorkolme}{rgb}{0.0,0.3,0.1}

\usepackage{accents}
\usepackage{physics}
\usepackage{tensor}
\usepackage{cancel}
\usepackage{mathrsfs}

\newcommand{\nlinenotag}{\notag \\}
\newcommand{\pequiv}{\overset{\mathrm{P\ref{post:pdf:qcd}}}{\equiv}}

\newcommand{\beq}{\begin{equation}}
\newcommand{\eeq}{\end{equation}}

\newcommand{\Acal}{\mathcal{A}}
\newcommand{\Bcal}{\mathcal{B}}
\newcommand{\Ccal}{\mathcal{C}}

\newcommand{\Pcal}{\mathcal{P}}
\newcommand{\Ical}{\mathcal{I}}
\newcommand{\icalid}{\mathcal{I}^{(1)}}
\newcommand{\icalx}{\mathcal{I}^{(x)}}
\newcommand{\Kcal}{\mathcal{K}}
\DeclareMathOperator\Id{Id}
\newcommand{\bI}{\mathbf{I}}
\DeclareMathOperator\Li{Li}
\DeclareMathOperator\minimize{minimize}
\newcommand{\circumeq}{\mathrel{\widehat{=}}}

\newcommand{\bA}{\mathbf{A}}

\newcommand{\bB}{\mathbf{B}}

\newcommand{\bC}{\mathbf{C}}

\newcommand{\fnc}{\mathrm{f.n.c.}}
\newcommand{\tmc}{\mathrm{t.m.c.}}
\newcommand{\sumrulGLS}{\mathrm{GLS}}
\newcommand{\sumrulG}{\mathrm{G}}
\newcommand{\sumrulA}{\mathrm{A}}
\newcommand{\pdtbf}{\mathbf{f}}

\newcommand{\pdtwritout}{\begin{bmatrix} f_g(x,Q^2) \\ f_d(x,Q^2) \\ \vdots \\ f_{\bar t}(x,Q^2)\end{bmatrix}}
\newcommand{\pdtwritoutz}{\begin{bmatrix} f_g(z,Q^2) \\ f_d(z,Q^2) \\ \vdots \\ f_{\bar t}(z,Q^2)\end{bmatrix}}
\newcommand{\pdtwritoutzdot}{\begin{bmatrix} f_g(\cdot,Q^2) \\ f_d(\cdot,Q^2) \\ \vdots \\ f_{\bar t}(\cdot,Q^2)\end{bmatrix}}

\newcommand{\xbj}{{x_\mathrm{Bj}}}

\input{defs.tex}

\input{glossary.tex}
\makenoidxglossaries

\author{Henri H\"anninen}
\emailAdd{henri.j.hanninen@jyu.fi}
\affiliation{
Department of Mathematics and Statistics, University of Jyväskylä, \\
 P.O. Box 35, 40014 University of Jyv\"askyl\"a, Finland}

\title{Mathematical inverse problem for the world data inference of the parton distribution functions of the proton}


\abstract{
We show that the inference problem of constraining the parton distribution functions of the proton from deeply inelastic scattering data can be formulated as a linear tensor reconstruction inverse problem.
This means that instead of fitting model parameters to the world data, a reconstructive approach to solve a system of coupled linear functional integral equations can be formulated.
We leverage the mathematical structures of the integral equations defined by perturbative QCD and global analysis to pose the problem of global analysis as a coupled system of linear integral equations, and construct a proof-of-principle methodology to solve it.
To formulate this approach concretely, we review all next-to-leading order accuracy results for inclusive virtual photon, neutral current, charged current, heavy flavor production, and neutrino deep-inelastic lepton--proton scattering.
This mathematical methodology of inference opens a path towards a parametrization-bias-free extraction of the proton PDFs from the world data of deeply inelastic scattering in the spirit of indirect measurement employed in mathematical inverse problems, including robust estimation of uncertainties with reduced bias from model parametrization, while closely adhering to the established paradigm of perturbative QCD and global analysis.}

\begin{document} 
\allowdisplaybreaks
\maketitle
\flushbottom 

\section{Introduction} \label{sec:intro}

    \noindent
    The goal of this article is to give a mathematical answer to the question:
    \vspace{2mm}

    \noindent
    \begin{center}
    \begin{minipage}{0.95\linewidth}
    ``What would the \gls{world data} inference problem for the \glspl{parton distribution function} (PDFs) of the proton look like, if it was written in the mathematical language of inverse problems as an indirect measurement problem, as is done in medical imaging{?}''
    \end{minipage}    
    \end{center}
    \vspace{2mm}

    \noindent
    The fields of high-energy physics and medical imaging both strive to learn physical truths of phenomena not directly observable, yet in one field it is possible to solve the internal structure of the subject from data in an \textit{imaging} sense, whereas in the other it has seemed necessary to rely on fitting theoretically motivated ansatz model parametrizations. The underlying physics phenomena are of course completely distinct, but why would that rule out the use of analogous methodologies from one or the other? And why does medical imaging not \textit{need to use fitting} as a part of its methods?
    To give an initial vague answer, the underlying reason is that in medical imaging the mathematical theory of inference and physical experiment are constructed in unison to yield a robust general purpose methods to solve the directly unobservable from the measurement. For example, computed tomography reconstructs the three-dimensional structure of the target from a set of two-dimensional translucent projections \textit{without fitting}, which is enabled by the mathematical theory of the X-ray transform~\cite{JohnFritz:1938:xraytransform,Natterer:2001:mathematicalmethods-img-rec,Hansen:2021ct}, which is a concrete example of applying the core principle of inverse problems: harmonious development of physical theory, experimental setup, and mathematical theory of the inference problem.

    The answers to these questions are explored in the mathematical framework of inverse problems, which enables the precise formalization of the indirect measurement problems as mathematical questions, with the associated capability to pose questions about the existence, uniqueness, and accuracy of solutions, and the methods for the discovery thereof.
    Specifically, we will be considering the inference problem of constraining the \gls{parton distribution function}s of the proton from all available data, an inverse problem known in the high-energy physics community as \textit{\gls{global analysis}}. We strive to formalize the problem of \gls{global analysis} in the framework of inverse problems as faithfully as possible to be able to consider the mathematical nature of the inference problem while being as grounded as possible to the established literature of the field that has been active and of central importance for the field of high-energy physics at least since the 1980s~\cite{Gluck:1980cp, Duke:1983gd, Eichten:1984eu}; see Refs.~\cite{ParticleDataGroup:2024cfk,Gao:2017yyd:global-analysis-modern-rev,Ethier:2020way:global-analysis-modern-rev,Amoroso:2022eow:global-analysis-modern-rev,Huston:2023ofk:global-analysis-modern-rev,Blumlein:2012bf:dis-review,Sterman:1995:handbook-qcd} for reviews, and Refs.~\cite{NNPDF:2021njg,Hou:2019efy,Harland-Lang:2014zoa,Martin:2009iq,PDF4LHCWorkingGroup:2022cjn:pdf-comparison} for overview and technical detail on the present-day global analysis problem for the proton PDFs.

\begin{figure}
        \centering
        \begin{minipage}{0.49\linewidth}
        \includegraphics[width=0.99\linewidth]{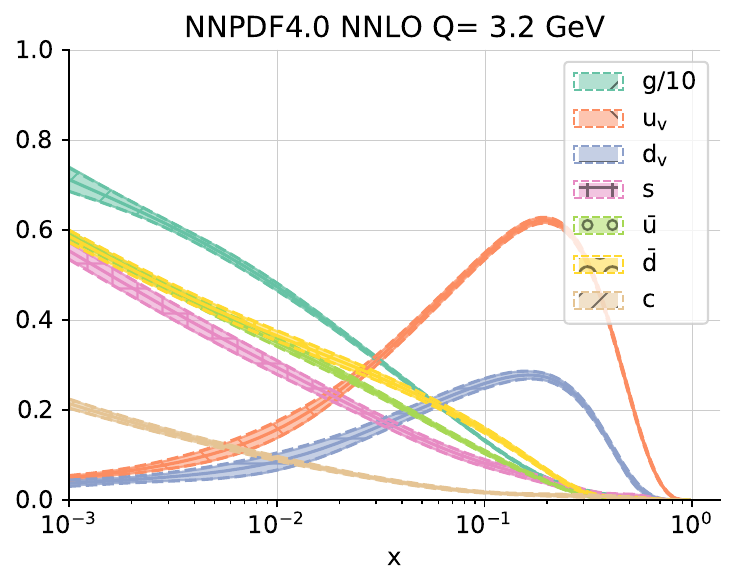}    
        \end{minipage}
        \begin{minipage}{0.49\linewidth}
        \includegraphics[width=0.99\linewidth]{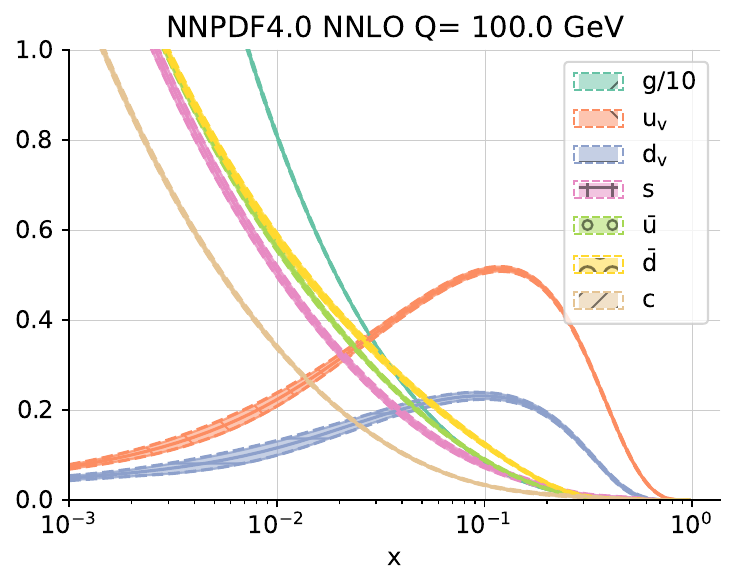}    
        \end{minipage}
        \caption{Illustration of the parton distribution functions of the proton at two momentum scales, $Q=3.2 \, \mathrm{\gls{GeV}}$ on the left, $Q=100 \, \mathrm{\gls{GeV}}$ on the right, as determined by global analysis by the NNPDF collaboration~\cite{NNPDF:2021njg}.
        The valence contributions from the up and down quarks, $u_v$ and $d_v$, show the $x$ and $Q^2$ scale dependence 
        of structure of the proton, which in the early days of modern elementary particle physics was interpreted as a simple bound state of two up and one down quark within the quark model~\cite{Gell-Mann:1961omu,Gell-Mann:1964ewy,Zweig:1964jf,Zweig:1964ruk,Neeman:1961jhl,Bjorken:1964gz}.
        All the rest are contributions from the so-called parton sea, which show the significance of the contributions from other partons than just the $uud$ to the structure of the proton.
        The fact that it has not been possible to calculate the PDFs from first principles within QCD highlights the fact that our collective understanding of the rich structure of the proton is, at least in part, limited by the accuracy of their inference in global analysis.
        Reproduced from Ref.~\cite{NNPDF:2021njg} by the NNPDF Collaboration, \href{https://doi.org/10.1140/epjc/s10052-022-10328-7}{\textit{Eur.Phys.J. C} \textbf{82}, 428 (2022)}, (\href{https://creativecommons.org/licenses/by/4.0/}{CC BY 4.0}).}
        \label{fig:nnpdf-pdfs-overview}
    \end{figure}

    The \gls{parton distribution function}s describe the rich energy and momentum scale dependent structure of the proton, and roughly speaking, they describe the probability of finding a given \gls{parton} in the target proton in the particle collision, which are quantified as functions of \gls{bjorkenx} and photon \gls{virtuality} $Q^2$ variables.
    Figure \ref{fig:nnpdf-pdfs-overview} illustrates the present day state of the art of the knowledge of the proton PDFs: the \glspl{valence quark} $u$ and $d$ have pronounced peaks at moderate $x$, whereas all the other lines from the ``background'' parton sea grow at small $x$, and comparing the left and right figures shows how the structure of the proton evolves in $Q^2$: increasing $Q^2$ corresponds to peering to shorter distance scales into the proton structure.
    \Gls{quantum chromodynamics} (QCD) is the part of the Standard Model that describes the strong nuclear interaction, and so enables precision calculations to relate the parton distribution functions to measurable observables in experiments~\cite{ParticleDataGroup:2024cfk,Ellis:1996mzs:qcd-collider-phys,Sterman:1995:handbook-qcd,Huston:2023ofk:global-analysis-modern-rev}.
    For our purposes in this work, it is sufficient to presume that each \gls{flavor} of \gls{quark} and \gls{anti}quark, and the \gls{gluon} have their own specific PDF $f_a(x,Q^2)$ and the task is to explore the mathematical tools required to consider solving them from measurements. In a moment we will consider briefly what type of regularity properties---such as degree of smoothness or integrability---they can be expected to possess.

    Before we begin formulating the approach of this work, a brief history of global analysis of the PDFs is in order to contextualize the foundations and goals of this approach in the long-established results of the field.
    The first global analyses of the PDFs by Gl\"uck \textit{et al.}, 1982~\cite{Gluck:1980cp}; Duke and Owens, 1984~\cite{Duke:1983gd}; Eichten, Quigg, Hinchliffe, and Lane, 1984~\cite{Eichten:1984eu} were widely used in computations of high-energy particle collision phenomena~\cite{Sterman:1995:handbook-qcd}.
    Substantial improvements in experiments and theory calculations have made the modern day global analysis a highly complex task of application of QCD and data analysis, which makes it quite a non-trivial result that the three major modern fits of the PDFs by the CT~\cite{Hou:2019efy}, MSHT~\cite{Bailey:2020ooq-MSHT}, and NNPDF~\cite{NNPDF:2021njg} collaborations have converged to agree to a substantial degree in their results~\cite{ParticleDataGroup:2024cfk}. It is hard to over-emphasize the importance of the role of the PDFs in modern high-energy physics, particle physics, and cosmology: be it the study of quark gluon plasma: the state of matter in the earliest moments of the universe~\cite{Heinz:2000bk:qgp-plasma}, search for new elementary particles such as the Higgs boson~\cite{ATLAS:2012yve:Higgs}, or for an as yet unknown particle as an answer to questions about dark matter~\cite{CMS:2016gox:dark,Alekhin:2015byh:dark,Hong:2017avi:dark}, the parton distribution functions are a central tool needed to compute high precision theory predictions for high-energy hadron collisions that enable these fields of modern physics to build on that foundation. 

    Historically the functional parametrization ansatz for the PDFs has had variations of the form~\cite{Martin:2009iq,Sterman:1995:handbook-qcd}:
    \begin{equation}
        \label{eq:fit-ansatz}
        xf(x) \approx Ax^\alpha (1-x)^\beta P(x), ~~ A>0, ~ \alpha<0, ~ \beta>0,
    \end{equation}
    where $A$ is an overall normalization, and $P(x)$ has been a smooth function $x$~\cite{Martin:2009iq,Sterman:1995:handbook-qcd}, or a functional output from a neural network~\cite{NNPDF:2021njg}.
    The functional ansatz is composed of two distinct parts: at small $x$ there is a theoretical prediction for the behavior $x f(x) \sim x^\alpha$, and at large $x$ another prediction prescribes the factor $(1-x)^\beta$.
    The sole unifying feature of these predictions is that they cannot be deduced within the framework of perturbative QCD~\cite{Ball:2016spl}, which is the most successful paradigm of QCD, and other theory pictures need to be employed. The small $x$ prediction originates from Regge theory of high-energy scattering~\cite{Regge:1959mz,Collins:1977jy:intro-to-regge,Ellis:1996mzs:qcd-collider-phys}, whereas the large $x$ prediction emerges from Brodsky--Farrar quark counting rules~\cite{Brodsky:1973kr}.
    Naively, the prediction at small $x$ reflects the physical expectation that the \glspl{sea quark} and \glspl{gluon} carry very small fractions of the parent proton longitudinal momentum, so the probability of probing them in the collision is most significant at small $x$. And on the other hand, at $x \to 1$ all PDFs vanish, since any individual parton in the proton has no possibility of carrying all of the longitudinal momentum of the whole proton.
    
    The validity of the assumptions made by the functional ansatz \eqref{eq:fit-ansatz} has been a topic of recent interest~\cite{Ball:2016spl,Carrazza:2021yrg:pdf-parametrization}, and the small-$x$ prediction from Regge theory is in some tension with the prediction of gluon saturation~\cite{Gelis:2010nm,Armesto:2022mxy:pdf-saturation}, where the growth of the gluon density should halt as a maximal density is reached at small $x$.
    These points are in the core of the motivation for this article: how can one perform the inference of the PDFs without assuming a functional ansatz parametrization such as given by Eq.~\eqref{eq:fit-ansatz}? Such an approach would give a complementary perspective of the problem to enable the assessment of the validity of the canonical ansatz. 
    Notably, from the perspective of this work, all of the discussed global analyses of the PDFs are model parameter fits---even with the application of neural networks for a more general manner of parameterizing the PDFs---especially in comparison with medical imaging, where the unknown quantity is mathematically solved from measurement \textit{without fitting any parameters}.

    The fundamental fact that any particle carrying color charge cannot exist freely---a phenomenon known as color confinement~\cite{ParticleDataGroup:2024cfk,Peskin:0201503972,Sterman:1995:handbook-qcd}---makes the inference of their physical properties and \glspl{interaction} an inherent inverse problem. Their study is further complicated by the following fundamental challenges:
    \begin{itemize}
        \item Fundamental inaccessibility. Free quarks or gluons cannot be observed, so the structure of hadronic bound states like the proton need to be studied. Random nature of the collider experiments and quantum effects introduce randomness to all observations, so many or all parton types contribute to the theoretical description of the measurements. For example all the partons contribute to the total probability of scattering which is quantified by the total cross section. The inference method needs to leverage different measurements simultaneously to have any contrast between the distributions of the partons in the proton.
        The solution to these challenges is to include all feasible physical processes in the global analysis.
        \item \Gls{renormalization} scheme dependence.
        \Gls{renormalization}~\cite{tHooft:1972tcz,tHooft:1973mfk,Weinberg:1973xwm,Collins:1984renormalization} is an essential step in the application of quantum field theories that connects the originally non-interacting free theory to the fully interacting theory that more accurately describes the real world. Basically, elementary particles are able to interact with themselves, which means that the original definitions of masses and charges in the free theory must be amended based on these dynamics, a process which is described by the renormalization procedure. This procedure mandates the choice of a ``scheme'' under which this process is systematically performed, and there is some freedom how to choose this scheme.
        The scheme must be fixed first before posing the inverse problem, after which the parton distribution functions become well-defined mathematical objects that can be solved for from data.
        \item Deep interdependence of unknown quantities. 
        The inference of the PDFs is complicated by other unknowns, such as the precise masses of the quarks and the strength of the strong nuclear force, both of which depend on the momentum scale of the \gls{interaction}, effects known as ``running''~\cite{ParticleDataGroup:2024cfk,CMS:2024trs:running-as,Deur:2016tte:running-as,Fusaoka:1998vc}.
        To simplify the initial construction of this approach, we assume all other variables than the PDFs to be known from experiment.
    \end{itemize}
    Our goal is to formulate the mathematics of solving the PDFs under these assumptions in as general manner as possible, i.e. as a functional optimization problem analogously to medical imaging.

    To construct this inverse problem methodology to solve the proton PDFs from the \gls{world data} in a \textit{functional sense}---in contrast with solving model parameters by fitting---we employ the following foundational postulates about the inference problem, and two additional weaker assumptions to make the construction of this initial approach more tractable:
    \begin{postulate}
        \label{post:pdf:global}
        The PDFs $f_a(x,Q^2), \, a \in \lbrace g,d,u,s,c,b,t,\bar d,\bar u,\bar s,\bar c,\bar b,\bar t \rbrace$ are only meaningful as whole and must be solved simultaneously from data. We will construct a mathematical structure to treat the PDF family as a unified object, and discuss the choice of a functional basis.
    \end{postulate}
    \begin{postulate}
        \label{post:pdf:smooth}
        The PDFs $f_a(x,Q^2)$ defined after \gls{renormalization} are sufficiently smooth and integrable functions $f_a(x,Q^2)$ to be solved from data in a functional sense. The experimental data are measurements of theoretically smooth quantities such as scattering cross sections, and the mathematical problem is to solve the smooth PDFs from a set of smooth observables. This is a generalization of the approach of defining an ansatz functional parametrization used in fitting. We use the $\overline{\mathrm{MS}}$ scheme. We discuss this assumption a bit further in App.~\ref{app:pdf-smooth}
    \end{postulate}
    \begin{postulate}
        \label{post:pdf:qcd}
        \Gls{perturbative QCD} and \gls{phenomenological} description define a physically accurate problem of \gls{global analysis} that can be taken as the starting point of defining a mathematical inverse problem to solve the PDFs from \gls{world data}.
        We take the diligently calculated and verified pQCD results available in the literature as the foundation of this approach, which we merely formalize as a mathematical inverse problem.
        \begin{itemize}
            \item \textup{We will refer back to this foundational assumption at steps of calculations, where we accept the established QCD theory result from literature, denoted by (P\ref{post:pdf:qcd}), and refer to the literature for details.}
        \end{itemize}
    \end{postulate}
    \begin{weakpostulate}
        \label{post:nonpert}
        All other \gls{non-perturbative} degrees of freedom than the proton PDFs are known be it either experimentally or phenomenologically. This includes parton masses, the running of the \gls{strong coupling} $\as(\mu^2)$, and the \gls{nuclear modification} factors $R^A(x,Q^2)$.
    \end{weakpostulate}
    \begin{weakpostulate}
        \label{post:massless}
        A parton must be massless to be described by a PDF. Due to this, we mainly consider the theory of high-energy scattering for massless quarks. However, \glspl{heavy quark} are of critical importance for the real global analysis problem, so we consider the mathematical structure of the inference problem with the inclusion of \glspl{heavy quark} both as observables, and how their systematical inclusion requires the use of a \gls{flavor number scheme}. Our approach largely follows the zero-mass variable \gls{flavor number scheme} (ZM-VFNS), which is contrasted with more general schemes in Sec.~\ref{sec:flavor-number-schemes}.
    \end{weakpostulate}
    \noindent
    On the whole, we expect that the solution functions of the inverse problem would resemble the solutions achieved via fitting of parameters shown in Fig.~\ref{fig:nnpdf-pdfs-overview}---given the high accuracy and agreement of the existing global analyses~\cite{ParticleDataGroup:2024cfk,PDF4LHCWorkingGroup:2022cjn:pdf-comparison}---while mathematically retaining the possibility that the inverse problem solutions could capture physical features that have not been included in the ansatz parametrizations.
    In the interest of developing a full picture of the mathematical structure of the inverse problem, we will also formally write the typically negligible bottom and top quark contributions in expressions where they typically would be neglected and left out due to vanishingly small contributions. We mainly focus on constructing this approach at \gls{leading order} (LO) and \gls{next-to-leading order} (NLO) accuracy in \gls{perturbative QCD}, but we do take a brief look beyond in Sec.~\ref{sec:nnlo}.

    The feature that this approach does not require an ansatz functional form with associated model parameters, such as Eq.~\eqref{eq:fit-ansatz}, reduces the bias on estimated uncertainties of the reconstructed data that can be caused by the use of such a parametrized fit. In Ref.~\cite{Hanninen:2025iuv} we apply the same principles employed in this work to the dipole amplitude inference problem relevant to small-$x$ physics, and showed that by randomly sampling replicas of the simulated dataset, we were able to reconstruct point-wise uncertainties of the reconstructed dipole amplitudes. The same boot-strapping method is general enough to be applied directly to define the uncertainties of the reconstructed PDFs.

    To end this section, let us collect the key assumptions and non-assumptions of the approach taken in this work.
    The framework assumes that:
    \begin{itemize}
        \item the PDFs must be solved simultaneously;
        \item the PDFs are sufficiently regular and integrable;
        \item \gls{perturbative QCD} implemented at a fixed order in $\as$ describes experimental \glspl{observable} adequately accurately;
        \item renormalization, factorization, and flavor number schemes are implemented systematically everywhere;
        \item all \glspl{observable} and theory constraints are explicitly expressed as a unified system of equations;
        \item the convolutional integral structure of the problem can be leveraged, which emerges universally at NLO accuracy;
        \item other non-perturbative inputs such as $\as(Q^2)$, quark masses, and nuclear modifications are known;
        \item \glspl{observable} are theoretically smooth quantities, and they are sampled by experimental data with statistically consistent uncertainties and correlations.
    \end{itemize}
    Then the reconstruction method imposes:
    \begin{itemize}
        \item QCD sum rules at all scales in $Q^2$;
        \item smoothness in $x$ and $Q^2$ are used to regularize the solutions, the latter of which is connected to consistency with the DGLAP evolution equation~\cite{ParticleDataGroup:2024cfk};
        \item compensation for theoretical effects not explicitly included in the forward operator construction, such as higher orders in $\as$, higher twist;
        \item optionally, positivity.
    \end{itemize}
    The method does \textit{not} necessitate:
    \begin{itemize}
        \item an explicit functional ansatz for the PDFs at an initial scale, or at any scale in $Q^2$;
        \item exact satisfaction of fixed-order DGLAP evolution by the solution PDFs;
        \item the existence or uniqueness of the reconstructed solution, which remains a mathematical question to be established for the chosen operators, domain, data, and constraints.
    \end{itemize}

    This article is structured as follows:
    In Sec.~\ref{sec:intro} we introduce the inverse problem of solving the parton distribution functions from particle collision data and key beats of its extensive history.
    In Sec.~\ref{sec:data} we describe the physical processes and experiments theoretically considered in this work, and in Secs.~\ref{sec:core}, \ref{sec:linear-intop-formalism} we lay out the mathematical principles that form the core of this approach and the changes they require us to implement in the classical \gls{implicit inverse problem} of global analysis.
    In Sec.~\ref{sec:forward-problem-lo} we proceed to rewrite the individual parts of the implicit global analysis problem into the required form for the \gls{explicit inverse problem}, and in Sec.~\ref{sec:tensor-ip-lo} we construct the explicit tensor inverse problem of the parton distributions at leading order accuracy in perturbative QCD.
    Then in Secs.~\ref{sec:forward-problem-nlo},~\ref{sec:tensor-ip-nlo} we follow this format to write parts of the next-to-leading order accuracy tensor reconstruction problem.
    Sections~\ref{sec:flavor-number-schemes} and~\ref{sec:nnlo} discuss important features of the full inference problem on a high-level, with the former discussing the compatibility of this approach with \glspl{flavor number scheme}, and the latter taking a look towards higher orders in perturbation theory and the factorization theorem of perturbative QCD.
    Mathematical feasibility of solving the integral equations underlying the inverse problem is briefly discussed in Sec.~\ref{sec:integral-eq-theory-wiener-hopf}.
    A discretization based method to solve the inverse problem in practice is constructed in Sec.~\ref{sec:discretization}, along with a discussion of the uncertainties of the reconstructed PDFs, and an example of applying the discretization methodology.
    Discussion and outlook are given in Sec.~\ref{sec:discuss}. In appendix \ref{app:full-global-data-operators} we give the full explicit form of the integral operators of global analysis of the PDFs.

    \textit{N.B.} We consider the topic of this article to be of interest to both high-energy physicists and inverse problems mathematicians, so we strive to cover topics in an accessible manner, and provide a glossary in Appendix \ref{sec:hep-nomenclature}. We refer to the literature on points of importance where the intermediate steps require advanced expertize in specific areas and we just apply the final results in the current work.

\subsection{Included deeply inelastic scattering observables and measurements} \label{sec:data}

    \begin{table}[th]
        \centering
        \setlength{\tabcolsep}{4pt}
        \begin{tabular}{llllll} \label{tab:processes}
            Process & \Glspl{parton} & \Gls{observable} & Experiment & Refs.~\\
            \toprule
            $e^{\pm} p \to e^{\pm} X$ & $q, \bar q, g$ &
            $\sigma_{r,\mathrm{NC}}^\pm \, (\approx \sigma_{r}^\gamma) , \, F_L^\mathrm{NC}, \, xF_3^\mathrm{NC}$ & HERA I+II, ZEUS &~\cite{Abramowicz:2015mha:HERAIInewcombined,Abramowicz:2015mha:HERAIInewcombined:hepdata.68951,H1:2008rkk:fl,ZEUS:2009nwk:fl,H1:2010fzx:fl, ZEUS:2002pdo:F3} 
            \\
            $e^+ p \to \bar{\nu} X$ & $q, \bar q$ &
            $\sigma_{r,\mathrm{CC}}^{\pm}$ & HERA I+II &~\cite{Abramowicz:2015mha:HERAIInewcombined} 
            \\
            $e^{\pm} p \to e^{\pm} c\bar{c}X, e^{\pm} b\bar{b}X$ & $c, b, g$ &
            $\sigma_{r}^{c\bar c}, \, \sigma_{r}^{b\bar b}$ & HERA I+II &~\cite{Abramowicz:2018flt:HERAII-charm-bottom-combined} \\
            \midrule
            $\nu(\bar \nu) N \to \mu^- (\mu^+) X$ & $q, \bar q$ &
            $\sigma^{(\nu, \bar \nu)N}, F_2^{(\nu, \bar \nu) N}, \, xF_3^{(\nu, \bar \nu) N}$ & NuTeV, CHORUS &~\cite{NuTeV:2005wsg, NuTeV:2005:hepdata.11120, CHORUS:2005cpn,CHORUS:2005:hepdata.6187} \\
            \midrule
            $e p \to e X, e d \to e X$ & $q, \bar q, g$ & $F_L^\mathrm{NC}$ & JLab &~\cite{Tvaskis:2016uxm:jlabfl,JeffersonLabHallCE94-110:2004nsn;jlabfl} 
            \\
            $\mu p \to X, \mu d \to X$ & $q, \bar q$ & Gottfried s.r. $S_\mathrm{G}$, $F_L^\mathrm{NC}$ & NMC &~\cite{NewMuon:1996uwk,NewMuon:1996fwh} 
            \\
            $\nu(\bar \nu) N \to \mu^- (\mu^+) X$ & $q, \bar q$ & GLS s.r. $S_\mathrm{GLS}$ & CCFR/NuTeV &~\cite{Leung:1992yx:GLS,NuTeVCCFR:1995wdi:GLS,Kim:1998kia:GLS} \\
            $\nu(\bar \nu) D \to \mu^- (\mu^+) X$ & $q, \bar q$ & Adler s.r. $S_\mathrm{A}$ & WA25 (BEBC) &~\cite{WA25:1983qub:Adler-exp} \\
            \bottomrule
        \end{tabular}
        \caption{Particle scattering processes and associated observables considered in this work as measured---or at least in principle measurable---. We limit the scope to lepton--hadron deeply inelastic scattering processes~\cite{Blumlein:2012bf:dis-review}, which lead to linear inverse problems, limiting the considered processes to a small subset of all processes and data conventionally included in global analysis of the PDFs. Adapted from Ref.~\cite{10.1093/ptep/ptac097}, and extended with the sum  rule (s.r.) observables as they are also associated with linear integral operators, which enables their inclusion as additional constraints in the inverse problem framework. The sensitivity to partons $c,b$ is implied for the infinite momentum frame where PDFs can be defined, i.e. the zero mass limit.  For an excellent and comprehensive overview of the free proton world data see Ref.~\cite{NNPDF:2021njg}, and the references therein, as the measurements chosen here are merely a minute fraction of all published data that could be considered as a part of the world data.}
    \end{table}

    We consider in this work high-energy scattering observables that have strictly a linear dependence on the parton distribution functions, and do not require any other non-perturbative QCD input such as fragmentation functions or branching ratios~\cite{ParticleDataGroup:2024cfk}, and the running of the strong coupling is taken as known (assumption \ref{post:nonpert}). Either a non-linear dependence on the PDFs or a multiplicative dependence on another non-perturbative quantity would produce a non-linear inverse problem, which are left for future work.
    In practice this means that our focus is on \gls{lepton}-hadron \gls{deeply inelastic scattering} and its observables shown in Table \ref{tab:processes}.
    To make it explicit what we mean here by linearity, let $\lbrace \tilde f_a\rbrace_{a \in \lbrace g,d,u,\cdots,\bar t\rbrace}$ and $\lbrace \hat f_a\rbrace_{a \in \lbrace g,d,u,\cdots,\bar t\rbrace}$ be two sets of PDFs, then the observable $O$ is a \textit{linear functional} that acts on the PDFs if:
    \begin{equation}
        O(\alpha \tilde f_a + \beta \hat f_a) = \alpha O (\tilde f_a) + \beta O(\hat f_a), \quad \forall \, \alpha, \beta \in \mathbb{R}.
        \label{eq:linear-pdf-operator}
    \end{equation}
    The observables of lepton--hadron DIS~\cite{Blumlein:2012bf:dis-review} that have this useful property are the \gls{reduced cross section} $\sigma_r$, \gls{structure function}s $F_i$, and the quantum field theoretic sum rules $S$. Most intuitively the cross section quantity is related to the probability of a specific \gls{interaction} happening in a given collision of particles in a particle collider, and the structure function and sum rule quantities capture some information about the partonic structure of the nucleon, which is either a proton or a neutron, see Appendix \ref{sec:hep-nomenclature} for some additional detail.

    The mathematical structure of the inverse problem will be such that individual observables cannot have complete resolution of the parton distribution functions even under unphysically perfect circumstances, which requires the simultaneous leverage of multiple experiments. The experiments that are included in table \ref{tab:processes} fall broadly into two categories: beam-beam collider experiments, HERA, and fixed target scattering experiments NMC, CCFR/NuTeV, and CHORUS.
    The former produce high-energy beams of \gls{lepton}s---electrons $e^-$, positrons $e^+$ or muons $\mu$---and beams of protons or nuclei that are then directed to collide within particle detectors.
    The fixed target experiments on the other hand collided a beam of neutrinos or antineutrinos with a dense stationary target of iron or lead. Neutrino beams are produced by large particle colliders, such as the Large Hadron Collider (LHC) or TeVatron, as decay products of various particles---like muons---momentarily created in the collisions which are collimated with magnetic fields before decay. The produced high-energy neutrinos that form in the decay of the intermediate particles remain broadly collimated into a beam thanks to their small mass.
    These experiments opened a critical ``orthogonal'' perspective into the structure of the proton by leveraging the fact that neutrinos only interact with matter via the \gls{weak nuclear} interaction~\cite{ParticleDataGroup:2024cfk}, which is fundamentally different in its sensitivity to the flavors of partons within the proton than the \gls{electromagnetic} interaction dominating in the beam-beam experiments. However, the use of neutrino DIS data in global analysis faces the significant challenge, that the nucleons in the target nuclei are not free, and the structure of the proton in the nucleus receives \gls{nuclear modification}s to the parton distributions. As the present work is about theoretical methodology, we will discuss the feasibility of the inclusion of this data in the inverse problem.

    We will briefly introduce the theoretical picture of each scattering process and its significance in the broader global analysis inverse problem in their respective sections. Importantly, we will recollect all relevant formulae from theory calculations that are needed for our construction, as we are of the opinion that the precise equations that are used should be made abundantly clear, especially here where their exact mathematical structure is of central importance. 
    Our core goal is to keep the constructed inverse problem approach firmly grounded in the real experiments and available world data, while expecting that future experiments like the upcoming Electron-Ion Collider (EIC)~\cite{Accardi:2012qut:EIC,AbdulKhalek:2021gbh:EIC}, Electron-ion collider in China (EicC)~\cite{Anderle:2021wcy:EicC}, and possibly also the considered Large Hadron Electron Collider (LHeC)~\cite{LHeCStudyGroup:2012zhm,LHeC:2020van}, will supplement the deeply inelastic scattering world data with higher fidelity and precision, hopefully enabling the novel---perhaps delicate---mathematical methods considered in this work.

\section{Core principles and a coupled system of linear inverse problems} \label{sec:core}

    Consider the inference of two unknown functions $f(x)$ and $g(x)$ from two distinct measurements $b_1(q)$ and $b_2(q)$---which are functions of an unspecified variable $q$---that are only sensitive to the sum and difference of the functions $f(x)$ and $g(x)$:
    \begin{align}
        b_1(q) & \coloneqq \int K_1(x, q) (f(x) + g(x)) \ud x, \\
        b_2(q) & \coloneqq \int K_2(y, q) (f(y) - g(y)) \ud y.
    \end{align}
    To solve the above coupled system of inverse problems for $f$ and $g$, we formalize the integrals as linear integral operators $\Kcal_i$:
    \begin{equation}
        b_i(q) = \int K_i(x, q) (f(x)\pm g(x))\ud x
        \eqqcolon \mathcal{K}_{i}(f \pm g) = \mathcal{K}_{i}(f) \pm \mathcal{K}_{i}(g). 
    \end{equation}
    It will be useful to stack the unknown functions into a vector, which allows us to write the above system as a single linear system of functional equations:
    \begin{equation}\label{eq:main-ip}
        \begin{bmatrix}
            \mathrm{b}_1 \\
            \mathrm{b}_2
        \end{bmatrix}
        =
        \begin{bmatrix}
            \Kcal_1 & \Kcal_1 \\
            \Kcal_2 & -\Kcal_2
        \end{bmatrix}
        \begin{bmatrix}
            f \\
            g
        \end{bmatrix}.
    \end{equation}
    This functional matrix structure is of course extensible for more unknown functions and measurements, and enables the consideration of existence and uniqueness of solutions with mathematical rigor in the inverse problems framework.
    We will apply these principles of solving coupled inverse problems to the global data inference problem of the proton PDFs, to construct an approach that does not use fitting of model parameters, broadly analogously to the approach constructed in Ref.~\cite{Hanninen:2025iuv}, where we applied similar principles to the inference of the dipole scattering amplitude in the dipole picture. On a high level, this approach is also a generalization of that of Ingelman and R\"uckl~\cite{Ingelman:1987zv} who considered solving PDFs from neutral and charged current DIS measurements as linear systems of equations, or even more generally of the Rosenbluth separation~\cite{Rosenbluth:1950yq} and Sachs' form factor methods~\cite{Ernst:1960zza,Sachs:1962zzc}, however here we construct the methodology to be able to include more observables, and higher order corrections in perturbative QCD as linear integral operators. This approach is also related at least in spirit to that of Refs.~\cite{Lappi:2023lmi:phys-pdf-basis,Lappi:2024dvv:phys-pdf-basis}, but instead of eliminating the PDFs from the problem, the task is to leverage functional methods to solve them from physical observables.
    
    If the integral operators $\Kcal_i$ have inverse operators $\Kcal_i^{-1}$, the system \eqref{eq:main-ip} has the explicit solution 
    \begin{align}
        f(x) & = \frac{1}{2} \Big[ (\Kcal^{-1}_1 b_1)(x) + (\Kcal^{-1}_2 b_2)(x) \Big]
        \\
        g(x) & = \frac{1}{2} \Big[ (\Kcal^{-1}_1 b_1)(x) - (\Kcal^{-1}_2 b_2)(x) \Big],
    \end{align}
    i.e. the mathematical structures defined by the definition of the observables $b_1$ and $b_2$ can be leveraged to solve the unknown functions directly from measurement \textit{without any modeling and fitting of the unknowns}.
    If the invertibility of the operators is limited for example by the availability of data in the measurements $b_i(q)$, i.e. exact inverses do not mathematically exist, inverse problems methods such as discretization and regularization can be employed~\cite{clason2021regularizationinverseproblems, Hansen:2021ct}, as is done in numerous practical applications like medical imaging, and as we applied to the dipole amplitude inference problem~\cite{Hanninen:2025iuv}. For example, computed tomography is mathematically based on the inversion of the X-ray transform~\cite{JohnFritz:1938:xraytransform}, which is the abstract reference point and inspiration to construct this approach. Another example is the image denoising inverse problem~\cite{Goyal:2019:imagedenoising}, which is a tractable and solvable problem---within limits of course---even though the operation of blurring an image is an information destructive process. Even if the linear operators we construct here do not have exact inverses, something might still be achievable within the framework of inverse problems.

    To make the mathematical analogue to computed tomography slightly more tangible, we consider the X-ray integral transform by F. John~\cite{JohnFritz:1938:xraytransform}
    \begin{equation}
        Xf(\theta,x) \coloneqq \int_{-\infty}^\infty f(x + t \theta) \ud t,
        \label{eq:xray-transform}
    \end{equation}
    where $x \in \mathbb{R}^3$ is a spatial coordinate, $f(x): \mathbb{R}^3 \mapsto \mathbb{R}$ is a function that describes the subject structure, $\theta \in S^2$ is a unit direction in $\mathbb{R}^3$, and $t$ parametrizes a straight line going through $(x, \theta)$. This X-ray transform, or just ray transform, is the mathematical foundation of computed tomography~\cite{Natterer:2001:mathematicalmethods-img-rec,Hansen:2021ct}, where partially transparent projections of a subject are used to reconstruct the three-dimensional internal structure of the subject. The measurement is identified with the X-ray transform of the unknown internal structure, which then defines the inverse problem of solving the unknown structure from this measurement of the transform. In this work we follow this perspective, and identify some of the standard high-energy physics scattering \gls{observable}s such as cross sections as measured constraints of some integral equations, which we will need to identify from the standard theory calculations available in the literature. The inverse problem of global analysis then is concerned with the methods of solving the a priori unknown parton distribution functions from these integral equations defined by the observables.
    And just as in computed tomography where mathematical methods are applied to solve the internal structure without fitting, our task is to formulate a mathematical framework to enable an analogous approach of reconstruction of the proton PDFs.

\section{Linear integral operator formalism} \label{sec:linear-intop-formalism}

    We apply the above linear operator formalism to the \gls{hadronic tensor} $W_{\mu\nu}$, which arises in the quantum field theoretic calculation of physical \gls{observable}s such as the \gls{reduced cross section} when one observes that the mathematical structures are amenable to a factorization that separates the probe and target particle degrees of freedom~\cite{ParticleDataGroup:2024cfk,Ellis:1996mzs:qcd-collider-phys,Peskin:0201503972}. 
    We choose the hadronic tensor as the fundamental object to reconstruct, because the parton distribution functions arise from the structure functions $F_i$, which are originally defined as functional components of the hadronic tensor, and so the functional linearity originates from the structure of the hadronic tensor. Our aim is to develop a full inverse problems theory picture of QCD, where the parton distribution functions are the universal unknown to be solved from data, and to this end we must describe how to systematically restructure the mathematical expressions of observables to facilitate the solution of the PDFs.
    The unpolarized hadronic tensor is
    \begin{equation}
        W_{\mu\nu} \coloneqq \left( \frac{q_\mu q_\nu}{q^2} - g_{\mu\nu} \right) F_1(x,Q^2) + \left( p_\mu - \frac{p \cdot q}{q^2}q_\mu \right)\left(p_\nu - \frac{p \cdot q}{q^2}q_\nu \right) F_2(x,Q^2)
    \end{equation}
    where $p_\mu, \, q_\nu$ are four-momenta of the interacting parton and force mediating virtual \gls{boson}, and the structure functions $F_i$ are defined as
    \begin{equation} \label{eq:wilson-convol}
        F_i(x,Q^2) \coloneqq \qty(\sum_a C_i^a \otimes f_a)(x,Q^2),
    \end{equation}
    where $C_i^a$ are so-called Wilson coefficient or convolution functions~\cite{ParticleDataGroup:2024cfk}, and the convolution operation $\otimes$ is defined as:
    \begin{equation}
    \label{eq:convolution-C-def}
        (C \otimes f)(x) \coloneqq \int_x^1 C(y) f\left( \frac{x}{y} \right) \frac{\ud y}{y}.
    \end{equation}
    Equation \eqref{eq:wilson-convol} defines a linear functional acting on the parton distribution functions $f_a$, which we will formalize in a more concrete way that will enable us to leverage the linearity as we construct the \gls{explicit inverse problem}.
    To treat the PDFs $f_a$ as a unified family of functions, we define a \gls{parton distribution tensor} as
    \begin{equation}
    \label{eq:pdt-definition}
            \mathbf{f}^{(n_f+1)}(x,Q^2) \coloneqq 
            \begin{bmatrix}
                f_g(x,Q^2) & f_{q_1}(x,Q^2) & \dots & f_{q_{n_f}}(x,Q^2) &
                f_{\bar q_1}(x,Q^2) & \dots & f_{\bar q_{n_f}}(x,Q^2)
            \end{bmatrix}^T
    \end{equation}
    on which the Wilson coefficient function convolutions act as an integral operator with a row-matrix structure:
    \begin{equation}
        \mathcal{C}_i \coloneqq
        \begin{bmatrix}
            C_i^g & C_i^d & \cdots & C_i^{t} & C_i^{\bar d} & \cdots & C_i^{\bar t}
        \end{bmatrix},
    \end{equation}
    where each element is an integral operator with the kernel $C_i^a$, and
    which operates on $\pdtbf$ with the rules of an inner product to reproduce the standard sum of convolutions:
    \begin{equation}
        \qty(\mathcal{C}_i \pdtbf)(x,Q^2) = \qty(\sum_a C_i^a \otimes f_a)(x,Q^2).
    \end{equation}
    With this we perform a second factorization to separate the parton distributions from the hadronic tensor, and rewrite the hadronic tensor as a linear integral operator valued row matrix that acts on the \gls{parton distribution tensor} $\pdtbf$:
    \begin{equation}
        W_{\mu\nu}(x,Q^2) = \left\lbrace \left( \frac{q_\mu q_\nu}{q^2} - g_{\mu\nu} \right) \mathcal{C}_1 + \left( p_\mu - \frac{p \cdot q}{q^2}q_\mu \right)\left(p_\nu - \frac{p \cdot q}{q^2}q_\nu \right) \mathcal{C}_2 \right\rbrace
        \pdtbf.
    \end{equation}
    Here we see that, by leveraging the linearity of this functional operation, we can explicitly factorize the definition of the parton distribution functions $f_a$ from the hadronic tensor $W_{\mu\nu}$. We will push this extraction through the standard program of perturbative QCD calculations to be able to redefine the conventional expressions of cross sections and observables as functional operators that act on the \gls{parton distribution tensor} $\pdtbf$.
    The goal is to economically apply this inverse problems approach to established results available in the literature, rather than perform the same calculations again in a slightly different form, and luckily this is fairly straightforward. This shift in perspective enables us to define linear parts of the global analysis problem as linear integral equations of the \gls{parton distribution tensor}, and to solve for the PDFs from data without fitting.

    To concisely give an overview of how this reformulation of the global analysis inference problem proceeds in practice, we refer back to the toy model of a coupled inverse problem given in Eq.~\eqref{eq:main-ip}. Following the silhouette of the toy inverse problem, we will write the inverse problem for the proton PDFs such that the \gls{parton distribution tensor} $\pdtbf$ takes the place of the vector of unknown functions to be solved, and linear observables such as cross sections, structure functions, and sum rule data take the place of the placeholder datasets $b_i$. We write this as the matrix valued integral equation:
    \begin{equation}
        \label{eq:ip-general-linear}
        O^\mathrm{WD}(x,Q^2,y)
        =
        \qty(\mathbf{G}^{lp \to l' X}
        \pdtbf)(x,Q^2,y),
    \end{equation}
    where the world data $O^\mathrm{WD}$ is comprised of the linear observables, the \gls{parton distribution tensor} $\mathbf{f}$ was defined in Eq.~\eqref{eq:pdt-definition}, and the forward ``global data operator'' of lepton--proton DIS $\mathbf{G}^{lp \to l' X}$ is a linear integral operator that we will construct in practice observable by observable in the next section.
    The full form of $\mathbf{G}^{lp \to l' X}$ is then constructed at leading order in Sec.~\ref{sec:tensor-ip-lo}, and in a more complete formulation in Sec.~\ref{sec:tensor-ip-nlo}.
    To write the inverse problem in the most general and extensible manner, in the remainder of this article we will be working with the \gls{parton distribution tensor} including all six flavors of quarks and anti-quarks defined as
    \begin{equation}
    \label{eq:pdt-6-definition}
        \mathbf{f} \coloneqq \pdtbf^{(6+1)} \equiv
        \begin{bmatrix}
        f_g&
        f_d &
        f_u &
        f_s &
        f_c &
        f_b &
        f_t &
        f_{\bar d}&
        f_{\bar u}&
        f_{\bar s}&
        f_{\bar c} &
        f_{\bar b} &
        f_{\bar t}
        \end{bmatrix}^T,
    \end{equation}
    which is also known as the canonical or flavor basis~\cite{Candido:2022tld:EKO} that spans the functional space of the PDFs. The choice of this basis is not unique~\cite{Candido:2022tld:EKO}, we merely adopt the canonical basis as the most commonly used basis, which will make it more straightforward to collect the expressions from the literature in one shared basis, which is needed to write the global analysis problem down into one unified system of inverse problems.

\section{Linear forward problems of deeply inelastic scattering at LO in pQCD}
\label{sec:forward-problem-lo}

    The \gls{leading order} (LO) inference problem acts as an introductory toy-model of the full problem. The most useful aspect of the LO problem is that it more simply shows the surface level importance of different observables and how they give critical separation power to the world data inference problem. However, to develop the actual methodology and mathematical inverse problem framing of the real world data inference problem, we will need to go to next-to-leading order accuracy to see the proper mathematical structures emerge, which will then persist at higher orders in the perturbative expansion.

    \subsection{Sum rules}

    We begin our proper formulation of the literature results into constraints of integral equations from the so-called sum rules~\cite{Blumlein:2012bf:dis-review}. In basic terms, they are integral constraints on the PDFs that arise from QCD theory and encode some fundamental properties~\cite{ParticleDataGroup:2024cfk,Sterman:1995:handbook-qcd,Peskin:0201503972} of a hadron, in this case the proton. There is the additional nuance that some of the sum rules are expected to be mathematically perfectly satisfied by reality, such as conservation of total momentum for the constituents of the proton, whereas some others we will discuss receive non-trivial higher-order contributions in perturbative QCD that change the value of the sum rule from the ``naive'' theoretical value. Our task is to formulate both kinds of sum rule constraints as conditions to be fulfilled by the solutions to our inverse problem.
    We will differentiate these by calling the former the basic sum rules, whereas the latter will be called the ``flavor structure'' sum rules.
    The basic rules describe physically trivial facts such as ``the elements of the proton collectively carry all of its momentum'' whereas the latter describe non-trivial deviations in the structure of the proton that arise from the rich QCD dynamics between the constituents of the proton~\cite{Sterman:1995:handbook-qcd}.
    
    We will first consider the basic rules: total momentum conservation and flavor number conservation rules, which mandate the conservation of momentum as discussed above, and the fact that the proton at rest is a bound state of $uud$ \glspl{valence quark}. The total momentum conservation ($\tmc$) rule is expressed as:
    \begin{align}
        \int_0^1 \sum_{a\in\lbrace g,d,u,..\rbrace} x f_a(x, Q^2) \ud x 
        = \int_0^1 x & \big[ 
            f_g(x, Q^2)
            + f_d(x, Q^2) + f_u(x, Q^2) + f_s(x, Q^2) 
            \notag \\[-12pt]
            &\, + f_c(x, Q^2) + f_b(x, Q^2) + f_t(x, Q^2) 
            \notag \\
            &\, + f_{\bar d}(x, Q^2) + f_{\bar u}(x, Q^2) + f_{\bar s}(x, Q^2) 
            \notag \\
            &\, + f_{\bar c}(x, Q^2)  + f_{\bar b}(x, Q^2) + f_{\bar t}(x, Q^2) 
        \big] \ud x \equiv 1.
    \end{align}
    The net flavor number conservation sum rule is more simply stated for each flavor of quark individually, and in practical terms states that each flavor of quark $a$ should have on average an excess of $N_a$ quarks compared to antiquarks, i.e. net count of $2$ up quarks, $1$ down quark, and so on:
    \begin{align}
        \int_0^1 f_u(x, Q^2) - f_{\bar u}(x, Q^2) \ud x & = N_u \equiv 2
        \\
        \int_0^1 f_d(x, Q^2) - f_{\bar d}(x, Q^2) \ud x & = N_d \equiv 1
        \\
        \int_0^1 f_a(x, Q^2) - f_{\bar a}(x, Q^2) \ud x & = N_a \equiv 0, \quad a \in \lbrace s, c, b, t \rbrace.
    \end{align}
    The last identity states that no other quark flavors are believed to exists as inherent valence quarks within the proton at rest, although this has been challenged recently by the possible existence of a charm quark contribution to the valence quark composition of the proton~\cite{NNPDF:2022qks}.
    The total momentum conservation rule and the quark net number rules for each flavor should hold exactly at all momentum scales $Q^2$.
    If one were to look for evidence for intrinsic charm~\cite{Brodsky:1980pb:Hoyer-intrinsic-charm,Ball:2015dpa:intrinsic-charm-dis}, i.e. the presence of a charm quark contribution to the proton wavefunction, then the charm PDF $f_c$ is allowed to be non-zero below the charm production scale $\mu = m_c$. If it is disallowed, the PDF should be exactly zero below the production scale, where dynamical production of charm enables it to contribute; more careful examination is left for future work.

    Next, we move on to the ``flavor structure'' sum rules considered in this work, which are the Gross--Llewellyn-Smith (GLS)~\cite{Gross:1969jf}, Adler~\cite{Adler:1965ty}, and Gottfried~\cite{Gottfried:1967kk} sum rules~\cite{Sterman:1995:handbook-qcd,ParticleDataGroup:2024cfk}.
    The GLS sum rule counts the expected average number of valence quarks in a nucleon at a fixed $Q^2$, which is three in the quark-parton model at all $Q^2$:
    \begin{align}
        S_\sumrulGLS(Q^2) 
        & \coloneqq \int_0^1 \frac{x F_3^{\nu N}(x,Q^2)}{2x} \ud x 
        \notag \\
        & \pequiv \int_0^1  f_d(x,Q^2) + f_u(x,Q^2) - f_{\bar d}(x,Q^2) - f_{\bar u}(x,Q^2) \ud x ~~ \big(= 3\big).
    \end{align}
    However, this sum rule has been experimentally observed~\cite{Leung:1992yx:GLS,NuTeVCCFR:1995wdi:GLS,Kim:1998kia:GLS} not to be exactly satisfied in this form, i.e. it fails to be exactly equal to three always. In the perturbative QCD formalism this is described as the right hand side constant receiving scale dependent corrections~\cite{Sterman:1995:handbook-qcd}.
    In this inverse problem approach we leverage this measured sum rule as an observable constraint, which means that we can forgo these perturbative corrections, and rather we will work with an inference problem where the data $S_\sumrulGLS$ and the unknown PDFs are both non-perturbative, which will be made concrete in the next section.

    The Adler sum rule~\cite{Adler:1965ty} leverages the isospin symmetry between the proton and the neutron to count the excess of up quarks compared to down quarks in the proton, which is one in the quark-parton model~\cite{Sterman:1995:handbook-qcd}:
    \begin{align}
        S_\sumrulA(Q^2)
        & \coloneqq \int_0^1 \frac{F_2^{\nu n}(x,Q^2) - F_2^{\nu p}(x,Q^2)}{2x} \ud x 
        \notag \\
        & \pequiv
        \int_0^1  f_u(x,Q^2) - f_{\bar u}(x,Q^2) - \qty(f_d(x,Q^2) - f_{\bar d}(x,Q^2)) \ud x
    \end{align}
    where in the derivation the isospin symmetry is used to replace the neutron $u_n(x),d_n(x)$ PDFs with those of the proton $d_p(x),u_p(x)$~\cite{Sterman:1995:handbook-qcd}, respectively. The value of this sum rule has been measured $S_\sumrulA(Q^2)$ at varying scales $Q^2$~\cite{WA25:1983qub:Adler-exp}, which we will formulate as a scale dependent experimental constraint for the inverse problem.

    The Gottfried sum rule~\cite{Gottfried:1967kk} is sensitive to the number of anti-up and anti-down quarks in the proton~\cite{Sterman:1995:handbook-qcd} in the sense that if their contributions are identical, i.e. $\bar u \equiv \bar d$, then a correction term vanishes. The Gottfried sum rule is written~\cite{Sterman:1995:handbook-qcd}
    \begin{align}
        S_G(Q^2) 
        & \coloneqq
        \int_0^1 \frac{F_2^p(x,Q^2) - F_2^n(x,Q^2)}{x}\ud x 
        \notag \\
        & \pequiv
        \frac{1}{3} \int_0^1 f_u(x,Q^2) + f_{\bar u}(x,Q^2) - \qty(f_d(x,Q^2) + f_{\bar d}(x,Q^2)) \ud x 
    \end{align}
    where if $\bar d \equiv \bar u$ one gets the naive value of $\frac{1}{3}$. The Gottfried sum rule has been measured~\cite{NewMuon:1991hlj,NewMuon:1996uwk}, and rules out the assumption $\bar u = \bar d$ in the proton, i.e. the measurement of $S_G(Q^2)$ provides experimental evidence for flavor asymmetry in the nucleon sea~\cite{Sterman:1995:handbook-qcd}.
    
    In the next section we will write these definitions of sum rule constraints as integral operators which connect the PDFs to the measurement constraints.

    \subsection{Observables as linear functional operators}
    \label{sec:obs-as-linear-ops}

    We formalize the sum rules discussed in the previous section---as well as other important observables---as linear integral operators that act on the parton distribution functions to leverage the fact that they satisfy the functional linearity defined in Eq.~\eqref{eq:linear-pdf-operator}.
    Explicitly, our formalism can incorporate any sum rule that can be written as a linear operator:
    \begin{equation}
        S(Q^2) \equiv \int_0^1 L(\pdtbf(x,Q^2)) \ud x,
    \end{equation}
    where $L$ is a linear functional composed with the parton distribution function tensor $\pdtbf(x,Q^2)$, and $S(Q^2)$ is the theoretically prescribed or experimentally observed value of the sum rule.

    We can write the $S_\sumrulGLS$, $S_\sumrulA$, $S_\sumrulG$ sum rules as linear integral operators that act on the \gls{parton distribution tensor} $\pdtbf$ by defining:
    \begin{align}
        \Ical^{(1)}[f] & \coloneqq \int_0^1 f(x) \ud x,
        \\
        \Ical^{(x)}[f] & \coloneqq \int_0^1 x f(x) \ud x,
        \\
        \label{eq:fwdop-sumrule-fnc}
        \Ical_\mathrm{f.n.c.}[\pdtbf] & \coloneqq 
            \begin{bmatrix}
                0 &
                \icalid & 0 & 0 & 0 & 0 & 0 &
                -\icalid & 0 & 0 & 0 & 0 & 0
                \\
                0 &
                0 & \icalid & 0 & 0 & 0 & 0&
                0 & -\icalid & 0 & 0 & 0 &0
                \\
                0 &
                0 & 0 & \icalid & 0 & 0 & 0&
                0 & 0 & -\icalid & 0 & 0 &0
                \\
                0 &
                0 & 0 & 0 & \icalid & 0 &0&
                0 & 0 & 0 & -\icalid & 0 &0
                \\
                0 &
                0 & 0 & 0 & 0 & \icalid &0&
                0 & 0 & 0 & 0 & -\icalid &0
                \\
                0 &
                0 & 0 & 0 & 0 & 0 & \icalid &
                0 & 0 & 0 & 0 & 0 & -\icalid
            \end{bmatrix}
            \pdtbf
        \\
        \label{eq:fwdop-sumrule-tmc}
        \Ical_\mathrm{t.m.c.}[\pdtbf] & \coloneqq 
            \begin{bmatrix}
                \icalx &
                \icalx & \icalx & \icalx & \icalx & \icalx & \icalx &
                \icalx & \icalx & \icalx & \icalx & \icalx & \icalx
            \end{bmatrix}
            \pdtbf
        \\
        \label{eq:fwdop-sumrule-GLS}
        \Ical_\mathrm{GLS}[\pdtbf] & \coloneqq 
            \begin{bmatrix}
                0 &
                \icalid & \icalid & 0 & 0 & 0 & 0 &
                -\icalid & -\icalid & 0 & 0 & 0 & 0
            \end{bmatrix}
            \pdtbf
        \\
        \label{eq:fwdop-sumrule-adler}
        \Ical_\mathrm{A}[\pdtbf] & \coloneqq 
            \begin{bmatrix}
                0 &
                -\icalid & \icalid & 0 & 0 & 0 & 0 &
                \icalid & -\icalid & 0 & 0 & 0 & 0
            \end{bmatrix}
            \pdtbf
        \\
        \label{eq:fwdop-sumrule-gottfried}
        \Ical_\mathrm{G}[\pdtbf] & \coloneqq 
            \frac{1}{3}
            \begin{bmatrix}
                0 &
                -\icalid & \icalid & 0 & 0 & 0 & 0 &
                -\icalid & \icalid & 0 & 0 & 0 & 0
            \end{bmatrix}
            \pdtbf.
    \end{align}
    With these integral operators the sum rules can be expressed as:
    \begin{align}
        S_\mathrm{f.n.c.}(Q^2) &= \qty(\Ical_\mathrm{f.n.c.} \pdtbf)(Q^2) = 
        \int_0^1 
        \begin{bmatrix}
            f_d(x, Q^2) - f_{\bar d}(x, Q^2) \\
            f_u(x, Q^2) - f_{\bar u}(x, Q^2) \\
            f_s(x, Q^2) - f_{\bar s}(x, Q^2) \\
            f_c(x, Q^2) - f_{\bar c}(x, Q^2) \\
            f_b(x, Q^2) - f_{\bar b}(x, Q^2) \\
            f_t(x, Q^2) - f_{\bar t}(x, Q^2)
        \end{bmatrix}
        \ud  x
        =
        \begin{bmatrix}
            N_d \\ N_u \\ N_s \\ N_c \\ N_b \\ N_t
        \end{bmatrix}
        =
        \begin{bmatrix}
            1 \\ 2 \\ 0 \\ 0 \\ 0 \\ 0
        \end{bmatrix}
        \\
        S_\mathrm{t.m.c.}(Q^2) &= \qty(\Ical_\mathrm{t.m.c.} \pdtbf)(Q^2) = 1
        \\
        S_\mathrm{GLS}(Q^2)  &= \qty(\Ical_\mathrm{GLS} \pdtbf)(Q^2),
        \\
        S_\mathrm{A}(Q^2)  &= \qty(\Ical_\mathrm{A} \pdtbf)(Q^2),
        \\
        S_\mathrm{G}(Q^2)  &= \qty(\Ical_\mathrm{G} \pdtbf)(Q^2),
    \end{align}
    where the flavor and total momentum conservation rules must be satisfied at all $Q^2$, and last three as sum rules of the ``flavor symmetry'' kind act as inverse problems constraints with the experimental observations of $S_i(Q^2)$, as discussed in the previous section and in Sec.~\ref{sec:data}.

    Cross sections $\sigma$ and structure functions $F_i$ will be formulated in this linear integral operator formalism as well. However, they can have more varied form as integral equations.    
    In general, we will need to work with the following forms of linear integral operators:
    \begin{align}
        O^\sigma_\mathrm{LO}(x,Q^2,y) & =  L(x,Q^2,y) \pdtbf(x,Q^2), \\
        \label{eq:integral-operator-sigma-obs-first-kind}
        O^\sigma_{\mathrm{1st}}(x,Q^2,y) & = \int_x^1 L(z, x, y, Q^2) \pdtbf(z,Q^2) \ud z ,  \\
        \label{eq:integral-operator-sigma-obs-second-kind}
        O^\sigma_{\mathrm{2nd}}(x,Q^2,y) & = \lambda(x, Q^2, y) \pdtbf(x,Q^2) + \int_x^1 L(z, x, y, Q^2) \pdtbf(z,Q^2) \ud z ,  \\
        \label{eq:integral-operator-F-obs-first-kind}
        O^F_{\mathrm{1st}}(x,Q^2) & = \int_x^1 L(z,x,Q^2) \pdtbf(z,Q^2) \ud z ,  \\
        O^F_{\mathrm{2nd}}(x,Q^2) & = \lambda(x, Q^2) \pdtbf(x,Q^2) + \int_x^1 L(z,x,Q^2) \pdtbf(z,Q^2) \ud z ,
    \end{align}
    where the first three are of ``$\sigma$'' type, and the last two of ``$F$'' type, for cross sections and structure functions, respectively. In the following sections we will write various well-known observables in this formalism, first in leading order in perturbative QCD, then moving onto the more involved and interesting next-to-leading order, and finally considering the case of higher orders in Sec.~\ref{sec:nnlo}.

    Strictly speaking, the cross section observables we consider are differential cross sections, i.e. they are derivatives of a proper ``integrated cross section''. However, to simplify our notation, we will write the double-differential cross sections as functions, as already used above:
    \begin{equation}
        \sigma^i(x,Q^2,y) \coloneqq \frac{\ud^2\sigma^i}{\ud x \ud y} \equiv \frac{\ud^2\sigma^i}{\ud x \ud y}\qty(x,Q^2 = Q^2(x,y,s), y),
    \end{equation}
    where we have written out the functional arguments of the differential cross section function. In this framing, it is natural that as one computes integrated cross sections as an integral over $x$, or any other variable, said variable is integrated out, reducing the differential function into a partially integrated one.

    One final detail to fix before moving to discuss the experimental observables is that of the choice of the \gls{flavor number scheme} (FNS). The \gls{world data} of DIS spans multiple orders in magnitude all the way from $Q^2 \sim O(1 \, \mathrm{\gls{GeV}})$ to $Q^2 \gg 10^3 \, \mathrm{\gls{GeV}}$, which means that the physical theory has to address data where first at low $Q^2$ it is energetically impossible to probe some flavors of quarks, all the way up to so high momentum scales where all the masses of the quarks are negligibly small in comparison to the momentum scale of the deeply inelastic scattering.
    These transitions between qualitatively different regimes turn out to be theoretically non-trivial, and we will discuss more general solutions in Sec.~\ref{sec:flavor-number-schemes}. As stated in the assumption~\ref{post:massless}, we will mainly work in the simple but consistent zero-mass variable \gls{flavor number scheme}, which always treats quarks as massless and determines which flavors can contribute based on a simple momentum threshold rule that models the transition from energetically impossible to possible. Specifically, all observables $O(Q^2)$ are treated in the ZM-VFNS as defined piece-wise~\cite{Barontini:2024xgu:flavor-number-scheme-names-and-intro}: 
    \begin{equation}
        O^\mathrm{ZM-VFNS}\qty(Q^2) \coloneqq
        \begin{cases}
            O^{(3)}\qty(Q^2) \quad \mathrm{if}~ Q^2 < m_c^2 \\
            O^{(4)}\qty(Q^2) \quad \mathrm{if}~ m_c^2 \leq Q^2 < m_b^2 \\
            O^{(5)}\qty(Q^2) \quad \mathrm{if}~ m_b^2 \leq Q^2 < m_t^2 \\
            O^{(6)}\qty(Q^2) \quad \mathrm{if}~ m_t^2 \leq Q^2 ,
        \end{cases}
    \end{equation}
    where $O^{(n_f)}$ refers to the theory prescription of the physical observable $O$ in the approximation of $n_f$ massless quarks, and $m_c$, $m_b$, and $m_t$ are the masses of the charm, bottom, and top quarks, respectively\footnote{
    The masses of the quarks~\cite{ParticleDataGroup:2024cfk} are 
    $m_d \approx 0.0047 \,\mathrm{\gls{GeV}}/c^2$, 
    $m_u \approx 0.0022 \,\mathrm{\gls{GeV}}/c^2$, 
    $m_s \approx 0.094 \,\mathrm{\gls{GeV}}/c^2$ ($\overline{\mathrm{MS}}$ masses at $\mu = 2 \gev$), 
    $m_c \approx 1.27 \,\mathrm{\gls{GeV}}/c^2$, 
    $m_b \approx 4.18 \,\mathrm{\gls{GeV}}/c^2$ ($\overline{\mathrm{MS}}$ masses at $\mu = m_q$), 
    $m_t \approx 172.6 \,\mathrm{\gls{GeV}}/c^2$ (from direct measurements), 
    which compared to the mass of the proton $m_p \approx 0.938272 \,\mathrm{\gls{GeV}}/c^2 \approx 1.67 \times 10^{-27} \,\mathrm{kg}$ can be separated into three very light, and three (very) heavy quarks. We consider all physical observables at $Q^2 > m_p^2$, which means the first three flavors of quark can be always approximated as massless.
    }. This means that all quarks are treated equally, and the kinematics of the scattering dictate whether a given flavor of quark can contribute to the interaction. In this picture, below the production threshold of a given heavier flavor, it nor its PDFs exists. We will in practice write the forward operators based on the perturbative QCD for the full case of $n_f \equiv 6$, i.e. we include all contributions in the concrete expressions. To then get a prescription for any $O^{(n_f)}$ for $n_f \leq 5$, one explicitly sets the elements of the forward operators for the heavier flavors to zero. This treatment would also necessitate the world data to be split into these same regimes by $Q^2$, but we will not write that out explicitly for this proof of principle formulation. Rather, we are interested for the capability of implementing a more realistic flavor number scheme in this formalism, which is discussed in Sec.~\ref{sec:flavor-number-schemes}, as the corrections from a non-zero mass $m$5 of a quark can be significant around the scale $Q^2 \approx m^2$~\cite{Barontini:2024xgu:flavor-number-scheme-names-and-intro}. To have a toy-model of a proper flavor number scheme, we will adapt the following:
    \begin{equation}
        O^\mathrm{3cb+ZM-VFNS}\qty(Q^2) \coloneqq
        \begin{cases}
            O^{\mathrm{FFNS3cb}}\qty(Q^2) &\mathrm{if}~ Q^2 < m_b^2 \\
            O^{(5)}\qty(Q^2) &\mathrm{if}~ m_b^2 \leq Q^2 < m_t^2 \\
            O^{(6)}\qty(Q^2) &\mathrm{if}~ m_t^2 \leq Q^2 ,
        \end{cases}
        \label{eq:toy-model-flavor-number-scheme-F3cb+ZM}
    \end{equation}
    which at large $Q^2$ just takes the zero mass approximation approach, but at low $Q^2$ implements the FFNS3cb scheme~\cite{Barontini:2024xgu:flavor-number-scheme-names-and-intro}, which includes mass effects for charm and bottom while neglecting two-mass effects. Both charm and bottom are produced from three light flavors. The motivation to work with such a toy-model is to be able to consider the implementation of the more realistic---and complex---perturbative QCD theory prescriptions in this inverse problems formalism. Fully implemented this scheme could perhaps even be considered as moderately realistic, however, in this initial work we will not consider the inclusion of heavy mass effect anywhere else than for the experimental production of heavy flavors in DIS: all other observables are taken to be purely in the zero mass approximation, which especially at low $Q^2$ is unfortunately dissatisfactory in real phenomenological QCD analysis of HERA data, for example. Lastly, we note that this FNS partitioning of the kinematic regimes by $Q^2$ also affects the running of the strong coupling $\as(Q^2)$ as that depends on the number of light flavors, and needs to be part of the systematic implementation of the FNS within each kinematic regime.

    \subsection{Inclusive electromagnetic DIS} \label{sec:lo-dis}

    The first of the canonical high-energy scattering observables---cross sections---we write in the linear integral operator formalism is the total cross section for electron--proton DIS~\cite{Blumlein:2012bf:dis-review}, which quantifies the \gls{bjorkenx} and $Q^2$ dependent total scattering probability in a scattering mediated by the electromagnetic force~\cite{Bjorken:1969ja}. The process of an electron $(e^-)$ or a positron $(e^+)$ scattering off a proton $p$ via the exchange of a virtual photon $\gamma^*$ and producing the experimental observation of the outgoing probe lepton and any final state $X$ is denoted as
    \begin{equation} \label{eq:process:dis-incl}
        e^\pm +p \to e^\pm + X.
    \end{equation}
    The classical \gls{leading order} prescription for the \gls{reduced cross section} of this process is~\cite{Abramowicz:2015mha:HERAIInewcombined, ParticleDataGroup:2024cfk, Blumlein:2012bf:dis-review, Sterman:1995:handbook-qcd, Peskin:0201503972,Griffiths:1987tj}:
    \begin{align} \label{eq:cs:dis-inclusive}
        \sigma^{\gamma}_r(x,Q^2) \pequiv{} & 
        xy^2 F^{\gamma}_1(x,Q^2) + (1-y) F^{\gamma}_2(x,Q^2) ,
    \end{align}
    where the equivalence indicates that we take this physical theory result as our starting point for writing the inverse problem, $\gamma$ specifies that the scattering is mediated by the photon,
    and $y \coloneqq \frac{Q^2}{xs} \in (0,1)$ is known as inelasticity. Leading order refers to leading contribution proportional to $\as^0$ to the observable in the perturbative expansion in \gls{strong coupling} $\as$; we will return to higher order corrections in later sections.
    Lastly, the proton structure functions $F_i^\gamma$ that appear are defined as:
    \begin{align}
        \label{eq:def-f2}
        F^{\gamma}_2(x, Q^2) & \coloneqq \sum_a e_a^2 x f_{a}(x, Q^2),
        \\
        F^{\gamma}_L(x, Q^2) & \coloneqq F^{\gamma}_2(x, Q^2) - 2 xF^{\gamma}_1(x, Q^2),
    \end{align}
    where $e_a$ is the \gls{fractional charge} of the quark flavor $a$, and $L$ refers to longitudinal. At leading order $F_L^\gamma \equiv 0$, known as the Callan--Gross relation, allows one to relate the total cross section directly to $F_2^\gamma$:
    \begin{align} \label{eq:cs:dis-inclusive-callan-gross}
        \sigma^{\gamma}_r(x,Q^2,y) ={} &
        \qty( \frac{y^2}{2} + (1-y) ) \sum_a e_a^2 x f_{a}(x, Q^2).
    \end{align}
    Here we can see fundamental issue of resolving between the parton \glspl{flavor} in the proton: in the leading picture the total cross section observable is functionally identically dependent of the PDFs $f_a(x,Q^2)$, and additional observables are a mathematical necessity to achieve any resolution between the flavors.

    We define the \gls{leading order} \gls{inclusive} \gls{reduced cross section} \gls{forward operator} based on Eq.~\eqref{eq:cs:dis-inclusive-callan-gross} as:
    \begin{align}
        \label{eq:fwdop-incldis-lo}
        \bA^{\gamma}_r (x,Q^2,y) \coloneqq{}
        &
        \left(\frac{xy^2}{2} + x (1-y) \right)
        \begin{bmatrix}
            0 &
            e_d^2 \mathrm{Id} & 
            e_u^2 \mathrm{Id} &
            e_s^2 \mathrm{Id} &
            e_c^2 \mathrm{Id} &
            e_b^2 \mathrm{Id} &
            e_t^2 \mathrm{Id} &
            e_{\bar d}^2 \mathrm{Id} &
            \cdots &
            e_{\bar t}^2 \mathrm{Id}
        \end{bmatrix},
    \end{align}
    which is a row-matrix operator, where the different parton contributions differ only by their squared fractional charge $e_a^2$, and the gluon does not contribute for a variety of reasons~\cite{Peskin:0201503972,Sterman:1995:handbook-qcd}, with having no electric charge being one. The identity maps $\Id$ map the PDFs onto themselves, and are written to make it explicit that the forward operator is a functional operator, which is an important distinction, especially once we consider more non-trivial operations.

    To demonstrate the action of the total cross section forward operator defined in Eq.~\eqref{eq:fwdop-incldis-lo}, we compute:
    \begin{align}
        \sigma^{\gamma}_r (x,Q^2,y) & \eqqcolon \qty(\bA^{\gamma}_r \pdtbf)(x,Q^2,y)
        \notag \\
        &= 
        \left(\frac{xy^2}{2} + x (1-y) \right)
        \begin{bmatrix}
            0 & e_d^2 \mathrm{Id} & e_u^2 \mathrm{Id} & \cdots & e_{\bar t}^2 \mathrm{Id}
        \end{bmatrix}
        \pdtwritout
        \nlinenotag
        & =
        \left(\frac{xy^2}{2} + x (1-y) \right)
        \sum_a e_a^2 f_a(x,Q^2),
    \end{align}
    which is the classical tree-level expression for the inclusive reduced cross section~\cite{ParticleDataGroup:2024cfk,Griffiths:1987tj} given in Eq.~\eqref{eq:cs:dis-inclusive-callan-gross}. The above construction exactly reproduces the canonical result, as intended. We will elaborate in Secs.~\ref{sec:tensor-ip-lo} and~\ref{sec:discretization} why this reformulation  of the classical results has practical implications for solving the global analysis inverse problem without fitting of model parameters.

    To illustrate the construction process for another observable, and to include some sensitivity to the gluon distribution $f_g$ in the toy-model leading order inverse problem, we note that the Altarelli--Martinelli (AM) equation allows one to relate the longitudinal structure function $F_L^\gamma$ to $F_2^\gamma$ and the gluon distribution $G(y, Q^2) \equiv f_g(y, Q^2)$~\cite{Altarelli:1978FL,Cooper-Sarkar:1997pqx,Tvaskis:2016uxm:jlabfl}\footnote{Refs.~\cite{Cooper-Sarkar:1997pqx} and~\cite{Tvaskis:2016uxm:jlabfl} differ with a factor of $z$ in the gluon contribution term, we use the expression from the former.}:
    \begin{equation}
    \label{eq:fl-lodis}
        F_L^{\gamma}(x,Q^2)
        \approx
        F_L^{\gamma,\mathrm{AM}}(x,Q^2) \pequiv \frac{\as}{\pi}
        \int_x^1 \frac{\ud z}{z} \qty(\frac{x}{z})^2 
        \left \lbrace
            F_2(z, Q^2) + 2 \sum_a e_a^2 \qty(1 - \frac{x}{z}) z f_g(z,Q^2)
        \right \rbrace.
    \end{equation}
    We can see that $F_L^\gamma \propto \as$ so strictly at $\as^0$ order in pQCD one should have $F_L^\gamma \equiv 0$. However, there is no technical limitation for including observables at mixed orders in the $\as$ expansion, which we show by constructing the forward operator for $F_L^\gamma$ given by Eq.~\eqref{eq:fl-lodis}.
    To write the forward operator, we use the definition of $F_2^\gamma$ in \eqref{eq:def-f2}, and define two integral operators:
    \begin{align}
        \Ical_{L,q}^\mathrm{AM} [f] \coloneqq {} &
        \int_x^1 \frac{\ud z}{z} \qty(\frac{x}{z})^2 f(z), 
        \\
        \Ical_{L,g}^\mathrm{AM} [f] \coloneqq {} &
        \int_x^1 \frac{\ud z}{z} \qty(\frac{x}{z})^2 \qty(1 - \frac{x}{z}) z f(z), 
    \end{align}
    with which we can write the forward operator for $F_L^\gamma$:
    \begin{align}
        \label{eq:fwdop-incldis-fl-lo}
        \bB_L^\mathrm{AM} (x,Q^2) \coloneqq
        &
        \frac{\as}{\pi}
        \begin{bmatrix}
            2 \undertilde{e}^2 \Ical_{L,g}^\mathrm{AM} &
            e_d^2 x \Ical_{L,q}^\mathrm{AM} & 
            \cdots &
            e_t^2 x \Ical_{L,q}^\mathrm{AM} &
            e_{\bar d}^2 x \Ical_{L,q}^\mathrm{AM} &
            \cdots &
            e_{\bar t}^2 x \Ical_{L,q}^\mathrm{AM}
        \end{bmatrix},
    \end{align}
    where we can now explicitly see the replacement of the identity $\Id$ with the integral operators $\Ical_{L,q}^\mathrm{AM}$ and $\Ical_{L,g}^\mathrm{AM}$, and we defined a fractional charge vector:
    \begin{equation}
        \label{eq:e-undertilde-frac-charge-vec}
        \undertilde{e} \coloneqq [0, e_d, e_u, \dots, e_t, e_{\bar d}, e_{\bar u},\dots, e_{\bar t}]    
    \end{equation}
    for the convenience of writing the squared fractional charge sum $\sum_a e_a^2 \equiv \undertilde{e}^2$ appearing in the gluon contribution term, which could also expressed in terms of the average squared charge as $\undertilde{e}^2 \equiv n_f \langle e^2 \rangle$, and which also makes it explicit that $\undertilde{e}^2$ is a function of the number of flavors $n_f$ changes, which becomes relevant in Sec.~\ref{sec:flavor-number-schemes}. With the forward operator $\bB_L^\mathrm{AM}$, one computes the observable structure function by acting on the \gls{parton distribution tensor} $\pdtbf$:
    \begin{equation*}
        F^{\gamma,\mathrm{AM}}_L(x,Q^2) = \qty(\bB_L^\mathrm{AM} \pdtbf) (x,Q^2).
    \end{equation*}

    Next we move on to observables sensitive to more distinct physical processes, which enable the framework of global analysis to build resolving power between the different flavors of partons in the proton.

    \subsection{Neutral current DIS} \label{sec:lo-nc-dis}

    Next we move on from the purely electromagnetic picture of the electron--proton scattering, and consider a more complete picture of DIS, where the interaction can also be mediated by the weak nuclear force\footnote{In complete generality, lepton--proton DIS can also be mediated by the exchange of a Higgs boson~\cite{Ma:1974cj:dis-scalar-higgs, Soar:2009yh:dis-scalar-higgs, Ellis:1975ap}, but that is not considered in this work where our focus is on the most well-understood and measured observables.}. 
    This more comprehensive picture is known as neutral current (NC) DIS, and the purely electromagnetic case from the previous section is a part of the full neutral current picture. 
    This electrically neutral weak nuclear interaction is mediated by the $Z$ boson. The $Z$ boson has the mass $M_Z = 91.1880 \pm 0.0020 \, \mathrm{\gls{GeV}}$~\cite{ParticleDataGroup:2024cfk}, which is heavier than the energy equivalent of an iron nucleus, or approximately $97$ times the mass of the proton. This heavy mass of the boson means that it cannot be produced in low-$Q^2$ collisions, and at higher and higher $Q^2$ its contribution becomes more significant: at $Q^2 \sim M_Z^2$ the $Z$-exchange contribution is of comparable scale with the electromagnetic interaction mediated by the virtual photon $\gamma^*$~\cite{ZEUS:2002pdo:F3}, and is therefore essential to include in the world data inference problem.

    The neutral current deeply inelastic scattering 
    \begin{equation}
        e^\pm p \to e^\pm X : ~ Z + q \to q',
    \end{equation}
    with the sub-process of the $Z$ boson interacting with a parton $q$, has the \gls{observable} \gls{reduced cross section}~\cite{ParticleDataGroup:2024cfk,Abramowicz:2015mha:HERAIInewcombined,ZEUS:2002pdo:F3}:
    \begin{equation}
        \label{eq:nc-dis-crosssection}
        \sigma^{\pm}_{r,\mathrm{NC}}(x,Q^2,y)
        \pequiv
            F_2^{\pm,\mathrm{NC}}(x,Q^2) \mp \frac{Y_-}{Y_+} x F_3^{\pm,\mathrm{NC}}(x,Q^2) - \frac{y^2}{Y_+} F^{\pm,\mathrm{NC}}_L(x,Q^2)
        ,
    \end{equation}
    where the sign of the superscript refers to the charge of the electron probe $e^\pm$.
    The measurement of the neutral current cross section with both electron $e^-$ and positron $e^+$ probes enables one to directly extract the structure function $F_3$ from experimental data~\cite{ZEUS:2002pdo:F3}:
    \begin{equation}
        xF_3^\mathrm{NC}(x,Q^2) \simeq \frac{Y_+}{2Y_-} \left( \sigma_{r,\mathrm{NC}}^{-}(x,Q^2,y) - \sigma_{r,\mathrm{NC}}^{+}(x,Q^2,y) \right),
    \end{equation}
    which makes it a physical observable, perhaps analogously to the Rosenbluth separation method~\cite{Rosenbluth:1950yq}\footnote{The above subtraction of cross sections exactly cancels out the dominant electromagnetic contributions $F_i^\gamma$, but a more technical treatment would account for the sub-dominant interference contributions $F_2^{\gamma Z}$ and $F_L^{\gamma Z}$, which have lepton sign dependent weak vectorial couplings and so do not cancel out. We do not go into these details here as the subtracted quantity $\sigma_{r,\mathrm{NC}}^{-} - \sigma_{r,\mathrm{NC}}^{+}$ will not be a linearly independent observable from the reduced cross sections $\sigma^{\pm}_{r,\mathrm{NC}}$.}.
    The inelasticity variable is defined $Y_\pm \coloneqq 1 \pm (1-y)^2$, the mass of the target has been neglected in the $\frac{M^2}{Q^2} \to 0$ limit, and the $\pm$ takes $+$ for an electron $e^-$, and a $-$ for a positron $e^+$~\cite{ParticleDataGroup:2024cfk}.
    The neutral current structure functions can be written in terms of a purely photonic component, a Z boson component, and their interference~\cite{ParticleDataGroup:2024cfk}:
    \begin{align}
        F_2^{\pm,\mathrm{NC}}(x,Q^2) & \pequiv
        F_2^\gamma (x,Q^2)
        - \qty(g_V^{e,\pm} \pm \lambda g_A^{e,\pm}) \eta_{\gamma Z} F_2^{\gamma Z}(x,Q^2)
        \notag \\ 
        & \qquad + \qty(\qty(g_V^{e,\pm})^2 + \qty(g_A^{e,\pm})^2 \pm 2 \lambda g_V^{e,\pm} g_A^{e,\pm}) \eta_Z F_2^Z(x,Q^2),
        \notag \\
        \label{eq:nc-f2-lo}
        & \overset{\sum_\lambda}{=}
        F_2^\gamma (x,Q^2)
        - g_V^{e,\pm} \eta_{\gamma Z} F_2^{\gamma Z} (x,Q^2)
        + \qty(\qty(g_V^{e,\pm})^2 + \qty(g_A^{e,\pm})^2) \eta_Z F_2^Z(x,Q^2),
    \end{align}
    with the same decomposition holding for $F_L^\mathrm{NC}$, the sum over the helicities $\lambda \in \lbrace -1 ,1 \rbrace$ eliminates the terms proportional to $\lambda$ since we are considering the unpolarized case, and
    \begin{align}
        xF_3^{\pm,\mathrm{NC}}(x,Q^2) & \pequiv
        - \qty(g_A^{e,\pm} \pm \lambda g_V^{e,\pm}) \eta_{\gamma Z} x F_3^{\gamma Z} (x,Q^2)
        \nonumber \\
        & \qquad
        + \qty(2 g_V^{e,\pm} g_A^{e,\pm} \pm \lambda \qty(\qty(g_V^{e,\pm})^2 + \qty(g_A^{e,\pm})^2)) \eta_Z x F_3^Z(x,Q^2)
        \notag \\
        \label{eq:nc-f3-lo}
        & \overset{\sum_\lambda}{=}
        - g_A^{e,\pm} \eta_{\gamma Z} x F_3^{\gamma Z} (x,Q^2)
        + 2 g_V^{e,\pm} g_A^{e,\pm} \eta_Z x F_3^Z(x,Q^2)
        ,
    \end{align}
    where $g_V^{e,\pm} \coloneqq \pm(\frac{1}{2} - 2 \sin^2 \theta_W) \approx \pm 0.038$, $\sin^2 \theta_W \approx 0.231$, $g_A^{e,\pm} \coloneqq \pm \frac{1}{2}$, where the $\pm$ is assigned the same sign as the electric charge of the lepton $e^\pm$.
    The structure function $xF_3^\mathrm{NC}$ has been measured by the ZEUS collaboration~\cite{ZEUS:2002pdo:F3}, and so we consider it as an observable, as discussed in Sec.~\ref{sec:data}, and the value of the squared sine of the weak mixing angle $\theta_W$ is determined experimentally~\cite{ParticleDataGroup:2024cfk}, having approximately the value $\sin^2 \theta_W \approx 0.231$, as is the mass $M_Z$ of the $Z$ boson.
    The factors $\eta_j$---which are related to the relative differences between the photon $\gamma$ and $Z$ as bosons mediating the interaction~\cite{ParticleDataGroup:2024cfk}---are
    \begin{align*}
        \eta_\gamma & \coloneqq 1, \\
        \eta_{\gamma Z}(Q^2) & \coloneqq \frac{G_F M_Z^2}{2 \sqrt{2} \pi \aem} \frac{Q^2}{Q^2 + M_Z^2}, \\
        \eta_Z(Q^2) & \coloneqq \eta_{\gamma Z}(Q^2)^2,
    \end{align*}
    where $G_F \approx 1.166 \times 10^{-5} \, \mathrm{\gls{GeV}}^{-2}$ is the Fermi constant~\cite{ParticleDataGroup:2024cfk}.
    These decompositions hold generally in perturbative QCD. At leading order the partonic description of the $\gamma$, $\gamma Z$, and $Z$ contributions are~\cite{ParticleDataGroup:2024cfk}:
    \begin{align}
        F_2^{\gamma}(x,Q^2) & = x \sum_{a, a\neq g} e_a^2 \qty(f_a(x,Q^2) + f_{\bar a}(x,Q^2)),
        \label{eq:f2-nc-gamma}
        \\
        F_2^{\gamma Z}(x,Q^2) & = x \sum_{a, a\neq g} 2 e_a g_V^a \qty(f_a(x,Q^2) + f_{\bar a}(x,Q^2)),
        \label{eq:f2-nc-gammaZ}
        \\
        F_2^{Z}(x,Q^2) & = x \sum_{a, a\neq g} \qty( \qty(g_V^a)^2 + \qty(g_A^a)^2) \qty(f_a(x,Q^2) + f_{\bar a}(x,Q^2)),
        \label{eq:f2-nc-Z}
        \\
        F_3^{\gamma}(x,Q^2) & = 0,
        \\
        F_3^{\gamma Z}(x,Q^2) & = x \sum_{a, a\neq g} 2 e_a g_A^a \qty(f_a(x,Q^2) - f_{\bar a}(x,Q^2)),
        \label{eq:f3-nc-gammaZ}
        \\
        F_3^{Z}(x,Q^2) & = x \sum_{a, a\neq g} 2 g_V^a g_A^a \qty(f_a(x,Q^2) - f_{\bar a}(x,Q^2)),
        \label{eq:f3-nc-Z}
    \end{align}
    where $g_V^a \coloneqq \pm \frac{1}{2} - 2e_a \sin^2 \theta_W$ and $g_A^a \coloneqq \pm \frac{1}{2}$, with $\pm$ taking $+$ for the $u$-type quarks $(u,c,t)$, and $-$ for the $d$-type $(d,s,b)$ quarks\footnote{This flavor dependent sign of the vectorial coupling $g_V^a$ means that the numerical values are approximately $\approx+0.19$ for the up type quarks, and $\approx-0.35$ for the down type.}.
    Here we see that the electromagnetically mediated inclusive DIS discussed in the previous section is a strict subsection of the full neutral current picture, here being the $\gamma$ contribution.
    With the above we can write the full neutral current structure function $F_2^\mathrm{NC}$ defined by Eq.~\eqref{eq:nc-f2-lo} for a lepton probe as:
    \begin{equation}
        \label{eq:fwdop-nc-Z-LO-f2}
        F_2^{\pm,\mathrm{NC}}(x,Q^2) =
        x \begin{bmatrix}
            0 &
            \phi^{\pm}_{2,d} & \cdots & \phi^{\pm}_{2,\bar t}
        \end{bmatrix}
        \pdtwritout
        \eqqcolon
        \qty(\bB^{\pm,\mathrm{NC}}_{2} \pdtbf)(x, Q^2) 
        ,
    \end{equation}
    where we defined a helicity $\lambda$ averaged effective electro-weak charge:
    \begin{align}
        \label{eq:lo-nc-phi2}
        \phi^{\pm}_{2,a} \equiv \phi^{\pm}_{2,a}(Q^2) \coloneqq {} &
            \sum_{\lambda=-1,1} \bigg( e_a^2 
            - 2 e_a g_V^a \qty(g_V^{e,\pm} \pm \lambda g_A^{e,\pm}) \eta_{\gamma Z}(Q^2) 
            \notag \\
            & \qquad + \qty( \qty(g_V^a)^2 + \qty(g_A^a)^2) \qty(\qty(g_V^{e,\pm})^2 + \qty(g_A^{e,\pm})^2 \pm 2 \lambda g_V^{e,\pm} g_A^{e,\pm}) \eta_Z(Q^2) \bigg)
            \notag \\
            ={} & e_a^2 
            - 2 e_a g_V^a g_V^{e,\pm} \eta_{\gamma Z}(Q^2)
            + \qty( \qty(g_V^a)^2 + \qty(g_A^a)^2) \qty(\qty(g_V^{e,\pm})^2 + \qty(g_A^{e,\pm})^2) \eta_Z(Q^2),
    \end{align}
    which reproduces the helicity averaged definition given by Eq.~\eqref{eq:nc-f2-lo}. The helicity averaging is done to match the unpolarized real world data from the combined measurement by H1 and ZEUS collaborations~\cite{Abramowicz:2015mha:HERAIInewcombined}. We write similarly for $xF_3^\mathrm{NC}$:
    \begin{equation}
        \label{eq:fwdop-nc-Z-LO-f3}
        x F_3^{\pm,\mathrm{NC}}(x,Q^2) =
        x \begin{bmatrix}
            0 &
            \phi^{e^\pm}_{3,d} & \cdots & - \phi^{e^\pm}_{3,\bar t}
        \end{bmatrix}
        \pdtwritout
        \eqqcolon
        \qty(\bB^{\pm,\mathrm{NC}}_{3} \pdtbf) (x, Q^2)
        ,
    \end{equation}
    where we note the change of sign for the antiquark terms, and use
    \begin{align}
    \label{eq:lo-nc-phi3}
        \phi^{\pm}_{3,a} \equiv \phi^{\pm}_{3,a}(Q^2) \coloneqq {} &
            \sum_{\lambda=-1,1} \bigg( - 2 e_a g_A^a \qty(g_A^{e,\pm} \pm \lambda g_V^{e,\pm}) \eta_{\gamma Z} 
            \notag \\
            & + 2 g_V^a g_A^a \qty(2 g_V^{e,\pm} g_A^{e,\pm} \pm \lambda \qty(\qty(g_V^{e,\pm})^2 + \qty(g_A^{e,\pm})^2)) \eta_Z \bigg)
            \notag \\
            ={} & - 2 e_a g_A^a g_A^{e,\pm} \eta_{\gamma Z} 
            + 2 g_V^a g_A^a \qty(2 g_V^{e,\pm} g_A^{e,\pm}) \eta_Z,
    \end{align}
    which by definition reproduces \eqref{eq:nc-f3-lo}.

    We will omit discussing $F_L^\mathrm{NC}$ here again, since we will need to return to include it properly when we move up to next-to-leading order accuracy in the Sec.~\ref{sec:forward-problem-nlo}.
    Instead, we finish this section by writing the forward operator for the neutral current reduced cross section of Eq.~\eqref{eq:nc-dis-crosssection} with the LO accuracy identity $F_L^\mathrm{NC} \equiv 0$. Therefore, it is written in terms of the operators for $F_{2,3}^\mathrm{NC}$:
    \begin{align}
        \bA^{\pm}_{r,\mathrm{NC}}(x,Q^2,y)
        & \coloneqq
        \left[
            Y_+ \bB_2^{\pm,\mathrm{NC}}(x,Q^2) \mp Y_- \bB_3^{\pm,\mathrm{NC}}(x,Q^2) 
        \right]
        \notag \\
        &=
        \begin{bmatrix}
            0 &
            Y_+ \phi^{\pm}_{2,d} \mp Y_- \phi^{\pm}_{3,d} 
            & \cdots &Y_+ \phi^{\pm}_{2,\bar t} \pm Y_- \phi^{\pm}_{3,\bar t} 
        \end{bmatrix},
        \label{eq:fwdop-nc-crosssec-lo}
    \end{align}
    which by construction reproduces the definition of the neutral current reduced cross section Eq.~\eqref{eq:nc-dis-crosssection} when operating on $\pdtbf(x,Q^2)$.


    \subsection{Charged current DIS} \label{sec:lo-cc-dis}

    The name charged current (CC) reflects the substantial step away from the previous sections that electric charge is transferred in the interaction. Charged current DIS is the process 
    \begin{equation}
        \label{eq:process:dis-charged-current}
        e^-(e^+) +p \to \nu (\bar \nu) + X,
    \end{equation}
    where the lepton flavor changes when the force carrier $W^+$ or $W^-$ gauge boson is emitted. This means that the incoming electron or positron scatters by emitting a $W^\pm$ boson, and turns into a neutrino $\nu$ or an antineutrino $\bar \nu$, and the electrically charged boson interacts with the constituents of the proton. The reverse is possible as well, which we will cover in Sec.~\ref{sec:lo-neutrino-dis}.
    With the $e^\pm$ probe the identification of these events is straightforward as the probe vanishes, and in the final state the corresponding neutrino escapes undetected.

    The charged current DIS reduced cross sections are defined as~\cite{Abramowicz:2015mha:HERAIInewcombined}:
    \begin{align}
        \label{eq:cc-dis-crosssection}
        \sigma_{r,\mathrm{CC}}^{\pm}(x,Q^2,y)
        \pequiv
            & \frac{Y_+}{2} F_2^{W^\pm}(x,Q^2) \mp \frac{Y_-}{2} x F_3^{W^\pm}(x,Q^2) 
            - \frac{y^2}{2} F^{W^\pm}_L(x,Q^2)
        ,
    \end{align}
    where 
    $Y_\pm \coloneqq 1 \pm (1-y)^2$.
    Due to the way the $W^\pm$ couples only to specific flavors of quarks and antiquarks, i.e. $W^+$ interacting with $ d, s, b, \bar u , \bar c, \bar t$, and $W^-$ with $u,c,t, \bar d, \bar s, \bar b$, the structure functions appear in the charged current cross section in the form~\cite{Ingelman:1987zv,Cooper-Sarkar:1997pqx,ParticleDataGroup:2024cfk}:
    \begin{align}
        \label{eq:cc-f2-lo-W+}
        F^{W^+}_2(x,Q^2) &\coloneqq 
            x 
            \qty(f_d(x,Q^2) + f_s(x,Q^2) + f_b(x,Q^2) + f_{\bar u}(x,Q^2) + f_{\bar c}(x,Q^2) + f_{\bar t}(x,Q^2))
            ,
        \\
        \label{eq:cc-f3-lo-W+}
        x F^{W^+}_3(x,Q^2) &\coloneqq 
            x
            \qty(f_d(x,Q^2) + f_s(x,Q^2) + f_b(x,Q^2) - f_{\bar u}(x,Q^2) - f_{\bar c}(x,Q^2) - f_{\bar t}(x,Q^2)),
        \\
        \label{eq:cc-f2-lo-W-}
        F^{W^-}_2(x,Q^2) &\coloneqq 
            x 
            \qty(f_u(x,Q^2) + f_c(x,Q^2) + f_t(x,Q^2) + f_{\bar d}(x,Q^2) + f_{\bar s}(x,Q^2) + f_{\bar b}(x,Q^2))
            ,
        \\
        \label{eq:cc-f3-lo-W-}
        x F^{W^-}_3(x,Q^2) &\coloneqq 
            x
            \qty(f_u(x,Q^2) + f_c(x,Q^2) + f_t(x,Q^2) - f_{\bar d}(x,Q^2) - f_{\bar s}(x,Q^2) - f_{\bar b}(x,Q^2)).
    \end{align}
    For mathematical completeness we have written back in the heavy quark contributions for $b,t$ and used the Cabibbo--Kobayashi--Maskawa matrix unitarity relations to sum over the final state partons, following Ref.~\cite{Ingelman:1987zv}. However, since the above theory result is for massless partons~\cite{Ingelman:1987zv}, it would be more accurate to only include the light quarks $u,d,s$ here, and implement the heavy quark contributions to the charged current DIS in a separate framework, for example as in Ref.~\cite{Risse:2025smp:heavy-cc}. We discuss heavy quarks in more detail in their specific section. To adhere more closely to Refs.~\cite{Ingelman:1987zv,Cooper-Sarkar:1997pqx}, we will limit ourselves to leading order in $\as$ and take $F^{W^\pm}_L \equiv 0$ for the remainder of this section.

    The crux of the above discussion is that measurements of $e^+p$ and $e^-p$ charged current DIS are two distinct observables, and will need to be correspondingly written into their own forward operators. Writing the above structure functions, the probe specific cross sections become
    \begin{align}
        \sigma_{r,\mathrm{CC}}^{+}(x,Q^2,y) ={} &
        x f_{\bar u}(x,Q^2) + x f_{\bar c}(x,Q^2) + x f_{\bar t}(x,Q^2) 
        \notag \\
        & + (1-y)^2 (x f_d(x,Q^2) + x f_s(x,Q^2) + x f_b(x,Q^2)),
        \\
        \sigma_{r,\mathrm{CC}}^{-}(x,Q^2,y) ={} &
        x f_u(x,Q^2) + x f_c(x,Q^2) + x f_t(x,Q^2) 
        \notag \\
        & + (1-y)^2 (x f_{\bar d}(x,Q^2) + x f_{\bar s}(x,Q^2) + x f_{\bar b}(x,Q^2)).
    \end{align}

    We can now write the above cross sections in to the forward operator form, which become:
    \begin{align}
        \label{eq:fwdop-cc-lo}
        \mathbf{A}_{r,\mathrm{CC}}^{+} (y) & \coloneqq
        x
        \begin{bmatrix}
            0 & 
            (1-y)^2 \mathrm{Id} & 0 &
            (1-y)^2 \mathrm{Id} & 0 &
            (1-y)^2 \mathrm{Id} & 0 &
            0 & \Id &
            0 & \Id &
            0 & \Id
        \end{bmatrix},
        \\
        \label{eq:fwdop-cc-lo2-eminus}
        \mathbf{A}_{r,\mathrm{CC}}^{-} (y) & \coloneqq
        x
        \begin{bmatrix}
            0 &
            0 & \Id & 
            0 & \Id &
            0 & \Id&
            (1-y)^2 \mathrm{Id} & 0 &
            (1-y)^2 \mathrm{Id} & 0 &
            (1-y)^2 \mathrm{Id} & 0
        \end{bmatrix},
    \end{align}
    where we observe that the Bjorken-$x$ is completely factorized into the PDFs, and the operators only depend on $y$, and they reproduce the original equations when operating on the \gls{parton distribution tensor} $\pdtbf$ defined in Eq.~\eqref{eq:pdt-definition}.

    \subsection{Heavy quark production in DIS} \label{sec:lo-heavyq}

    The quarks are typically divided by mass into two groups, those that are so light that they can very reasonably be considered massless $(d,u,s)$, and those that are not and are called the heavy quarks: charm, bottom, and top $(c,b,t)$. The previous sections have worked in the ``massless'' limit where all the parton masses are exactly zero. However, to build a more physically accurate picture of the high-energy collisions, one must amend the theory for the notable masses of the heavier quarks, which compared to the mass of the proton $m_p$ are $m_c \approx 1.36 m_p$, $m_b \approx 4.46 m_p$, and $m_t \approx 184 m_p$ for the charm, bottom, and top quarks~\cite{ParticleDataGroup:2024cfk}. The heavy quarks are so massive that particle remnants of their decay can be identified in experiment, which enables a more direct separation of the events where they are produced, which allows one to define observables that are more sensitive to the specific heavy quark PDFs. Another cause of intrigue is the question of whether the wavefunction of the proton has an intrinsic contribution from the charm quark, which is a long-standing open question~\cite{Brodsky:1980pb:Hoyer-intrinsic-charm,Martin:2009iq,NNPDF:2022qks}, which would be a significant discovery about the fundamental structure of the proton as the charm quark is $\sim 36\%$ \textit{more massive} than the proton itself~\cite{ParticleDataGroup:2024cfk}. Allowing for intrinsic charm in this formalism is in principle possible, for example based on Ref.~\cite{Ball:2015dpa:intrinsic-charm-dis}, but is left for future work.
    
    We base the following exposition of heavy quark physics in DIS on Ref.~\cite{Klein2012}. The heavy quark production process in DIS for charm $c$ (bottom $b$) quarks is:
    \begin{equation}
        e^\pm + p \to e^\pm + c \bar c \, (b \bar b) + X,
    \end{equation}
    which means that the experiment identifies outgoing particles that can be associated with a $c\bar c$ or $b \bar b$ quark pair having been produced in the original scattering of the lepton and proton. In principle top quark could be produced as well, but it would require incredibly high energy collisions, beyond what has been reached in lepton--proton colliders so far, however the next generation Electron Ion--Collider~\cite{Accardi:2012qut:EIC,AbdulKhalek:2021gbh:EIC} is expected to reach high enough energies to study top quark related physics~\cite{Jiang:2025frv:topq:eic}. In the rest of this section we will focus on the charm and bottom quarks, in line with Ref.~\cite{Klein2012}, since in the kinematic regime of the best currently available DIS data from the HERA particle collider~\cite{Abramowicz:2018flt:HERAII-charm-bottom-combined}, charm is the most predominantly produced of the heavy quarks, with the heavier bottom being produced less frequently, but at detectable rates~\cite{Abramowicz:2018flt:HERAII-charm-bottom-combined}. The theory picture is also only considered for the neutral current case, since that is the only available data currently~\cite{Abramowicz:2018flt:HERAII-charm-bottom-combined}, but in principle charged current production of heavy quarks could be considered as an observable as well.

	In this section we will write down the leading order expressions for the observable of heavy quark production in DIS without taking reference of the \gls{flavor number scheme} that one would use as a part of the global data inference of the PDFs. Here we focus on the integral equation form of the observables, and in Sec.~\ref{sec:flavor-number-schemes} we will continue more comprehensively the discussion of quark \glspl{flavor number scheme}, and their implementation within this inverse problems approach.
	However, in the interest of not leaving the heavy quark observables into a partially defined state, for the time being we will consider as a toy model of the physics a simplistic framing based on the fixed \gls{flavor number scheme} (FFNS)~\cite{Buza:1996wv-heavyq-nlo,Klein2012}. This means that for the production of charm quark, we will assume three active light quarks described by PDFs, $n_f =3$, which contribute to the production of charm as a heavy inactive flavor. And similarly, and more naively, we assume for bottom production observables that charm can be taken to be light, i.e. that $m_c \ll m_b$, in which case we have four light flavors $n_f = 4$, and bottom is produced as a heavy flavor. The number of light flavors $n_f$ will appear as a variable in the following expressions due to the complicating detail that the $Q^2$ scale of the measurement will affect which flavors can and cannot be considered as light.
	The assumption that charm mass is negligible in comparison to that of bottom is mostly reasonable but not always valid, and would require a systematic inclusion of corrections in the transition regime in FFNS, or an alternative variable \gls{flavor number scheme}~\cite{Klein2012}, which we return to in Sec.~\ref{sec:flavor-number-schemes}.

    HERA data of heavy quark production is for the reduced cross section~\cite{Abramowicz:2018flt:HERAII-charm-bottom-combined}:
    \begin{equation}
        \label{eq:sigma-red-qqbar}
        \sigma_{r}^{Q\bar Q}(x,Q^2,y) \pequiv
        F^{Q \bar Q}_2(x,Q^2) - \frac{y^2}{1+(1-y)^2} F^{Q \bar Q}_L(x,Q^2)
        ,
    \end{equation}
    where $Q\bar Q$ refers to the produced heavy quark-antiquark pair, and the measurement has been done for the charm and bottom quarks.
    The heavy quark structure functions at leading order are~\cite{Klein2012,Witten:1975bh,Babcock:1977fi,Shifman:1977yb,Leveille:1978px,Gluck:1980cp}
    \begin{equation}
        F^{Q \bar Q}_{2,L} (x, Q^2, m^2) \pequiv e_Q^2 \as \int_{ax}^1 \frac{\ud z}{z} H^{(1)}_{g,(2,L)} \left( \frac{x}{z}, \frac{m^2}{Q^2} \right) f_g(n_f, z, Q^2),
        \label{eq:f2-fl-heavyqq-lo}
    \end{equation}
    where $m$ is the mass and $e_Q$ the fractional charge of the heavy quark $Q$, $n_f$ is the number of light quark flavors, and $a\coloneqq 1 + 4 \frac{m^2}{Q^2}$. The number of light flavors can vary for example depending on whether the charm quark is considered as light compared to the heaviest included quark in the problem.
    The count of order of expansion is here understood process specifically, i.e. order-$\as$ is here the leading contribution to heavy quark production, whereas for the inclusive case it is next-to-leading order in the $\as$ expansion. The coefficient functions are~\cite{Klein2012}:
    \begin{align}
        H^{(1)}_{g,2}\qty(\tau, \frac{m^2}{Q^2}) & \pequiv
        8 T_F \Bigg \lbrace
        v \left[ - \frac{1}{2} + 4 \tau (1-\tau) + 2 \frac{m^2}{Q^2} \tau (\tau-1) \right]
        \notag \\
        &
        \quad \quad \quad + \left[ - \frac{1}{2} + \tau - \tau^2 + 2 \frac{m^2}{Q^2} \tau (3\tau-1) + 4 \frac{m^4}{Q^4}\tau^2 \right] \ln\left( \frac{1-v}{1+v} \right)
        \Bigg \rbrace,
        \\
        H^{(1)}_{g,L}\qty(\tau, \frac{m^2}{Q^2}) & \pequiv 
        16 T_F \left[ \tau (1-\tau) v + 2 \frac{m^2}{Q^2} \tau^2 \ln\left( \frac{1-v}{1+v} \right) \right],
    \end{align}
    where $T_F = \frac{1}{2}$, and the center of mass system velocity $v$ of the heavy quark is
    \begin{equation}
        v \equiv v(m^2, Q^2, \tau) \coloneqq \sqrt{1- \frac{4 m^2 \tau}{Q^2 (1-\tau)}},
    \end{equation}
    and $\tau$ is just the name for the first variable of the coefficient functions $H$.
    Only the gluon PDF $f_g$ contributes at this order, and one must go to the next order in the perturbative expansion to see contributions from all partons.

    To write the forward operators, we define the integral operators:
    \begin{equation}
        \mathcal{H}^{(1)}_{g,(2,L)}[f]\left( x, \frac{m^2}{Q^2} \right) \coloneqq
        \as \int_{ax}^1 H^{(1)}_{g,(2,L)} \left( \frac{x}{z}, \frac{m^2}{Q^2} \right) f(z) \frac{\ud z}{z}.
    \end{equation}
    With these, the heavy quark production forward operators for the structure functions at leading order are:
    \begin{equation} \label{eq:fwdop-heavyQQ-LO-BQQ}
        \bB^{Q \bar Q}_{2,L} (x, Q^2, m^2) \coloneqq 
        \begin{bmatrix}
        e_Q^2 \mathcal{H}^{(1)}_{g,(2,L)} \left( x, \frac{m^2}{Q^2} \right) &
        0 & 0& 0& 0& 0& 0&
        0 & 0& 0& 0& 0 &0
        \end{bmatrix},
    \end{equation}
    which gives the operator for charm and bottom cases by selecting the corresponding heavy quark mass $m$, fractional charge $e_Q$.
    The forward operator for the reduced cross section \eqref{eq:sigma-red-qqbar} is then just a linear combination of the structure function operators:
    \begin{equation}
        \label{eq:fwdop-heavyQQ-LO-AQQ}
        \bA^{Q \bar Q}_{r}(x,Q^2,y,m^2) \coloneqq \bB^{Q \bar Q}_2(x,Q^2,m^2) - \frac{y^2}{1+(1-y)^2} \bB^{Q \bar Q}_L(x,Q^2,m^2).
    \end{equation}
    We define shorthands for the charm and bottom production cross sections:
    \begin{align}
        \label{eq:fwdop-heavyQQ-LO-Acc}
        \bA^{c \bar c}_{r}(x,Q^2,y) \coloneqq {} & \bA^{Q \bar Q}_{r} (x, Q^2, y, m_c^2),
        \\
        \label{eq:fwdop-heavyQQ-LO-Abb}
        \bA^{b \bar b}_{r}(x,Q^2,y) \coloneqq {} & \bA^{Q \bar Q}_{r} (x, Q^2, y, m_b^2),
    \end{align}
    which by construction reproduce their corresponding observables when operating on $\pdtbf$.

    \subsection{Neutrino deeply inelastic scattering} \label{sec:lo-neutrino-dis}

    Neutrino deeply inelastic scattering ($\nu$DIS) is in practical terms the reverse of the charged current lepton-proton DIS: instead of the charged lepton probe scattering off the proton, a high-energy neutrino scatters off the target via the weak nuclear interaction and producing an outgoing charged lepton detected in the experiment\footnote{Neutral current by $Z$ boson exchange is possible as well, but in that case the (anti-)neutrino does not turn into a charged lepton that can be detected in an experiment, and instead after scattering the neutrino escapes undetected. We limit the discussion only to charged current.}. $\nu$DIS complements the previously discussed charged lepton-proton observables in a critical way, since the electrically neutral neutrino and antineutrino can only interact via the weak force, which gives them a unique sensitivity to the flavor structure of the target. Specifically, the neutrino $\nu$ can only interact with the $d,s,\bar u$, and $\bar c$ quarks, while the antineutrino $\bar \nu$ interacts with $u,c,\bar d$, and $\bar s$ quarks~\cite{ParticleDataGroup:2024cfk,Sterman:1995:handbook-qcd,Peskin:0201503972,Griffiths:1987tj}.
    $\nu$DIS was first described by Bjorken and Paschos~\cite{Bjorken:1969in:neutrinoDIS}, and Llewellyn-Smith~\cite{LlewellynSmith:1971uhs:neutrinoDIS}.

    The high theoretical impact unfortunately comes with highly non-trivial experimental challenges, such as the production of a tightly collimated high-energy beam of neutrinos, which cannot be refocused after production, or the fact that they are so weakly interacting with anything, that a single neutrino needs a lead target light-years thick to have a $50\%$ chance of interacting~\cite{Griffiths:1987tj}. The latter challenge means in practice that neutrino scattering experiments need to have very large detectors, which have been made from materials such as lead~\cite{CHORUS:2005cpn}, iron~\cite{NuTeV:2005wsg}, and others such as argon~\cite{ArgoNeuT:2011bms,ArgoNeuT:2014rlj,MicroBooNE:2018xad,MicroBooNE:2019nio,MicroBooNE:2023foc}, carbon~\cite{T2K:2013nor:carbon,T2K:2018lnf:carbon}, and water~\cite{T2K:2019dgm:water,NINJA:2020gbg:water}, as reviewed in Ref.~\cite{ParticleDataGroup:2024cfk}.
    
    The use of dense material targets introduces the theoretical complication, that the neutrinos are not interacting with free protons anymore, and rather they scatter off protons and neutrons bound within nuclei. This compounds the problem in two ways: the structure of the proton is altered by being bound within the nucleus, and the partonic structure of the neutron is not the same as that of a proton. The latter issue is canonically amended thanks to the isospin symmetry between the proton and neutron, which allows one to relate the PDFs of the neutron directly those of the proton~\cite{Sterman:1995:handbook-qcd,ParticleDataGroup:2024cfk,Peskin:0201503972}, which we will do here as well. The former problem is more non-trivial, since the nuclear modifications that the PDFs receive in the nucleus---also known as the EMC effect---stem from fundamentally non-perturbative QCD effects, originally discovered by the European Muon Collaboration~\cite{EuropeanMuon:1983wih}, when they observed that an iron nucleus target in DIS cannot simply be considered to be composed of free protons and neutrons.

    In theory, the nuclear modifications of the PDFs are described by a flavor $a$ dependent and nucleus $A$ specific smooth functions $R_a^A$ defined as~\cite{Eskola:2016oht,Eskola:2021nhw}:
    \begin{equation}
        f_a^{p /A}(x, Q^2) \coloneqq R_a^{p /A}(x,Q^2) f_a(x, Q^2),
    \end{equation}
    where $f_a$ is the regular PDF of the free proton we have been discussing so far. The inference of the nuclear modification factors $R_a^{p /A}(x,Q^2)$ from data is an inverse problem in its own right and of comparable complexity to the inference of the free proton PDFs themselves~\cite{Eskola:2016oht,Eskola:2021nhw,Paakkinen:2025pcq}. 
    From the mathematical inverse problems perspective, the introduction of the multiplicative modifications $R_a^{p /A}(x,Q^2)$ turns the problem into a non-linear inverse problem, which goes beyond the scope of the current work. For the simplicity of the initial construction of this approach we will essentially assume, as stated in A\ref{post:nonpert}, that the nuclear modifications are known well enough from experiment, and trust that the neutrino DIS experiments have extracted the neutrino-proton cross sections and structure functions systematically correctly. This enables the inclusion of the neutrino DIS observables as linear problems into the global data inference problem, when the only unknown family of functions are the free proton PDFs.

    The core observable of the neutrino DIS experiments is the differential cross section $\sigma^{\nu, \bar \nu}(x,Q^2,y)$~\cite{ParticleDataGroup:2024cfk,Cooper-Sarkar:1997pqx,Formaggio:2012cpf:neutrino-dis-review,NuTeV:2005wsg,CHORUS:2005cpn}, which is for a proton target~\cite{ParticleDataGroup:2024cfk}:
    \begin{equation}
        \sigma^{\nu(\bar \nu)p}(x,Q^2,y) \pequiv
        \frac{4 \pi \aem^2}{x y Q^2} \frac{1}{2} \qty(\frac{G_F M_W^2}{4 \pi \aem} \frac{Q^2}{Q^2+M_W^2})^2
        \sigma_r^{\nu(\bar \nu)p}(x,Q^2,y),
    \end{equation}
    where $\aem$ is the \gls{fine structure constant}, $G_F$ is the Fermi constant~\cite{ParticleDataGroup:2024cfk}, $M_W$ is the mass of the $W^\pm$ boson, and where we have defined the analog of the HERA \gls{reduced cross section} for a presentation more similar to the other observables, which reads:
    \begin{align}
        \label{eq:lo-nudis-nup-crosssection}
        \sigma_r^{\nu(\bar \nu)p}(x,Q^2,y) \coloneqq
        Y_+ F^{\nu(\bar \nu) p}_2(x,Q^2) - y^2 F^{\nu(\bar \nu) p}_L(x,Q^2) \pm Y_-x F^{\nu(\bar \nu) p}_3(x,Q^2) 
        ,
    \end{align}
    where the $\pm$ is $+$ for $\nu$, and $-$ for $\bar \nu$, and $Y_\pm \coloneqq 1 \pm (1-y)^2$.
    We note here that for a proton target the neutrino DIS structure functions $F^{\nu(\bar \nu) p}_i$ are exactly the same\footnote{Aside from the varying conventions of including, or not including, a factor of $2$ in the definition of the structure functions. We follow the convention of Ref.~\cite{ParticleDataGroup:2024cfk}, where a factor of $2$ is kept in $F_i^{W^\pm}$.} as the charged current structure functions $F^{W^\pm}_i$ seen in Sec.~\ref{sec:lo-cc-dis}: the incoming lepton probe scatters via a $W^\pm$ boson exchange\footnote{The correspondence is $F^{\nu p} \equiv F^{W^+}$ and $F^{\bar \nu p} \equiv F^{W^-}$, since the neutral neutrino $\nu_e$ emits a $W^+$ and turns into an electron $e^-$ in the process, and electrical charge is conserved. Correspondingly for the anti-neutrino, which turns into a positron $e^+$.}, and the QCD picture of the scattering from the target does not care which lepton is incoming or outgoing. However, with real world experiments a practical difference arises, since neutrino DIS experiments use nuclear targets, which contain neutrons.
    The neutrino--neutron scattering cross section is sensitive to the \textit{neutron} structure functions $F_i^n$:
    \begin{align}
        \sigma_r^{\nu(\bar \nu)n}(x,Q^2,y) \pequiv
        Y_+ F^{\nu(\bar \nu) n}_2(x,Q^2) - y^2 F^{\nu(\bar \nu) n}_L(x,Q^2) \pm Y_-x F^{\nu(\bar \nu) n}_3(x,Q^2) 
        ,
    \end{align}
    which we will relate to those of the proton with the isospin symmetry $u \leftrightarrow d$ and $\bar u \leftrightarrow \bar d$~\cite{Cooper-Sarkar:1997pqx}. The structure functions are, purely in terms of the proton PDFs $f_a(x,Q^2)$:
    \begin{align}
        F^{\nu p}_2(x,Q^2) & = 2x [f_d + f_s + f_{\bar u} + f_{\bar c}],
        \\
        x F^{\nu p}_3(x,Q^2) & = 2 x [f_d + f_s - f_{\bar u} - f_{\bar c}],
    \end{align}
    \begin{align}
        F^{\nu n}_2(x,Q^2) & = 2x [f_u + f_s + f_{\bar d} + f_{\bar c}],
        \\
        x F^{\nu n}_3(x,Q^2) & = 2 x [f_u + f_s - f_{\bar d} - f_{\bar c}],
    \end{align}
    \begin{align}
        F^{\bar \nu p}_2(x,Q^2) & = 2x [f_u + f_c + f_{\bar d} + f_{\bar s}],
        \\
        x F^{\bar \nu p}_3(x,Q^2) & = 2 x [f_u + f_c - f_{\bar d} - f_{\bar s}],
    \end{align}
    \begin{align}
        F^{\bar \nu n}_2(x,Q^2) & = 2x [f_d + f_c + f_{\bar u} + f_{\bar s}],
        \\
        x F^{\bar \nu n}_3(x,Q^2) & = 2 x [f_d + f_c - f_{\bar u} - f_{\bar s}],
    \end{align}
    where we have again suppressed the arguments of the PDFs: $f_a \equiv f_a(x,Q^2)$. This assumes no top or bottom quarks in the target nucleon, which is reasonable at the $Q^2$ scales of the experiments, and that the collision energy is above the threshold for charm production~\cite{Cooper-Sarkar:1997pqx}.

    For an isoscalar target---atomic nucleus with equal number of protons and neutrons: $N_p = N_n$---one needs to average the proton and neutron contributions, since the experimental probe is as likely to hit either a proton or a neutron:
    \begin{align}
        \label{eq:nudis-f2N}
        F_2^{\nu N} &\coloneqq \frac{F_2^{\nu p}+F_2^{\nu n}}{2} 
        = x \big[ f_u + f_d + 2 f_s + f_{\bar u} + f_{\bar d} + 2 f_{\bar c} \big],
        \\
        \label{eq:nudis-f3N}
        x F_3^{\nu N} &\coloneqq \frac{x F_3^{\nu p}+x F_3^{\nu n}}{2} 
        = x \big[ f_u + f_d + 2 f_s - f_{\bar u} - f_{\bar d} - 2 f_{\bar c}\big],
        \\
        \label{eq:barnudis-f2N}
        F_2^{\bar \nu N} &\coloneqq \frac{F_2^{\bar \nu p}+F_2^{\bar \nu n}}{2} 
        = x \big[ f_u + f_d + 2 f_c + f_{\bar u} + f_{\bar d} + 2 f_{\bar s} \big],
        \\
        \label{eq:barnudis-f3N}
        x F_3^{\bar \nu N} &\coloneqq \frac{x F_3^{\bar \nu p}+x F_3^{\bar \nu n}}{2} 
        = x \big[ f_u + f_d + 2 f_c - f_{\bar u} - f_{\bar d} - 2 f_{\bar s}\big],
    \end{align}
    where again the arguments are suppressed for clarity. This averaging is also done at the cross section level, which gives for the isoscalar neutrino--nucleon DIS the cross section: 
    \begin{equation}
        \sigma^{\nu(\bar \nu)N}(x,Q^2,y)
        \coloneqq \frac{\sigma^{\nu(\bar \nu)p}(x,Q^2,y) + \sigma^{\nu(\bar \nu)n}(x,Q^2,y)}{2}.
        \label{eq:lo-nudis-isoscalar-dis-crosssection}
    \end{equation}

    We write first the forward operators for the nucleon structure functions in neutrino DIS, as the experiments discussed in Sec.~\ref{sec:data} have also extracted $F_2$, $x F_3$ as observables, though with some model dependency. As forward operators they are: 
    \begin{align}
        \label{eq:fwdop-nudis-f2-nu}
        \bB^{\nu N}_{2,\lo}(x) & = x 
        \begin{bmatrix}
            0 &
            \Id & \Id & 2 \Id & 0 & 0 & 0 &
            \Id & \Id & 0 & 2 \Id & 0 & 0
        \end{bmatrix},
        \\
        \label{eq:fwdop-nudis-f3-nu}
        \bB^{\nu N}_{3,\lo}(x) & = x 
        \begin{bmatrix}
            0 &
            \Id & \Id & 2\Id & 0 & 0 & 0 &
            - \Id & - \Id & 0 & -2\Id & 0 & 0
        \end{bmatrix},
        \\
        \label{eq:fwdop-nudis-f2-barnu}
        \bB^{\bar \nu N}_{2,\lo}(x) & = x 
        \begin{bmatrix}
            0 &
            \Id & \Id & 0 & 2 \Id & 0 & 0 &
            \Id & \Id & 2 \Id & 0 & 0 & 0
        \end{bmatrix},
        \\
        \label{eq:fwdop-nudis-f3-barnu}
        \bB^{\bar \nu N}_{3,\lo}(x) & = x 
        \begin{bmatrix}
            0 &
            \Id & \Id & 0 & 2 \Id  & 0 & 0 &
            - \Id & - \Id & -2\Id & 0 & 0 & 0
        \end{bmatrix},
    \end{align}
    which reproduce the definitions \eqref{eq:nudis-f2N}, \eqref{eq:nudis-f3N} when operating on the PDFs $\pdtbf$.
    With these the forward operator for the isoscalar neutrino--nucleon cross section can be defined:
    \begin{align}
        \label{eq:fwdop-nudis-nuN-cs-isoscalar}
        \mathbf{A}_r^{\nu N} (x,Q^2,y) 
        & \coloneqq
        \left[ Y_+ \bB^{\nu N}_2(x) 
        +
        Y_-x \bB^{\nu N}_3(x) \right]
        \\
        & =
        \begin{bmatrix}
            0 &
            \Id & \Id & 2\Id & 0 & 0 & 0 &
            (1-y)^2 \Id & (1-y)^2 \Id & 0 & 2(1-y)^2  \Id & 0 & 0
        \end{bmatrix},
        \notag
    \end{align}
    and similarly for the antineutrino--nucleon cross section:
    \begin{align}
        \label{eq:fwdop-nudis-antinuN-cs-isoscalar}
        \mathbf{A}_r^{\bar \nu N} (x,Q^2,y) 
        & \coloneqq
        \left[ Y_+ \bB^{\bar \nu N}_2(x) 
        -
        Y_-x \bB^{\bar \nu N}_3(x) \right]
        \\
        & =
        \begin{bmatrix}
            0 &
            (1-y)^2 \Id & (1-y)^2 \Id & 0 & 2(1-y)^2 \Id & 0 & 0 &
            \Id & \Id & 2\Id & 0 & 0 & 0
        \end{bmatrix},
        \notag
    \end{align}
    where for both cross section forward operators we set $F_L \equiv 0$ at this order in $\as$, in accordance with the literature~\cite{ParticleDataGroup:2024cfk,Cooper-Sarkar:1997pqx}. These forward operators enable us to include the original neutrino and antineutrino DIS scattering cross section data as a part of the formulation of the global data inference problem, as well as---at least in principle---the inclusion of the neutrino DIS extraction of the structure functions $F_2$ and $xF_3$ as a form of data to constrain the reconstructed PDFs. However, since the extraction of these structure functions requires modeling assumptions of the PDFs, they are not as clean as observables, as the original measured scattering cross sections.

\section{Parton distribution function reconstruction inverse problem at LO} \label{sec:tensor-ip-lo}

    We are now ready to collect the re-written leading order accuracy observables and forward operators constructed in Sec.~\ref{sec:forward-problem-lo}. We write the full leading order accuracy linear continuous tensor reconstruction problem for the proton parton distribution functions roughly in the form of the prototype inverse problem of Eq.~\eqref{eq:main-ip}:
    \begin{equation}
    \label{eq:tensor-ip-lo}
        \mathbf{O}^\mathrm{WD} \coloneqq
        \begin{bmatrix}
            \mathbf{O}^\sigma \\
            \mathbf{O}^F \\
            \mathbf{O}^S
        \end{bmatrix} = \mathbf{G}^{lp\to l' X}_\mathrm{LO} \pdtbf,
    \end{equation}
    where the unified forward global data operator $\mathbf{G}$ is constructed by stacking the individual forward operators, and datasets of each kind have been kept in their own block sections $\mathbf{O}^i$, within which each dataset is a sorted list of data points, which we will construct explicitly in Sec.~\ref{sec:discretization}. 
    The cross section observable block is constructed from the forward operators defined in Eqs.~\eqref{eq:fwdop-incldis-lo}, \eqref{eq:fwdop-nc-crosssec-lo}, \eqref{eq:fwdop-cc-lo}, \eqref{eq:fwdop-cc-lo2-eminus}, \eqref{eq:fwdop-heavyQQ-LO-Acc}, \eqref{eq:fwdop-heavyQQ-LO-Abb}, \eqref{eq:fwdop-nudis-nuN-cs-isoscalar}, \eqref{eq:fwdop-nudis-antinuN-cs-isoscalar}:
    \begin{equation}
        \label{eq:tensor-ip-observable-block-cs}
        \mathbf{O}^\sigma \coloneqq
        \begin{bmatrix}
            \sigma^\gamma_r(x,Q^2,y) \\
            \sigma^{+}_{r,\mathrm{NC}}(x,Q^2,y) \\
            \sigma^{-}_{r,\mathrm{NC}}(x,Q^2,y) \\
            \sigma^{+}_{r,\mathrm{CC}}(x,Q^2,y) \\
            \sigma^{-}_{r,\mathrm{CC}}(x,Q^2,y)\\
            \sigma^{c\bar c}_{r}(x,Q^2,y)\\
            \sigma^{b\bar b}_{r}(x,Q^2,y)
            \\ \sigma_r^{\nu N}(x,Q^2,y)
            \\ \sigma_r^{\bar \nu N}(x,Q^2,y)
        \end{bmatrix}
        =
        \begin{bmatrix}
            \bA^\gamma_r (x,Q^2,y) \\
            \bA^{+}_{r,\mathrm{NC}}(x,Q^2,y) \\
            \bA^{-}_{r,\mathrm{NC}}(x,Q^2,y) \\
            \bA^{+}_{r,\mathrm{CC}} (x,Q^2,y) \\
            \bA^{-}_{r,\mathrm{CC}} (x,Q^2,y)
            \\ \bA^{c \bar c}_{r}(x,Q^2,y)
            \\ \bA^{b \bar b}_{r}(x,Q^2,y)
            \\ \bA^{\nu N}_r (x,Q^2,y)
            \\ \bA^{\bar \nu N}_r (x,Q^2,y)
        \end{bmatrix}
        \pdtbf,
    \end{equation}
    where the purely electromagnetic inclusive process of Eq.~\eqref{eq:fwdop-incldis-lo} is included only for illustrative purposes, and is superseded by the more complete picture of neutral current DIS. We also note that all above cross section observables are implicitly also functions of $s$ via the variable $y = \frac{Q^2}{xs}$.
    The structure function part is composed from the forward operators in Eqs.~\eqref{eq:fwdop-incldis-fl-lo}, \eqref{eq:fwdop-nc-Z-LO-f3}, \eqref{eq:fwdop-nudis-f2-nu}, \eqref{eq:fwdop-nudis-f3-nu}, \eqref{eq:fwdop-nudis-f2-barnu}, \eqref{eq:fwdop-nudis-f3-barnu}: 
    \begin{equation}
        \label{eq:tensor-ip-observable-block-F}
        \mathbf{O}^F \coloneqq
        \begin{bmatrix}
            F^{\gamma,\mathrm{AM}}_L (x,Q^2) \\
            xF_3^\mathrm{NC} (x,Q^2) \\
            F_2^{\nu N} (x,Q^2) \\
            xF_3^{\nu N} (x,Q^2) \\
            F_2^{\bar \nu N} (x,Q^2) \\
            xF_3^{\bar \nu N} (x,Q^2)
        \end{bmatrix}
        =
        \begin{bmatrix}
            \bB^{\mathrm{AM}}_{L} (x, Q^2) \\
            \bB_3^\mathrm{NC} (x,Q^2)\\
            \bB^{\nu N}_{2} (x)\\
            \bB^{\nu N}_{3}(x)\\
            \bB^{\bar \nu N}_{2} (x)\\
            \bB^{\bar \nu N}_{3}(x)
        \end{bmatrix}
        \pdtbf,
    \end{equation}
        and the sum rule part is from Eqs.~\eqref{eq:fwdop-sumrule-fnc}, \eqref{eq:fwdop-sumrule-tmc}, \eqref{eq:fwdop-sumrule-GLS}, \eqref{eq:fwdop-sumrule-adler}, \eqref{eq:fwdop-sumrule-gottfried}:
    \begin{equation}
        \label{eq:ip-sumrule-part}
        \mathbf{O}^S \coloneqq
        \begin{bmatrix}
            S_\mathrm{f.n.c.}(Q^2)
            \\ S_\mathrm{t.m.c.}(Q^2)
            \\ S_\mathrm{GLS}(Q^2)
            \\ S_\mathrm{A}(Q^2)
            \\ S_\mathrm{G}(Q^2)
        \end{bmatrix}
        =
        \begin{bmatrix}
            \Ical_\mathrm{f.n.c.}
            \\ \Ical_\mathrm{t.m.c.}
            \\ \Ical_\mathrm{GLS}
            \\ \Ical_\mathrm{A}
            \\ \Ical_\mathrm{G}
        \end{bmatrix}
        \pdtbf.
    \end{equation}
    Equation \eqref{eq:tensor-ip-lo}, with the observable blocks defined by Eqs.~\eqref{eq:tensor-ip-observable-block-cs}, \eqref{eq:tensor-ip-observable-block-F}, and \eqref{eq:ip-sumrule-part}, now defines the tensor reconstruction inverse problem for the proton PDFs at leading order accuracy, where each measurement specific block-element problem is linear, analogously to the dipole amplitude reconstruction inverse problem~\cite{Hanninen:2025iuv}. In total, there are $25$ rows of observable constraints, and $13$ columns of parton integral operators in each row, which as a whole form a coupled system of linear and integral equations, roughly in analogy with Eq.~\eqref{eq:main-ip}, with the main difference at LO being that all the element operators are not yet proper integral operators and only functional identity maps. When we move to NLO accuracy in the next section, all the theory prescriptions of the observables become integral equations.

    The global data forward operator $\mathbf{G}^{lp\to l' X}_\mathrm{LO}$ as defined by Eq.~\eqref{eq:tensor-ip-lo} can be written in the general form of the matrix valued integral operator written in \eqref{eq:main-ip}, which we have written out explicitly in appendix \ref{app:full-global-data-operators} due to its extensive size. By construction, operating with $\mathbf{G}^{lp\to l' X}_\mathrm{LO}$ on the \gls{parton distribution tensor} $\pdtbf$ reproduces the standard expressions of the included cross section, structure function, and sum rule observables. Equation \eqref{eq:tensor-ip-lo} is therefore written into the form of an explicit inverse problem for which one can with mathematical rigor make statements about the existence and uniqueness of solutions, and to consider solving the unknown functional information encoded by $\pdtbf$ from the world data without fitting model parameters. We note that Eq.~\eqref{eq:tensor-ip-lo} is implicitly only written for the case $n_f=6$, and therefore does not yet implement a proper \gls{flavor number scheme} as discussed in Sec.~\ref{sec:obs-as-linear-ops}; we will consider the full picture with a FNS in Sec.~\ref{sec:tensor-ip-nlo}. We will discuss the construction of a solution method in Sec.~\ref{sec:discretization}.

    A critical detail that we have not yet addressed is the role of the Dokshitzer--Gribov--Lipatov--Altarelli--Parisi (DGLAP) evolution equations~\cite{Dokshitzer:1977sg,Lipatov:1974qm,Gribov:1972ri,Altarelli:1977zs}. The DGLAP equations are a coupled system of integro-partial differential equations for the PDFs, which give a precise QCD theory prescription for the momentum scale $Q^2$ dependence of the PDFs~\cite{Martin:2008cn:dglap}.
    The DGLAP equation enables one to consider the initial condition of the evolution of the PDFs to be the non-perturbative functional input at some scale $Q^2_0$ in fits and comparisons with data, and the DGLAP equation predicts the behavior of the PDFs at all $Q^2 > Q^2_0$. This means that the global analyses of the PDFs have fit parameters describing the PDFs at the initial scale to the world data.
    
    In this inverse problems approach, the role of the DGLAP equation is shifted in the sense that the PDFs are solved from the world data defined by the problem in Eq.~\eqref{eq:tensor-ip-lo}, and we must then take care that the reconstructed solutions satisfy, at least within the accuracy of the perturbative order, the DGLAP equations. This is done with a PDE constraint~\cite{Biegler-pde-opti,leugering2014trends:pde-opti} in the reconstruction algorithm that numerically solves Eq.~\eqref{eq:tensor-ip-lo}.
    A general way to express a functional optimization problem with a PDE constraint is, in the optimization problem formalism employed in Ref.~\cite{Hanninen:2025iuv}:
    \begin{equation}
        \min_{\pdtbf, \mathbf{y}} \left|\left| \mathbf{O}^\mathrm{WD} - \mathbf{G}^{lp\to l' X}_\mathrm{LO} \pdtbf \right|\right|^2_2 + \lambda \left|\left| \mathbf{y} \right|\right|^2_2,
    \end{equation}
    where the first term is the ``data fitting'' term, and the latter is the PDE constraint term with a control variable $\mathbf{y}$, which quantifies the accuracy at which the reconstructed $\pdtbf$ satisfies the DGLAP equation, and $\lambda$ is a regularization parameter which acts as a relative weight between the minimization terms. This means that the above optimization problem is understood to be solved under the dynamical constraint:
    \begin{equation}
        \mathbf{y} \coloneqq \frac{\partial}{\partial \log Q^2} \pdtbf - \mathbf{P}_\mathrm{DGLAP} \otimes \pdtbf,
        \label{eq:pde-constraint-dglap}
    \end{equation}
    where the kernel $\mathbf{P}_\mathrm{DGLAP}$ is implied by the DGLAP equations for the gluon $f_g$ and quarks $f_q$~\cite{Martin:2008cn:dglap}. This constraint encodes the theoretical ideal that for PDFs which satisfy the DGLAP equations exactly at all $Q^2$, one has $\mathbf{y} \equiv 0$, which is by construction enforced in conventional global analysis, since the numerical solution of the DGLAP equations is built into the framework.
    In this approach this expectation is challenged in the sense that the world data is not available at all $Q^2$, and there is experimental uncertainty, so one cannot expect the reconstruction from real experimental data to perfectly satisfy the DGLAP evolution. Another practical reason is that the full DGLAP equation is derived within perturbative QCD, which means that in practice the evolution equation that can be implemented is an approximation of the all-order evolution equation implied by QCD. So even if we expect that the experiment would precisely agree with the complete all-orders form of the DGLAP evolution, that is not available to implement in practice. Ultimately this means, that we in general expect the reconstruction to approximately satisfy the DGLAP equation at all $Q^2$, which would be an opportunity to extract information about the $Q^2$-evolution of the PDFs from the world data, without explicitly requiring the agreement with the DGLAP equation. This perspective is formulated in the same vein as with the small-$x$ evolution of the dipole amplitude in Ref.~\cite{Hanninen:2025iuv}.

    For completeness, let us write the DGLAP equations for the \gls{parton distribution tensor} $\pdtbf$ defined in Sec.~\ref{sec:linear-intop-formalism}. It is most convenient to begin with the general form of the DGLAP equation~\cite{ParticleDataGroup:2024cfk,Moch:2004pa:dglap-general-nnnlo,Vogt:2004mw:dglap-general-nnnlo}:
    \begin{equation}
        \label{eq:dglap}
        \frac{\partial}{\partial \log Q^2} f_i(x,Q^2) = \frac{\as(Q^2)}{2\pi} \sum_j \int_x^1 P_{ij}(z,Q) f_j\left(\frac{x}{z}, Q^2 \right) \frac{\ud z}{z},
    \end{equation}
    where $i,j$ are the flavor indices of the partons, and the kernels $P_{ij}$ are at LO~\cite{Martin:2008cn:dglap}:
    \begin{align}
        P_{qq}(z) & \coloneqq \frac{4}{3} \frac{1+z^2}{(1-z)_+} + 2 \delta(1-z),
        \\
        P_{qg}(z) & \coloneqq \frac{1}{2} \left( z^2 + (1-z)^2 \right),
        \\
        P_{gq}(z) & \coloneqq \frac{4}{3} \frac{1+(1-z)^2}{z},
        \\
        P_{gg}(z) & \coloneqq 6 \left( \frac{1-z}{z} + \frac{z}{(1-z)_+}  + z(1-z) \right) + \left( \frac{11}{2} - \frac{n_f}{3} \right) \delta(1-z).
    \end{align}
    To write these in the integral operator formalism, we define the notation:
    \begin{equation}
        \mathcal{P}_{ij}[f] \coloneqq \frac{\as(Q^2)}{2\pi} \int_x^1 P_{ij}\qty(\frac{x}{z},Q) f(z, Q^2 ) \frac{\ud z}{z}
    \end{equation}
     also used in Ref.~\cite{Martin:2008cn:dglap} in a slightly different form.
     With this we can recast Eq.~\eqref{eq:dglap} in the form of a coupled system of integro-differential equations:
     \begin{equation}
         \frac{\partial}{\partial \log Q^2} \pdtbf(x,Q^2)
         =
         \frac{\partial}{\partial \log Q^2} \pdtwritout
         =
         \begin{bmatrix}
             \Pcal_{gg} & \Pcal_{gq} & \cdots & \Pcal_{gq} \\
             \Pcal_{qg} & \Pcal_{qq} & \cdots & \Pcal_{qq} \\
            \vdots & \vdots & \ddots & \vdots \\
             \Pcal_{qg} & \Pcal_{qq} & \cdots & \Pcal_{qq} \\
         \end{bmatrix}
         \pdtwritout
         \eqqcolon
         \mathbf{P}_\mathrm{DGLAP} \pdtbf,
     \end{equation}
    which is of the form used in Eq.~\eqref{eq:pde-constraint-dglap}. However, it is known that the DGLAP equation has a significantly simpler structure in a different basis of representation of the PDFs than the flavor basis we have considered so far. Instead of ``bare flavors'' the evolution equations structurally prefer a basis where the PDFs are represented in terms of the gluon and so-called singlet and non-singlet states~\cite{Moch:2004pa:dglap-general-nnnlo,Vogt:2004mw:dglap-general-nnnlo}, where the equation~\eqref{eq:dglap} has the block-diagonal representation:
    \begin{equation}
        \mathbf{P}_\mathrm{DGLAP}^{\{g,s,ns\}} = 
        \begin{bmatrix}
            \mathcal{P}_{gg} & \mathcal{P}_{gq} & 0 & 0 & 0 & 0 & 0 & 0 & 0 & 0 & 0 & 0 & 0 \\
            \mathcal{P}_{qg} & \mathcal{P}_{qq} & 0 & 0 & 0 & 0 & 0 & 0 & 0 & 0 & 0 & 0 & 0 \\
            0 & 0 & \mathcal{P}_{\mathrm{ns}}^\mathrm{v} & 0 & 0 & 0 & 0 & 0 & 0 & 0 & 0 & 0 & 0 \\
            0 & 0 & 0 & \mathcal{P}_{\mathrm{ns}}^+ & 0 & 0 & 0 & 0 & 0 & 0 & 0 & 0 & 0 \\
            0 & 0 & 0 & 0 & \mathcal{P}_{\mathrm{ns}}^+ & 0 & 0 & 0 & 0 & 0 & 0 & 0 & 0 \\
            0 & 0 & 0 & 0 & 0 & \mathcal{P}_{\mathrm{ns}}^+ & 0 & 0 & 0 & 0 & 0 & 0 & 0 \\
            0 & 0 & 0 & 0 & 0 & 0 & \mathcal{P}_{\mathrm{ns}}^+ & 0 & 0 & 0 & 0 & 0 & 0 \\
            0 & 0 & 0 & 0 & 0 & 0 & 0 & \mathcal{P}_{\mathrm{ns}}^+ & 0 & 0 & 0 & 0 & 0 \\
            0 & 0 & 0 & 0 & 0 & 0 & 0 & 0 & \mathcal{P}_{\mathrm{ns}}^- & 0 & 0 & 0 & 0 \\
            0 & 0 & 0 & 0 & 0 & 0 & 0 & 0 & 0 & \mathcal{P}_{\mathrm{ns}}^- & 0 & 0 & 0 \\
            0 & 0 & 0 & 0 & 0 & 0 & 0 & 0 & 0 & 0 & \mathcal{P}_{\mathrm{ns}}^- & 0 & 0 \\
            0 & 0 & 0 & 0 & 0 & 0 & 0 & 0 & 0 & 0 & 0 & \mathcal{P}_{\mathrm{ns}}^- & 0 \\
            0 & 0 & 0 & 0 & 0 & 0 & 0 & 0 & 0 & 0 & 0 & 0 & \mathcal{P}_{\mathrm{ns}}^-
        \end{bmatrix} \, ,
        \label{eq:P-operator-dglap-singlet-basis}
    \end{equation}
    where we have included all 6 flavors and the gluon, and
    for which the PDF basis is defined in terms of the flavor PDFs by
    \begin{align}
        f_g &= f_g, 
        \\
        \label{eq:pdf-basis:singlet}
        q_s &= \sum_i^{n_f} f_{q_i} + f_{ \bar q_i},
        \\
        \label{eq:pdf-basis:non-singlet-v}
        q_\mathrm{ns}^\mathrm{v} &= \sum_i^{n_f} f_{q_i} - f_{ \bar q_i}
        \\
        \label{eq:pdf-basis:non-singlet-pm}
        q_{\mathrm{ns},ij}^\pm &= f_{q_i} \pm f_{ \bar q_i} - \qty(f_{q_j} \pm f_{ \bar q_j}),
    \end{align}
    which give, for example, the basis
    $\{ f_g, \allowbreak q_s, \allowbreak q_\mathrm{ns}^\mathrm{v}, \allowbreak q^+_{\mathrm{ns},du}, \allowbreak q^+_{\mathrm{ns},ds}, \allowbreak q^+_{\mathrm{ns},dc}, \allowbreak q^+_{\mathrm{ns},db}, \allowbreak q^+_{\mathrm{ns},dt}, \allowbreak q^-_{\mathrm{ns},du}, \allowbreak q^-_{\mathrm{ns},ds}, \allowbreak q^-_{\mathrm{ns},dc}, \allowbreak q^-_{\mathrm{ns},db}, \allowbreak q^-_{\mathrm{ns},dt} \}$, which are needed to span the flavor space of thirteen partons. The simplification of the matrix operator of the DGLAP equation in the singlet-basis in contrast with the regular flavor basis makes it evident that the use of a carefully selected basis is essential in terms of computational implementation~\cite{Candido:2022tld:EKO}. It has been determined that Eq.~\eqref{eq:P-operator-dglap-singlet-basis} is in fact an all-orders accurate formulation of the DGLAP equation~\cite{Moch:2004pa:dglap-general-nnnlo,Vogt:2004mw:dglap-general-nnnlo}, and starting at order $\as^2$ all non-trivial contributions to the coefficients are contributing, i.e. the elements $\Pcal_i$ shown above become distinct. The perturbative corrections only affect the given elements which are of the general form~\cite{Moch:2004pa:dglap-general-nnnlo,Vogt:2004mw:dglap-general-nnnlo}
    \begin{equation}
        \mathcal{P}_i = \sum_{n=0} \qty(\frac{\as(Q^2)}{4\pi})^{n+1} P_i^{(n)},
    \end{equation}
    whereas in the flavor basis all the elements of the matrix would receive corrections at every order in $\as^n$. Explicit expressions for the coefficients in~\eqref{eq:P-operator-dglap-singlet-basis} can be found in Refs.~\cite{Moch:2004pa:dglap-general-nnnlo,Vogt:2004mw:dglap-general-nnnlo}.

\section{Linear forward problems of DIS at NLO in pQCD} \label{sec:forward-problem-nlo}

    In this section we consider how the inverse problem changes once we include \gls{next-to-leading order} (NLO) accuracy corrections, i.e. the first sub-leading contributions in the $\as$ expansion, to the observables. There are a number of significant reasons to move to NLO accuracy in pQCD, which in some sense make the leading order problem a practice version of the proper mathematical problem that only manifests at NLO. From the physical perspective the higher order corrections are critical, since they are what produce the parton sea in the parton model, which means that to quantitatively infer the distributions of the sea quarks, antiquarks, and gluons, one must develop the QCD theory picture at least at NLO accuracy~\cite{ParticleDataGroup:2024cfk,Peskin:0201503972,Sterman:1995:handbook-qcd}. Second important reason is that the NLO and higher order corrections are not negligibly small, which is perhaps reflected in the fact that the parton sea only becomes ``detectable'' at NLO.
    The QCD theory calculations also begin to need the additional step of renormalization, a step in the quantum field theory that we just take as a technical detail that must be done in a consistent scheme (P\ref{post:pdf:smooth}). Specifically, we will only work within the $\overline{\mathrm{MS}}$ scheme~\cite{ParticleDataGroup:2024cfk}, which is handled by collecting the results from the literature that have been derived in the $\overline{\mathrm{MS}}$ scheme, which is the most commonly used renormalization scheme for perturbative QCD~\cite{ParticleDataGroup:2024cfk}.
    Lastly, a mathematical reason essential for the construction of this inverse problem approach is that the first corrections beyond LO qualitatively change the form of the linear equations that we must solve: the identity map contributions receive perturbative corrections, which turn the problems into linear integral equations of the first or second kind, although luckily this first step in the perturbative order seems to be the only instance where this structural change happens. This will be relevant when we construct the method to solve the full problem in Sec.~\ref{sec:discretization}, which we will do with the higher order pQCD corrections in mind.

    In this section we will limit the discussion to key changes that stem from the move to NLO accuracy with very limited discussion of the physics, which was introduced in more detail in Sec.~\ref{sec:forward-problem-lo}. Our main focus is to collect from the literature the expressions for the \glspl{observable} needed to write the \gls{world data} inference problem at NLO accuracy, and rewrite them in the integral operator formalism.
    
    We will be working with coefficient function $C$ convolutions is defined as
    \begin{equation}
        (C \otimes f)(x) \coloneqq \int_x^1 C(y) f\left( \frac{x}{y} \right) \frac{\ud y}{y}.
    \end{equation}
    At points it will be convenient to change the integration variable to $z\coloneqq \frac{x}{y}$ and work with the convolutions as:
    \begin{equation}
    \label{eq:convolution-C-cov-z}
        (C \otimes f)(x) = \int_x^1 C\qty(\frac{x}{z}) f(z) \frac{\ud z}{z}
    \end{equation}
    as this will facilitate the explicit factorization of the PDFs from the forward operators. We will consider these convolution integral equations as integral linear operators under the notation:
    \begin{equation}
        \label{eq:linear-integral-operator}
        \Ccal[f](x) \coloneqq (C \otimes f)(x),
    \end{equation}
    with the implication that each Wilson coefficient function $C(z)$ implicitly defines such an integral operator, which will be denoted by $\Ccal$.

    \subsection{Inclusive electromagnetic massless DIS at NLO accuracy} \label{sec:nlo-inclusive-gamma-dis}

    The NLO structure functions for inclusive DIS were originally calculated by Altarelli, Ellis, and Martinelli~\cite{Altarelli:1978id}, reviewed in~\cite{Furmanski:1981cw}, and systematically re-derived as a part of higher order calculations in~\cite{Zijlstra:1992qd,Moch:1999eb}, and they are in the $\overline{\mathrm{MS}}$ scheme:
    \begin{align}
        F_2^\gamma(x,Q^2) \pequiv{} &
        \sum_a e_a^2 \int_x^1 x f_{a}\qty(\frac{x}{z}, Q^2) \left( \delta(1-z) + \frac{\alpha_s}{4\pi}c^{(1)}_{2,q}(z)\right) \frac{\ud z}{z}
        \notag \\ &
        + \sum_a e_a^2 
        \int_x^1 x f_{g}\qty(\frac{x}{z}, Q^2) \frac{\alpha_s}{4\pi} c^{(1)}_{2,g}(z) \frac{\ud z}{z}
        \label{eq:conventional-nlo-f2}
        \\
        F_L^\gamma(x, Q^2) \pequiv{} &
        \frac{\alpha_s}{4\pi} \sum_a e_a^2 \int_x^1 x f_{a}\qty(\frac{x}{z}, Q^2) c^{(1)}_{L,q}(z) \frac{\ud z}{z}
        \notag \\ &
        + \frac{\alpha_s}{4\pi} \sum_a e_a^2 \int_x^1  x f_{g} \qty(\frac{x}{z}, Q^2) c^{(1)}_{L,g}(z) \frac{\ud z}{z},
        \label{eq:conventional-nlo-fl}
    \end{align}
    where $e_a$ is the fractional charge of parton $a$, and the coefficient functions are~\cite{Moch:1999eb}
    \begin{align}
        c_{2,q}^{(1)}(z) & = \cf \bigg\lbrace \frac{9}{2} + \frac{5}{2}z - 2 \qty(\frac{3}{4} + \ln(z) - \ln(1-z) ) \qty(\frac{2}{1-z} - 1 - z)
        \notag \\
        & \quad \qquad - \qty(9 + 4 \zeta(2)) \delta(1-z) \bigg\rbrace_+,
        \\
        c_{2,g}^{(1)}(z) & = 6 - 2 \qty(4 + \ln(z) - \ln(1-z)_+) \qty(1 -2z + 2z^2 ), 
        \\
        c_{L,q}^{(1)}(z) & = 4 \cf z,
        \\
        c_{L,g}^{(1)}(z) & = 8 z(1-z),
    \end{align}
    where the color factor $\cf \equiv \frac{N_c^2-1}{N_c} = \frac{4}{3}$ is a constant of QCD~\cite{ParticleDataGroup:2024cfk,Peskin:0201503972}, and where $N_c$ is the number of color charges in the theory, i.e. three in standard QCD. The superscript ${}^{(1)}$ refers to the order in the perturbative $\as^n$ expansion, i.e. $n=1$ at NLO, and $\zeta(s)$ is the Riemann zeta function~\cite{Riemann1859}. We have intentionally left the factors of $n_f$ out compared to Refs.~\cite{Moch:1999eb,Vermaseren:2005qc} as we have not written the sum over the quark charges in terms of their average charge $\langle e^2 \rangle$, which is a function of the number of included flavors, and instead have the sum written in for now.
    The plus-distributions---objects denoted with $(q(x))_+$---that appear in the coefficient functions $c_{q,2}$ are defined with:
    \begin{equation}
        \int_0^1 a(z)_+ f(z) \ud z \coloneqq \int_0^1 a(z) \left( f(z) - f(1) \right) \ud z.
    \end{equation}
    However, one must be diligent in their application as they appear in the observable expressions in the form of the convolution \eqref{eq:convolution-C-def}, which when written out becomes:
    \begin{align}
        (c_+ \otimes f)(x) \coloneqq {} & \int_x^1 c_+(z) f\qty(\frac{x}{z}) \frac{\ud z}{z}
        \\ ={} &
        \int_x^1 C\qty(\frac{x}{z}) \left( f(z) - \frac{x}{z} f(x) \right) \frac{\ud z}{z}
        - f(x) \int_0^x C(z) \ud z 
        \\
        \eqqcolon &
        \int_x^1 C\qty(\frac{x}{z}) \int_x^1 \left( \delta(z'-z) - \frac{x}{z} \delta(z'-x) \right) f(z') \ud z' \frac{\ud z}{z} - f(x) c(x),
    \end{align}
    where we defined $c(x) \coloneqq \int_0^x C(z) \ud z$, and introduced a second integral with delta functions to write the integral regulated by the plus-distribution back into a integral linear operator form. The introduction of the second integration also makes it explicit in the method section (Sec.~\ref{sec:discretization}) that this type of an integral operator will require an additional step in the construction.


    The observable reduced cross section of electromagnetic DIS at NLO accuracy in the $\overline{\mathrm{MS}}$ scheme is exactly the same quantity as given in Eq.~\eqref{eq:cs:dis-inclusive}, but what changes is the theory prescription of the structure functions $F_i^\gamma$. Writing it out in terms of the NLO accuracy $F_2^\gamma$ and $F_L^\gamma$ from above, it is:
    \begin{align}
        \sigma^\gamma_r(x, Q^2,y)
        ={} & 
        \sum_a e_a^2 xf_{a}(x,Q^2) 
        \notag \\*
        &+ \sum_a e_a^2 \int_x^1 x f_{a} \qty(\frac{x}{z}, Q^2)
        \frac{\alpha_s}{4\pi} \Bigg( c_{2,q}^{(1)}(z)_+ 
        - \frac{y^2}{Y_+} c_{L,q}^{(1)}(z)\Bigg) \frac{\ud z}{z}
        \notag \\
        & + \sum_a e_a^2 \int_x^1 x f_{g} \qty(\frac{x}{z}, Q^2)
        \frac{\alpha_s}{4\pi} \Bigg( c_{2,g}^{(1)}(z)_+ 
        -\frac{y^2}{Y_+} c_{L,g}^{(1)}(z) \Bigg) \frac{\ud z}{z}
        ,
        \label{eq:sigma-incl-nlo}
    \end{align}
    where $Y_+ = (1-y)^2+1$ in the first two terms the sum over parton flavors $a$ includes the sum over $f_a$, whereas in the last term the sum only concerns the squared fractional charges $e_a^2$.

    We will now write the cross section and structure functions in the linear operator formalism with an operator of type $\lambda_1 \mathrm{Id} + \lambda_2 \mathcal{I}$, where the second term is new compared to the LO case. First, let us define for convenience, using the change of variable of Eq.~\eqref{eq:convolution-C-cov-z}:
    \begin{align}
        \mathcal{I}_{2-L,q}[f](x, Q^2, y) & \coloneqq x \Id \circ f + \int_x^1 x f \qty(z, Q^2)
        \frac{\alpha_s}{4\pi} \Bigg( c_{2,q}^{(1)}\qty(\frac{x}{z})_+ 
        - \frac{y^2}{Y_+} c_{L,q}^{(1)}\qty(\frac{x}{z}) \Bigg) \frac{\ud z}{z}
        \label{eq:nlo-incl-integralops1}
        \\
        \mathcal{I}_{2-L,g}[f](x, Q^2, y) & \coloneqq \int_x^1 x f \qty(z, Q^2)
        \frac{\alpha_s}{4\pi} \Bigg( c_{2,g}^{(1)}\qty(\frac{x}{z})_+ 
        -\frac{y^2}{Y_+} c_{L,g}^{(1)}\qty(\frac{x}{z}) \Bigg) \frac{\ud z}{z},
        \label{eq:nlo-incl-integralops2}
    \end{align}
    which thanks to renormalization are integrable operations on a smooth function $f$ and as such are of the form of integral operators discussed in Sec.~\ref{sec:obs-as-linear-ops}, specifically cross section integral operators of second~\eqref{eq:integral-operator-sigma-obs-second-kind} and first kind~\eqref{eq:integral-operator-sigma-obs-first-kind}.
    With these we can write
    \begin{align}
        \sigma^\gamma_{r,\nlo}(x, Q^2, y)
        & = 
        \begin{bmatrix}
            \undertilde{e}^2 \mathcal{I}_{2-L,g} &
            e_d^2 \Ical_{2-L,q} &
            \cdots &
            e_{\bar t}^2 \Ical_{2-L,q}
        \end{bmatrix}
        \pdtwritoutzdot
        \notag
        \\
        & \eqqcolon
        \qty(\bA^\gamma_{r,\nlo}
        \pdtbf)(x,Q^2,y),
        \label{eq:fwdop-incdis-nlo}
    \end{align}
    where we defined the linear inclusive DIS operator at NLO accuracy and used the fractional charge vector $\undertilde{e}$ defined in Eq.~\eqref{eq:e-undertilde-frac-charge-vec}.
    In Sec.~\ref{sec:tensor-ip-nlo} we will return to discuss how this integral operator structure enables the use of reconstructive methods, once we are ready to collect all the components of the full global data inverse problem.

    To write the forward operator for $F_L^\gamma$ as defined at NLO by the Eq.~\eqref{eq:conventional-nlo-fl} we define the integral operators:
    \begin{align}
        \mathcal{I}_{L,q}[f](x, Q^2) & \coloneqq
        \frac{\alpha_s}{4\pi} \int_x^1 c_{L,q}^{(1)}\qty(\frac{x}{z}) x f\qty(z, Q^2) \frac{\ud z}{z} ,
        \label{eq:nlo-incl-fl-integralops1}
        \\
        \mathcal{I}_{L,g}[f](x, Q^2) & \coloneqq
        \frac{\alpha_s}{4\pi} \int_x^1 c_{L,g}^{(1)}\qty(\frac{x}{z}) x f\qty(z, Q^2) \frac{\ud z}{z} ,
        \label{eq:nlo-incl-fl-integralops2}
    \end{align}
    which are structure function integral operators of the first kind~\eqref{eq:integral-operator-F-obs-first-kind}.
    With these the forward operator is written as
    \begin{align}
        F_L^{\gamma}(x, Q^2) &=
        \qty(\bB^\gamma_L \pdtbf) (x,Q^2) 
        \coloneqq
        \begin{bmatrix}
            \undertilde{e}^2 \mathcal{I}_{L,g} &
            e_d^2 \Ical_{L,q} &
            \cdots &
            e_{\bar t}^2 \Ical_{L,q}
        \end{bmatrix}
        \pdtwritoutzdot ,
        \label{eq:fwdop-incdis-nlo-fl}
    \end{align}
    where we can observe that the integral equation structure is distinct from Eq.~\eqref{eq:fwdop-incdis-nlo} due to the missing ``$\Id$'' terms.

    \subsection{Neutral current massless DIS at NLO accuracy}
    \label{sec:nlo-nc}

    We move on to discuss the full neutral current picture of DIS at NLO accuracy only examining the form of the forward operators, and refer the reader to Sec.~\ref{sec:lo-nc-dis} where the physics was considered in more detail. Recall the definition for the total NC cross section from Eq.~\eqref{eq:nc-dis-crosssection}, which is the primary observable quantity:
    \begin{equation}
        \label{eq:nc-dis-crosssection-nlo}
        \sigma^{\pm}_{r,\mathrm{NC}}(x,Q^2,y)
        \pequiv
        F_2^{\pm,\mathrm{NC}}(x,Q^2) \mp \frac{Y_-}{Y_+} x F_3^{\pm,\mathrm{NC}}(x,Q^2) - \frac{y^2}{Y_+} F^{\pm,\mathrm{NC}}_L(x,Q^2).
    \end{equation}
    In this section we collect the QCD corrections for massless quarks to the structure functions $F_2^{\pm,\mathrm{NC}}$, $F_3^{\pm,\mathrm{NC}}$, and $F_L^{\pm,\mathrm{NC}}$. These were originally derived in Ref.~\cite{Bardeen:1978yd:nc}, reviewed in~\cite{Sterman:1995:handbook-qcd}, with more modern re-calculations being found in Refs.~\cite{Moch:1999eb, Vermaseren:2005qc}. We need the Wilson coefficients that in the massless approximation at NLO hold for all of the electro-weak bosons $V \in \{ \gamma, Z, W^\pm \}$:
    \begin{align}
        c_{2,q}^{(1)}(z) & = \cf \bigg\lbrace \frac{9}{2} + \frac{5}{2}z - 2 \qty(\frac{3}{4} + \ln(z) - \ln(1-z) ) \qty(\frac{2}{1-z} - 1 - z)
        \notag \\
        & \quad \qquad - \qty(9 + 4 \zeta(2)) \delta(1-z) \bigg\rbrace_+,
        \label{eq:coeff-C2q}
        \\
        \label{eq:coeff-C3q}
        c_{3,q}^{(1)}(z) & = c_{2,q}^{(1)}(z) - 2 \cf(1+z),
        \\
        c_{2,g}^{(1)}(z) & = 
        6 - 2 \qty(4 + \ln(z) - \ln(1-z)) \qty(1 -2z + 2z^2 ) 
        ,
        \\
        c_{L,q}^{(1)}(z) & = 4 \cf z,
        \\
        c_{L,g}^{(1)}(z) & = 8 
        z(1-z),
        \label{eq:coeff-CLg}
    \end{align}
    where we again have left the factors of $n_f$ out intentionally from the gluonic coefficients, as we will express the sum over flavors and electro-weak charges differently yet equivalently.
    With $V \equiv \gamma$, these are exactly the same as discussed in the previous section, other than the new coefficient $c_{3,q}$.
    To express the NC structure functions $F_2^{\gamma Z}$, $F_2^{Z}$, $F_3^{\gamma Z}$, and $F_3^{Z}$ in terms of the PDFs, we apply the one-loop correction structure of Eq.~\eqref{eq:conventional-nlo-f2} to the structure functions defined in Eqs.~\eqref{eq:f2-nc-gammaZ}, \eqref{eq:f2-nc-Z}, \eqref{eq:f3-nc-gammaZ}, and \eqref{eq:f3-nc-Z}.
    Equations~\eqref{eq:nc-f2-lo} and~\eqref{eq:nc-f3-lo} are general all-order formulae for the structure function contributions from $\gamma$, $Z$, and $\gamma Z$ exchanges or interferences, and are assumed to define the electro-weak contributions to the structure functions.

    Our starting point to write out the full neutral current case at NLO accuracy is Eq.~\eqref{eq:fwdop-nc-Z-LO-f2}, which spells out the leading form of the electro-weak factorization. Equations~\eqref{eq:f2-nc-gamma}--\eqref{eq:f2-nc-Z} spell out how the replacement of the photon interaction by $\gamma Z$ or $Z$ changes the sensitivity of the observable to the PDFs depending on the interaction of the boson with the quarks. Putting this together, the full NC observable sums over the parton flavors and the electro-weak contributions $\gamma$, $\gamma Z$, $Z$, and we want to write the experimental observable as a single matrix-form operator acting on the universal PDF tensor.
    To this end, first recall the helicity-averaged effective squared parity-preserving electro-weak charge $\phi^{\pm}_{2,a}$ we defined in Eq.~\eqref{eq:lo-nc-phi2}:
    \begin{align}
        \notag
        \phi^{\pm}_{2,a} =
            e_a^2 
            - 2 e^a g_V^a \qty(g_V^{e,\pm}) \eta_{\gamma Z}(Q^2)
            & + \qty( \qty(g_V^a)^2 + \qty(g_A^a)^2) \qty(\qty(g_V^{e,\pm})^2 + \qty(g_A^{e,\pm})^2) \eta_Z(Q^2),
    \end{align}
    and the effective parity-breaking counterpart of Eq.~$\phi^{\pm}_{3,a}$~\eqref{eq:lo-nc-phi3}:
    \begin{align}
        \notag
        \phi^{\pm}_{3,a} =
            - 2 e^a g_A^a g_A^{e,\pm} \eta_{\gamma Z}(Q^2) 
            + 2 g_V^a g_A^a \qty(2 g_V^{e,\pm} g_A^{e,\pm}) \eta_Z(Q^2).
    \end{align}
    At NLO accuracy for NC, the gluonic contribution is an analogous sum over the charges of the electro-weakly interacting quarks, so to write the full NC contribution from the gluon PDF, one has to replace the sum over the squared fractional charges of the quarks by the sum over the effective electro-weak charges $\phi^{\pm}_{2,a}$, which we denote by\footnote{Despite appearances, these are just functions of $Q^2$. For practical purposes it is a real-valued function of $Q^2$ predicted by electro-weak theory.}:
    \begin{align}
        \phi^\pm_{g} \coloneqq {} & \sum_{a \in \lbrace d,\dots, \bar t\rbrace}
            e_a^2 - 2 e^a g_V^a g_V^{e,\pm} \eta_{\gamma Z}(Q^2)
            + \qty( \qty(g_V^a)^2 + \qty(g_A^a)^2) \qty(\qty(g_V^{e,\pm})^2 + \qty(g_A^{e,\pm})^2) \eta_Z(Q^2).
        \label{eq:nlo-nc-gluon-eff-charge}
    \end{align}
    This arises as one sums over all exchanged vector bosons and all quark flavors, the order of which can be interchanged at NLO in the massless quark approximation. With the effective charges \eqref{eq:lo-nc-phi2} and \eqref{eq:nlo-nc-gluon-eff-charge} and the coefficient functions \eqref{eq:coeff-C2q}--\eqref{eq:coeff-CLg} we can then write the structure functions $F_2^\mathrm{NC}$ and $F_L^\mathrm{NC}$ in the matrix integral operator form:
    \begin{align}
        F_{2, \mathrm{NC}}^{\pm,\nlo}(x, Q^2) &=
        \begin{bmatrix}
            \phi^\pm_{g} \Ccal_{2,g} &
            \phi^{\pm}_{2,d} \Ccal_{2,q} &
            \cdots &
            \phi^{\pm}_{2,\bar t} \Ccal_{2,q}
        \end{bmatrix}
        \pdtwritoutzdot
        \coloneqq
        \qty(\bB_{2, \mathrm{NC}}^{\pm,\nlo} \pdtbf)(x,Q^2),
        \label{eq:fwdop-nlo-nc-f2}
    \end{align}
    and
    \begin{align}
        F_{L, \mathrm{NC}}^{\pm,\nlo}(x, Q^2) &=
        \begin{bmatrix}
            \phi^\pm_g \Ccal_{L,g} &
            \phi^{\pm}_{L,d} \Ccal_{L,q} &
            \cdots &
            \phi^{\pm}_{L,\bar t} \Ccal_{L,q}
        \end{bmatrix}
        \pdtwritoutzdot
        \coloneqq
        \qty(\bB_{L, \mathrm{NC}}^{\pm,\nlo} \pdtbf)(x,Q^2),
        \label{eq:fwdop-nlo-nc-fl}
    \end{align}
    where ${\phi}^\pm_L \equiv {\phi}^\pm_2$, and we are using the integral operator notation of Eq.~\eqref{eq:linear-integral-operator} for the coefficient functions $c_i$.
    Similarly, noting that the only non-zero coefficient is the quark contribution $c_{3,q}^{(1)}$, we write for the parity breaking $F_3$:
    \begin{align}
        F_{3, \mathrm{NC}}^{\pm,\nlo}(x, Q^2) &=
        \begin{bmatrix}
            0 &
            \phi^{\pm}_{3,d} \Ccal_{3,q} &
            \cdots &
            -\phi^{\pm}_{3,\bar t} \Ccal_{3,q}
        \end{bmatrix}
        \pdtwritoutzdot
        \coloneqq
        \qty(\bB_{3, \mathrm{NC}}^{\pm,\nlo} \pdtbf)(x,Q^2),
        \label{eq:fwdop-nlo-nc-f3}
    \end{align}
    for which we note that the sign changes for the antiquark contributions according to Eqs.~\eqref{eq:f3-nc-gammaZ} and~\eqref{eq:f3-nc-Z}.
    With the above structure function forward operators $\bB_i$ it is now convenient to write the experimental observable given in Eq.~\eqref{eq:nc-dis-crosssection-nlo} into a forward operator at NLO accuracy:
    \begin{align}
        \label{eq:fwdop-nlo-nc-cs}
        \sigma^{\pm, \nlo}_{r,\mathrm{NC}}(x,Q^2,y)
        & =
        \Bigg[ \bB_{2, \mathrm{NC}}^{\pm,\nlo}(x,Q^2) \mp \frac{Y_-}{Y_+} x \bB_{3, \mathrm{NC}}^{\pm,\nlo}(x,Q^2) - \frac{y^2}{Y_+} \bB_{L, \mathrm{NC}}^{\pm,\nlo}(x,Q^2)
        \Bigg] \pdtbf
        \notag
        \\
        & = 
        \begin{bmatrix}
            \phi^\pm_{g} \qty( \Ccal_{2,g} - \frac{y^2}{Y_+} \Ccal_{L,g}) &
            \Ccal_{q,d}^\mathrm{NC} &
            \cdots &
            \Ccal_{q,\bar t}^\mathrm{NC}
        \end{bmatrix}
        \pdtwritoutzdot
        \notag
        \\
        & \eqqcolon
        \qty(\bA^{\pm, \nlo}_{r,\mathrm{NC}}
        \pdtbf)(x,Q^2,y),
    \end{align}
    where we defined the combined integral operator for the quarks:
    \begin{equation}
        \Ccal_{q,a}^\mathrm{NC} \coloneqq \phi^{\pm}_{2,a} \qty( \Ccal_{2,q} - \frac{y^2}{Y_+} \Ccal_{L,q} ) \mp \frac{x Y_-}{Y_+} \phi^{\pm}_{3,a} (-1)^{\delta_{a \bar a}} \Ccal_{3,q},
    \end{equation}
    which is understood as a single unified convolution integral, and the abuse of notation $(-1)^{\delta_{a \bar a}}$ is intended to produce $+1$ for quarks, and $-1$ for anti-quarks, which results in definite signs in the expressions when written out explicitly for all flavors. Equations~\eqref{eq:fwdop-nlo-nc-f3} and~\eqref{eq:fwdop-nlo-nc-cs} now define the forward operators of interest with corresponding real world data available.

    \subsection{Charged current massless DIS at NLO}

    In this section we follow the approach of the previous section, and directly move on to write the NLO accuracy results of the experimental observables which were discussed in more detail in Sec.~\ref{sec:lo-cc-dis}. Experimentally it is possible to isolate the experiment for the exchange of either $W^+$ or $W^-$ based on the incoming probe lepton and the identification of the final state, which means that we must write the theory result for these specific observables to consider the structure of the mathematical inverse problem. 

    The perturbative QCD results for the charged current Wilson coefficients are known well beyond NLO~\cite{Moch:2007rq:cc-nnlo,Moch:2008fj:cc-nnlo,Moch:2007gx:cc-nnlo-neutrino,Davies:2016ruz:cc-nnlo-neutrino,Davies:2016bwb:cc-dis-davies-phd}, but for the sake of simplicity of the present approach we limit the discussion to the classical results derived at NLO accuracy~\cite{Bardeen:1978yd:nc}.
    The experimentally measured reduced charged current cross section is~\cite{Abramowicz:2015mha:HERAIInewcombined}
    \begin{equation}
        \label{eq:cc-dis-reduced-crosssection}
        \sigma_{r,\mathrm{CC}}^{\pm}(x,Q^2,y)
        \pequiv
            \frac{Y_+}{2} F_2^{W^\pm}(x,Q^2) \mp \frac{Y_-}{2} x F_3^{W^\pm}(x,Q^2) 
            - \frac{y^2}{2} F^{W^\pm}_L(x,Q^2)
        ,
    \end{equation}
    where the sign of the observable refers to that of the charged probe $e^\pm$, or equivalently the $W^\pm$ boson. Our task is to write the NLO accuracy results for the CC structure functions $F^{W^\pm}$. The coefficient functions that we need are given in Eqs.~\eqref{eq:coeff-C2q}--\eqref{eq:coeff-CLg}.
    As we write the NLO contributions, we need to be aware that $W^+$ only interacts with ``down'' and ``anti-up'' flavors $(d,s,b,\bar u, \bar c, \bar t)$ and $W^-$ interacts with their counterparts $(u,c,t,\bar d, \bar s, \bar b)$~\cite{ParticleDataGroup:2024cfk}, which makes the other half of the quark flavors ``invisible'' to the $W^\pm$ at leading and next-to-leading order.

    Using the strategy of the previous section, we introduce the NLO corrections to the leading order CC structure functions of Eqs.~\eqref{eq:cc-f2-lo-W+}--\eqref{eq:cc-f3-lo-W-}\footnote{As discussed in the LO CC section, for the inclusive cross section \textit{in the massless approximation} we can leverage the unitarity relations of the CKM matrix whereby we can reorder the sum over CKM rows to get unities. If we were to consider heavy quarks in the charged current picture, we would have to account for flavor mixing into heavy flavors using the CKM matrix. We leave these complications out of the scope of this work, only noting that this has been considered in detail for example in Refs.~\cite{Gao:2021fle:heavy-cc,Risse:2025smp:heavy-cc}. However, in principle the CKM matrix is just a linear mapping acting on the PDFs in the flavor space, and as such should be possible to include in this framework.}:
    \begin{align}
        F_{i,\nlo}^{W^+}(x, Q^2) &=
        \begin{bmatrix}
            \frac{n_f}{2} \Ccal_{i,g} &
            \Ccal_{i,q} & 0 & \Ccal_{i,q} & 0 & \Ccal_{i,q} & 0 &
            0 & \Ccal_{i,q} & 0 & \Ccal_{i,q} & 0 & \Ccal_{i,q}
        \end{bmatrix}
        \pdtbf
        \notag \\
        &
        \coloneqq
        \qty(\bB_{i,\nlo}^{W^+} \pdtbf)(x,Q^2),
        \label{eq:fwdop-nlo-cc-f2fl-w+}
    \end{align}
    and
    \begin{align}
        F_{i,\nlo}^{W^-}(x,Q^2) & = 
        \begin{bmatrix}
            \frac{n_f}{2} \Ccal_{i,g} &
            0 & \Ccal_{i,q} & 0 & \Ccal_{i,q} & 0 & \Ccal_{i,q} & 
            \Ccal_{i,q} & 0 & \Ccal_{i,q} & 0 & \Ccal_{i,q}& 0
        \end{bmatrix}
        \pdtbf
        \notag \\
        &
        \eqqcolon 
        \qty(\bB_{i,\nlo}^{W^-} \pdtbf)(x,Q^2),
        \label{eq:fwdop-nlo-cc-f2fl-w-}
    \end{align}
    for $i \in \lbrace2,L\rbrace$, and where we have introduced a factor of $n_f/2$ for the gluonic contribution since only half of the normally possible quark flavors produced from the gluon can interact with the $W^\pm$ boson, and we had not already included a factor of $n_f$ in the gluonic coefficient function as discussed in the previous section.
    The parity breaking $F_3^{W^\pm}$ become: 
    \begin{align}
        F_{3,\nlo}^{W^+}(x,Q^2) & = 
        \begin{bmatrix}
            0 &
            \Ccal_{3,q} & 0 & \Ccal_{3,q} & 0 & \Ccal_{3,q} & 0 &
            0 & - \Ccal_{3,q} & 0 & - \Ccal_{3,q} & 0 & - \Ccal_{3,q}
        \end{bmatrix}
        \pdtbf
        \notag \\ 
        & \eqqcolon 
        \qty(\bB_{3,\nlo}^{W^+} \pdtbf)(x,Q^2),
        \label{eq:fwdop-nlo-cc-f3-w+}
        \\
        F_{3,\nlo}^{W^-}(x,Q^2) & = 
        \begin{bmatrix}
            0 &
            0 & \Ccal_{3,q} & 0 & \Ccal_{3,q} & 0 & \Ccal_{3,q} & 
            - \Ccal_{3,q} & 0 & - \Ccal_{3,q} & 0 & - \Ccal_{3,q}& 0
        \end{bmatrix}
        \pdtbf
        \notag \\ 
        & \eqqcolon 
        \qty(\bB_{3,\nlo}^{W^-} \pdtbf)(x,Q^2),
        \label{eq:fwdop-nlo-cc-f3-w-}
    \end{align}
    where again the LO contribution is included as the delta-function term in the definition of $C_{3,q}$ given in Eq.~\eqref{eq:coeff-C3q}.
    With the above forward operators for the structure functions we can write the reduced CC cross section observable in the forward operator form:
    \begin{align}
        \label{eq:fwdop-nlo-cc-csecred}
        \sigma_{r,\mathrm{CC}}^{\pm,\nlo}(x,Q^2,y)
        & =
        \left[\frac{Y_+}{2} \bB_{2,\nlo}^{W^\pm}(x,Q^2) \mp \frac{Y_-}{2} x \bB_{3,\nlo}^{W^\pm}(x,Q^2) 
        - \frac{y^2}{2} \bB^{W^\pm}_{L,\nlo}(x,Q^2) \right] \pdtbf
        \notag \\
        & \eqqcolon
        \qty(\bA^{\pm, \nlo}_{r,\mathrm{CC}}
        \pdtbf)(x,Q^2,y),
    \end{align}
    which by construction reproduces Eq.~\eqref{eq:cc-dis-reduced-crosssection}. Writing this out more explicitly:
    \begin{align}
        \bA^{+, \nlo}_{r,\mathrm{CC}}(x,Q^2,y) & = 
        \begin{bmatrix}
            \frac{n_f Y_+}{2}\qty( \Ccal_{2,g} - \frac{y^2}{Y_+} \Ccal_{L,g}) &
            \Ccal^{W^+}_{q,d} & 0 & \Ccal^{W^+}_{q,s} & 0 & \Ccal^{W^+}_{q,b} & 0 &
            0 & \Ccal^{W^+}_{q, \bar u} & 0 & \Ccal^{W^+}_{q, \bar c} & 0 & \Ccal^{W^+}_{q, \bar t}
        \end{bmatrix}
        ,
        \label{eq:fwdop-nlo-cc-A-w+}
        \\
        \bA^{-, \nlo}_{r,\mathrm{CC}}(x,Q^2,y) & = 
        \begin{bmatrix}
            \frac{n_f Y_+}{2}\qty( \Ccal_{2,g} - \frac{y^2}{Y_+} \Ccal_{L,g}) &
            0 & \Ccal^{W^-}_{q,u} & 0 & \Ccal^{W^-}_{q,c} & 0 & \Ccal^{W^-}_{q,t} &
            \Ccal^{W^-}_{q, \bar d} & 0 & \Ccal^{W^-}_{q, \bar s} & 0 & \Ccal^{W^-}_{q, \bar b} & 0
        \end{bmatrix} ,
        \label{eq:fwdop-nlo-cc-A-w-}
    \end{align}
    where we defined the integral operator:
    \begin{align}
        \Ccal_{q,a}^{W^\pm}[f] & \coloneqq \frac{Y_+}{2} \Ccal_{2,q}[f] \mp (-1)^{\delta_{a \bar a}} \frac{Y_-}{2} x \Ccal_{3,q}[f] - \frac{y^2}{2} \Ccal_{L,q}[f]
        \\
        & = \int_x^1 \qty(\frac{Y_+}{2} c_{2,q}\qty(\frac{x}{z}) \mp (-1)^{\delta_{a \bar a}} \frac{Y_-}{2} x c_{3,q}\qty(\frac{x}{z}) - \frac{y^2}{2} c_{L,q}\qty(\frac{x}{z})) f(z) \frac{\ud z}{z}
        ,
        \label{eq:fwdop-nlo-cc-Cq}
    \end{align}
    where the delta function sets the correct sign for the antiquark contributions for the coefficient $c_3$.
    The result is very similar for $W^-$, the only change would be the actively visible flavors to the boson; both are shown in appendix~\ref{app:full-global-data-operators}. We also note that the exactly zero operator elements seen here are due to the finite order of perturbation theory used here, and beginning at the following next-to-next-to-leading order one would see non-zero contributions of order $\mathcal{O}(\as^2)$.

    \subsection{Heavy quark production in DIS at NLO}

    We will examine the NLO corrections to heavy flavor production observables in DIS, and their shape in the larger mathematical structure of the inverse problem in a kind of a toy model isolation. Systematic inclusion of one-loop corrections from heavy flavors requires their comprehensive implementation into all observables included in global analysis, including a choice of a \gls{flavor number scheme} as discussed in Sec.~\ref{sec:obs-as-linear-ops}. In this work we limit the examination of these corrections only to the heavy flavor production observables. Roughly in the sense that charm and bottom production measurements are taken to be independent observables, which can be given an approximate forward operator in a fixed-flavor number scheme: both charm and bottom are assumed to be produced from three light flavors, a scheme known as FFNS3cb~\cite{Barontini:2024xgu:flavor-number-scheme-names-and-intro}.
    Due to the deep integration of the heavy quark mass effects into various parts of the perturbative QCD calculations, this re-framing of the heavy quark effects is very much a toy model of the complete problem. But due to the real importance of the heavy quark effects in global analysis, we consider it crucial that this framework can handle their inclusion at least in principle, and return to discuss more general formulations of flavor number schemes in Sec.~\ref{sec:flavor-number-schemes}.

    To write the NLO accuracy corrections to heavy flavor production observables of Tab.~\ref{tab:processes}, we will have to collect the contributions from a few sources in the literature. Specifically, we will recall the modern re-calculations of the quark non-singlet~\cite{Blumlein:2016xcy:Hq-heavy-wilson-nonsinglet} and pure singlet contributions~\cite{Blumlein:2019qze:Hq-heavy-wilson}, which are written in an analytic form for the former, and in an almost analytic form in terms of iterated integrals representing special functions containing elliptic integrals and polylogarithms~\cite{Remiddi:1999ew}. The last of the $\as^2$ order contributions, that of the gluon singlet state, is unfortunately still not known even in the latter form, and so we will describe it according to the Refs.~\cite{Laenen:1992zk-heavyq-nlo,Buza:1995ie:orignal-heavy-nlo,Buza:1996wv-heavyq-nlo} which gives the contribution in an implicit integral form, which was numerically implemented in Ref.~\cite{Riemersma:1994hv-heavyq-nlo}, and later verified in an independent calculation and implementation in Refs.~\cite{Hekhorn:2018ywm:phd-ref-article1,Hekhorn:2019nlf-felix-phd}.

    The order $\as^2$ result for the inclusive production heavy of a massive quark\footnote{The label $Q$ often used for the ``big'' heavy quark might unfortunately be confused with the virtuality squared $Q^2$ of the virtual boson. These have nothing to do with each other; the former is a label for the heaviest flavor index, and the latter is the momentum scale of the scattering process. Due to this, some references prefer to use the label $H$ for the heavy quark, or the upright $\mathrm{Q}$.} $Q$ in DIS from $n_f$ light flavors is of the form~\cite{Klein2012}
    \begin{align}
    F_{i}^{Q\overline{Q}}(x, n_f \!+\! 1, Q^2, m^2) \pequiv{} & \sum_{k=1}^{n_f} e_k^2 L_{i,q}^{\text{NS}} \left( x, \frac{Q^2}{m^2}, \frac{m^2}{\mu^2} \right) 
    \! \otimes \! \left[ f_k(x, \mu^2, n_f) + f_{\overline{k}}(x, \mu^2, n_f) \right] \nonumber \\
    &+ e_Q^2 \biggl[ H_{i,q}^{\text{PS}} \left( x, \frac{Q^2}{m^2}, \frac{m^2}{\mu^2} \right) \otimes \Sigma(x, \mu^2, n_f) \nonumber \\
    &\qquad+ H_{i,g}^{\text{S}} \left( x, \frac{Q^2}{m^2}, \frac{m^2}{\mu^2} \right) \otimes f_g(x, \mu^2, n_f) \biggr],
    \quad i \in \lbrace 2,L \rbrace,
    \label{eq:nlo-heavyq-structure-fun-Fi}
    \end{align}
    where $f_k$ are the PDFs of the light quarks, and $\Sigma$ is the singlet PDF:
    \begin{equation}
        \Sigma(x, Q^2, n_f) = \sum_{l=1}^{n_f} \left[ f_l(x, Q^2) + f_{\overline{l}}(x, Q^2) \right],
    \end{equation}
    where the sum over the light flavors is $a \in \lbrace d,u,s \rbrace$ for both charm and bottom production in the FFNS3cb scheme. In this initial approach we neglect the regime, where they both should be considered massive.

    The so-called Compton contributions $L_{i,q}^{\text{NS}}$ are~\cite{Blumlein:2016xcy:Hq-heavy-wilson-nonsinglet,Buza:1995ie:orignal-heavy-nlo}
    \begin{align}
     L_{L,q}^{\text{NS},(2),C}\left(z,\frac{Q^2}{m^2},\frac{m^2}{\mu^2}\right)
     &=
     a_s^2 \cf T_F \Biggl\{
     96 \frac{z^3}{\xi^2} \Biggl[\ln\left(\frac{1+\sqrt{1-\frac{4z}{(1-z)\xi}}}{1-\sqrt{1-\frac{4z}{(1-z)\xi}}}\right)
     \ln\left(\frac{1+\sqrt{1-\frac{4z}{\xi}}}{1-\sqrt{1-\frac{4z}{\xi}}}\right)
    \nonumber\\ &
    +2 \Biggl[
    - \Li_2\left(\frac{(1-z)\left(1+\sqrt{1-\frac{4z}{(1-z)\xi}}\right)}{1+\sqrt{1-\frac{4z}{\xi}}} \right)
    + \Li_2\left(\frac{1- \sqrt{1-\frac{4z}{\xi}}}{1 +\sqrt{1-\frac{4z}{(1-z)\xi}}} \right)
    \nonumber\\ &
    + \Li_2\left(\frac{1 - \sqrt{1-\frac{4z}{(1-z)\xi}}}
                      {1 + \sqrt{1-\frac{4z}{\xi}}}\right)
    - \Li_2\left(\frac{1 + \sqrt{1-\frac{4z}{(1-z)\xi}}}
                      {1 + \sqrt{1-\frac{4z}{\xi}}}\right)
     \Biggr] \Biggr]
    \nonumber\\ &
    - \frac{16}{3 \xi} (22 z^2 - z\xi) \sqrt{1 - \frac{4z}{
    \xi}}
    \ln\left(\frac{
     \sqrt{1-\frac{4z}{\xi}} + \sqrt{1-\frac{4z}{(1-z)\xi}}}
    {\sqrt{1-\frac{4z}{\xi}} - \sqrt{1-\frac{4z}{(1-z)\xi}}}\right)
    \nonumber\\ &
    -\frac{8}{9(1-z) \xi} \sqrt{1-\frac{4z}{(1-z)\xi}}\left[60 z -478 z^2 +372 z^3 \right. 
    \nonumber\\ &
    \left.
    - \xi (6 - 31 z + 25 z^2)\right]
    +\ln\left(\frac{1+\sqrt{1-\frac{4z}{(1-z)\xi}}}{1-\sqrt{1-\frac{4z}{(1-z)\xi}}} \right)
    \nonumber\\ &
    \times
    \left[ \frac{32(6 z^2 - 9z +2) z^2}{(1-z)^2 \xi^2} 
    + 96 \frac{z^3}{\xi^2} \ln \left(\frac{1-z}{z^2}\right)\right]
    \Biggr\}
    \theta\left(\frac{\xi}{\xi+4} - z\right)
    \label{eq:FLR}
    \end{align}
    and
    \begin{align}
    L_{2,q}^{\text{NS},(2),C}\left(z,\frac{Q^2}{m^2},\frac{m^2}{\mu^2}\right)
     &=
     a_s^2\cf T_F \Biggl\{
     \left[-\frac{16 z^2}{\xi^2(1-z)} \left(1-9z+9z^2\right) + \frac{4}{3} \frac{1+z^2}{1-z} \right]
    \nonumber\\ &
    \times  \Biggl[ \Biggl[
     \ln\left(\frac{1+\sqrt{1-\frac{4z}{\xi}}}{1-\sqrt{1-\frac{4z}{\xi}}}\right)
    + \ln\left(\frac{1-z}{z^2}\right) \Biggr]
     \ln\left(\frac{1+\sqrt{1-\frac{4z}{(1-z)\xi}}}{1-\sqrt{1-\frac{4z}{(1-z)\xi}}}\right)
    \nonumber\\ &
    +2 \Biggl[
    - \Li_2\left(\frac{(1-z)\left(1+\sqrt{1-\frac{4z}{(1-z)\xi}}\right)}{1+\sqrt{1-\frac{4z}{\xi}}} \right)
    + \Li_2\left(\frac{1- \sqrt{1-\frac{4z}{\xi}}}{1 +\sqrt{1-\frac{4z}{(1-z)\xi}}} \right)
    \nonumber\\ &
    + \Li_2\left(\frac{1 - \sqrt{1-\frac{4z}{(1-z)\xi}}}
                      {1 + \sqrt{1-\frac{4z}{\xi}}}\right)
    - \Li_2\left(\frac{1 + \sqrt{1-\frac{4z}{(1-z)\xi}}}
                      {1 + \sqrt{1-\frac{4z}{\xi}}}\right)
     \Biggr] \Biggr]
    \nonumber\\ &
    + \frac{8}{9(1-z)\xi}\left[26 z -168 z^2 +188 z^3 - (8 - 6z+17z^2) \xi\right]
    \sqrt{1-\frac{4z}{\xi}}
    \nonumber\\ & \times
    \ln\left(\frac{
     \sqrt{1-\frac{4z}{\xi}} + \sqrt{1-\frac{4z}{(1-z)\xi}}}
    {\sqrt{1-\frac{4z}{\xi}} - \sqrt{1-\frac{4z}{(1-z)\xi}}}\right)
    +\frac{2}{27 (1-z)^2 \xi} \sqrt{1-\frac{4z}{(1-z) \xi}}
    \nonumber\\ & \times
    \left[ -1702 z +9516 z^2 -14260 z^3
    +6456 z^4  \right.
    \nonumber\\ & \left.
    +(223-709 z +1108 z^2 -622 z^3) \xi\right]\nonumber\\ &
    -\frac{4}{3(1-z)^3\xi^2}\left[6z^2(-15+70z-90z^2+36z^3)-\xi^2(1-3z^2+2z^3)\right]\nonumber\\ &
    \times\ln\left(\frac{1+\sqrt{1-\frac{4z}{(1-z)\xi}}}{1-\sqrt{1-\frac{4z}{(1-z)\xi}}}\right)
    \Biggr\}\theta\left(\frac{\xi}{\xi+4} - z\right),
    \end{align}
    where $\xi \coloneqq \frac{Q^2}{m^2}$, and $T_F= \frac{1}{2}$ is a QCD color factor.
    The pure singlet contributions for $i\in \lbrace L, 2 \rbrace$~\cite{Blumlein:2019qze:Hq-heavy-wilson}
    \begin{equation}
        H_{i,q}^\mathrm{PS} \equiv H_{i,q}^\mathrm{PS}\left(z,\frac{Q^2}{m^2},\frac{m^2}{\mu^2}\right)
    \end{equation}
    are given in appendix~\ref{app:heavy-flavor-wilson-coeffs} Eqs.~\eqref{eq:heavy-wilson-HLq} and~\eqref{eq:heavy-wilson-H2q} due to their size.
    
    The last term we need is $H_{i,g}^S$, i.e. the gluon singlet contribution, which is only known as a complicated integral form.
    In the original Ref.~\cite{Buza:1995ie:orignal-heavy-nlo}, the order $\as^2$ gluon contributions are given in the compact implicit integral form as:
    \begin{align}
      H_{L,g}^{(2)} \left(z, \frac{Q^2}{m^2}, \frac{\mu^2}{m^2} \right) &= A_{Qg}^{(1)} \left( \frac{\mu^2}{m^2} \right) \otimes C_{L,q}^{(1)} \left( z, \frac{Q^2}{\mu^2} \right) + C_{L,g}^{(2)} \left( z, \frac{Q^2}{\mu^2} \right) , 
      \\
      H_{2,g}^{(2)} \left( z, \frac{Q^2}{m^2}, \frac{\mu^2}{m^2} \right) &= A_{Qg}^{(2)} \left( \frac{\mu^2}{m^2} \right) + A_{Qg}^{(1)} \left( \frac{\mu^2}{m^2} \right) \otimes C_{2,q}^{(1)} \left( z, \frac{Q^2}{\mu^2} \right) + C_{2,g}^{(2)} \left( z, \frac{Q^2}{\mu^2} \right) ,
    \end{align}
    where $A_i$ are so-called operator matrix elements, and $C_i$ are the massless Wilson coefficient functions familiar from the previous sections. However, this is not how they are in practice computed in theory--data comparisons such as global analysis, and instead one has to have a numerical implementation, which is discussed for example in Refs.~\cite{Hekhorn:2018ywm:phd-ref-article1,Hekhorn:2019nlf-felix-phd}. Given the extensive empirical numerical use they have gone through in theory computations, from the perspective of this work we will just assume that they are sufficiently nice integration kernels that are only available numerically, for example in the implementation~\cite{Hekhorn:heavy-quark-yadism-leproHQ-zenodo} of Ref.~\cite{Hekhorn:2019nlf-felix-phd}. Writing down a formulation of the integral definition would need to go into details about specifics about the numerical implementation of integration and regularization, which go beside the point of this theory review, and so we refer the reader to Ref.~\cite{Hekhorn:2019nlf-felix-phd}.
    From the practical point of view, we take the coefficient functions $H_{i,g}^S$ as being available numerically from the \texttt{LeProHQpy} package~\cite{Hekhorn:heavy-quark-yadism-leproHQ-zenodo,Hekhorn:2019nlf-felix-phd}:
    \begin{align}
        H_{L,g}^{(2)} \left(z, \frac{Q^2}{m^2} \right) &\circumeq \texttt{LeProHQ.cg1("FL", "VV", Q\^{}2/m\^{}2, eta)} ,
        \\
        H_{2,g}^{(2)} \left( z, \frac{Q^2}{m^2} \right) &\circumeq \texttt{LeProHQ.cg1("F2", "VV", Q\^{}2/m\^{}2, eta)} ,
    \end{align}
    where \texttt{"VV"} refers to the vectorial-vectorial coupling of pure virtual photon exchange we are now limited to, and $\texttt{eta} \circumeq \eta \coloneqq \frac{\hat s-4m^2}{4m^2} \equiv \frac{Q^2}{4m^2}\qty(\frac{z}{x}-1)-1$ is the distance from heavy flavor $Q$ production threshold~\cite{Hekhorn:2019nlf-felix-phd,Hekhorn:heavy-quark-yadism-leproHQ-zenodo}\footnote{Writing out the partonic invariant mass squared for the parton with momentum fraction $z$: $\hat s \coloneqq (z P + q)^2 = 2z(P\cdot q) -Q^2 = z(s+Q^2)-Q^2 = z \frac{Q^2}{x} - Q^2$.}.

    According to Refs.~\cite{Hekhorn:2019nlf-felix-phd,Klein2012}, we have included the term(s) $L_i^\mathrm{NS}$ as it ever-so-slightly is a closer definition of the experimental observable~\cite{Forte:2010ta:FONLL,Hekhorn:2019nlf-felix-phd}, where the heavy quark production is tagged regardless of its production mechanism. This is in contrast with the standard choice used in global analyses where it is not included, and the definition for the observable of heavy flavor production is taken to be the explicit probing of the heavy flavor by the virtual boson. These definitions differ by whether the virtual boson interacts with light flavors directly, which is disallowed by the explicit requirement of probing the heavy flavor. In the alternative definition the heavy flavor can be produced before or after the boson interacts with a light flavor. From our perspective the inclusion of the $L_i^\mathrm{NS}$ term would seem reasonable, so we keep it for completeness in our inspection, but our higher-level point is to merely demonstrate that it can be done in this inverse problems approach. We acknowledge that there is a non-trivial physical definition to be made in regards to the inclusion of the contribution, and strive to accommodate the possibility of taking either choice within our inverse problems formalism.
    
    It now remains to write the experimental observables defined in Eq.~\eqref{eq:nlo-heavyq-structure-fun-Fi}, and the reduced cross sections using them, in the forward operator form. For the gluon contribution we simply have:
    \begin{equation}
        \mathcal{H}_{i,g,Q}^{\text{S}}[f](x, Q^2, m^2) \coloneqq 
        e_Q^2 \int_{ax}^1 H_{i,g}^{\text{S}} \left( \frac{x}{z}, \frac{Q^2}{m^2}, \frac{m^2}{\mu^2} \right) f(z, Q^2) \frac{\ud z}{z},
    \end{equation}
    where $a \coloneqq 1 + 4 \frac{m^2}{Q^2}$~\cite{Klein2012}, and we set the scale of the PDFs as $\mu^2 \equiv Q^2$. For the quarks we define:
    \begin{equation}
        \Ical^{L+H}_{i,q,Q} [f] (x, Q^2, m^2)
        \coloneqq
        \int_{ax}^1 \left \lbrace
        e_q^2 L_{i,q}^{\text{NS}} \left( \frac{x}{z}, \frac{Q^2}{m^2}, \frac{m^2}{\mu^2} \right)
        + e_Q^2 H_{i,q}^{\text{PS}} \left( \frac{x}{z}, \frac{Q^2}{m^2}, \frac{m^2}{\mu^2} \right)
        \right \rbrace f(z,Q^2) \frac{\ud z}{z},
    \end{equation}
    where $q$ in $e_q$ acts also as the flavor index of $\Ical^{L+H}_{i,q,Q}$, otherwise the kernel terms are independent of the flavor index, and similarly for the fractional charge $e_Q$ of the heavy flavor $Q$.
    With these integral operators we are ready to write the expressions for the experimental observables $\sigma^{Q \bar Q}_r$ for charm and bottom production reduced cross sections. For charm we assume $n_f=3$ light flavors, and the produced charm is the heavy flavor $(Q \equiv c)$, so:
    \begin{align}
        & F_{i}^{c \bar c} (x, n_f+1 = 4 , Q^2, m_c^2) = \qty(\bB_{i, \nlo}^{c \bar c} \pdtbf)(x,Q^2, m_c)
        \\
        & \qquad \coloneqq
        \begin{bmatrix}
            \mathcal{H}_{i,g,c}^{\text{S}} &
            \Ical^{L+H}_{i,d,c} &
            \Ical^{L+H}_{i,u,c} &
            \Ical^{L+H}_{i,s,c} &
            0 &
            0 & 0 &
            \Ical^{L+H}_{i,\bar d,c} &
            \Ical^{L+H}_{i,\bar u,c} &
            \Ical^{L+H}_{i,\bar s,c} &
            0 &
            0 & 0
        \end{bmatrix}
        \pdtwritoutzdot
    \end{align}
    In FFNS3cb scheme, bottom production is written completely analogously as produced from three light flavors, and only the mass of the produced heavy flavor is changed:
    \begin{align}
        & F_{i}^{b \bar b} (x, n_f+1 = 4 , Q^2, m_b^2) = \qty(\bB_{i, \nlo}^{b \bar b} \pdtbf)(x,Q^2,m_b)
        \\
        & \quad \coloneqq
        \begin{bmatrix}
            \mathcal{H}_{i,g,b}^{\text{S}} &
            \Ical^{L+H}_{i,d,b} &
            \Ical^{L+H}_{i,u,b} &
            \Ical^{L+H}_{i,s,b} &
            0 &
            0 & 0 &
            \Ical^{L+H}_{i,\bar d,b} &
            \Ical^{L+H}_{i,\bar u,b} &
            \Ical^{L+H}_{i,\bar s,b} &
            0 &
            0 & 0
        \end{bmatrix}
        \pdtwritoutzdot
    \end{align}

    With the above forward operators we write the operators for the heavy flavor reduced cross sections $\sigma_r = F_2 - \frac{y^2}{Y_+} F_L$:
    \begin{equation}
        \bA_{r,\nlo}^{c \bar c}(x,Q^2,y, m_c) = \bB_{2, \nlo}^{c \bar c} (x,Q^2, m_c) - \frac{y^2}{Y_+} \bB_{L, \nlo}^{c \bar c} (x,Q^2, m_c)
        \label{eq:fwdop-nlo-hq-cc}
    \end{equation}
    for charm production reduced cross section $\sigma_r^{c \bar c}$, and
    \begin{equation}
        \bA_{r,\nlo}^{b \bar b}(x,Q^2,y, m_b) = \bB_{2, \nlo}^{b \bar b} (x,Q^2, m_b) - \frac{y^2}{Y_+} \bB_{L, \nlo}^{b \bar b} (x,Q^2, m_b)
        \label{eq:fwdop-nlo-hq-bb}
    \end{equation}
    for bottom $\sigma_r^{b \bar b}$. According to assumption~\ref{post:nonpert}, we assume the heavy quark masses $m_c$ and $m_b$ to be known, and as such we consider them as a part of the external input that goes into the prescription of the forward operators, like with $\as(Q^2)$. This builds in an implicit---and naive---flavor number scheme into the above forward operators, and a more realistic implementation would need to describe all the forward operators in a consistent more accurate \gls{flavor number scheme}, which are touched on in Sec.~\ref{sec:flavor-number-schemes}. This naive \gls{flavor number scheme} is perhaps the closest to the fixed flavor-number scheme (FFNS)~\cite{Buza:1996wv-heavyq-nlo,Klein2012}, where the number of flavors is a constant and only the most heavy flavor is considered heavy, and the others are considered massless. We apply this prescription on the level of the observable cross section, and not the whole global analysis problem, which means that for charm production we considered charm as the heavy flavor, and similarly for bottom where the only heavy flavor is bottom. This is somewhat of a toy model of the application of the FFNS framework, as there is a physical regime of scattering at low $Q^2$, where both charm and bottom masses can be non-negligible.

    \subsection{Neutrino DIS at NLO for massless quarks}

        We will write the NLO corrections to the isoscalar target neutrino DIS cross section observable of Eq.~\eqref{eq:lo-nudis-isoscalar-dis-crosssection} by leveraging the NLO corrections to charged current DIS, and presuming that they can be taken to be the same for both proton and neutron by assuming isospin symmetry. We note that results up to NNLO~\cite{Zijlstra:1992kj}, and even NNNLO accuracy~\cite{Moch:2007gx:cc-nnlo-neutrino,Blumlein:2022gpp:nnnlo-f2-f3,Davies:2016bwb:cc-dis-davies-phd} are known, and that a more comprehensive implementation would require the consideration of heavy flavor production discussed in the previous section, as experimental data is specifically available for heavy flavor production in $\nu$DIS~\cite{NuTeV:2005wsg}.

        Taking Eq.~\eqref{eq:lo-nudis-nup-crosssection}, but writing it in terms of the NLO CC structure functions $F_i^{W^\pm}$ it becomes clear how to implement the one-loop QCD corrections in the case of a proton target in $\nu$DIS, since in the partonic picture of high-energy scattering the interaction with the parton does not care where the interaction carrying boson originated:
        \begin{align}
        \sigma^{\nu(\bar \nu)p}_r(x,Q^2,y) =
        Y_+ F^{W^\pm}_2(x,Q^2) - y^2 F^{W^\pm}_L(x,Q^2) \pm Y_-x F^{W^\pm}_3(x,Q^2) 
        .
        \end{align}
        Accounting for neutrons in isoscalar nuclear targets goes as discussed in Sec.~\ref{sec:lo-neutrino-dis}, i.e. by isospin symmetry, which is written out at LO in Eq.~\eqref{eq:lo-nudis-isoscalar-dis-crosssection}. However, there the isospin symmetry is implicitly assumed to be precisely one-to-one, i.e. a nuclear target with the same number of protons and neutrons. This is not always exactly satisfied in real world experiment\footnote{For example, a water molecule target in $\nu$DIS~\cite{T2K:2019dgm:water} has an oxygen with 8 protons and 8 neutrons, and two hydrogen which are just a proton each, which is an imbalance of 10 to 8 in isoscalarity.}, and in the general case with a nuclear target of mass number $A$, proton number $Z$, and neutron number $A-Z$, the quasi-isoscalar PDFs are taken to be the weighted average of the PDFs, which can be expressed as~\cite{Candido:2024rkr:yadism,yadism:manual:online}:
        \begin{equation}
            \begin{bmatrix}
                C'_d \\
                C'_u
            \end{bmatrix}
            =
            \frac{1}{A} \begin{bmatrix}
                Z & A-Z \\
                A-Z & Z
            \end{bmatrix}
            \begin{bmatrix}
                C_d \\
                C_u
            \end{bmatrix}
            \coloneqq
            \mathbf{w}_{\! A,Z}\begin{bmatrix}
                C_d \\
                C_u
            \end{bmatrix}
            ,
        \end{equation}
        where the averaging is incorporated into the implementation of the coefficient functions, which we will also want to leverage here. We are working with the full flavor basis representation, so we need to extend the above weighting to act on the full set of PDFs without acting on the other flavors:
        \begin{equation}
            \bC' = \begin{bmatrix}
                1 & \\
                & \mathbf{w}_{A,Z} & \\
                & & \bI^{4{\times}4} \\
                & & & \mathbf{w}_{A,Z} & \\
                & & & & \bI^{4{\times}4}
            \end{bmatrix}
            \bC 
            \eqqcolon {\mathbf{W}}_{\! \! A,Z} \bC,
        \end{equation}
        where it is made explicit that both $d,u$ and $\bar d, \bar u$ are mixed, and it is implied that the upper and lower corners are purely zeroes. The order of the flavor basis used is $g,d,u,...,t,\bar d, \bar u,...,\bar t$. The unities map the unaffected flavors onto themselves.
        While this isoscalarity mixing of the PDFs itself has nothing to do with the NLO corrections we are primarily focused on, we include this detail as we strive towards accurate description of real world data in our inverse problem formalism. In practice this is fortunately just one additional step after we have clarified the NLO accuracy coefficient functions $C_i$, on which we can operate on the weighted average matrix given above.

        Writing the NLO accuracy forward operators for the neutrino DIS observables is conveniently done with the forward operators defined in the NLO CC section in Eqs.~\eqref{eq:fwdop-nlo-cc-f2fl-w+}--\eqref{eq:fwdop-nlo-cc-f3-w-}. For a proton target we simply have:
        \begin{align}
        \sigma^{\nu(\bar \nu)p}_r(x,Q^2,y) & =
        \left[ Y_+ \bB^{W^\pm}_2(x,Q^2) - y^2 \bB^{W^\pm}_L(x,Q^2) \pm Y_-x \bB^{W^\pm}_3(x,Q^2) \right] \pdtbf
        \notag \\
        & \eqqcolon
        \qty(\bA^{\nu(\bar \nu)p}_{r,\nlo} \pdtbf) (x,Q^2).
        \label{eq:fwdop-nlo-nudis-p}
        \end{align}
        We then need to write the observable for a general mixed-isoscalar nuclear target ${}^A_Z\mathrm{N}$ of mass number $A$ and proton number $Z$:
        \begin{align}
        \sigma_r^{\nu(\bar \nu){}^A_Z\mathrm{N}}(x,Q^2,y)
        & =
        \qty({\mathbf{W}}_{\! \! A,Z} \bA^{\nu(\bar \nu)p}_{r,\nlo} \pdtbf) (x,Q^2)
        \eqqcolon
        \qty(\bA^{\nu(\bar \nu){}^A_Z\mathrm{N}}_{r,\nlo} \pdtbf) (x,Q^2).
        \label{eq:fwdop-nlo-nudis-isoscalarmixing}
        \end{align}
        For concreteness, we expand the forward operator:
        \begin{align}
            & {\mathbf{W}}_{\! \! A,Z} \bA^{\nu(\bar \nu)p}_{r,\nlo}
            \notag \\*
            & \quad =
            \begin{bmatrix}
                \Ccal_g ^{\nu(\bar \nu)p} &
                \Ccal_{d}^{\nu(\bar \nu){}^A_Z\mathrm{N}} &
                \Ccal_{u}^{\nu(\bar \nu){}^A_Z\mathrm{N}} &
                \Ccal_s ^{\nu(\bar \nu)p} &
                \cdots &
                \Ccal_t ^{\nu(\bar \nu)p} &
                \Ccal_{\bar d}^{\nu(\bar \nu){}^A_Z\mathrm{N}} &
                \Ccal_{\bar u}^{\nu(\bar \nu){}^A_Z\mathrm{N}} &
                \Ccal_{\bar s} ^{\nu(\bar \nu)p} &
                \cdots &
                \Ccal_{\bar t} ^{\nu(\bar \nu)p}
            \end{bmatrix}
            \label{eq:fwdop-nlo-nudis-isoscalarmixing-WA}
        \end{align}
        where we use the short-hands:
        \begin{align}
            \Ccal_{g}^{\nu(\bar \nu)p}[f]
            & \coloneqq
            \frac{Y_+}{2}\qty( \Ccal_{2,g}[f] - \frac{y^2}{Y_+} \Ccal_{L,g}[f]),
            \\
            \Ccal_{d(\bar d)}^{\nu(\bar \nu){}^A_Z\mathrm{N}}[f]
            & \coloneqq
            \frac{Z}{A}\Ccal_{q,d(\bar d)}^{\nu(\bar \nu)p}[f] +\frac{A-Z}{A}\Ccal_{q, u(\bar u)}^{\nu(\bar \nu)p} [f],
            \\
            \Ccal_{u(\bar u)}^{\nu(\bar \nu){}^A_Z\mathrm{N}}[f]
            & \coloneqq
            \frac{A-Z}{A} \Ccal_{q,d(\bar d)}^{\nu(\bar \nu)p}[f] + \frac{Z}{A} \Ccal_{q,u (\bar u)}^{\nu(\bar \nu)p}[f],
            \\
            \Ccal_{q}^{\nu(\bar \nu)p}[f]
            & \coloneqq
            \Ccal_{q,a}^{\nu(\bar \nu)p}[f],
        \end{align}
        where we use the same definitions of integral operators \eqref{eq:fwdop-nlo-cc-Cq} as given in the NLO CC section, but with the sign of the $F_3$-term changed:
        \begin{align}
        \Ccal_{q,a}^{\nu(\bar \nu)p}[f] & \coloneqq \frac{Y_+}{2} \Ccal_{2,q}[f] \pm (-1)^{\delta_{a \bar a}} \frac{Y_-}{2} x \Ccal_{3,q}[f] - \frac{y^2}{2} \Ccal_{L,q}[f]
        \\
        & = \int_x^1 \qty(\frac{Y_+}{2} C_{2,q}\qty(\frac{x}{z}) \pm (-1)^{\delta_{a \bar a}} \frac{Y_-}{2} x C_{3,q}\qty(\frac{x}{z}) - \frac{y^2}{2} C_{L,q}\qty(\frac{x}{z})) f(z) \frac{\ud z}{z}
        .
        \label{eq:fwdop-nlo-nudis-Cq}
    \end{align}
        The coefficients for all other flavors than $u,d$ are just the proton observable coefficients since gluon and the other flavors of quarks are not affected by the isoscalar mixing. The mixing for the anti-quarks $\bar d, \bar u$ is the same as for $d, u$. This gives the forward operators for nuclear target $\nu$DIS observables tabulated in Tab.~\ref{tab:processes}. Writing the same for $x F_i^{\nu N}$ goes similarly, remembering that $\nu \sim W^+$ and $\bar \nu \sim W^-$, which means that the forward operators are almost the same as the corresponding CC operators, but with isoscalar mixing from the nuclear structure:
        \begin{align}
        F_{i,\nlo}^{\nu{}^A_Z\mathrm{N}}(x, Q^2) &=
        \begin{bmatrix}
            \frac{1}{2} \Ccal_{i,g} &
            \frac{Z}{A}\Ccal_{i,q} &  \frac{A-Z}{A}\Ccal_{i,q} & \Ccal_{i,q} & 0 & \Ccal_{i,q} & 0 &
            \frac{A-Z}{A}\Ccal_{i,q} & \frac{Z}{A}\Ccal_{i,q} & 0 & \Ccal_{i,q} & 0 & \Ccal_{i,q}
        \end{bmatrix}
        \pdtbf
        \notag \\
        &
        \coloneqq
        \qty(\bB_{i,\nlo}^{\nu{}^A_Z\mathrm{N}} \pdtbf) (x,Q^2),
        \label{eq:fwdop-nlo-nudis-f2fl-nu}
        \\
        F_{i,\nlo}^{\bar \nu{}^A_Z\mathrm{N}}(x,Q^2) & = 
        \begin{bmatrix}
            \frac{1}{2} \Ccal_{i,g} &
            \frac{A-Z}{A}\Ccal_{i,q} & \frac{Z}{A}\Ccal_{i,q} & 0 & \Ccal_{i,q} & 0 & \Ccal_{i,q} & 
            \frac{Z}{A}\Ccal_{i,q} &  \frac{A-Z}{A}\Ccal_{i,q} & \Ccal_{i,q} & 0 & \Ccal_{i,q}& 0
        \end{bmatrix}
        \pdtbf
        \notag \\
        &
        \eqqcolon \qty(\bB_{i,\nlo}^{\bar \nu{}^A_Z\mathrm{N}} \pdtbf) (x,Q^2),
        \label{eq:fwdop-nlo-nudis-f2fl-nubar}
    \end{align}
    for $i \in \lbrace 2, L \rbrace$, and
    \begin{align}
        xF_3^{\nu{}^A_Z\mathrm{N}}(x,Q^2) & = 
        \begin{bmatrix}
            0 &
            \frac{Z}{A} \Ccal_{3,q} & \frac{A-Z}{A}\Ccal_{3,q} & \Ccal_{3,q} & 0 & \Ccal_{3,q} & 0 &
            -\frac{A-Z}{A}\Ccal_{3,q} & -\frac{Z}{A} \Ccal_{3,q} & 0 & - \Ccal_{3,q} & 0 & - \Ccal_{3,q}
        \end{bmatrix}
        \pdtbf
        \notag \\
        &
        \coloneqq \qty(\bB_{3,\nlo}^{\nu{}^A_Z\mathrm{N}} \pdtbf) (x,Q^2),
        \label{eq:fwdop-nlo-nudis-f3-nu-w+}
        \\
        xF_3^{\bar \nu{}^A_Z\mathrm{N}}(x,Q^2) & = 
        \begin{bmatrix}
            0 &
            \frac{A-Z}{A}\Ccal_{3,q} & \frac{Z}{A}\Ccal_{3,q} & 0 & \Ccal_{3,q} & 0 & \Ccal_{3,q} & 
            -\frac{Z}{A} \Ccal_{3,q} & -\frac{A-Z}{A}\Ccal_{3,q} & - \Ccal_{3,q} & 0 & - \Ccal_{3,q}& 0
        \end{bmatrix}
        \pdtbf
        \notag \\
        &
        \coloneqq \qty(\bB_{3,\nlo}^{\bar \nu{}^A_Z\mathrm{N}} \pdtbf)(x,Q^2),
        \label{eq:fwdop-nlo-nudis-f3-barnu-w-}
    \end{align}
    for the parity-breaking $F_3$. The key difference with these compared to those of the CC case is the isoscalar mixing between the up and down flavors.
    With these we have all the necessary forward operators required to include $\nu$DIS in the coupled inverse problem at NLO accuracy.

\section{Parton distribution function reconstruction inverse problem at NLO} \label{sec:tensor-ip-nlo}

    In the previous section we reviewed the literature of NLO accuracy description of experimental observables of lepton--proton DIS. In this section we gather them into the coupled inference problem of \gls{global analysis}, where the task is to reconstruct the PDFs concurrently from the available \gls{world data}. In contrast with the LO accuracy toy problem discussed in Sec.~\ref{sec:tensor-ip-lo}, at NLO accuracy all elements of the forward operator become linear integral operators, and the full world data inference problem can be written into the form of a matrix valued integral equation. It would appear that the matrix valued integral equation has some novel features not inspected mathematically previously, as the integral kernel elements can contain plus-distributions, and take the form of different kinds of integral equations. To the best of our knowledge, mathematical theory for matrix valued integral equations has seemed to assume the overall matrix to have common properties~\cite{Polyanin2008handbook:integral-equations,Moslehian:2023:matrix-operator-equations-applications-integral-equations,GohbergKrein1958:wiener-hopf-systems}, whereas now especially the Wilson coefficient convolutions and sum rule integral equation constraints appear quite distinct.
        
    At NLO accuracy the world data inference problem of the PDFs becomes a coupled system of implied integral equations:
    \begin{equation}
    \label{eq:tensor-ip-nlo}
        \mathbf{O}^\mathrm{WD} \coloneqq
        \begin{bmatrix}
            \mathbf{O}^\sigma \\
            \mathbf{O}^F \\
            \mathbf{O}^S
        \end{bmatrix} = 
        \int\mathbf{G}^{lp\to l' X}_\mathrm{NLO} \pdtbf,
    \end{equation}
    where the integral implies that the world data forward operator acting on the PDFs $\pdtbf$ requires the element-wise convolution of the forward operator with the PDFs. The observable modalities of the cross  sections $\mathbf{O}^\sigma$, structure functions $\mathbf{O}^F$, and sum rules $\mathbf{O}^S$ are defined analogously with the LO case.
    At \gls{next-to-leading order}, the cross section observable modality is constructed from the forward operators defined by Eqs.~\eqref{eq:fwdop-incdis-nlo}, \eqref{eq:fwdop-nlo-nc-cs}, \eqref{eq:fwdop-nlo-cc-csecred}, \eqref{eq:fwdop-nlo-hq-cc}, \eqref{eq:fwdop-nlo-hq-bb}, and~\eqref{eq:fwdop-nlo-nudis-isoscalarmixing}; suppressing here all labels specifying ``NLO'' for clarity:
    \begin{equation}
        \label{eq:tensor-ip-observable-block-cs-nlo}
        \mathbf{O}^\sigma \coloneqq
        \begin{bmatrix}
            \sigma^\gamma_r(x,Q^2,y) \\
            \sigma^{+}_{r,\mathrm{NC}}(x,Q^2,y) \\
            \sigma^{-}_{r,\mathrm{NC}}(x,Q^2,y) \\
            \sigma_{r,\mathrm{CC}}^{+}(x,Q^2,y) \\
            \sigma_{r,\mathrm{CC}}^{-}(x,Q^2,y)\\
            \sigma^{c\bar c}_{r}(x,Q^2,y)\\
            \sigma^{b\bar b}_{r}(x,Q^2,y)
            \\ \sigma_r^{\nu {}^A_Z \mathrm{N}}(x,Q^2,y)
            \\ \sigma_r^{\bar \nu {}^A_Z\mathrm{N}}(x,Q^2,y)
        \end{bmatrix}
        =
        \int
        \begin{bmatrix}
            \bA_{r}^{\gamma} (x,Q^2,y) \\
            \bA_{r,\mathrm{NC}}^{+}(x,Q^2,y) \\
            \bA_{r,\mathrm{NC}}^{-}(x,Q^2,y) \\
            \bA_{r,\mathrm{CC}}^{+} (x,Q^2,y) \\
            \bA_{r,\mathrm{CC}}^{-} (x,Q^2,y)
            \\ \bA_{r}^{c \bar c}(x,Q^2,y)
            \\ \bA_{r}^{b \bar b}(x,Q^2,y)
            \\ \bA_r^{\nu(\bar \nu){}^A_Z\mathrm{N}} (x,Q^2,y)
            \\ \bA_r^{\nu(\bar \nu){}^A_Z\mathrm{N}} (x,Q^2,y)
        \end{bmatrix}
        \pdtbf(\cdot,Q^2).
    \end{equation}
    The structure function observable modality constructed with the Eqs.~\eqref{eq:fwdop-incdis-nlo-fl}, \eqref{eq:fwdop-nlo-nc-f3}, \eqref{eq:fwdop-nlo-nudis-f2fl-nu}, \eqref{eq:fwdop-nlo-nudis-f2fl-nubar}, \eqref{eq:fwdop-nlo-nudis-f3-nu-w+}, and~\eqref{eq:fwdop-nlo-nudis-f3-barnu-w-}:
    \begin{equation}
        \label{eq:tensor-ip-observable-block-F-nlo}
        \mathbf{O}^F \coloneqq
        \begin{bmatrix}
            F_L^\gamma (x,Q^2) \\
            xF_3^\mathrm{NC} (x,Q^2) \\
            F_2^{\nu N} (x,Q^2) \\
            xF_3^{\nu N} (x,Q^2) \\
            F_2^{\bar \nu N} (x,Q^2) \\
            xF_3^{\bar \nu N} (x,Q^2)
        \end{bmatrix}
        =
        \int
        \begin{bmatrix}
            \bB^{\gamma}_{L} (x, Q^2) \\
            \bB_{3, \mathrm{NC}}^{\pm} (x,Q^2)\\
            \bB_{2}^{\nu{}^A_Z\mathrm{N}} (x)\\
            \bB_{3}^{\nu{}^A_Z\mathrm{N}}(x)\\
            \bB_{2}^{\bar \nu{}^A_Z\mathrm{N}} (x)\\
            \bB_{3}^{\bar \nu{}^A_Z\mathrm{N}}(x)
        \end{bmatrix}
        \pdtbf(\cdot,Q^2),
    \end{equation}
        and the sum rule modality is formed from Eqs.~\eqref{eq:fwdop-sumrule-fnc}, \eqref{eq:fwdop-sumrule-tmc}, \eqref{eq:fwdop-sumrule-GLS}, \eqref{eq:fwdop-sumrule-adler}, \eqref{eq:fwdop-sumrule-gottfried}:
    \begin{equation}
        \label{eq:ip-sumrule-part-nlo}
        \mathbf{O}^S \coloneqq
        \begin{bmatrix}
            S_\mathrm{t.m.c.}(Q^2)
            \\ S_\mathrm{f.n.c.}(Q^2)
            \\ S_\mathrm{GLS}(Q^2)
            \\ S_\mathrm{A}(Q^2)
            \\ S_\mathrm{G}(Q^2)
        \end{bmatrix}
        =
        \int
        \begin{bmatrix}
            \Ical_\mathrm{t.m.c.}
            \\ \Ical_\mathrm{f.n.c.}
            \\ \Ical_\mathrm{GLS}
            \\ \Ical_\mathrm{A}
            \\ \Ical_\mathrm{G}
        \end{bmatrix}
        \pdtbf(\cdot,Q^2).
    \end{equation}
    The sum rule modality is exactly the same as at leading order in Eq.~\eqref{eq:ip-sumrule-part}, since the forward operators that define the sum rules do not get perturbative corrections themselves, thanks to the rules being defined as integral constraints directly on the PDFs. The perturbative corrections would go to the expected values of the observables that is compared against experimental data $\mathbf{O}^S$, so since we directly use the experimental data as input, we forgo these corrections. We note that the fundamental sum rules such as the total momentum conservation rule are also included in the sum rule modality as a theoretical constraint, and it does not receive any perturbative corrections either. In that case the experimental datapoints are replaced by the precise theoretical constant given by the sum rule, for example $S_\tmc(Q^2) \equiv 1$ at all $Q^2$.

    To make the integral equation nature of Eq.~\eqref{eq:tensor-ip-nlo} apparent, let us bring the integral over $z$ out:
    \begin{align}
    \label{eq:tensor-ip-nlo-integral}
        \mathbf{O}^\mathrm{WD} & = \int_0^1 \mathbf{G}^{lp\to l' X}_\mathrm{NLO}(z,x,Q^2,y) \pdtbf(z,Q^2) \frac{\ud z}{z}
        \notag \\
        \Leftrightarrow
        \begin{bmatrix}
            \sigma^\gamma_r(x,Q^2,y)
            \\
            \vdots
            \\
            F^\gamma_L (x,Q^2)
            \\
            \vdots
            \\
            S_\mathrm{t.m.c.}(Q^2)
            \\
            \vdots
        \end{bmatrix}
        & = \int_0^1 \begin{bmatrix}
            \undertilde{e}^2 \mathcal{I}_{2-L,g} \theta_x & e_d^2 \Ical_{2-L,q} \theta_x & \cdots &  e_{\bar t}^2 \Ical_{2-L,q} \theta_x
            \\
            \vdots& \ddots & \cdots & \vdots
            \\
            \undertilde{e}^2 \Ical_{L,g} \theta_x & e_d^2 \Ical_{L,q} \theta_x & \cdots & e_{\bar t}^2 \Ical_{L,q} \theta_x \\
            \vdots & \ddots & \cdots & \vdots \\
            z^2 & z^2 & \cdots & z^2 \\
            \vdots & \ddots & \cdots & \vdots
        \end{bmatrix} \pdtwritoutz
        \frac{\ud z}{z},
    \end{align}
    where we include the sum rules under the same operation, and the implied integration of the integral operators $\Ical$ is replaced by the unified integration over the full matrix problem. The full form of the matrix operator is given in Eq.~\eqref{eq:app:full-fwdop-G-nlo} in appendix~\ref{app:full-global-data-operators}. The above implies that some of the forward operators must be amended with a step function $ \theta_x \coloneqq \theta(z-x)$ which keeps the integration regime consistent with the physical definition, be it $z_\mathrm{min} = x$ for massless parton DIS, or other like seen with the heavy quark production case. 

    It is time to return to the matter of the implementation of a full \gls{flavor number scheme} as outlined in Sec.~\ref{sec:obs-as-linear-ops}. Equation~\eqref{eq:tensor-ip-nlo-integral}---aside from the heavy flavor production observables---is written in the zero mass approximation for $n_f=6$ flavors, i.e. for $Q^2 > m_t^2$. To formalize the inverse problem in the regimes of lower $Q^2$, we need to partition both the world data $O^\mathrm{WD}$ and forward operator $\mathbf{G}^{lp\to l' X}_\mathrm{NLO}$ by the transitions in the number of light flavors as given by the $Q^2$ thresholds defined by the FNS given in Eq.~\eqref{eq:toy-model-flavor-number-scheme-F3cb+ZM}. With an implemented FNS, the world data inference problem Eq.~\eqref{eq:tensor-ip-nlo-integral} becomes:
    \begin{equation}
    \label{eq:tensor-ip-nlo-integral-fns}
        \mathbf{O}^{\mathrm{WD}}_\mathrm{FNS} \coloneqq
        \begin{bmatrix}
            \mathbf{O}^{\mathrm{WD}}_{Q^2<m_b^2} \\
            \mathbf{O}^{\mathrm{WD}}_{m_b^2 \leq Q^2<m_t^2} \\
            \mathbf{O}^{\mathrm{WD}}_{m_t^2 \leq Q^2}
        \end{bmatrix} = 
        \int
        \begin{bmatrix}
            \mathbf{G}^{lp\to l' X}_{Q^2<m_b^2} \\ 
            \mathbf{G}^{lp\to l' X}_{m_b^2 \leq Q^2<m_t^2} \\
            \mathbf{G}^{lp\to l' X}_{m_t^2 \leq Q^2}
        \end{bmatrix}
        \pdtbf
        \eqqcolon
        \int
        \mathbf{G}^{lp\to l' X}_\mathrm{FNS}
        \pdtbf,
    \end{equation}
    where it is implied that \textit{all included observables} of a partitioned $\mathbf{G}^{lp\to l' X}_{Q^2 \in I_i}$ should be written using the prescribed number of available parton flavors and their associated masses. In the used toy-model example~\eqref{eq:toy-model-flavor-number-scheme-F3cb+ZM} in the lowest interval $Q^2<m_b^2$, the scheme describes charm and bottom with masses, whereas in the higher two partitions, all flavors are massless and only the number of flavors changes based on the kinematic possibility of top production. Instead of repeating all expressions explicitly for this toy-model construction, we will discuss in Sec.~\ref{sec:flavor-number-schemes} how the more advanced flavor number schemes differ at the observable description level, but ultimately this partitioning of the world data into distinct regimes is largely done in this manner.

    Equation~\eqref{eq:tensor-ip-nlo-integral-fns} now expresses a coupled integral equation formulation of the problem of global analysis of the PDF from DIS observables and sum rules, and is the culmination of this mathematical inverse problems approach: the problem of global analysis of the PDFs is given a mathematical form in terms of elementary structures which can be leveraged to solve the inference problem without fitting a model ansatz for the PDFs. And by construction, the elements of the kernel matrix are explicitly related to the canonical pQCD results from the literature.
    This matrix integral equation form for example enables one to consider the rank of the operator matrix, and in that sense the uniqueness of solutions that one can attain at a given fixed order in perturbative QCD. At higher order in the $\as$ expansion the varying flavor dependent operator elements begin to receive more flavor specific corrections~\cite{Blumlein:2022gpp:nnnlo-f2-f3,Moch:2007rq:cc-nnlo}, so it is possible the rank of the problem in the cross section and structure function modalities improves at higher order. A foundation question for the problem of global analysis in general is, how much of the PDFs it is possible to reconstruct from data without any modeling assumptions made of the specific form of the PDFs.
    
    In a mathematical sense this integral equation formulation puts the Wilson coefficient functions into the same role as that of the kernel of the X-ray transform~\eqref{eq:xray-transform}. The mathematical structure changes qualitatively from the LO case of Sec.~\ref{sec:tensor-ip-lo}, since there all of the row operators were not of the integral form, whereas at NLO and beyond the problem has this form. This seems to be a new type of a coupled system of integral equations problem, since the cross section and structure function observables have distinctly different structure than the sum rule equations, and the high-level problem is to solve them simultaneously. In Sec.~\ref{sec:integral-eq-theory-wiener-hopf} we briefly examine from a mathematical perspective the solution of the former kind of integral equation.
    A general method to solve an inverse problem of the form of Eq.~\eqref{eq:tensor-ip-nlo-integral} is constructed in Sec.~\ref{sec:discretization}.

    To conclude this section, we formally state the optimization problem of global analysis as the system of integral equations~\eqref{eq:tensor-ip-nlo-integral-fns} that constrain the PDFs via experimental data, and the DGLAP system of PDEs~\eqref{eq:P-operator-dglap-singlet-basis} which the solutions of the former must satisfy within perturbative accuracy:
    \begin{equation}
    \begin{split}
    \label{eq:pde-dglap-constrained-world-data-optimization}
        &\underset{\pdtbf, \mathbf{y}}{\minimize} ~
        \norm{ \mathbf{O}^\mathrm{WD}_\mathrm{FNS} -
        \int\mathbf{G}^{lp\to l' X}_\mathrm{FNS} \pdtbf }_{L^2(\Omega)}^2 + \lambda \norm{\mathbf{y}}_{L^2(\Omega)}^2
        ,
        \\
        &\text{subject to}~
        \mathbf{y} \coloneqq
        \frac{\partial}{\partial \log Q^2} \pdtbf(x,Q^2) - \mathbf{P}^\mathrm{DGLAP}_\mathrm{FNS} \pdtbf
        ,
    \end{split}
    \end{equation}
    where $\lambda$ is a regularization parameter which is basically a relative weight between the two minimization terms.
    Equation~\eqref{eq:pde-dglap-constrained-world-data-optimization} is a functional optimization problem for the PDFs $\pdtbf$ and the control variable $\mathbf{y}$, which controls how accurately a reconstructed solution $\tilde \pdtbf$ of the system~\eqref{eq:pde-dglap-constrained-world-data-optimization} satisfies the DGLAP equation, i.e. $\mathbf{y}=\mathbf{y}(x,Q^2)\equiv0 \, \forall \, x,Q^2$ would correspond to perfect agreement. 
    The FNS-unified evolution operator $\mathbf{P}^\mathrm{DGLAP}_\mathrm{FNS}$ has a flavor number scheme implicitly built in: within each $Q^2$-regime $I_i$ of the FNS $\lbrace I_i\rbrace_{i = \lbrace1,2,3\rbrace}$ of~\eqref{eq:toy-model-flavor-number-scheme-F3cb+ZM} the number of light flavors $n_f$ changes, so the number of active flavors contributing to DGLAP evolution also changes. This means that the DGLAP evolution equation Eq.~\eqref{eq:P-operator-dglap-singlet-basis} has to be written in each regime $I_i$ separately, and the full constrained problem then implies that the specific DGLAP equation of FNS-regime $I_i$ has to be satisfied by the reconstructed solution $\tilde \pdtbf$ within perturbative accuracy.
    Since the DGLAP equation has to be implemented at a fixed order in perturbative QCD, the solution reconstructed from real physical data is not expected to perfectly satisfy the perturbatively imperfect theory. The former term is the so-called ``data fitting'' term, which minimizes the theory prediction of the reconstruction $\tilde \pdtbf$ against the experimental world data, which is the analogue of for example the $\chi^2$ test used in fitting to quantify the quality of fit.
    
    We note here that it is completely possible---if not even preferable---to separate the theoretical sum rule constraints into a separate optimization term with its own weight $\lambda_2$ and control parameter $\mathbf{w} \coloneqq \mathbf{O}^S - \mathbf{G}^S \pdtbf$ constructed from the total momentum and flavor number conservation rules, for example. This enables stricter enforcement of the theoretical sum rule constraints, whereas the data has both experimental and forward operator uncertainty, which means that the data optimization term is not expected to be minimized as precisely as the sum rules should be satisfied.
    
    The as yet undetermined domain of the problem $\Omega$ is an open set in which the optimization problem is defined, which would be a set that encompasses the datapoints included in the world data in the experimental variables $x$, $Q^2$, $y$, and $s$. Equation~\eqref{eq:pde-dglap-constrained-world-data-optimization} now defines a functional optimization problem in the same vein as used in many applied fields~\cite{leugering2014trends:pde-opti,Biegler-pde-opti}, and a perfect optimizer $\tilde \pdtbf^*$ would yield exactly zero from both terms, i.e. it would perfectly predict the data, whilst also perfectly satisfying the DGLAP equation. A proof of the existence and uniqueness of such a solution---within perturbative accuracy---of the problem~\eqref{eq:pde-dglap-constrained-world-data-optimization} would be a promising indication for the viability of practical reconstruction of the PDFs from real world data. Such a proof is left for future work.

\section{Functional linearity beyond NLO and collinear factorization} \label{sec:nnlo}

	In this section we first concretely examine the preservation of the functional linearity at next-to-next-to-leading order (NNLO) in pQCD, and then in the following subsection examine more generally the origin of the functional linearity at higher orders in perturbative QCD. Lastly, in the third subsection we take a brief look at an even more general case, and whether the linearity is preserved even then. Applicability of this inverse problems approach at higher orders in the perturbative expansion of QCD is critical, since the state-of-the-art of global analysis of the PDFs is at next-to-next-to-next-to-leading order (NNNLO, N$^3$LO) accuracy~\cite{Gao:2017yyd:global-analysis-modern-rev,Ethier:2020way:global-analysis-modern-rev,Amoroso:2022eow:global-analysis-modern-rev,Huston:2023ofk:global-analysis-modern-rev}.

	\subsection{Inclusive DIS at next-to-next-to-leading order}


    We will overview the broad mathematical shape of the theory side of the inference problem for \gls{inclusive} electromagnetic DIS to conclude that the present approach applies at NNLO accuracy~\cite{Zijlstra:1992qd,vanNeerven:1991nn,Zijlstra:1991qc,SanchezGuillen:1990iq,Moch:1999eb,Moch:2004xu,Vermaseren:2005qc,Blumlein:2022gpp:nnnlo-f2-f3} as well, which is in fact due to the collinear factorization theorem of perturbative QCD, which we then consider in the following subsection. To see this, we write the general expression for the inclusive DIS reduced cross section~\cite{Moch:1999eb,Moch:2004xu,Vermaseren:2005qc}:
    \begin{align}
        \sigma^\gamma_r(x, Q^2,y)
        \pequiv {} & 
        \sum_a e_a^2 \int_x^1 x f_{a} \qty(\frac{x}{z}, Q^2)
        \Bigg( C_{2,q}(z)_+ 
        - \frac{y^2}{Y_+} C_{L,q}(z)\Bigg) \frac{\ud z}{z}
        \notag \\
        & + \sum_a e_a^2 \int_x^1 x f_{g} \qty(\frac{x}{z}, Q^2)
        \Bigg( C_{2,g}(z)_+ 
        -\frac{y^2}{Y_+} C_{L,g}(z) \Bigg) \frac{\ud z}{z}
        ,
        \label{eq:sigma-incl-nnlo}
    \end{align}
    where the coefficient functions are written as expansions in $\as$, so up to order $\as^n$ ($N^nLO$):
    \begin{equation}
        \label{eq:coeff-pqcd-expansion-as}
        C_{i,a}(z) \coloneqq \sum_{m=0}^n \left(\frac{\as}{4 \pi}\right)^m c^{(m)}_{i,a}(z).
    \end{equation}
    Thus to write the observable reduced cross section $\sigma^\gamma_r$ at NNLO accuracy, the coefficient functions $c^{(m)}_{i,a}(z)$ are needed for $a \in \lbrace g,q \rbrace$, $i \in \lbrace 2,L \rbrace$, and $m \in \lbrace 0,1,2 \rbrace$. The lower order coefficients for $m=0$ and $m=1$ have been discussed in Sec.~\ref{sec:nlo-inclusive-gamma-dis}, whereas the $m=2$ coefficients are given in appendix~\ref{app:nnlo} due to their extensive size. However, we note that this mathematical inverse problem approach exists at all because the coefficient functions are known, and as their mathematical structure is of significance for higher order application of this methodology, we record all coefficients from the literature. As we develop the world data inference problem formalism, we are mindful of the precise mathematical structures that are present in the problem.

    To write Eq.~\eqref{eq:sigma-incl-nnlo} in the forward operator form, we define the integral operators
    \begin{align}
        \Ccal_q[f] & \coloneqq \int_x^1 x f (z, Q^2) 
            \Bigg( C_{2,q}\qty(\frac{x}{z})_+ - \frac{y^2}{Y_+} C_{L,q}\qty(\frac{x}{z})\Bigg) \frac{\ud z}{z},
        \\
        \Ccal_g[f] & \coloneqq \int_x^1 x f (z, Q^2) 
            \Bigg( C_{2,g}\qty(\frac{x}{z})_+ - \frac{y^2}{Y_+} C_{L,g}\qty(\frac{x}{z})\Bigg) \frac{\ud z}{z}.
    \end{align}
    With these the general expression for the reduced cross section~\eqref{eq:sigma-incl-nnlo} is at an arbitrary order in the perturbative expansion:
    \begin{align}
        &\sigma^\gamma_r(x, Q^2,y) =
        \begin{bmatrix}
            \undertilde{e}^2 \Ccal_{g} &
            e_d^2 \Ccal_{q} &
            \cdots &
            e_{\bar t}^2 \Ccal_{q}
        \end{bmatrix}
        \pdtwritoutzdot
        \eqqcolon
        \qty(\bA^{\gamma,(n)}_r \pdtbf) (x, Q^2,y),
    \end{align}
    where $\undertilde{e}^2$ was defined in Eq.~\eqref{eq:e-undertilde-frac-charge-vec}.
    Here we can see that the corrections in the $\as$-expansion just become correction terms to the elements of the forward integral operator, and that the overall form is the same we saw at NLO in Eq.~\eqref{eq:fwdop-incdis-nlo}.

    A key motivation to collect explicitly the coefficient functions in a unified form from the literature is that their properties as integral operator kernels are of significance for the mathematical foundations of this approach. It is also relevant for the mathematical inverse problems approach to have an explicit relationship between the unknown quantities and the experimental data so that the structure of the mathematical relationship can be analyzed and leveraged to solve the unknown quantities from the data with bespoke methods.
    Some of the coefficient functions $C_i(z)$ exhibit singularities for example at $z \to 1$, which can either be useful or a challenge depending on the nature of the singularity~\cite{hochstadt1989integraleq,kress1989linear-inteq-methods-quadrature}. Precise mathematical inspection of these features is out of the scope of this work, which is more in the vein of collecting the necessary results in a unified form, to enable this work in the future.
    These details fortunately are not necessary for the application of the general purpose solution algorithm constructed in Sec.~\ref{sec:discretization}.

\subsection{Collinear factorization and functional linearity}

	To consider the general applicability of this integral equation inverse problems approach to global analysis beyond NLO, we seek to identify the origin of the functional linearity. The cross section observables that we have considered are known to be linearly dependent on the proton structure functions $F_2^b$, $F_L^b$, and $F_3^b$, for the boson $b \in \lbrace \gamma, Z, W^\pm\rbrace$. For these type of linear observables we can then consider whether the structure functions $F_i^b$ are linear integral operators for the PDFs generally. This is provided by the collinear factorization, which states that the structure functions can be generally factorized as~\cite{Collins:1987pm:collinear-factorization,Collins:1989gx:collinear-factorization-long}
	\begin{equation} \label{eq:thrm:coll-factorization}
        F_i^b(x,Q^2) \pequiv \sum_a \int_x^1 C_{i,a}^b\left(\frac{x}{z}, \frac{Q^2}{\mu^2}, \as(\mu^2)\right) f_{a,H}(z,\mu^2) \frac{\ud z}{z} + O(1 \, \mathrm{\gls{GeV}} /Q),
    \end{equation}
    where $C_{ia}$ are integration kernels known as Wilson coefficients, DIS coefficients, or hard scattering functions, and $f_{a,H}$ is the PDF for parton flavor $a$ in a \gls{hadron} $H$. The Wilson coefficients $C^b_{ia}$ are specific to the physical process, and they can exhibit varying divergences at least as the function of $\frac{x}{z}$. The final term implying corrections on the scale of $1 \, \mathrm{\gls{GeV}} / Q$ come from higher order terms in the so-called twist expansion~\cite{Braun:2022gzl:rev-twist,Schienbein:2007gr:target-mass-corrections}, which we briefly overview in Sec.~\ref{sec:twist-tmc}, and $\mu$ is the factorization scale which can be chosen to be $\mu^2 \equiv Q^2$ for DIS~\cite{Collins:1987pm:collinear-factorization,Collins:1989gx:collinear-factorization-long}, and which is typically chosen to be always exactly $Q^2$ in global analysis for all processes. 
	For a more technical discussion why this hold, we refer to Refs.~\cite{Collins:1987pm:collinear-factorization,Collins:1989gx:collinear-factorization-long}. Its surface level interpretation is that scattering off hadronic targets is dominated by short distance interaction with localized partons, and generally all flavors of partons contribute, although the dependence of which flavors do depends on the practical details of the scattering, such as the energy scale and nature of the probe.

    The linear functional structure of the collinear factorization~\eqref{eq:thrm:coll-factorization} means that the approach of inference developed in this work is broadly applicable to observables of this form, for example to the lepton-proton DIS processes considered. And the general structure of the perturbative expansion of the coefficient functions in Eq.~\eqref{eq:coeff-pqcd-expansion-as} means that the theory can be implemented at any order in $\as$. Observables of this form can be included in the coupled system of integral equations formulated in Sec.~\ref{sec:tensor-ip-nlo}, and therefore at any order in perturbative QCD fall within the scope of the solution algorithm discussed in Sec.~\ref{sec:discretization}.

    For completeness, let us write the collinear factorization of $F_i$ into the forward operator form, for which we set $\mu^2 \equiv Q^2$ and define:
    \begin{equation}
        \Ccal^b_{i,a}[f](x,Q^2) \coloneqq \int_x^1 C^b_{i,a}\left(\frac{x}{z}, \as(Q^2)\right) f(z,Q^2) \frac{\ud z}{z}.
    \end{equation}
    With this one can write the leading twist collinear factorization \eqref{eq:thrm:coll-factorization} into the row-form integral operator:
    \begin{equation}
        F^b_i(x,Q^2) = 
        \begin{bmatrix}
            \Ccal^b_{i,g} &
            \Ccal^b_{i,d} &
            \cdots &
            \Ccal^b_{i,\bar t}
        \end{bmatrix}
        \pdtwritoutzdot
        \eqqcolon
        \qty(\bB^b_i \pdtbf) (x,Q^2),
    \end{equation}
    which is of the linear form of Eq.~\eqref{eq:ip-general-linear}. Since this factorized form stems directly from the collinear factorization \eqref{eq:thrm:coll-factorization}, it holds to all orders in the perturbative expansion in $\as$, and the corrections affect the individual elements $\Ccal_{i,a}$. As long as renormalization produces coefficient functions that are integrable kernels in a distributional sense for the parton distribution functions, this integral operator approach is applicable.
    This would seem to be a promising indication of the feasibility of the application of this approach at any order in perturbative QCD and the practical usage of existing theory calculations after this straightforward re-factorization process.
    In Sec.~\ref{sec:discretization} we construct a general purpose methodology for the numerical solution of these types of integral equation inverse problems using multiple measurement modalities simultaneously.

\subsection{Functional linearity beyond leading twist: target mass corrections} \label{sec:twist-tmc}

    So far we have only considered the high-$Q^2$ limit of the forward operators in perturbative QCD, also known as leading twist. So-called higher twist effects are perturbative corrections suppressed by powers of $\left(\frac{1}{Q^2}\right)^n$. An important class of physical effects are the target mass corrections (TMCs)~\cite{Schienbein:2007gr:target-mass-corrections} which arise as corrections on the order $\left(\frac{M^2}{Q^2}\right)^n$, where $M$ is the mass of the target like the proton. In a realistic inference of the PDFs from real data, these corrections can have a substantial effect: for example at $x = 0.8$ the TMC effect is $\sim 30 \%$ at $Q^2 = 5 \, \mathrm{\gls{GeV}}^2$~\cite{Schienbein:2007gr:target-mass-corrections}. Thus, also the TMC effects become essential to consider, as one is striving for the highest accuracy with NNLO and N$^3$LO corrections in perturbative QCD. We limit this initial examination only to the TMC effects, as the full spectrum of higher twist physics~\cite{Braun:2022gzl:rev-twist} is much too broad to discuss in this work.

    To inspect how TMCs affect the mathematical structure of the inference problem, we will consider as a toy-model representative of TMCs their implementation to charm quark production in neutrino--nucleon DIS discussed in Ref.~\cite{Schienbein:2007gr:target-mass-corrections}. Taking only leading contribution to $F_{2,c}$, the observable is written with TMC as:
    \begin{align}
        F_{2,c}^\mathrm{TMC}(x,Q^2) \simeq {} &
        \frac{2x^2}{\bar \xi r^3}\left(1 + \frac{m_c^2}{Q^2}\right)^2 f_s(\bar \xi)
        + \frac{12 M^2 x^3}{Q^2 r^4} \left(1 + \frac{m_c^2}{Q^2}\right) \int_{\bar \xi}^1 \frac{f_s(u)}{u} \ud u
        \notag \\
        &
        + \frac{24 M^4 x^4}{Q^4 r^5} \int_{\bar \xi}^1 \int_{u}^1 \frac{f_s(v)}{v}\ud v \ud u,
        \label{eq:f2-tmc}
    \end{align}
    where the shorthands used are:
    \begin{align}
        \xi & = \xi(x,Q^2,M^2) \coloneqq \frac{2x}{1+ \sqrt{1+4x^2 \frac{M^2}{Q^2}}},
        \\
        \bar \xi & = \bar \xi(x,Q^2,M^2) \coloneqq R_{ij} \xi(x,Q^2,M^2) \approx \left(1 + \frac{m_c^2}{Q^2}\right) \xi(x,Q^2,M^2),
        \\
        r & = r(x,Q^2,M^2) \coloneqq \sqrt{1 + \frac{4 x^2 M^2}{Q^2}}.
    \end{align}
    It will be convenient to perform a standard change of integration order to reduce the double integral of the last term into a single one~\cite{Schienbein:2007gr:target-mass-corrections}:
    \begin{equation*}
        \int_{\bar \xi}^1 \int_u^1 \frac{f(v)}{v} \ud v \ud u = \int_{\bar \xi}^1 (v-\bar \xi) \frac{f(v)}{v} \ud v.
    \end{equation*}
    We define the integral operations:
    \begin{align}
        \Ical_{\bar \xi}[f](\bar \xi) & \coloneqq \int_0^1 \delta(z-\bar \xi) f(z) \ud z,
        \\
        \Ical_u[f](z) & \coloneqq \int_{z}^1 \frac{f(u)}{u} \ud u,
        \\
        \Ical_v[f](z) & \coloneqq \int_{z}^1 (v - \bar \xi) \frac{f(v)}{v} \ud v.
    \end{align}
    With these we rewrite Eq.~\eqref{eq:f2-tmc} as
    \begin{align}
        F_{2,c}^\mathrm{TMC}(x,Q^2) ={} &
        \Bigg[ \frac{2x^2}{\bar \xi r^3}\left(1 + \frac{m_c^2}{Q^2}\right)^2 \Ical_{\bar \xi}
        + \frac{12 M^2 x^3}{Q^2 r^4} \left(1 + \frac{m_c^2}{Q^2}\right) \Ical_{\bar \xi} \Ical_u
        + \frac{24 M^4 x^4}{Q^4 r^5} \Ical_{\bar \xi} \Ical_v \Bigg] f_s(\cdot),
    \end{align}
    where the notation $f_s(\cdot)$ intentionally leaves the parameter implicit, as that goes into the definition of the integral operators above, and will be practically handled by the discretization method in Sec.~\ref{sec:discretization}. This has been merely a brief introductory glance into the mathematical structure of the target mass corrections and how they fit into the overall inverse problem approach to global analysis, but this would seem like a positive indication for the feasibility of the approach even in practice. The high-level picture is that the TMCs introduce a new linear functional operation on top of the perturbative QCD corrections we have discussed. Systematical implementation of TMCs necessitates the introduction of these corrections to all observables, but that discussion is beyond the scope of this work.

    \section{Wilson coefficient convolutions as Wiener--Hopf integral equations}
    \label{sec:integral-eq-theory-wiener-hopf}


    A foundational question about the viability of this approach of global data inference of the PDFs that we have not examined so far is that of the existence of solutions for the Wilson coefficient function integral equations \eqref{eq:wilson-convol} defined in Eq.~\eqref{eq:convolution-C-cov-z}. We will inspect this existential question in the light of Wiener--Hopf equations~\cite{Wiener1931:wiener-hopf-original,Hopf1934:wiener-hopf-original,Kisil:2021:wiener-hopf-methods, Lawrie2007:wiener-hopf-review, Corduneanu:1973:wiener-hopf}, which can be of first or second-kind:
    \begin{align}
        \int_0^\infty k(x-y) f(y) \ud y & = g(x), \quad x \in (0,\infty),
        \label{eq:wiener-hopf-first}
        \\
        f(x) - \int_0^\infty k(x-y) f(y) \ud y & = g(x), \quad x \in (0,\infty),
        \label{eq:wiener-hopf-second}
    \end{align}
    where $k(x-y)$ is a known difference kernel, $g(x)$ is a function known from measurement, and $f$ is the unknown function to be solved. These are of interest because precise understanding of existence and analytical solution methods are known, but they can depend on the nature of the difference kernel. A standard result is that if the kernel is $L^1$-integrable, it suffices to verify that the Fourier transform of the difference kernel does not go to unity anywhere. See Refs.~\cite{Kisil:2021:wiener-hopf-methods, Lawrie2007:wiener-hopf-review, Corduneanu:1973:wiener-hopf} for reviews on the topic.
    
    We will inspect the NLO DIS Wilson coefficient convolutions as Wiener--Hopf equations to preliminarily evaluate whether they can be analytically solvable as integral equations, even on the parton flavor level of the problem.
    The convolution equations---where the convolution $C\otimes f$ plays the role of the function known from data---are generally of the form:
    \begin{equation}
        (C \otimes f)(x) = \int_x^1 C\qty(\frac{x}{z})f(z) \frac{\ud z}{z}.
    \end{equation}
    If we do a change of variable:
    \begin{equation}
        \frac{x}{z} = e^{\ln{x/z}} = e^{\ln(x) - \ln(z)} \eqqcolon e^{\tilde x - \tilde z}
    \end{equation}
    we can change the dependence of the kernel function on the ratio of the arguments $x,z$ into a difference of their logarithms $\tilde x - \tilde z$, which is of the form seen in the Wiener--Hopf equations~\eqref{eq:wiener-hopf-first} and~\eqref{eq:wiener-hopf-second}.
    With this change the Wilson coefficient convolution becomes:
    \begin{equation}
        (C \otimes f)(x) = \int_x^1 C(e^{\tilde x - \tilde z})f(z) \frac{\ud z}{z}
    \end{equation}
    which when we change the integration variable to $\tilde z = \ln z$, can be written as:
    \begin{equation}
        (C \otimes f)(x) = -\int_0^{\ln(x)} \tilde C({\tilde x - \tilde z}) \tilde f({\tilde z}) \ud \tilde z,
    \end{equation}
    where for $f$ we defined $\tilde f(\tilde z) \coloneqq f(e^{\tilde z})$, and similarly for $C(z)$.
    This is almost a Wiener--Hopf equation of the first-kind~\eqref{eq:wiener-hopf-first}, but instead of limit at $\infty$, the upper limit $\ln(x)$ is finite.
    The first coefficient functions of interest are given in Eqs.~\eqref{eq:coeff-C2q}--\eqref{eq:coeff-CLg},
    which when written after the logarithmic change of variable $z = e^{\tilde z}$ are:
    \begin{align}
        \tilde C_{2,q}^{(1)}({\tilde x - \tilde z}) & = \cf \bigg\lbrace \frac{9}{2} + \frac{5}{2}e^{\tilde x - \tilde z} - 2 \qty(\frac{3}{4} + \ln(e^{\tilde x - \tilde z}) - \ln(1-e^{\tilde x - \tilde z}) ) \qty(\frac{2}{1-e^{\tilde x - \tilde z}} - 1 - e^{\tilde x - \tilde z})
        \notag \\
        & \qquad - \qty(9 + 4 \zeta(2)) \delta(1-e^{\tilde x - \tilde z}) \bigg\rbrace_+,
        \\
        \tilde C_{3,q}^{(1)}({\tilde x - \tilde z}) & = C_{2,q}^{(1)}(e^{\tilde x - \tilde z}) - 2 \cf(1+e^{\tilde x - \tilde z}),
        \\
        \tilde C_{2,g}^{(1)}({\tilde x - \tilde z}) & = \left\lbrace 6 - 2 \qty(4 + \ln(e^{\tilde x - \tilde z}) - \ln(1-e^{\tilde x - \tilde z})) \qty(1 -2e^{\tilde x - \tilde z} + 2e^{2(\tilde x - \tilde z)} ) \right\rbrace,
        \\
        \tilde C_{L,q}^{(1)}({\tilde x - \tilde z}) & = 4 \cf e^{\tilde x - \tilde z},
        \\
        \tilde C_{L,g}^{(1)}({\tilde x - \tilde z}) & = 8 e^{\tilde x - \tilde z}(1-e^{\tilde x - \tilde z}).
    \end{align}
    Since at least parts of these Wilson coefficient functions acting as the difference kernel need to be taken as distributions, applicability or generalization of existing proofs of existence of solutions is to be verified in future work. Systems of Wiener--Hopf equations have been investigated~\cite{GohbergKrein1958:wiener-hopf-systems}, as has equations with non-integrable kernels~\cite{Geleg:1966:wiener-hopf-nonint-kernel}, both of which would appear relevant to the present coupled system of equations. From the point of view of this work, this framing in the context of solvable integral equations is part of the motivation to develop this mathematical inverse problems approach. The standard approach to solve the Wiener--Hopf equations \eqref{eq:wiener-hopf-first}, \eqref{eq:wiener-hopf-second} would inspect the Fourier transforms of the above difference kernels $\tilde C(\tilde z)$, but this is left for a separate publication.
    Hopefully existence and uniqueness can be rigorously established, which would support confidence in any potential reconstructions achieved from real data using the methodology constructed in Sec.~\ref{sec:discretization}. Alternatively, if for example the uniqueness of solutions would fail, it would be of interest from the perspective of interpreting existing fit based global analyses if it became possible to characterize the linearly independent solutions of the same integral equation.
    The goal of this inverse problems approach is not to solve the PDFs analytically from data, but rather to establish existence and uniqueness of solutions when computationally solved using the experimental data as input.

\section{Flavor number schemes and functional linearity} \label{sec:flavor-number-schemes}
	
    \Glspl{flavor number scheme} are theoretical frameworks that solve the problem of simultaneously handling regimes of data where the number of light quarks can change as a function of the probe scale as described in Sec.~\ref{sec:obs-as-linear-ops}. Technically, a parton can be associated with a PDF only at the massless limit, which is in practice sufficiently satisfied by data when $\frac{m_a^2}{Q^2} \ll 1$ for flavor $a$. This means that in the global analysis inverse problem one is faced with the issue that the number of parton flavors that satisfy $\frac{m_a^2}{Q^2} \ll 1$ changes with $Q^2$. 
	This can perhaps be thought a bit like having physical phases of the system at different $Q^2$-scales: at low $Q^2$ only the lightest quarks are light enough to embody a PDF, and at higher and higher $Q^2$ scale each heavier quark flavor will in turn begin to sufficiently satisfy  $\frac{m_a^2}{Q^2} \ll 1$ and ``melt'' to begin to contribute as a part of the PDF family and $Q^2$ evolution.
	The theoretical approaches that construct consistent systematic frameworks for the description of this multi-scale physics are known as \glspl{flavor number scheme}, and they are given overviews in Refs.~\cite{Klein2012,Martin:2009iq,NNPDF:2021njg,Hekhorn:2019nlf-felix-phd,Forte:2010ta:FONLL}.
    
    Established frameworks---known as general-mass variable  \glspl{flavor number scheme} (GM-VFNS)---to describe this multi-scale dynamics are FONLL~\cite{Forte:2010ta:FONLL}, S-ACOT~\cite{Kramer:2000hn:s-acot}, ACOT-$\chi$~\cite{Tung:2001mv:acot-chi,Kretzer:2003it:acot-chi}, VFNS~\cite{Ablinger:2026xza:nnnlo-vfns}, and TR~\cite{Thorne:1997ga:mscheme-tr,Thorne:1997uu:mscheme-tr,Thorne:2006qt:mscheme-tr}, all of which have the objective of systematically addressing how to simultaneously describe these regimes of varying number of evolving light quark PDFs.
    The goal of this work is not to develop a new \gls{flavor number scheme}, but rather to develop a mathematical formalism of solving the PDFs from world data while being general enough to be amenable to any of the established  \glspl{flavor number scheme} that seek to capture the rich QCD dynamics in their effective theoretical prescriptions of the physics.
    
    We will consider the implementation of a \gls{flavor number scheme} in this inverse problems formalism using FONLL scheme~\cite{Forte:2010ta:FONLL} as a concrete example of the procedure. FONLL has the nice feature that the perturbative expansion in coupling can be systematically handled with an arbitrary order of resummation of large logarithms. However, thanks to the functional linearity stemming from the collinear factorization~\cite{Collins:1987pm:collinear-factorization}, this approach should be applicable with any \gls{flavor number scheme}. Our discussion will cover at a high-level the considerations that are taken in a given scheme, but a comprehensive systematic implementation of a \gls{flavor number scheme} would mandate including the heavy quark mass effects in all of the forward operators discussed in the previous sections---if a quark flavor is at some momentum scale considered heavy, then its contribution to all forward operators is heavy at that scale, even to those where light flavors are primarily produced---and so is left for future work. An economical possibility would be to leverage a robust computational implementation of the coefficient functions such as Yadism~\cite{Candido:2024rkr:yadism} to handle the generation of the forward operators.

    In the FONLL scheme~\cite{Forte:2010ta:FONLL} one defines generally for all structure functions $F$:
    \begin{align}
        \label{eq:F-fonll}
        F^\mathrm{FONLL}(x,Q^2) & \coloneqq F^{(d)}(x,Q^2) + F^{(n_l)}(x,Q^2), \\
        F^{(d)}(x,Q^2) & \coloneqq F^{(n_l+1)}(x,Q^2) - F^{(n_l,0)}(x,Q^2),
    \end{align}
    where $F^\mathrm{FONLL}$ is the effective multi-scale prescription, $F^{(n_l)}$ is the theory picture with $n_l$ light flavors, and $F^{(d)}$ is a ``difference term'' where the double counting of light contributions is subtracted from the picture with $n_l+1$ flavors. They are~\cite{Forte:2010ta:FONLL}:
    \begin{align}
        F^{(n_l)}(x,Q^2) & \coloneqq 
        x \int_x^1 \frac{\ud z}{z} \sum_{i=q,\bar q, g} B_i \left( \frac{x}{z}, \frac{Q^2}{m_i^2}, \as^{(n_l+1)}(Q^2) \right) f_i^{(n_l +1)}(z,Q^2)
        , \\
        F^{(n_l+1)}(x,Q^2) & \coloneqq
        x \int_x^1 \frac{\ud z}{z} \sum_{i=q,\bar q, g} A_i^{(n_l+1)} \left( \frac{x}{z}, \log \frac{Q^2}{m_i^2}, \as^{(n_l+1)}(Q^2) \right) f_i^{(n_l +1)}(z,Q^2)
        , \\
        F^{(n_l,0)}(x,Q^2) & \coloneqq 
        x \int_x^1 \frac{\ud z}{z} \sum_{i=q,\bar q, g} B_i^{(0)} \left( \frac{x}{z}, \frac{Q^2}{m_i^2}, \as^{(n_l+1)}(Q^2) \right) f_i^{(n_l +1)}(z,Q^2).
    \end{align}
    Precise definitions and discussion of the coefficient functions $B_i$, $A_i^{(n_l+1)}$, and $B_i^{(0)}$ are given in Ref.~\cite{Forte:2010ta:FONLL}, but for our purposes it is sufficient to consider them as unspecified integration kernels. Especially, we highlight that the \gls{flavor number scheme} approximation of the cross-regime expression for the general structure function $F^\mathrm{FONLL}$ in Eq.~\eqref{eq:F-fonll} is a linear integral equation for the PDFs $\pdtbf$, which we will take advantage of. If we define the integral operators
    \begin{align}
        \mathcal{B}_i[f] & \coloneqq x \int_x^1 \frac{\ud z}{z} B_i \left( \frac{x}{z}, \frac{Q^2}{m_i^2}, \as^{(n_l+1)}(Q^2) \right) f(z,Q^2), \\
        \mathcal{A}_i[f] & \coloneqq x \int_x^1 \frac{\ud z}{z} A_i^{(n_l+1)} \left( \frac{x}{z}, \log \frac{Q^2}{m_i^2}, \as^{(n_l+1)}(Q^2) \right) f(z,Q^2), \\ 
        \mathcal{B}_i^{(0)}[f] & \coloneqq x \int_x^1 \frac{\ud z}{z} B_i^{(0)} \left( \frac{x}{z}, \frac{Q^2}{m_i^2}, \as^{(n_l+1)}(Q^2) \right) f_i(z,Q^2),
    \end{align}
    we can express the above general structure functions in the row matrix form:
    \begin{align}
        F^{(n_l)}(x,Q^2) & \equiv \begin{bmatrix}
        \Bcal_g & \Bcal_d & \cdots & \Bcal_{\bar t}
        \end{bmatrix} \pdtbf, \\
        F^{(n_l+1)}(x,Q^2) & \equiv \begin{bmatrix}
        \Acal_g & \Acal_d & \cdots & \Acal_{\bar t}
        \end{bmatrix} \pdtbf, \\
        F^{(n_l,0)}(x,Q^2) & \equiv \begin{bmatrix}
        \Bcal_g^{(0)} & \Bcal_d^{(0)} & \cdots & \Bcal_{\bar t}^{(0)}
        \end{bmatrix} \pdtbf, \\
        F^\mathrm{FONLL}(x,Q^2) & \equiv \begin{bmatrix}
        \Acal_g + \Bcal_g - \Bcal_g^{(0)} & \Acal_d + \Bcal_d -\Bcal_d^{(0)} & \cdots & \Acal_{\bar t} + \Bcal_{\bar t} - \Bcal_{\bar t}^{(0)}
        \end{bmatrix} \pdtbf.
    \end{align}
    A realistic implementation of a \gls{flavor number scheme} will have to work with at least three regimes of flavor number: at low $Q^2$ the three lightest $u,d,s$ flavors are light, and $c,b,t$ are heavy, at intermediate $Q^2$, i.e. $m_c \lesssim Q \lesssim m_b$, there is four light and two heavy flavors, and at moderately high $Q^2$ $(m_b \lesssim Q \lesssim m_t)$ all but top can be considered as light, perhaps neglecting some non-trivial transition regime dynamics where a flavor is smoothly transitioning from massive to light. For example, the NNPDF global analysis~\cite{NNPDF:2021njg} uses the FONLL variable \gls{flavor number scheme} with the maximum of five flavors, i.e. top is always heavy, which is reasonable with historical and present day experimental data. The question of what to do at so low $Q^2$ that the light flavors would need to be considered as massive is well beyond the scope of this work and is not considered.

    In conclusion, thanks to the functional linearity of DIS that originates from the collinear factorization, this inverse problems approach is fully compatible with the existing theory picture of  \glspl{flavor number scheme}. In a future work we consider the practical numerical implementation of this approach in at least one of the  \glspl{flavor number scheme} discussed above.

\section{A method to solve the global data inference inverse problem}
    \label{sec:discretization}

    We will construct a systematic method that reduces the continuous functional global analysis inverse problem into a discrete block-form matrix-vector inverse problem of the form $O^\mathrm{WD} = \bA \pdtbf$, where $O^\mathrm{WD}$ is a carefully arranged list of the world data measurement points, $\bA$ is a forward operator matrix, and $\pdtbf$ will be a block vector structure encoding the discretized information about the PDFs $f_a(x,Q^2)$ for each flavor $a$ over discretized nodes in $x$ and $Q^2$. This discrete linear inverse problem is then solvable computationally with standard inverse problems methods, which we demonstrated for the dipole amplitude inverse problem in Ref.~\cite{Hanninen:2025iuv}, and discuss in the context of this problem later in this section. We note that on a surface level this discretization might share similarities with the one used in Refs.~\cite{Barontini:2023vmr:pineline-fktable-discretization,DelDebbio:2021whr:fktable} as we are constructing a discretization for the same problem after all. However, this approach has distinct features, as we specifically leverage the higher level integral equation structures of the problem, and use standard techniques from mathematical inverse problems to reduce the multi-dimensional discretized table multiplication into just a matrix-vector multiplication, which at least in principle can be solved with standard methods, and critically, does not require a parametrized fit ansatz. Technically, the discretization we will construct is broader than the FK-tables of Refs.~\cite{Barontini:2023vmr:pineline-fktable-discretization,DelDebbio:2021whr:fktable}, as the FK-tables are constructed for each parton flavor separately, and at fixed $x$, whereas we will construct a discrete forward operator that encompasses all observables of the world data for all flavors at all $x$, $Q^2$, and $y$, i.e. the entire problem of global analysis is reduced in to a matrix-vector multiplication problem.
    
    The high-level structure of the approach is that the world data $O^\mathrm{WD}$ is split into block vectors of observables of cross section type $O^\sigma(x,Q^2,y)$, structure function type $O^F(x,Q^2)$, and sum rule type $O^S(Q^2)$. Analogously the parton distribution functions are discretized and ``stacked'' into a block vector form, and then the integral operators discussed in previous sections need to be discretized, here with quadrature rule approximation, in a way that respects the block structure of both $\pdtbf$ and $O^\mathrm{WD}$.

    
    We begin this construction by considering two types of integral equations that emerge for a given PDF $f_a$ depending on the observable $O(x)=O(\sum_a f_a(x))$ considered:
    \begin{align}
        O_a(x) & = \int_x^1 C\qty(\frac{x}{z}) f_a(z) \frac{\ud z}{z},
        \label{eq:vie1}
        \\
        O_a(x) & = \lambda_a f_a(x) + \int_x^1 C\qty(\frac{x}{z}) f_a(z) \frac{\ud z}{z},
        \label{eq:vie2}
    \end{align}
    where $a$ is the flavor of the parton of $f_a$, and also denotes its isolated contribution to the ``quasi-observable'' $O_a(x)$ for which $\sum_a O_a(x) \coloneqq O(x)$, i.e. the full physical observable.
    Above, the former is a Volterra integral equation of the first kind, and the latter is an inhomogeneous Volterra integral equation of the second kind, both of which have their bespoke mathematical theories of existence and discovery of solutions~\cite{Volterra1,Volterra2,Volterra3,Volterra4,Volterra5,brunner2017volterra}. Here we will however only consider a very general purpose methodology that can straightforwardly accommodate both cases, since our full problem will include both kinds simultaneously. For example, in Eq.~\eqref{eq:sigma-incl-nlo} both are seen: quarks contribute at LO so they fulfill the integral equation of the second kind, whereas the gluon distribution $f_g$ only appears at NLO so it fulfills an equation of the first kind.
    We will first consider the method to solve a hypothetical inverse problem for an isolated PDF from either type of integral equation, after which we move to extend the method to solve real observables from the stacked tensor problem which includes all flavors simultaneously.

    Natural sets of nodes---discrete points at which data is measured or reported---is defined for $x$ and $Q^2$ by the available data, which we take as the baseline ranges and discretization points for the variables in the problem. However, numerical accuracy likely requires to take a refinement of the partition on $x$ to enable sufficiently accurate integration over $z$.

    To construct a method to solve the coupled integral equations that form the problem of the global analysis of the PDFs, we apply the Nyström quadrature method~\cite{Nystrom:1930-nystrommethod-quadrature,kress1989linear-inteq-methods-quadrature}. The first step is to approximate the integral in Eqs.~\eqref{eq:vie1} and \eqref{eq:vie2} with a quadrature rule:
    \begin{equation}
        \mathcal{Q}_n \coloneqq \sum_{j=1}^n \omega_j \frac{f(z_j)}{z_j} \approx \int_x^1 f(z) \frac{\ud z}{z},
        \label{eq:quadrature-Qn-zint}
    \end{equation}
    with the nodes $z_1, \dots, z_n \in [0,1]$ and weights $\omega_1, \dots, \omega_n$, which can depend on $n$, and for which $\omega_j = 0$ if $z_j < x$. Then we define the discretized integral operator with
    \begin{equation}
        (\mathcal{C}_n f)(x) \coloneqq \mathcal{Q}_n\qty(C\qty(\frac{x}{\cdot}) f) 
        \coloneqq \sum_{j=1}^n \omega_j \frac{C\qty(\frac{x}{z_j})}{z_j} f(z_j)
        \label{eq:wilson-Cn-by-quadrature}
    \end{equation}
    for continuous $f \in C(0,1]$ and $x \in (0,1]$.
    The Eqs.~\eqref{eq:vie1}, \eqref{eq:vie2} then become
    \begin{align}
        O_a & = C_n f_a^{(n)},
        \\
        O_a & = \lambda_a f_a^{(n)} + C_n f_a^{(n)}.
    \end{align}
    To then fully discretize these, we take $x$ on the measured points $x_k \in \lbrace x_1, \dots, x_m \rbrace$ at the fixed $Q^2$:
    \begin{align}
        O_a(x_k) & = \sum_{j=1}^n \omega_j \frac{C\qty(\frac{x_k}{z_j})}{z_j} f_a^{(n)}(z_j),
        \\
        O_a(x_k) & = \lambda_a f_a^{(n)}(x_k) + \sum_{j=1}^n \omega_j \frac{C\qty(\frac{x_k}{z_j})}{z_j} f_a^{(n)}(z_j).
    \end{align}
    We can then compact this by defining $\undertilde f \coloneqq (f^{(n)}(z_1), \dots, f^{(n)}(z_j))$ and
    \begin{align}
        M_{jk} & \coloneqq \omega_j \frac{C\qty(\frac{x_k}{z_j})}{z_j}, ~~ 1 \leq j \leq n, ~~ 1 \leq k \leq m,
        \\
        \undertilde O & \coloneqq \qty(O(x_1), \dots, O(x_m))^T.
    \end{align}
    With these the now fully discretized equations can be expressed as:
    \begin{align}
        \undertilde O_a & = M \undertilde f_a,
        \\
        \undertilde O_a & = \qty( \lambda_a \mathbf{I} + M) \undertilde f_a,
    \end{align}
    where $\mathbf{I} \equiv \delta_{x_k, z_j}$, which might be notable if real data and numerical accuracy of the quadrature enforce $m < n$. This is the fully discretized problem for the contribution of the flavor $a$ to observable $O$. We now need to include the contributions from all the flavors.
    To do this, we recall the one-form or covector formulation of the forward operator that acts on the \gls{parton distribution tensor} discussed in Sec.~\ref{sec:linear-intop-formalism}, which in practice turns the flavor identifier $a$ into a summation index as the operator elements act on the PDT elements:
    \begin{align}
        \undertilde O & = M^a \undertilde f_a \coloneqq \mathbf{M} \pdtbf,
        \label{eq:vie1-fstack}
        \\
        \undertilde O & = \qty( \lambda_a \mathbf{I} + M^a) \undertilde f_a \coloneqq (\mathbf{\lambda I + M}) \pdtbf,
        \label{eq:vie2-fstack}
    \end{align}
    where we now identify each operator $M^a$ by its associated flavor as that will be the case for the real problem.
    
    The flavor summed problems in Eqs.~\eqref{eq:vie1-fstack}, \eqref{eq:vie2-fstack} have been written implicitly at fixed $Q^2$ and $y$, which we need to amend to be able to leverage the observables of \gls{world data} for all bins of $Q^2$ and $y$.
    Since $Q^2$ nor $y$ are not integrated over in the considered observables, they are essentially ``inert'' from the perspective of the linear operation. This means that we can compose the full problem for each $Q^2$ and $y$ using the so-called ``stacked form'' used in regularizing discrete inverse problems, see for example Sec.~5.1.1 in Ref.~\cite{hansen1998rank:stacked-form}, where we combine the individual sub-problems into a larger block-matrix block-vector problem.
    
    To initiate the stacked form, we write out the PDT discretized for a fixed $Q^2=Q^2_1$, assuming an ordered list of nodes $Q^2_1, \dots, Q^2_N$:
    \begin{equation}
    \pdtbf_n(Q^2_1) =
        \begin{bmatrix}
            g(z_1, Q^2_1) \\
            g(z_2, Q^2_1) \\
            \vdots \\
            g(z_n, Q^2_1) \\
            d(z_1, Q^2_1) \\
            \vdots \\
            d(z_n, Q^2_1) \\
            \vdots \\ \vdots \\
            \bar t(z_1, Q^2_1) \\
            \vdots \\
            \bar t(z_n, Q^2_1)
        \end{bmatrix}
    \end{equation}
    This flavor stacked PDT allows the computation of observables at $Q^2_1$. To cover all bins in $Q^2$ we stack these as block vectors:
    \begin{equation}
    \pdtbf_{N} =
        \begin{bmatrix}
            \pdtbf_n(Q^2_1) \\
            \pdtbf_n(Q^2_2) \\
            \vdots \\
            \pdtbf_n(Q^2_N)
        \end{bmatrix}.
        \label{eq:pdt-discr-Q-stack}
    \end{equation}
    To write the operator that acts on $\pdtbf_{N}$, we need to keep track of the flavor $a$ of the operator, and the discretized variables $z$, $x$, and $Q^2$, so we define the element
    \begin{equation}
        \bA_{a, \nu ml} \coloneqq A_a(z_\nu,x_m,Q^2_l).
    \end{equation}
    With this we write the operator that acts to yield the measurement of observable $O(x_m,Q^2_l)$:
    \begin{equation}
        \bA_{ml} 
        \coloneqq 
        \begin{bmatrix}
            \bA_{g,1ml} & \bA_{g,2ml} & \cdots & \bA_{g,{n}ml} & 
            \bA_{d,1ml} & \cdots & \bA_{d,{n}ml} & 
            \cdots
            \bA_{\bar t,1ml} & \cdots & \bA_{\bar t,{n}ml}
        \end{bmatrix},
    \end{equation}
    where $n$ is the last index of $z_\nu$. With this then, by construction:
    \begin{equation}
        O(x_m,Q^2_l) = \bA_{ml} \pdtbf_{n}(Q^2_l) = \sum_a \sum_\nu A_{a,\nu ml} f_{a}(z_\nu,Q^2_l).
    \end{equation}
    To cover the full dataset of $O(x_m,Q^2_n)$ over sets of points $x_m \in \lbrace x_1 , \dots, x_M\rbrace$ and $Q^2_n \in \lbrace Q^2_1 , \dots, Q^2_N\rbrace$ we stack these operations into a block matrix:
    \begin{equation}
        O_{MN} \coloneqq
        O_{\substack{1\leq m \leq M,\\ 1\leq n \leq N}}
        =
        \begin{bmatrix}
            O(x_1, Q^2_1) \\
            O(x_1, Q^2_2) \\
            \vdots \\
            O(x_1, Q^2_N) \\
            O(x_2, Q^2_1) \\
            \vdots \\
            O(x_2, Q^2_N) \\
            \vdots \\
            O(x_M, Q^2_1) \\
            \vdots \\
            O(x_M, Q^2_N)
        \end{bmatrix}
        =
        \begin{bmatrix}
            \bA_{11} & 0 & 0 & \cdots &  0\\
            0 & \bA_{12} & 0 & \cdots & \vdots \\
            \vdots & 0 & \ddots & 0 &0  \\
            0 & 0 & \cdots & 0 & \bA_{1N} \\
            \bA_{21} & 0 & \cdots & 0 &  0\\
            \, 0_{\vdots} & \ddots & 0 & \cdots & 0 \\
            0 & 0 & \cdots & 0 & \bA_{2N} \\
            \vdots & \ddots & \vdots & \vdots & \vdots \\ 
            \bA_{M1} & 0 & 0 & \cdots &  0\\
            \, 0_{\vdots} & \ddots & 0 & \cdots & 0 \\
            0  & 0 & \cdots & 0 & \bA_{MN}
        \end{bmatrix}
        \begin{bmatrix}
            \pdtbf_l(Q^2_1) \\
            \pdtbf_l(Q^2_2) \\
            \vdots \\
            \pdtbf_l(Q^2_N)
        \end{bmatrix}
        =
        \bA_{MN} \pdtbf_N
        \label{eq:obs-F-class-discr-ip}
    \end{equation}
    If the observable is measured as a function of other variables such as $y$, that must be stacked similarly from the $(x, Q^2)$ constructions $O_{MN}$ at fixed $y$, with the nodes $y_1, \dots , y_K$:
    \begin{equation}
        O(x,Q^2,y)=
        \begin{bmatrix}
            O_{MN}(y_1) \\
            \vdots \\
            O_{MN}(y_K)
        \end{bmatrix}
        =
        \begin{bmatrix}
            \bA_{MN}(y_1) \\
            \vdots \\
            \bA_{MN}(y_K)
        \end{bmatrix}
        \pdtbf_N
        \eqqcolon
        \bA_{MNK} \pdtbf_N,
        \label{eq:obs-sigma-class-discr-ip}
    \end{equation}
    where the stacking now happens only in the forward operator, since $\pdtbf_N$ is not dependent on the external variable $y$. 
    We note that with the real world data all the bins might have different number of datapoints but this is just a bookkeeping problem, which was already implemented for testing of the dipole amplitude reconstruction approach~\cite{Hanninen:2025iuv} with HERA data, albeit with only one observable~\cite{oma4:invdip-github, oma-talk-2025:invdip-qcd-seminar}. It is more practical and straightforward in computation to dynamically bookkeep the sizes of sets of datapoints, than to very generally write them out as above, which here assumes uniform bin sizes in the data.

    The above construction assumed that the integrals of Eqs.~\eqref{eq:vie1} and~\eqref{eq:vie2} are well-defined as given. However, the integration kernels of the physical problem can contain plus-distributions, which will need specific attention in the discretization process. Consider the convolution with a plus-distribution:
    \begin{align}
        O_a(x) & = \int_x^1 C_+\qty(\frac{x}{z}) f_a(z) \frac{\ud z}{z}
        \notag \\
        & = \int_x^1 C\qty(\frac{x}{z}) \int_x^1 \left( \delta(z'-z) - \frac{x}{z} \delta(z'-x) \right) f(z') \ud z' \frac{\ud z}{z} - f(x) c(x),
        \label{eq:plus-distr-doubleint}
    \end{align}
    with $c(x) \coloneqq \int_0^x C(z) \ud z$. The issue is that the kernel $C_+$ is singular at $\frac{x}{z} \to 1$, and furthermore, it is not an $L^1([x,1])$ function, i.e. its integral without the test function $f$ acting as a regulator does not converge. The plus-distributions are one approach to systematically solve these challenges, and as it is largely the standard method, we adjust the method to be compatible with the physical theory including these distributions.
    Above we have broken the non-locally acting plus-distribution into a double integration that separates the non-locality from the test function $f$ so that we can formulate the discretized operator for the plus-distribution convolution.

    We will form the practical discretization based on Eq.~\eqref{eq:plus-distr-doubleint}, and to that end we have to introduce a new stack at the bottom of the stacking form, i.e. as the first step of the integral discretization. The role of this stack layer is to perform the task of the Dirac delta-functions to pick the correct non-local values from the discretized PDF $f(z_j)$. Denoting the variable of the innermost stack by $z'_{j'}$, we write the first discretization:
    \begin{equation}
        \int_x^1 \left( \delta(z'-z) - \frac{x}{z} \delta(z'-x) \right) f(z') \ud z' 
        \approx \left( \delta_{z'_{j'} z_j} - \frac{x_m}{z_j} \delta_{z'_{j'} x_m} \right) f(z'_{j'}),
    \end{equation}
    which is just matrix multiplication in the discretized problem that outputs a vector discretized in $z_j$ what can be inserted into the place of $f(z_j)$ in the above construction of the discretization stack. Lastly, the new additional term with $c(x)$ is analogous to the identity terms seen previously, and the factor $c(x)$ is computable analytically or at least numerically, which is standard practice for example in the implementation of NNPDF global analysis~\cite{NNPDF:2021njg,Candido:2024rkr:yadism}. The introduction of a full new stack in the discretization is of course somewhat unsatisfactory from the perspective of practical numerical implementation, since it increases the size of the discretized problem fairly substantially. This is hopefully something that can be remedied with a more elegant solution in the future.


    Now, the last step is to construct the full problem of global analysis with all of its observables described in sections \ref{sec:tensor-ip-lo} and \ref{sec:tensor-ip-nlo}. The observables considered fall in three classes: cross section type, measured as functions of $(x,Q^2,y)$, structure function type depending on $(x,Q^2)$, and sum rule type, which are either functions of $Q^2$ or theoretically fixed constants for all $Q^2$. 
    Each observable can have different nodes for $x$, $Q^2$, and $y$ due to independent experimental setups and physics, i.e. each discretized sub-problem of the global analysis is presumable to be of different discretized dimensions.
    Let us for generality list these classes of observables as $O^\sigma_i(x,Q^2,y)$, $O^F_i(x,Q^2)$, and $O^S_i(Q^2)$, with the index $i$ referring to an ordered list of included datasets of that class, with the number of datasets as $N_\sigma$, $N_F$, and $N_S$.
    The construction of the discrete problems of classes $\sigma$ and $F$ are given by Eqs.~\eqref{eq:obs-sigma-class-discr-ip} and \eqref{eq:obs-F-class-discr-ip}, respectively.
    To write the sum rule observables we return to Eq.~\eqref{eq:pdt-discr-Q-stack} and implement a quadrature to approximate the integral over $x$ that is the core difference of the sum rules to the other observables:
    \begin{equation}
        \mathcal{X}_{n}(f_a) \coloneqq \sum_{j=1}^{n} \tilde \omega_j f_a(x_j) \approx \int_0^1 f_a(x) \ud x,
        \label{eq:quad-x-int-0to1}
    \end{equation}
    with the nodes $x_j \equiv z_j \in [0,1]$ same as in Eq.~\eqref{eq:quadrature-Qn-zint}, and weights $\tilde \omega_1, \dots, \tilde \omega_n$. With this we write the approximations of the sum rules~\eqref{eq:fwdop-sumrule-fnc}, \eqref{eq:fwdop-sumrule-tmc}, \eqref{eq:fwdop-sumrule-GLS}, \eqref{eq:fwdop-sumrule-adler}, and~\eqref{eq:fwdop-sumrule-gottfried}. To be able to apply the quadrature $\mathcal{X}_n$ directly to the $Q^2$-stacked $\pdtbf_N$, which was also flavor and $x$ stacked first, we need to take care of the meta structure of the operating matrix. To this end, we define the matrix that operates on $\pdtbf$ at a fixed $Q^2_l$:
    \begin{equation}
        \mathcal{X}_{n}^\fnc \coloneqq
        \begin{bmatrix}
                0 &
                \mathcal{X}_n & 0 & 0 & 0 & 0 & 0 &
                -\mathcal{X}_n & 0 & 0 & 0 & 0 & 0
                \\
                0 &
                0 & \mathcal{X}_n & 0 & 0 & 0 & 0&
                0 & -\mathcal{X}_n & 0 & 0 & 0 &0
                \\
                0 &
                0 & 0 & \mathcal{X}_n & 0 & 0 & 0&
                0 & 0 & -\mathcal{X}_n & 0 & 0 &0
                \\
                0 &
                0 & 0 & 0 & \mathcal{X}_n & 0 &0&
                0 & 0 & 0 & -\mathcal{X}_n & 0 &0
                \\
                0 &
                0 & 0 & 0 & 0 & \mathcal{X}_n &0&
                0 & 0 & 0 & 0 & -\mathcal{X}_n &0
                \\
                0 &
                0 & 0 & 0 & 0 & 0 & \mathcal{X}_n &
                0 & 0 & 0 & 0 & 0 & -\mathcal{X}_n
            \end{bmatrix}.
    \end{equation}
    The sum rule needs to be satisfied for all $Q^2$, so we just stack these constraints for each $Q^2_l$ and arrive at an inverse problem for $\pdtbf_N$ defined in Eq.~\eqref{eq:pdt-discr-Q-stack}:
    \begin{equation}
        S_\fnc \equiv 
        \begin{bmatrix}
            S_\fnc(Q^2_1) \\
            \vdots \\
            S_\fnc(Q^2_N)
        \end{bmatrix}
        =
        \begin{bmatrix}
            \mathcal{X}_{n}^\fnc & \cdots & \mathcal{X}_{n}^\fnc
        \end{bmatrix}
        \begin{bmatrix}
            \pdtbf_n(Q^2_1) \\
            \pdtbf_n(Q^2_2) \\
            \vdots \\
            \pdtbf_n(Q^2_N)
        \end{bmatrix}
        \eqqcolon
        \mathbf{X}_{n, N}^\fnc \pdtbf_N,
    \end{equation}
    where now the block matrix and vector structures takes care of enforcing the sum rule constraint for all $Q^2$.

    To implement the remaining sum rules~\eqref{eq:fwdop-sumrule-tmc}, \eqref{eq:fwdop-sumrule-GLS}, \eqref{eq:fwdop-sumrule-adler}, and~\eqref{eq:fwdop-sumrule-gottfried}, we follow the above formula and define another quadrature
    \begin{equation}
        \mathcal{Y}_{n}(f_a) \coloneqq \sum_{j=1}^{n} \tilde \omega_j x_j f_a(x_j) \approx \int_0^1 x f_a(x) \ud x.
    \end{equation}
    To have a quadrature matrix that can operate on $\pdtbf_n$, we need to stack $\mathcal{Y}_n$ for each flavor:
    \begin{equation}
        \mathbf{Y}_n = \begin{bmatrix}
            \mathcal{Y}_n & (\cdots)^{2N_f+1-2} & \mathcal{Y}_n
        \end{bmatrix},
    \end{equation}
    which acts as:
    \begin{equation}
        \mathbf{Y}_n\pdtbf = \begin{bmatrix}
            \mathcal{Y}_n & (\cdots)^{2N_f+1-2} & \mathcal{Y}_n,
        \end{bmatrix}
        \pdtwritout
        =
        \sum_{a=g}^{\bar t} \mathcal{Y}_n f_a
        \approx
        \sum_{a=g}^{\bar t} \int_0^1 x f_a(x).
    \end{equation}
    It is now straightforward to write the total momentum conservation rule:
    \begin{equation}
        S_\tmc  =
        \begin{bmatrix}
             \mathbf{Y}_{n} & \cdots & \mathbf{Y}_{n}
        \end{bmatrix}
        \begin{bmatrix}
            \pdtbf_n(Q^2_1) \\
            \pdtbf_n(Q^2_2) \\
            \vdots \\
            \pdtbf_n(Q^2_N)
        \end{bmatrix}
        \eqqcolon
        \mathbf{Y}_{n, N}^\tmc \pdtbf_N.
        \label{eq:fwdop-sumrule-tmc-discr}
    \end{equation}
    The remaining three sum rules depend on specific linear combinations of the PDFs, which we write as using the quadrature rule defined in Eq.~\eqref{eq:quad-x-int-0to1}:
    \begin{align}
        \mathcal{X}_n^\mathrm{GLS} & \coloneqq 
        \begin{bmatrix}
                0 &
                \mathcal{X}_n & \mathcal{X}_n & 0 & 0 & 0 & 0 &
                -\mathcal{X}_n & -\mathcal{X}_n & 0 & 0 & 0 & 0
            \end{bmatrix},
        \\
        \mathcal{X}_n^\mathrm{A} & \coloneqq 
            \begin{bmatrix}
                0 &
                -\mathcal{X}_n & \mathcal{X}_n & 0 & 0 & 0 & 0 &
                \mathcal{X}_n & -\mathcal{X}_n & 0 & 0 & 0 & 0
            \end{bmatrix},
        \\
        \mathcal{X}_n^\mathrm{G} & \coloneqq 
        \frac{1}{3}\begin{bmatrix}
                0 &
                -\mathcal{X}_n & \mathcal{X}_n & 0 & 0 & 0 & 0 &
                -\mathcal{X}_n & \mathcal{X}_n & 0 & 0 & 0 & 0
            \end{bmatrix},
    \end{align}
    with which the sum rules become:
    \begin{align}
        S_{\mathrm{GLS}}(Q^2) & = 
        \begin{bmatrix}
            S_\mathrm{GLS}(Q^2_1) \\
            \vdots \\
            S_\mathrm{GLS}(Q^2_N)
        \end{bmatrix} 
        = 
        \begin{bmatrix}
            \mathcal{X}_n^\mathrm{GLS} & \cdots & \mathcal{X}_n^\mathrm{GLS}
        \end{bmatrix}
            \pdtbf_N
        \eqqcolon \mathbf{X}_{n,N}^\mathrm{GLS} \pdtbf_N,
        \\
        S_\mathrm{A}(Q^2) & = 
        \begin{bmatrix}
            S_\mathrm{A}(Q^2_1) \\
            \vdots \\
            S_\mathrm{A}(Q^2_N)
        \end{bmatrix} 
        = 
        \begin{bmatrix}
            \mathcal{X}_n^\mathrm{A} & \cdots & \mathcal{X}_n^\mathrm{A}
        \end{bmatrix}
            \pdtbf_N
        \eqqcolon \mathbf{X}_{n,N}^\mathrm{A} \pdtbf_N,
        \\
        S_\mathrm{G}(Q^2) & = 
        \begin{bmatrix}
            S_\mathrm{G}(Q^2_1) \\
            \vdots \\
            S_\mathrm{G}(Q^2_N)
        \end{bmatrix} 
        = 
        \begin{bmatrix}
            \mathcal{X}_n^\mathrm{G} & \cdots & \mathcal{X}_n^\mathrm{G}
        \end{bmatrix}
            \pdtbf_N
        \eqqcolon \mathbf{X}_{n,N}^\mathrm{G} \pdtbf_N.
    \end{align}
    With these, we have now formulated all of the sum rule constraints for the discretized inverse problem.


    Now we can finally write the full global data inverse problem with all the observable classes discussed above, and in Sec.~\ref{sec:data}:
    \begin{equation}
        O^\mathrm{WD} = 
        \begin{bmatrix}
            O^\sigma_1(x,Q^2,y) \\
            \vdots \\
            O^\sigma_{N_\sigma}(x,Q^2,y) \\
            O^F_1(x,Q^2) \\
            \vdots \\
            O^F_{N_F}(x,Q^2) \\
            O^S_1(Q^2) \\
            \vdots \\
            O^S_{N_S}(Q^2)
        \end{bmatrix}
        =
        \begin{bmatrix}
            \bA_{M_1 N_1 K_1} \\
            \vdots \\
            \bA_{M_{N_\sigma} N_{N_\sigma} K_{N_\sigma}} \\
            \bA^F_{M_1 N_1} \\
            \vdots \\
            \bA^F_{M_{N_F} N_{N_F}} \\
            \bA^S_{N_1} \\
            \vdots \\
            \bA^S_{N_S} 
        \end{bmatrix}
        \pdtbf_N,
    \end{equation}
    where for generality each sub-problem has individual dataset dimensions $M_j, N_j, K_j$, reflecting the dataset dependent availability of measurements at a given $x$, $Q^2$, or $y$.
    The \gls{parton distribution tensor} $\pdtbf_N$ is understood to be discretized on nodes of $x$ and $Q^2$ that cover all possible available measurement points included in the world data $O^\mathrm{WD}$. If a given dataset $O_j^\tau$ does not include a given node in $\hat x$ or $\hat Q^2$, the forward operator for that dataset is constructed such that the contribution from $\pdtbf(\hat x,Q^2)$ and $\pdtbf(x,\hat Q^2)$ is zero, i.e. the element of the operator is identically zero.
    This construction is intentionally done on a high-level to leave it open for example at which order in pQCD this is to be implemented in practice, and the order at least technically could depend on the observable as well, whether that is reasonable from the perspective of QCD is a matter of nuance and outside the scope of this work. This construction is done within one fixed regime of a full flavor number scheme, which we discussed in Sec.~\ref{sec:obs-as-linear-ops}. This discretization has to be done separately for each distinct regime of $n_f$ with their own associated forward operators, and then the problem with a fully implemented FNS is constructed by stacking the independent $n_f$ specific world data forward operators, and corresponding subsets of the world data binned by $Q^2$ as defined by the FNS.

    Discretization as a method to solve the inverse problem of course has the side effect, that instead of the original continuous inverse problem, one is solving a finite dimensional vector space inverse problem, which needs to be constructed as a proxy to the original problem.
    The above problem is of the linear form $O^\mathrm{WD} = \mathbf{A}\pdtbf$, which can be at least in principle solved by general purpose methods such as Tikhonov--Phillips regularization~\cite{Tikhonov:1943,Tikhonov:1963,Tikhonov:1977,Phillips:1962,Hoerl:1962}, Cimmino's~\cite{Cimmino:1938} or Kaczmarz's~\cite{Kaczmarz:1937, Gordon:1970:ART} methods, which are broadly applied in real world applications of inverse problems in fields such as medical imaging~\cite{Hansen:2021ct}.
    A key aspect of the regularization method is the selection of a regularization algorithm that enforces some expected properties of the recovered solution, i.e. here one might expect the solution to be a discretization of the smooth and integrable PDFs $f_a(x,Q^2)$, instead of having discontinuities or other weaker regularity properties.
    These aspects of the practical solution of a linear discrete problem of the form $O^\mathrm{WD} = \mathbf{A}\pdtbf$ were discussed in the context of the inference of the dipole scattering amplitude in Ref.~\cite{Hanninen:2025iuv}, where we applied these methods to reconstruct the dipole amplitude from simulated reduced cross section data.
    The practical numerical implementation, closure testing, and evaluation of feasibility of application this method to real world data are left for future work.

    \subsection{Uncertainties of the reconstructed solutions to the inverse problem}


    A fundamental requirement for any framework of inference for the PDFs is the propagation and quantification of uncertainties from all sources and their impact on the inferred PDFs. The uncertainty quantification workflows of the major global analyses discussed in Sec.~\ref{sec:intro} are established and robust, however, they might carry some bias from the functional parametrization ansatz \eqref{eq:fit-ansatz}, which this inverse problems approach seeks to amend. In general the framework needs to be able to incorporate both experimental uncertainties associated with each measurement, and theoretical uncertainties from various approximations such as a fixed order in pQCD, and even more fundamentally the accuracy of the collinear factorization theorem and so on.

    We construct the framework for handling experimental uncertainty following the general approach of the NNPDF collaboration, that is using the concept of randomly sampled replicas~\cite{NNPDF:2021njg}. Replica is a randomly sampled copy of a full set of measurement datapoints, where each measurement is given a randomly sampled representative within the experimentally determined uncertainties. This general purpose approach was applied in Ref.~\cite{Hanninen:2025iuv} to reconstruct the uncertainties of the reconstructed solutions to the linear inverse problem for the dipole amplitude, and since this formulation of the linear global analysis problem is on a high-level of the same form, the same approach applies. 
    A second fundamental source of uncertainty of the forward problem is the confidence in the accuracy of the physical theory describing the physics. A key source of theoretical uncertainty is that the calculations are performed at a finite order in perturbation theory. The handling of this type of uncertainty is a critical question for the approach. Conceptually, each forward operator can be given a theoretical uncertainty based on the analysis of the neglected higher order terms, and one then has to propagate the uncertainty of the forward operators into an additional uncertainty of the reconstructed solutions of the inverse problem. Perhaps the concept of replicas is also applicable to the forward operators, and one would have to sample the solutions both over samples of data replicas, and forward operator replicas to estimate the combined uncertainty for the solutions. In detail consideration and implementation of this is left for future work.


    \subsection{Computational difficulty of the reconstruction}

    To even roughly estimate the computational size of the problem, let us draw an analogue to image processing, and think of the value of a PDF $f_a(x_m, Q^2_n)$ at the node $(x_m, Q^2_n)$ as a ``pixel'', a sample of the original continuous data. Then each flavor $a$ is a distinct ``channel'' of data, perhaps analogously to different color channels of images. If we assume a discretization of $10^3$ nodes in both $x$ and $Q^2$, each channel describing the PDF $f_a$ on that set of nodes is $(10^3)^2=10^6$ pixels, i.e. one megapixel $1 \, \mathrm{Mpx}$, and thus the description of the full \gls{parton distribution tensor} would need $13 \, \mathrm{Mpx}$, which is of course a fairly sizable for a set of unknown data, but in comparison to the image processing problem, it is perfectly reasonable, and can be feasible to solve in practice thanks to the smoothness of the PDFs, which enables the use of differential operators in the regularization of the inverse problem~\cite{hanke1992regularization:large-ip}.
    
    Given the above, let us estimate the size of the full forward operator. Each row of the discretized global data forward operator $\mathbf{G}^{lp\to l' X}$ is then of the same size as $\pdtbf$, so given the above, $13 \, \mathrm{Mpx}$. The number of rows in $\mathbf{G}^{lp\to l' X}$ is then determined by the number of data points included in the world data $O^\mathrm{WD}$.
    If the global dataset were composed of ten measurements of observables of $10^3$ points each, the total problem of $10^4$ points would result in a forward operator of $10^4 \times 13 \, \mathrm{Mpx} = 1.3 \cdot 10^5 \, \mathrm{Mpx} = 130 \, \mathrm{Gpx}$.
    However, the present day global data taken as described in Sec.~\ref{sec:data} is much more limited than the above, with only the largest combined HERA dataset having more than $10^3$ points~\cite{Abramowicz:2015mha:HERAIInewcombined}, and the others have on the order of $\mathcal{O}(100)$ points or even fewer, as excellently tabulated in Ref.~\cite{NNPDF:2021njg} where they include a total of $2762$ DIS observable data points in their analysis. If we assume that the world data were to be of the order of $2500$ data points, and reduce the grid sizes to $200$ in $x$ and $100$ in $Q^2$, which is very close to the discretization used by the NNPDF collaboration~\cite{NNPDF:2021njg} for their resulting LHAPDF sets~\cite{Buckley:2014ana:LHAPDF}, the forward operator is reduced by a factor of $200$ down to $650 \, \mathrm{Mpx}$, i.e. samples of scalars. 
    Vast majority of the cells are exactly zero by construction, but for the sake of worst case estimation, let us assume that each cell would need to store a double-precision floating point value, which would take $8$ bytes (B). The operator matrix of $650 \cdot 10^6$ values would then use $650 \cdot 10^6 \times 8 \, \mathrm{B} \approx 5.2 \, \mathrm{GB}$ of memory, which is a completely workable size for a dataset. Unfortunately, the innermost discretization introduced for the plus-distributions in this initial method is quite costly as it increases the problem size by introducing a huge amount of zeroes as padding in the problem, though there might be some data savings to be had in the implementation of the padding made entirely of zeroes. Furthermore, since the computation of the forward operator element values do not depend on the values of the solution PDFs, it is sufficient to generate the forward operator once after fixing the phenomenological details such as parton masses, running coupling prescription and so on. This dramatically reduces the computational cost of solving the problem, as the physical observables do not require further numerical integration once the forward operator is generated, as we demonstrated with the dipole amplitude inverse problem~\cite{Hanninen:2025iuv}.
    Practical implementation and testing of this approach based on the framework of Ref.~\cite{Hanninen:2025iuv} is left for future work, but thanks to the smoothness of the PDFs, the problem should be solvable as a regularized inverse problem~\cite{hanke1992regularization:large-ip}.

    We emphasize that in this work we are primarily concerned with what is mathematically possible \textit{in principle}, and what the inverse problem of the global analysis of the proton PDFs looks like mathematically, when the explicit goal is to formulate a methodology not based on fitting model parameters. With future high-energy experiments providing ever more accurate data, the forward looking question is how does one apply mathematical methods of inference to that future data, without having to rely on model parametrizations, which is not done in medical, or other indirect, imaging either.

    \subsection{Example: discretized inverse problem for inclusive DIS at NLO}

    To illustrate the application of the above construction to real expressions of observables calculated in pQCD, we will take as a ``toy-model'' of the world data the inclusive DIS cross section, the longitudinal structure function $F_L^\gamma$, and the total momentum conservation sum rule. These observables form a toy problem for the reconstruction of the PDFs as an inverse problem:
    \begin{equation}
        \begin{bmatrix}
            \sigma^\gamma_r(x,Q^2,y) \\
            F_L^\gamma(x,Q^2) \\
            S_{\mathrm{t.m.c}}(Q^2)
        \end{bmatrix} 
        = 
        \int
        \begin{bmatrix}
            \mathbf{A}^\gamma_r(x,Q^2,y) \\
            \mathbf{B}_{L}^\gamma(x,Q^2) \\
            \Ical_{\mathrm{t.m.c}}
        \end{bmatrix} 
        \pdtbf(\cdot,Q^2),
    \end{equation}
    where the forward operators were defined in Eqs.~\eqref{eq:fwdop-incdis-nlo}, \eqref{eq:fwdop-incdis-nlo-fl}, and \eqref{eq:fwdop-sumrule-tmc}.
    To then apply the method of solving the problem constructed above, we must discretize the above sub-problems to arrive at the linear discrete inverse problem. The last of these observables we can directly take as discretized from Eq.~\eqref{eq:fwdop-sumrule-tmc-discr}.

    We could begin the illustration with the LO result for the inclusive cross section in Eq.~\eqref{eq:fwdop-incldis-lo}, but it is unfortunately almost unhelpful in its triviality: the observable is just a linear combination of the elements of the discretized $\pdtbf$, so one can just use the nodes of the measurement $x_1, \dots, x_M$ and be done. 
    We need to step up to NLO accuracy to really apply the quadrature and stacking steps: we write Eqs.~\eqref{eq:nlo-incl-integralops1}, \eqref{eq:nlo-incl-integralops2} using the Wilson coefficient function quadrature rule defined in Eq.~\eqref{eq:wilson-Cn-by-quadrature}, moving the discretization node count $n$ to superscript:
    \begin{align}
        C_{2-L,q} \otimes f_a & \approx 
        \mathcal{C}^{n}_{2-L,q} f_a(x)
        \label{eq:nlo-incl-integralops1-quad}
        \\
        C_{2-L,g} \otimes f_g & \approx
        \mathcal{C}^{n}_{2-L,g} f_g(x),
        \label{eq:nlo-incl-integralops2-quad}
    \end{align}
    where we defined all of the terms adjacent to the coefficient function to be absorbed for convenience:
    \begin{align}
        \qty(\mathcal{C}^{n}_{2-L,q}f)\qty(x) & \coloneqq 
        \frac{\alpha_s}{4\pi} \sum_{j=1}^n \frac{\omega_j}{z_j}
        \qty(c_{2,q}^{(1)}\qty(\frac{x}{z_j})_+ - \frac{y^2}{Y_+} c_{L,q}^{(1)}\qty(\frac{x}{z_j}))
        x f(z_j)
        \\
        \qty(\mathcal{C}^{n}_{2-L,g} f)\qty(x) & \coloneqq 
        \frac{\alpha_s}{4\pi} \sum_{j=1}^n \frac{\omega_j}{z_j} 
        \qty(c_{2,g}^{(1)}\qty(\frac{x}{z_j})_+ -\frac{y^2}{Y_+} c_{L,g}^{(1)}\qty(\frac{x}{z_j}))
        x f(z_j).
    \end{align}
    With these we can then write the discrete forward operator for the inclusive DIS cross section at NLO \eqref{eq:fwdop-incdis-nlo}:
        \begin{align}
        \bA^\gamma_{r,\nlo} & (x_\mu, y_\rho) =
        \begin{bmatrix}
            \undertilde{e}^2 \mathcal{C}^{n}_{2-L,g}(x_\mu) &
            e_d^2 \qty( x \Id + \mathcal{C}^{n}_{2-L,q}(x_\mu)) &
            \cdots &
            e_{\bar t}^2 \qty( x \Id + \mathcal{C}^{n}_{2-L,q}(x_\mu))
        \end{bmatrix}
    \end{align}
    where the identity is understood as $\Id \equiv \delta(x_\mu - z_j)$, which is the identity matrix when the sets of nodes are the same for $x$ and $z$. With these we can simply write for the whole inclusive cross section dataset, again using the shorthand\footnote{We elected to keep the dependence on $Q_j^2$ in the forward operator here for generality, as it is possible for $Q^2$ dependence to be present in the forward operator.} $\bA_{ijk} \coloneqq \bA(x_i, Q^2_j, y_k)$:
    \begin{align}
        &O^\gamma_r
        = 
        \begin{bmatrix}
            \sigma^\gamma_r(x_1, Q^2_1, y_1) \\
            \vdots \\
            \sigma^\gamma_r(x_M, Q^2_1, y_1) \\
            \sigma^\gamma_r(x_1, Q^2_2, y_1) \\
            \vdots \\
            \sigma^\gamma_r(x_M, Q^2_2, y_1) \\
            \vdots \\
            \sigma^\gamma_r(x_1, Q^2_N, y_1) \\
            \vdots \\
            \sigma^\gamma_r(x_M, Q^2_N, y_1) \\
            \sigma^\gamma_r(x_1, Q^2_1, y_2) \\
            \vdots \\
            \sigma^\gamma_r(x_M, Q^2_1, y_2) \\
            \vdots \\
            \sigma^\gamma_r(x_1, Q^2_N, y_K) \\
            \vdots \\
            \sigma^\gamma_r(x_M, Q^2_N, y_K) \\
        \end{bmatrix}
        =
        \begin{bmatrix}
            \bA_{111} \vphantom{\sigma^\gamma_r} & 0 & \cdots &  0\\
            \vdots & \vdots & \ddots & \vdots \\
            \bA_{M11}\vphantom{\sigma^\gamma_r} & 0 & \cdots & 0\\
            0 & \bA_{121}\vphantom{\sigma^\gamma_r} & \hspace{8mm} 0~~~\cdots &  0\\
            \vdots & \vdots & \vdots & \vdots \\
            0 & \bA_{M21}\vphantom{\sigma^\gamma_r} & \hspace{8mm} 0~~~\cdots & 0 \\
            \vdots & \ddots & \ddots &\vdots\\
            0 & \cdots & 0& \bA_{1N1}\vphantom{\sigma^\gamma_r} \\
            \vdots & \ddots & \vdots & \vdots \\
            0 & \cdots & 0 & \bA_{MN1}\vphantom{\sigma^\gamma_r} \\
            \bA_{112}\vphantom{\sigma^\gamma_r} & 0 & \cdots & 0 \\
            \vdots & \vdots & \ddots & \vdots \\
            \bA_{M12}\vphantom{\sigma^\gamma_r} & 0 & \cdots & 0 \\
            \vdots & \ddots & \ddots &\vdots\\
            0 & \cdots & 0& \bA_{1NK}\vphantom{\sigma^\gamma_r} \\
            \vdots & \ddots & \vdots & \vdots \\
            0 & \cdots & 0 & \bA_{MNK}\vphantom{\sigma^\gamma_r}
        \end{bmatrix}
        \begin{bmatrix}
            \pdtbf_l(Q^2_1) \\
            \pdtbf_l(Q^2_2) \\
            \vdots \\
            \pdtbf_l(Q^2_N)
        \end{bmatrix}
        \notag \\
        & \hspace{95mm}
        \eqqcolon
        \bA_{r,\nlo,(MNK)}^\gamma \pdtbf_N,
    \end{align}
    where $\bA_{r,\nlo,(MNK)}^\gamma$ is a block matrix of $M \cdot N \cdot K$ element matrices $\bA_{ijk}$. If a data point for some combination of $M,N,K$ does not exist in the dataset, the block row can be full of zeroes, or left out by indexing the data points by count instead of the sets of nodes of $x,Q^2,y$.

    Next we apply these steps to discretize the longitudinal structure function $F_L^\gamma$ at NLO accuracy from Eq.~\eqref{eq:fwdop-incdis-nlo-fl}. We approximate the integral operators defined in Eqs.~\eqref{eq:nlo-incl-fl-integralops1}, \eqref{eq:nlo-incl-fl-integralops2} with the quadrature rule:
    \begin{align}
        C_{L,q} \otimes f_a & \approx 
        \qty(\mathcal{C}^{n}_{L,q}f_a)\qty(x)
        \label{eq:nlo-incl-fl-integralops1-quad}
        \\
        C_{L,g} \otimes f_g & \approx
        \qty(\mathcal{C}^{n}_{L,g} f_g)\qty(x),
        \label{eq:nlo-incl-fl-integralops2-quad}
    \end{align}
    where we again defined the discretized convolutions $\mathcal{C}^{n}_{L,i}$:
    \begin{align}
        \qty(\mathcal{C}^{n}_{L,q}f)\qty(x) & \coloneqq 
        \frac{\alpha_s}{4\pi} \sum_{j=1}^n \frac{\omega_j}{z_j}
        c_{L,q}^{(1)}\qty(\frac{x}{z_j}) x f\qty(z_j),
        \\
        \qty(\mathcal{C}^{n}_{L,g} f)\qty(x) & \coloneqq 
        \frac{\alpha_s}{4\pi} \sum_{j=1}^n \frac{\omega_j}{z_j}
        c_{L,g}^{(1)}\qty(\frac{x}{z_j}) x f\qty(z_j).
    \end{align}
    With these the discretized forward operator for the longitudinal structure function is then
    \begin{align}
        \bB^\gamma_L (x_\mu) =
        \begin{bmatrix}
            \undertilde{e}^2 \mathcal{C}^{n}_{L,g}(x_\mu) &
            e_d^2 \mathcal{C}^{n}_{L,q}(x_\mu) &
            \cdots &
            e_{\bar t}^2 \mathcal{C}^{n}_{L,q}(x_\mu)
        \end{bmatrix},
    \end{align}
    and the discretized problem for the $F_L^\gamma$ dataset becomes, with the shorthand $\bB_{i} \coloneqq \bB^\gamma_L (x_i)$
    \begin{equation}
        F_L^\gamma(x, Q^2)
        =
        \begin{bmatrix}
            F_L^\gamma(x_1, Q^2_1) \\
            \vdots \\
            F_L^\gamma(x_M, Q^2_1) \\
            F_L^\gamma(x_1, Q^2_2) \\
            \vdots \\
            F_L^\gamma(x_M, Q^2_2) \\
            \vdots \\
            F_L^\gamma(x_1, Q^2_N) \\
            \vdots \\
            F_L^\gamma(x_M, Q^2_N)
        \end{bmatrix}
        =
        \begin{bmatrix}
            \bB_{1} \vphantom{F_L^\gamma} & 0 & \cdots &  0\\
            \vdots & \vdots & \ddots & \vdots \\
            \bB_{M}\vphantom{F_L^\gamma} & 0 & \cdots & 0\\
            0 & \bB_{1}\vphantom{F_L^\gamma} & \hspace{8mm} 0~~~\cdots &  0\\
            \vdots & \vdots & \vdots & \vdots \\
            0 & \bB_{M}\vphantom{F_L^\gamma} & \hspace{8mm} 0~~~\cdots & 0 \\
            \vdots & \ddots & \ddots &\vdots\\
            0 & \cdots & 0& \bB_{1}\vphantom{F_L^\gamma} \\
            \vdots & \ddots & \vdots & \vdots \\
            0 & \cdots & 0 & \bB_{M}\vphantom{F_L^\gamma} \\
        \end{bmatrix}
        \begin{bmatrix}
            \pdtbf_l(Q^2_1) \\
            \pdtbf_l(Q^2_2) \\
            \vdots \\
            \pdtbf_l(Q^2_N)
        \end{bmatrix}
        \eqqcolon
        \bB^\gamma_{L,(MN)}  \pdtbf_N,
    \end{equation}
    where the forward operator is a block matrix of $M \cdot N$ element matrices $\bB_i$.
    Lastly, the inclusion of the total momentum conservation sum rule is already written in Eq.~\eqref{eq:fwdop-sumrule-tmc-discr} with $\mathbf{Y}_{n, N}^\tmc$.
    With these we are ready to write the discretized ``toy-model'' global analysis problem for the PDFs, which becomes the following discretized system of linear equations:
    \begin{equation}
        \begin{bmatrix}
            \sigma^\gamma_r(x_m,Q^2_n,y_k)_{1\leq m \leq M, \, 1\leq n \leq N, \, 1 \leq k \leq K} \\
            F^\gamma_L(x_m,Q^2_n)_{1\leq m \leq M, \, 1\leq n \leq N} \\
            S_{\mathrm{t.m.c}}(Q^2_n)_{1\leq n \leq N}
        \end{bmatrix} 
        = 
        \begin{bmatrix}
            \bA_{r,\nlo,(MNK)}^\gamma \\
            \bB^\gamma_{L,(MN)} \\
            \mathbf{Y}_{n, N}^\tmc
        \end{bmatrix} 
        \pdtbf_N,
        \label{eq:discrete-global-analysis-toyproblem}
    \end{equation}
    where each of the block matrix operators $\bA$, $\bB$, and $\mathbf{Y}$ are by construction $N$ columns wide to match the discretization of $\pdtbf_N$. Equation \eqref{eq:discrete-global-analysis-toyproblem} casts the conventional data inference problem of global analysis of the PDFs into a discrete linear matrix-vector inverse problem, which can---at least in principle---be solved by numerical regularization methods broadly applied in inverse problems, and which we showed to be feasible for the inference of the dipole scattering amplitude as well~\cite{Hanninen:2025iuv}. We emphasize that this is merely a demonstration of the application of the discretization method, and our explicit goal is for this approach to be applicable to all physical processes discussed in Secs.~\ref{sec:forward-problem-lo}, \ref{sec:forward-problem-nlo}, and at any order in pQCD as outlined in Sec.~\ref{sec:nnlo}.

\section{Discussion and outlook} \label{sec:discuss}

    \subsection*{Overview}

    We have formulated the problem of global analysis of the parton distribution functions of the proton as a mathematical inverse problem with the express aim to discover a mathematical methodology of inference that does not rely on fitting model parameters to data. In this mathematical picture, the problem of global analysis becomes a system of coupled integral equations of the PDFs that are constrained by experimental observables, sum rules, and a PDE constraint from the DGLAP evolution equation. The closest similarity of this coupled integral equation framing in the existing QCD literature are the DGLAP evolution equations themselves. On the inverse problem side, a methodological analogy can be drawn with computed tomography, where an integral equation defined by the X-ray transform is solved to recover the internal structure of the subject from experimental data, which can be constrained by a PDE to leverage time evolution data in the reconstruction~\cite{Mang_2018:pde-constr-medical}.

    This approach is motivated by the fact that the PDFs have only been inferred from the world data as parametrized fits, which could mean that any features of the PDFs not included in the model parametrizations such as Eq.~\eqref{eq:fit-ansatz} would not be seen. The goal is to develop a mathematical methodology of inference not reliant on a parametrized ansatz that can \textit{complement existing global analyses} by enabling the testing of the validity of the ansatz itself, such as search for the small-$x$ asymptotics from world data implied by Regge theory without writing it into the ansatz.

    The formulation of the mathematical inverse problem required all of the exact theory equations in perturbative QCD that are needed to calculate predictions of the experimental observables. This necessitated the recollection of the fixed order results in a self-consistent form from the literature, so we have given an introductory review to the phenomenology of deeply inelastic scattering and its theory results in order to formulate the coupled problem. While numerous reviews of the problem at LO accuracy have been given, this is the first time that we are aware of that recollects the NLO accuracy results explicitly and writes out precisely how to compute experimental DIS observables at NLO accuracy in pQCD.
    Most or all of the observables have of course been presented individually, but one of the foundational points of this work is that they all share a unifying mathematical structure in global analysis which can be leveraged in this inverse problems approach.
    We reviewed the NLO theory results for electromagnetic, neutral current, charged current, heavy flavor production, and neutrino variants of DIS. We also gave a very brief overview of  \glspl{flavor number scheme}, target mass corrections, the structure of electromagnetic DIS at NNLO accuracy, and collinear factorization.

    \subsection*{Core principles and linearity}

    The construction of the mathematical inverse problem of global analysis---in the search of methodology that does not assume a model ansatz for the PDFs---required a shift in perspective to one that relies on the mathematical properties of the PDFs and the functional linearity of the overall inference problem. We chose the following as the foundational assumptions (P\ref{post:pdf:global}, P\ref{post:pdf:smooth}, P\ref{post:pdf:qcd}) of the approach to encode standard practice assumptions of global analyses of past and present in a mathematical way:
    \begin{itemize}
        \item The PDFs $f_a(x,Q^2)$ and observables $O^{\sigma,F}$ are sufficiently differentiable in $x$. With some exceptions they are also differentiable in $Q^2$.
        \item The PDFs are sufficiently regular to be integrable in the sense of Wilson coefficient convolutions, and sum rules. They do not have cancellations between each other that could be physical. This is essential for the interpretation of the sum rule constraints, for example.
        \item The PDFs need to be solved from the DIS world data simultaneously in a consistent theory scheme, which includes all choices regarding renormalization, factorization, and flavor number schemes.
    \end{itemize}
    Within the framing given by these assumptions of the problem, we selected compatible experimental observables based on their functionally linear dependence on the PDFs~\eqref{eq:ip-general-linear}, which were the electromagnetic, neutral current, charged current, heavy flavor production, and neutrino DIS.
    We condensed the usefulness of the functional linearity of the observables in the matrix valued integral equation inverse problem of Eq.~\eqref{eq:main-ip}, which gives a foundation for the solution of inference problems of this type without fitting model parameters.

    We rewrote canonical theory pQCD prescriptions of the observables up to NLO accuracy in the linear integral operator form, giving a unified presentation of all of the observables at NLO accuracy for the first time. This necessitated the factorization of the PDFs from the hadronic tensor discussed in Sec.~\ref{sec:linear-intop-formalism}. After this ``second factorization'' we were able to separate the non-perturbative PDFs $\pdtbf$ from the interactive parts of the hadronic tensor, which enabled us to write the expressions for the observables into so-called forward operators, which are linear integral operators that act on the PDFs to by-construction reproduce the established theory results.
    In this process we identified the Wilson coefficient convolutions as closely related to Wiener--Hopf integral equations, which are convolutional integral equations the kernels of which depend on the difference of the arguments, instead of their ratio.

    \subsection*{Global analysis as a coupled system of inverse problems}
    To formalize the full DIS world data inference problem, we then ``stacked'' the observable specific forward operators in the form of Eq.~\eqref{eq:main-ip}, which resulted in the construction of the world data forward operator in Secs.~\ref{sec:tensor-ip-lo} and~\ref{sec:tensor-ip-nlo}, which acts on the PDFs $\pdtbf$ to form a coupled system of integral equations defined by perturbative QCD. This system of integral equations must be complemented by the DGLAP evolution equations, which act as a PDE constraint which the solutions of the integral equation system must satisfy within perturbative accuracy. The matrix-valued formulation of the DGLAP equation is essentially exactly equivalent with the present formalism as is, aside from a small difference in notation: instead of taking a convolution of a kernel matrix with the PDFs, an integral operator valued matrix acts on the PDFs. This provides a pleasing duality between the DGLAP equation and the integral equation system defined the world data forward operator: the former describes the physics of the internal degrees of freedom of the PDFs, and the latter their connection to the experimental observables.

    The world data inference problem of coupled integral equations exhibits integral equations of different kinds, which can be grouped by their origins. The experimental observables, cross sections and structure functions, define integral equations for each parton flavor via the Wilson coefficient function convolutions, whereas the sum rules, some which act as experimental constraints and others as theoretical boundary conditions, form distinct kinds of integral equations for the PDFs. We identified the Wilson coefficient convolutions as Wiener--Hopf integral equations with kernels that in some cases are plus-distributions, which could be a path towards unique invertibility of the Wilson convolutions at least on the parton level by solving the convolution equations analytically. However, the ultimate goal is not to solve the PDFs analytically from data, but rather to prove existence and uniqueness of the solutions when computationally reconstructed from experimental data. The high-level problem then is the coupled problem of all these distinct integral equations, which seems to be a new kind of a mathematical inverse problem given the simultaneous use of completely distinct kinds of integral equations, all the while being under the constraint of the DGLAP equation.

    \subsection*{Higher-order perturbative corrections}
    Given the fact that the state of the art in modern global analysis is at N${}^3$LO, we considered in Sec.~\ref{sec:nnlo} the applicability of this approach at higher orders in perturbative QCD: first for electromagnetic DIS at NNLO, after which we concluded that the approach should be applicable to any observable that satisfies the collinear factorization. Since the higher order corrections affect only the Wilson coefficient functions, and this approach is built around the general structure of the Wilson coefficient convolution itself, the inverse problem can be formulated at any order in the $\as$ expansion. 
    
    Very briefly, we also considered the possibility of including target mass corrections in this formalism, which at least in principle should be possible thanks to their linear convolutional structure that acts on the Wilson convolutions. However, all higher twist corrections are not directly compatible in general, as the functional linearity provided by the collinear factorization can be broken.
    
    A foundational feature of this approach is also the total compatibility with existing literature of pQCD calculations of observables, which we postulated to form a mathematically formalizable inverse problem based on the feasibility and success of global analysis of the proton PDFs. The established theory results can be economically rewritten in this formalism; the only required step is the factorization of the PDFs from the convolutions in a unified flavor basis which separates the theory prescription of observables into linear forward operators that act on the PDFs.

    \subsection*{Existence and uniqueness of solutions of the integral equation system}
    In applications of inverse problems such as medical imaging, there are two broad avenues of development: methods for the solution of the inferred quantity from experimental data, and the theoretical understanding of how accurate and stable that solution is, whether it is the only possible solution to find, and if the solution depends continuously on perturbations in the measurement. The first of these is of course the path to making the application possible and useful in practice, whereas the theoretical foundations inform the interpretation of the result: if one establishes a mathematically rigorous proof that the inference problem has a mathematically unique solution within some definition of the problem, it allows the interpreter to trust what they see in the reconstructed image. Conversely, the characterization of all possible degenerate solutions informs what is not possible to be seen from the measured data.

    This discussion is of relevance to the present approach as we have primarily concerned ourselves with the question of how to formulate the extraction of the PDFs of the proton from DIS world data as a functional reconstruction problem, rather than a problem of optimizing parameters of a predefined ansatz. In the pursuit for an answer to this question we formalized the problem of global analysis as the functional optimization problem given in Eq.~\eqref{eq:pde-dglap-constrained-world-data-optimization}. However, stating the inference problem in this generalized form does not tell us anything of the existence or uniqueness of solutions for the problem.

    The problem stated in Eq.~\eqref{eq:pde-dglap-constrained-world-data-optimization} has two distinct pieces that constrain features of the sought-after solution for the PDFs: the system of integral equations mandate the prescription of the experimental data, whereas the DGLAP equation dictates how the solutions should evolve in $Q^2$.
    Let us first consider the former term; the latter will then act to eliminate unphysical solutions if any were to exist to the former alone. A candidate solution set of PDFs $\pdtbf$ must simultaneously satisfy the integral equation of each observable and sum rule constraint.
    As the constructed inverse problem is linear, there are only three distinct scenarios: either there are no solutions, there is one unique solution, or there are infinitely many solutions~\cite{kress1989linear-inteq-methods-quadrature}. We can perhaps eliminate the first scenario based on the empirical evidence that global analyses of the PDFs have discovered at least approximate solutions for the PDFs via parameter fitting. Thus the more critical question is concerned with the latter two possibilities: is there a single unique solution, or is there a non-trivial null space, which immediately causes infinitely many degenerate solutions? This will depend on the properties of the integral operator structure defined by the Wilson coefficient functions and by the structure formed by the observable operators included in the DIS world data; unfortunately this must be left for future work.

    The discussion above suggests that a necessary, though perhaps not sufficient\footnote{It might also be necessary that each integral operator element has a unique solution on the parton level for a unique solution to exist to the global problem, but on the other hand perhaps not, since the full problem necessitates the solution of a large number of integral equations simultaneously. So, it could be sufficient, that the distinct Wilson convolution integral operator elements collectively eliminate superfluous solutions of one another for the full global problem to have an unique solution, even if some sub-problems could have ambiguously more than one solution.},
    condition for the PDF inverse problem to admit a unique solution is that the global world-data forward operator be of full rank, or alternatively overdetermined by a set of theoretically consistent observables. For $n_f=6$, the proton structure is described by $2n_f+1=13$ independent PDFs, implying that at least thirteen linearly independent constraints are required to constrain all PDFs. If no such set can be constructed using DIS observables and sum rule constraints alone, then this would establish the existence of an unconstrained flavor subspace that cannot be determined from DIS data without introducing additional assumptions on the PDF parametrization or initial conditions.
    
    A candidate minimal set of thirteen constraints consists of the total momentum sum rule $S_\tmc$, the six flavor-number sum rules $S_\fnc$, and the experimental observables $\sigma_r^{\pm,\mathrm{NC}}$, $\sigma_r^{\pm,\mathrm{CC}}$, $\sigma_r^{cc}$, and $\sigma_r^{bb}$, provided the corresponding forward operators are linearly independent. In that case, the inclusion of neutrino DIS observables or additional sum rules would render the problem formally overdetermined, requiring compatibility through linear dependence among the constraints. Alternatively, replacing one of the $e^\pm$ neutral current observables with a neutrino DIS observable may provide a more diverse flavor basis while preserving full rank. Determining the linear independence of the available DIS observables, and thereby identifying an optimal set of thirteen constraints, remains an important topic for future study.
    
    It seems that there is a good possibility for it to be possible to construct a full rank world data forward operator.
    We also saw how the different observables receive nontrivial corrections at higher orders in the perturbative expansion in $\as$, which might indicate that the linear independence of the operators only improves at higher orders, similarly to the DGLAP equation receiving the most unique coefficients only at NNLO accuracy, which also improves the separation of the PDFs via the PDE constraint. We note that the uniqueness of solutions discussed here is in the sense of a fixed renormalization scheme, flavor number scheme, and other such choices.

    \subsection*{On computational implementation and uncertainties}
    To ready this methodology for a practical application to global analysis, we constructed a discretization-based algorithm for the solution of the world data inverse problem.
    The method leverages the stacking form technique to reduce the practical inverse problem to a matrix--vector reconstruction problem on the level of the full world data problem, including a flavor number scheme. 
    This reduced form of the linear inverse problem can at least in principle be solved with standard computational inverse problems methods such as Tikhonov regularization, which we earlier applied to the inference problem of the dipole amplitude in Ref.~\cite{Hanninen:2025iuv}.
    
    We estimated the expected computational difficulty of the problem, and found that a realistic implementation should be viable with similar discretization parameters as used in other global analyses. The computational implementation can use many of the same standard efficiency gains already used in global analysis, such as pre-computation of the coefficient functions~\cite{Candido:2024rkr:yadism}, which is basically the only truly demanding step in the construction of the forward operators. The reconstruction step is reduced to matrix-vector linear algebra which is highly efficient to perform in compute. The size of the computational problem can be simply reduced with the exclusion of the top and anti-top quarks, as the present day world data in DIS is not sensitive to top physics~\cite{Jiang:2025frv:topq:eic}.

    Since this methodology is developed as an alternative for fitting model parameters, the uncertainty estimation can be done in the same vein as randomly sampling replicas of the world data as is used by the NNPDF collaboration~\cite{NNPDF:2021njg}, which was also applied in Ref.~\cite{Hanninen:2025iuv} to the reconstruction formulation of the dipole amplitude inference problem. This step provides the bootstrapped uncertainties of the reconstructed solution based on the experimental uncertainties of the world data within a fixed theory prescription of the observables. To estimate the theoretical uncertainty caused by the truncated perturbative series, the framework of missing higher order uncertainties can be used~\cite{NNPDF:2024dpb:mhou} and will be relevant for the analysis of the results as the reconstructed PDFs will compensate for any physical effects not accounted for in the forward operators of the observables. 
    
    The novelty of this approach is that the reconstructed solutions nor their uncertainties are biased by a fit ansatz, because none is required. A comparison of the reconstructed solution and its uncertainties with the established global analyses of PDFs---as long as both use the same theory prescription of observables---could shine light on the question of whether the used fit parametrization is capturing all physics that the world data is sensitive to.

    Practical numerical implementation of the presented solution method has a few aspects to consider, all of which can be carried out by leveraging existing implementations. The generation of the forward operators requires the computation of the Wilson coefficient functions on discretized grids, which is implemented for example by yadism~\cite{Candido:2024rkr:yadism}, which also implements the heavy flavor effects in all observables and so would enable the relaxation of the second weak assumption about massless quarks~\ref{post:massless}. Yadism already leverages pre-computation of the coefficient functions on a discretized space without also performing the convolution with the PDFs which is very much aligned with the requirements of this approach, and so it seems like an ideal tested-and-verified computational implementation of DIS observables. 
    
    Second consideration is the implementation of the DGLAP constraint, which could be performed by applying the EKO package~\cite{Candido:2022tld:EKO} to compute the evolution of the reconstructed solutions to quantify the size of the PDE constraint term. Lastly, one needs an implementation for the stacking form reduction of the discretized global data problem, which can be implemented by extending the reconstruction code developed for the dipole amplitude inference problem~\cite{Hanninen:2025iuv,oma4:invdip-github}, for which we have implemented an analogous stacking form for the enforcement of the $x$-evolution of the reconstruction. This reconstruction pipeline will need to be robustly tested in a closure test with known PDF sets, where artificial world data is generated and the capability of the reconstruction algorithm is evaluated by its capability to recover the exactly known PDF fit ground truth used to generate the world data set. Randomly sampled replicas of the generated world data can also be used to test the bootstrapping of the uncertainties of the reconstructions in a setting where the ground truth PDFs are known. 
    
    The inclusion of the top quark can be eliminated to reduce the computational size of the initial problem, as it was kept in the formalism for completeness and for future outlook of the approach. However, with existing experimental energies the top contribution is vanishingly small and undetectable, and so exclusion is completely in line with standard practices in established global analyses. To exclude the top flavor, one just sets all top contributions in the forward operators to zero, or in the flavor space just cut them out of the basis completely, which helpfully reduces the problem size.

    \subsection*{On the analytical solution of the Wilson convolutions}
    The mathematical side of this work also opens a new avenue of inquiry in the inverse problems theory foundations of this approach. To build a more robust mathematical picture of the solution of the PDE constrained integral equation system, we need to consider the following questions: characterization of the existence and uniqueness of solutions to the parton level Wiener--Hopf integral equations defined by the Wilson coefficient functions, possibility of analytical solution of said equations when solutions exist, necessary and sufficient requirements for the existence of solution to the world data level coupled integral equation problem, and how the order of the perturbative expansion affects those criteria. Another aspect is the stability of solutions if they exist: can minute changes in the world data cause large changes in the reconstructed solution of the problem, and if so, the analysis of where and by how much this can happen. These questions are more tractable at NLO accuracy, but perhaps the overall structure endowed by the collinear factorization is enough to construct a more general theory that could incorporate higher order contributions in a general form.

    The presented solution method is very much a proof of principle brute force algorithm where very little of the integral equation structure is leveraged. More elegant algorithms are likely to be possible, which would leverage the mathematical theory developed for the plus-distribution Wiener--Hopf equations, and for the solution of systems of coupled integral equations. Such an improved algorithm could have an improved convergence rate to the solution of the problem, or even mathematically provable convergence---i.e. avoidance of local minima---as has been mathematically proven for example with the D-bar method for electrical impedance tomography~\cite{Mueller:2012:eit}. These considerations would be especially important in the scenario that the reconstructed PDFs would suggest features in the PDFs that have not been included in previous global analysis fit parametrizations.

    \subsection*{Novel aspects of the formalism}
    The inverse problem formalism developed for this work exhibits some novel features both from the perspective of high-energy physics and mathematical inverse problems. Aside from constructing the first of a kind method to solve the PDFs from DIS world data at NLO and higher order accuracy without fitting a parametrized ansatz, we formulated the observables into integral operators, which do not contain non-perturbative degrees of freedom. This was enabled by the ``second factorization'' in which we extracted the PDFs from the hadronic tensor by writing the hadronic tensor into a linear integral operator that acts on the PDFs. At a face value, the unified tensor of the PDFs $\pdtbf$ might seem analogous for example to a spinor, which has functional components that are part of a whole; is this factorization of the PDFs merely a mathematical convenience or something to interpret?
    
    Another new observation is the identification of the Wilson coefficient convolutions as closely related to Wiener--Hopf integral equations, for which solution methods have been constructed, and uniqueness of solutions has been characterized under varying conditions. This seems like a promising avenue to pursue in the development of the analytical solution theory of this approach. A novel feature of the integral equation framing are the plus-distributions seen in the coefficient functions, which have not been mathematically examined in the Wiener--Hopf equation context.

    From the inverse problems theory perspective, the problem of global analysis has some features that seem to not have been studied in the mathematical literature: the number of unknown functions to be inferred is sizable at $2 n_f +1 \in \lbrace9,11,13\rbrace$, and the fact that none of these unknowns are physically possible to be isolated and individually solved from data. All physical observables are only sensitive to their linear combinations, and all PDFs contribute in varying degree, and so a notable number of completely distinct physical observables with their own experiments are needed to achieve mathematical separation capability for the PDFs. The observables exhibit Wiener--Hopf integral equations of the first and second kind, and together with them one needs the QCD sum rules, which are different kinds of integrals from the observables, and prescribe fundamental boundary conditions of the proton such as the conservation of momentum, or the up and down flavor numbers of the proton. On top of all of this is the integro-PDE DGLAP evolution equation, which the reconstructed solutions should satisfy within the perturbative accuracy of the theory expansion, which gives a powerful physical constraint for the solutions, while being a technical challenge.

    \subsection*{Interpretation of the reconstruction}
    We have formulated the problem of global analysis in the mathematical language of indirect sensing, which might raise questions about its interpretation. Consider the Wigner quasiprobability distributions, which are a well-known observable in quantum mechanics~\cite{Wigner:1932,Hillery:Wigner-1984}, and even have been measured with photon~\cite{Lvovsky:2009:photon-wigner-tomo}, and phonon Fock states as reported in Nature~\cite{Chu2018-wigner-tomo}. It has been suggested that the PDFs are a suitable projection of a relativistic Wigner quasiprobability distribution of the generalized transverse momentum-dependent distributions~\cite{Ji:2003ak:pdf-wigner,Belitsky:2003nz:pdf-wigner,Lorce:2025aqp:pdfs-and-generalizations-wigner}. These connections perhaps raise the possibility that the PDFs---once defined within a systematic renormalization scheme---become observable as a whole when solved from data simultaneously in the most general manner, toward which this work opens a new path. At least in a generalized Wandzura–Wilczek approximation it has been shown that the Lorentz invariance relations of the PDFs are not violated~\cite{Metz:2008ib:pdf-lorentz-invariance-violation}, which is an important property of any relativistic observable.

    More concretely, the linear dependence of the DIS world data on the PDFs established in this work raises the question of whether it is possible to show that different observers of the DIS world data can agree on the reconstruction of the PDFs---or some related quantity---within an agreed upon system of schemes, such as $\overline{\mathrm{MS}}+\mathrm{FONLL}$ for example, which would be a step towards establishing the PDFs as a quantum mechanical \gls{observable}.
    If a general functional reconstruction of the PDFs would become possible via this approach, it would open the possibility of using such a reconstruction as a stepping stone to consider inferring other non-perturbative quantities from the world data in a functional sense, such as the running of the strong coupling, the hadronic fragmentation functions~\cite{ParticleDataGroup:2024cfk}, or nuclear modification factors without fitting model parameters. Perhaps this iterative process of inference could have an analogue drawn with the cosmic distance ladder~\cite{Camarena:2019rmj:cosmic-distance}, but instead of peering out into the universe, the task is to indirectly sense the non-perturbative QCD structure of the proton in ever finer resolution.

    \subsection*{Outlook}

    We have in this work constructed a first of a kind methodology to reformulate the inference problem of global analysis of the PDFs into a functional optimization problem defined by the established theory of perturbative QCD. Our core goal is to show that the solution of the PDFs from DIS world data is at least in principle possible without fitting model parameters. This inverse problem formalism of indirect sensing developed for perturbative QCD is very much aimed at future generations of high-energy physics experiments which will produce evermore high fidelity experimental data, which will hopefully enable the more advanced mathematical analysis methods developed in this work. 
    
    A generalized functional extraction of the PDFs from experimental data would enable the independent testing of the assumptions made in global analysis of the PDFs in the past, such as the choice of the functional form ansatz used in fit parametrization. This will hopefully shine complementary light on the initial condition of the PDFs, and perhaps help analyze the behavior of the PDFs at low $Q^2$ near the regime where the collinear factorization begins to falter and higher twist contributions become more prominent. Ultimately, we hope that such analyses would help to understand the non-perturbative structure of the proton in a more fundamental manner.

    To answer the question with which this article began, it seems that---in a mathematical sense---the global data inference of the proton PDFs is roughly analogous to solving thirteen totally overlapping computed tomography problems from datasets which are only sensitive to some limited linear combinations of the densities of the subject. And to overcome this fundamental challenge the approach is to simultaneously use most or all of the data from various experiments from around the globe, the data from some or all of which fall under limited data constraints associated with partial data inverse problems.
    On top of all of which is the challenge of accurately combining measurements from completely different experiments with their own distinct physics, methods of measurement, and uncertainties.
    So, in the mathematical sense that computed tomography is founded on solving an integral equation defined by the X-ray transform, the inverse problem of global analysis of the PDFs can be cast as a coupled system of linear integral equations of different kinds to solve thirteen unknown functions simultaneously.

\vspace{3mm}
\noindent
\textbf{Acknowledgments} 
The author is indebted to J. Ilmavirta for uncountably many discussions about the mathematical principles of inverse problems that helped the author to find the perspective of this work, and to F. Hekhorn for extremely helpful discussions and correspondence on QCD, DIS, and global analysis, and for helpful comments on the manuscript.
The author also thanks M. Kuha for helpful and encouraging discussions, and feedback on the manuscript; S. Siltanen for much-appreciated correspondence and manuscript feedback; K. Kansanen for insight on quantum mechanical observables; and J. Bl\"umlein for helpful correspondence.

The author is supported by the Research Council of Finland (Flagship of Advanced Mathematics for Sensing, Imaging and Modelling grant 359208), and the Vilho, Yrjö and Kalle Väisälä Foundation.

\bibliographystyle{JHEP-2modlong.bst}
\bibliography{refs}

\appendix

\begin{landscape}

\section{Full expressions for the DIS world data forward operator}
    \label{app:full-global-data-operators}
    We record here the full expression for the global data operator defined by Eq.~\eqref{eq:tensor-ip-lo}, which is the forward operator of the inverse problem for the thirteen unknown parton distribution functions of the proton to be inferred from data:
    {\footnotesize
    \begin{align}
        & \mathbf{G}^{lp\to l' X}_\mathrm{LO} (x,Q^2,y) =
        \notag \\
        & \begin{bmatrix}
            0 &
            a_{1}e_d^2 & 
            a_{1}e_u^2   &
            a_{1}e_s^2   &
            a_{1}e_c^2   &
            a_{1}e_b^2   &
            a_{1}e_t^2   &
            a_{1}e_{\bar d}^2   &
            a_{1}e_{\bar u}^2   &
            a_{1}e_{\bar s}^2   &
            a_{1}e_{\bar c}^2   &
            a_{1}e_{\bar b}^2   &
            a_{1}e_{\bar t}^2  
            \\
            0 &
            a_2^+(d) &
            a_2^+(u) &
            a_2^+(s) &
            a_2^+(c) &
            a_2^+(b) &
            a_2^+(t) &
            a_2^+(\bar d)&
            a_2^+(\bar u)&
            a_2^+(\bar s)&
            a_2^+(\bar c)&
            a_2^+(\bar b)&
            a_2^+(\bar t) 
            \\
            0 &
            a_2^-(d) &
            a_2^-(u) &
            a_2^-(s) &
            a_2^-(c) &
            a_2^-(b) &
            a_2^-(t) &
            a_2^-(\bar d)&
            a_2^-(\bar u)&
            a_2^-(\bar s)&
            a_2^-(\bar c)&
            a_2^-(\bar b)&
            a_2^-(\bar t) 
            \\
            0 & 
            (1-y)^2 & 0 &
            (1-y)^2 & 0 &
            (1-y)^2 & 0 &
            0 & 1 &
            0 & 1 &
            0 & 1
            \\
            0 &
            0 & 1 & 
            0 & 1 &
            0 & 1 &
            (1-y)^2 & 0 &
            (1-y)^2 & 0 &
            (1-y)^2 & 0
            \\ 
            e_c^2 \! \left[ \!\mathcal{H}^{(1)}_{g,2} \!-\! \frac{y^2}{Y_+} \mathcal{H}^{(1)}_{g,L}\! \right] &
            0 & 0& 0& 0& 0& 0&
            0 & 0& 0& 0& 0 &0
            \\ 
            e_b^2 \!\left[ \!\mathcal{H}^{(1)}_{g,2} \!-\! \frac{y^2}{Y_+} \mathcal{H}^{(1)}_{g,L}\!\right] &
            0 & 0& 0& 0& 0& 0&
            0 & 0& 0& 0& 0 &0
            \\ 
            0 &
            a_3 & a_3 & 2 a_3 & 0 & 0 & 0 &
            (1-y)^2 a_3 & (1-y)^2 a_3 & 0 & -2(1-y)^2 a_3 & 0 & 0
            \\ 
            0 &
            a_3 & a_3 & 0 & 2a_3 & 0 & 0 &
            (1-y)^2 a_3 & (1-y)^2 a_3 & -2(1-y)^2 a_3 & 0 & 0 & 0
            \\
            2 \frac{\as}{\pi} \undertilde{e}^2 \Ical^\mathrm{AM}_{L,g} &
            e_d^2 x \Ical^\mathrm{AM}_{L,q} & 
            e_u^2 x \Ical^\mathrm{AM}_{L,q} & 
            e_s^2 x \Ical^\mathrm{AM}_{L,q} & 
            e_c^2 x \Ical^\mathrm{AM}_{L,q} & 
            e_b^2 x \Ical^\mathrm{AM}_{L,q} & 
            e_t^2 x \Ical^\mathrm{AM}_{L,q} &
            e_{\bar d}^2 x \Ical^\mathrm{AM}_{L,q} &
            e_{\bar u}^2 x \Ical^\mathrm{AM}_{L,q} &
            e_{\bar s}^2 x \Ical^\mathrm{AM}_{L,q} &
            e_{\bar c}^2 x \Ical^\mathrm{AM}_{L,q} &
            e_{\bar b}^2 x \Ical^\mathrm{AM}_{L,q} &
            e_{\bar t}^2 x \Ical^\mathrm{AM}_{L,q}
            \\
            0&
            x \phi^{e^\pm}_{3,d} & x \phi^{e^\pm}_{3,u} & x \phi^{e^\pm}_{3,s}& 
            x \phi^{e^\pm}_{3,c} & x \phi^{e^\pm}_{3,b} & x \phi^{e^\pm}_{3,t}&
            -x \phi^{e^\pm}_{3,\bar d} & -x \phi^{e^\pm}_{3,\bar u} & -x \phi^{e^\pm}_{3,\bar s} &
            -x \phi^{e^\pm}_{3,\bar c} & -x \phi^{e^\pm}_{3,\bar b} & -x \phi^{e^\pm}_{3,\bar t}
            \\
            0 &
            x & x & 2 x & 0 & 0 & 0 &
            x & x & 0 & 2 x & 0 & 0
            \\
            0 &
            x & x & 2x & 0 & 0 & 0 &
            - x & - x & 0 & -2x & 0 & 0
            \\
            0 &
            x & x & 0 & 2 x & 0 & 0 &
            x & x & 2 x & 0 & 0 & 0
            \\
            0 &
            x & x & 0 & 2 x  & 0 & 0 &
            - x & - x & -2x & 0 & 0 & 0
            \\
                0 &
                \icalid & 0 & 0 & 0 & 0 & 0 &
                -\icalid & 0 & 0 & 0 & 0 & 0
                \\
                0 &
                0 & \icalid & 0 & 0 & 0 & 0&
                0 & -\icalid & 0 & 0 & 0 &0
                \\
                0 &
                0 & 0 & \icalid & 0 & 0 & 0&
                0 & 0 & -\icalid & 0 & 0 &0
                \\
                0 &
                0 & 0 & 0 & \icalid & 0 &0&
                0 & 0 & 0 & -\icalid & 0 &0
                \\
                0 &
                0 & 0 & 0 & 0 & \icalid &0&
                0 & 0 & 0 & 0 & -\icalid &0
                \\
                0 &
                0 & 0 & 0 & 0 & 0 & \icalid &
                0 & 0 & 0 & 0 & 0 & -\icalid
            \\
                \icalx &
                \icalx & \icalx & \icalx & \icalx & \icalx & \icalx &
                \icalx & \icalx & \icalx & \icalx & \icalx & \icalx
            \\ 
                0 &
                \icalid & \icalid & 0 & 0 & 0 & 0 &
                -\icalid & -\icalid & 0 & 0 & 0 & 0
            \\ 
            0 &
                -\icalid & \icalid & 0 & 0 & 0 & 0 &
                \icalid & -\icalid & 0 & 0 & 0 & 0
            \\ 
            0 &
                -\frac{1}{3}\icalid & \frac{1}{3}\icalid & 0 & 0 & 0 & 0 &
                -\frac{1}{3}\icalid & \frac{1}{3}\icalid & 0 & 0 & 0 & 0
        \end{bmatrix}
        \renewcommand*{\arraystretch}{1.0}
        \begin{matrix}
            \eqref{eq:fwdop-incldis-lo} \vphantom{\icalid_2}\\
            \eqref{eq:fwdop-nc-crosssec-lo} \rule{0pt}{0.12cm}\\
            \eqref{eq:fwdop-nc-crosssec-lo} \rule{0pt}{0.12cm}\\
            \eqref{eq:fwdop-cc-lo} \rule{0pt}{0.15cm}\\
            \eqref{eq:fwdop-cc-lo2-eminus} \rule{0pt}{0.13cm}\\
            \eqref{eq:fwdop-heavyQQ-LO-Acc} \rule{0pt}{0.45cm}\\
            \eqref{eq:fwdop-heavyQQ-LO-Abb} \rule{0pt}{0.45cm}\\
            \eqref{eq:fwdop-nudis-nuN-cs-isoscalar} \rule{0pt}{0.35cm}\\
            \eqref{eq:fwdop-nudis-antinuN-cs-isoscalar} \rule{0pt}{0.35cm}\\
            \eqref{eq:fwdop-incldis-fl-lo} \vphantom{\icalid}\\
            \eqref{eq:fwdop-nc-Z-LO-f3} \vphantom{\icalid}\\
            \eqref{eq:fwdop-nudis-f2-nu} \vphantom{\icalid}\\
            \eqref{eq:fwdop-nudis-f3-nu} \vphantom{\icalid}\\
            \eqref{eq:fwdop-nudis-f2-barnu} \vphantom{\icalid}\\
            \eqref{eq:fwdop-nudis-f3-barnu} \vphantom{\icalid}\\
            \eqref{eq:fwdop-sumrule-fnc} \vphantom{\icalid}\\
            \eqref{eq:fwdop-sumrule-fnc} \vphantom{\icalid}\\
            \eqref{eq:fwdop-sumrule-fnc} \vphantom{\icalid}\\
            \eqref{eq:fwdop-sumrule-fnc} \vphantom{\icalid}\\
            \eqref{eq:fwdop-sumrule-fnc} \vphantom{\icalid}\\
            \eqref{eq:fwdop-sumrule-fnc} \vphantom{\icalid}\\
            \eqref{eq:fwdop-sumrule-tmc} \vphantom{\icalid}\\
            \eqref{eq:fwdop-sumrule-GLS} \vphantom{\icalid}\\
            \eqref{eq:fwdop-sumrule-adler} \vphantom{\icalid}\\
            \eqref{eq:fwdop-sumrule-gottfried} \vphantom{\icalid}
        \end{matrix}
    \end{align}
    }
    \end{landscape}
    \newpage
    \begin{landscape}
    At NLO accuracy the global data operator, the elements of which are integral operators, from Eq.~\eqref{eq:tensor-ip-nlo} is:
    {\small
    \begin{align}
        & \mathbf{G}^{lp\to l' X}_\mathrm{NLO}(x,Q^2,y) =
        \notag \\
        & \begin{bmatrix}
            \undertilde{e}^2 \mathcal{I}_{2-L,g} &
            e_d^2 \Ical_{2-L,q} & 
            e_u^2 \Ical_{2-L,q} &
            e_s^2 \Ical_{2-L,q}  &
            e_c^2 \Ical_{2-L,q}  &
            e_b^2 \Ical_{2-L,q}  &
            e_t^2 \Ical_{2-L,q}  &
            e_{\bar d}^2 \Ical_{2-L,q}  &
            e_{\bar u}^2 \Ical_{2-L,q}  &
            e_{\bar s}^2 \Ical_{2-L,q}  &
            e_{\bar c}^2 \Ical_{2-L,q}  &
            e_{\bar b}^2 \Ical_{2-L,q}  &
            e_{\bar t}^2 \Ical_{2-L,q} 
            \\
            \phi^+_g \Ccal_{2-L,g} &
            \Ccal_{q,d}^{+,\mathrm{NC}} &
            \Ccal_{q,u}^{+,\mathrm{NC}} &
            \Ccal_{q,s}^{+,\mathrm{NC}} &
            \Ccal_{q,c}^{+,\mathrm{NC}} &
            \Ccal_{q,b}^{+,\mathrm{NC}} &
            \Ccal_{q,t}^{+,\mathrm{NC}} &
            \Ccal_{q,\bar d}^{+,\mathrm{NC}} &
            \Ccal_{q,\bar u}^{+,\mathrm{NC}} &
            \Ccal_{q,\bar s}^{+,\mathrm{NC}} &
            \Ccal_{q,\bar c}^{+,\mathrm{NC}} &
            \Ccal_{q,\bar b}^{+,\mathrm{NC}} &
            \Ccal_{q,\bar t}^{+,\mathrm{NC}}
            \\
            \phi^-_g \Ccal_{2-L,g} &
            \Ccal_{q,d}^{-,\mathrm{NC}} &
            \Ccal_{q,u}^{-,\mathrm{NC}} &
            \Ccal_{q,s}^{-,\mathrm{NC}} &
            \Ccal_{q,c}^{-,\mathrm{NC}} &
            \Ccal_{q,b}^{-,\mathrm{NC}} &
            \Ccal_{q,t}^{-,\mathrm{NC}} &
            \Ccal_{q,\bar d}^{-,\mathrm{NC}} &
            \Ccal_{q,\bar u}^{-,\mathrm{NC}} &
            \Ccal_{q,\bar s}^{-,\mathrm{NC}} &
            \Ccal_{q,\bar c}^{-,\mathrm{NC}} &
            \Ccal_{q,\bar b}^{-,\mathrm{NC}} &
            \Ccal_{q,\bar t}^{-,\mathrm{NC}}
            \\
            \frac{Y_+}{2}\Ccal_{2-L,g} &
            \Ccal^{W^+}_{q,d} & 0 & \Ccal^{W^+}_{q,s} & 0 & \Ccal^{W^+}_{q,b} & 0 &
            0 & \Ccal^{W^+}_{q, \bar u} & 0 & \Ccal^{W^+}_{q, \bar c} & 0 & \Ccal^{W^+}_{q, \bar t}
            \\
            \frac{Y_+}{2}\Ccal_{2-L,g} &
            0 & \Ccal^{W^-}_{q,u} & 0 & \Ccal^{W^-}_{q,c} & 0 & \Ccal^{W^-}_{q,t} &
            \Ccal^{W^-}_{q, \bar d} & 0 & \Ccal^{W^-}_{q, \bar s} & 0 & \Ccal^{W^-}_{q, \bar b} & 0
            \\ 
            \mathcal{H}_{r,g,c}^{\text{S}} &
            \Ical^{L+H}_{r,d,c} &
            \Ical^{L+H}_{r,u,c} &
            \Ical^{L+H}_{r,s,c} &
            0 &
            0 & 0 &
            \Ical^{L+H}_{r,\bar d,c} &
            \Ical^{L+H}_{r,\bar u,c} &
            \Ical^{L+H}_{r,\bar s,c} &
            0 &
            0 & 0
            \\ 
            \mathcal{H}_{r,g,b}^{\text{S}} &
            \Ical^{L+H}_{r,d,b} &
            \Ical^{L+H}_{r,u,b} &
            \Ical^{L+H}_{r,s,b} &
            0 &
            0 & 0 &
            \Ical^{L+H}_{r,\bar d,b} &
            \Ical^{L+H}_{r,\bar u,b} &
            \Ical^{L+H}_{r,\bar s,b} &
            0 &
            0 & 0
            \\ 
            \Ccal_g ^{\nu p} &
            \Ccal_{d}^{\nu{}^A_Z\mathrm{N}} &
            \Ccal_{u}^{\nu{}^A_Z\mathrm{N}} &
            \Ccal_s ^{\nu p} &
            \Ccal_c ^{\nu p} &
            \Ccal_b ^{\nu p} &
            \Ccal_t ^{\nu p} &
            \Ccal_{\bar d}^{\nu{}^A_Z\mathrm{N}} &
            \Ccal_{\bar u}^{\nu{}^A_Z\mathrm{N}} &
            \Ccal_{\bar s} ^{\nu p} &
            \Ccal_{\bar c} ^{\nu p} &
            \Ccal_{\bar b} ^{\nu p} &
            \Ccal_{\bar t} ^{\nu p}
            \\ 
            \Ccal_g ^{\bar \nu p} &
            \Ccal_{d}^{\bar \nu{}^A_Z\mathrm{N}} &
            \Ccal_{u}^{\bar \nu{}^A_Z\mathrm{N}} &
            \Ccal_s ^{\bar \nu p} &
            \Ccal_c ^{\bar \nu p} &
            \Ccal_b ^{\bar \nu p} &
            \Ccal_t ^{\bar \nu p} &
            \Ccal_{\bar d}^{\bar \nu{}^A_Z\mathrm{N}} &
            \Ccal_{\bar u}^{\bar \nu{}^A_Z\mathrm{N}} &
            \Ccal_{\bar s} ^{\bar \nu p} &
            \Ccal_{\bar c} ^{\bar \nu p} &
            \Ccal_{\bar b} ^{\bar \nu p} &
            \Ccal_{\bar t} ^{\bar \nu p}
            \\
            2 \frac{\as}{\pi} \undertilde{e}^2 \Ical_{L,g} &
            e_d^2 x \Ical_{L,q} &
            e_u^2 x \Ical_{L,q} & 
            e_s^2 x \Ical_{L,q} & 
            e_c^2 x \Ical_{L,q} & 
            e_b^2 x \Ical_{L,q} & 
            e_t^2 x \Ical_{L,q} &
            e_{\bar d}^2 x \Ical_{L,q} &
            e_{\bar u}^2 x \Ical_{L,q} &
            e_{\bar s}^2 x \Ical_{L,q} &
            e_{\bar c}^2 x \Ical_{L,q} &
            e_{\bar b}^2 x \Ical_{L,q} &
            e_{\bar t}^2 x \Ical_{L,q}
            \\
            0 &
            \phi^{\pm}_{3,d} \Ccal_{3,q} &
            \phi^{\pm}_{3,u} \Ccal_{3,q} &
            \phi^{\pm}_{3,s} \Ccal_{3,q} &
            \phi^{\pm}_{3,c} \Ccal_{3,q} &
            \phi^{\pm}_{3,b} \Ccal_{3,q} &
            \phi^{\pm}_{3,t} \Ccal_{3,q} &
            -\phi^{\pm}_{3,\bar d} \Ccal_{3,q} &
            -\phi^{\pm}_{3,\bar u} \Ccal_{3,q} &
            -\phi^{\pm}_{3,\bar s} \Ccal_{3,q} &
            -\phi^{\pm}_{3,\bar c} \Ccal_{3,q} &
            -\phi^{\pm}_{3,\bar b} \Ccal_{3,q} &
            -\phi^{\pm}_{3,\bar t} \Ccal_{3,q}
            \\
            \frac{1}{2} \Ccal_{2,g} &
            \frac{Z}{A}\Ccal_{2,q} &  \frac{A-Z}{A}\Ccal_{2,q} & \Ccal_{2,q} & 0 & \Ccal_{2,q} & 0 &
            \frac{A-Z}{A}\Ccal_{2,q} & \frac{Z}{A}\Ccal_{2,q} & 0 & \Ccal_{2,q} & 0 & \Ccal_{2,q}
            \\
            0 &
            \frac{Z}{A} \Ccal_{3,q} & \frac{A-Z}{A}\Ccal_{3,q} & \Ccal_{3,q} & 0 & \Ccal_{3,q} & 0 &
            -\frac{A-Z}{A}\Ccal_{3,q} & -\frac{Z}{A} \Ccal_{3,q} & 0 & - \Ccal_{3,q} & 0 & - \Ccal_{3,q}
            \\
            \frac{1}{2} \Ccal_{2,g} &
            \frac{A-Z}{A}\Ccal_{2,q} & \frac{Z}{A}\Ccal_{2,q} & 0 & \Ccal_{2,q} & 0 & \Ccal_{2,q} & 
            \frac{Z}{A}\Ccal_{2,q} &  \frac{A-Z}{A}\Ccal_{2,q} & \Ccal_{2,q} & 0 & \Ccal_{2,q}& 0
            \\
            0 &
            \frac{A-Z}{A}\Ccal_{3,q} & \frac{Z}{A}\Ccal_{3,q} & 0 & \Ccal_{3,q} & 0 & \Ccal_{3,q} & 
            -\frac{Z}{A} \Ccal_{3,q} & -\frac{A-Z}{A}\Ccal_{3,q} & - \Ccal_{3,q} & 0 & - \Ccal_{3,q}& 0
            \\
                0 &
                \icalid & 0 & 0 & 0 & 0 & 0 &
                -\icalid & 0 & 0 & 0 & 0 & 0
                \\
                0 &
                0 & \icalid & 0 & 0 & 0 & 0&
                0 & -\icalid & 0 & 0 & 0 &0
                \\
                0 &
                0 & 0 & \icalid & 0 & 0 & 0&
                0 & 0 & -\icalid & 0 & 0 &0
                \\
                0 &
                0 & 0 & 0 & \icalid & 0 &0&
                0 & 0 & 0 & -\icalid & 0 &0
                \\
                0 &
                0 & 0 & 0 & 0 & \icalid &0&
                0 & 0 & 0 & 0 & -\icalid &0
                \\
                0 &
                0 & 0 & 0 & 0 & 0 & \icalid &
                0 & 0 & 0 & 0 & 0 & -\icalid
            \\
                \icalx &
                \icalx & \icalx & \icalx & \icalx & \icalx & \icalx &
                \icalx & \icalx & \icalx & \icalx & \icalx & \icalx
            \\ 
                0 &
                \icalid & \icalid & 0 & 0 & 0 & 0 &
                -\icalid & -\icalid & 0 & 0 & 0 & 0
            \\ 
            0 &
                -\icalid & \icalid & 0 & 0 & 0 & 0 &
                \icalid & -\icalid & 0 & 0 & 0 & 0
            \\ 
            0 &
                -\frac{1}{3}\icalid & \frac{1}{3}\icalid & 0 & 0 & 0 & 0 &
                -\frac{1}{3}\icalid & \frac{1}{3}\icalid & 0 & 0 & 0 & 0
        \end{bmatrix}
        \renewcommand*{\arraystretch}{1.005}
        \begin{matrix}
            \eqref{eq:fwdop-incdis-nlo} \rule{0pt}{0.27cm}\\
            \eqref{eq:fwdop-nlo-nc-cs} \rule{0pt}{0.38cm}\\
            \eqref{eq:fwdop-nlo-nc-cs} \rule{0pt}{0.38cm}\\
            \eqref{eq:fwdop-nlo-cc-A-w+} \rule{0pt}{0.38cm}\\
            \eqref{eq:fwdop-nlo-cc-A-w-} \rule{0pt}{0.38cm}\\
            \eqref{eq:fwdop-nlo-hq-cc} \rule{0pt}{0.38cm}\\
            \eqref{eq:fwdop-nlo-hq-bb} \rule{0pt}{0.38cm}\\
            \eqref{eq:fwdop-nlo-nudis-isoscalarmixing-WA} \rule{0pt}{0.38cm}\\
            \eqref{eq:fwdop-nlo-nudis-isoscalarmixing-WA} \rule{0pt}{0.38cm}\\
            \eqref{eq:fwdop-incdis-nlo-fl} \vphantom{E_{\bar d}^2}\\
            \eqref{eq:fwdop-nlo-nc-f3} \vphantom{E_{\bar d}^2}\\
            \eqref{eq:fwdop-nlo-nudis-f2fl-nu} \vphantom{E_{\bar d}^2}\\
            \eqref{eq:fwdop-nlo-nudis-f3-nu-w+} \vphantom{E_{\bar d}^2}\\
            \eqref{eq:fwdop-nlo-nudis-f2fl-nubar} \vphantom{E_{\bar d}^2}\\
            \eqref{eq:fwdop-nlo-nudis-f3-barnu-w-} \vphantom{E_{\bar d}^2}\\
            \eqref{eq:fwdop-sumrule-fnc} \vphantom{E_{\bar d}^2}\\
            \eqref{eq:fwdop-sumrule-fnc} \vphantom{E_{\bar d}^2}\\
            \eqref{eq:fwdop-sumrule-fnc} \vphantom{E_{\bar d}^2}\\
            \eqref{eq:fwdop-sumrule-fnc} \vphantom{E_{\bar d}^2}\\
            \eqref{eq:fwdop-sumrule-fnc} \vphantom{E_{\bar d}^2}\\
            \eqref{eq:fwdop-sumrule-fnc} \vphantom{E_{\bar d}^2}\\
            \eqref{eq:fwdop-sumrule-tmc} \vphantom{E_{\bar d}^2}\\
            \eqref{eq:fwdop-sumrule-GLS} \vphantom{E_{\bar d}^2}\\
            \eqref{eq:fwdop-sumrule-adler} \vphantom{E_{\bar d}^2}\\
            \eqref{eq:fwdop-sumrule-gottfried} \vphantom{E_{\bar d}^2}
        \end{matrix}
        \label{eq:app:full-fwdop-G-nlo}
    \end{align}
    }
    \end{landscape}
    \noindent
    Each operator row has an equation reference that points to the definition of the operator elements in the main text.
    We use notations of Sec.~\ref{sec:forward-problem-lo} and defined the shorthands:
    \begin{align}
        a_1 & \coloneqq \frac{xy^2}{2} + x (1-y),
        \\
        a_2^\pm(a) & \coloneqq Y_+ \phi^{e^\pm}_{2,a} \mp Y_- \phi^{e^\pm}_{3,a},
        \\
        a_3 & \coloneqq \frac{G_F^2}{2\pi} sx.
    \end{align}
    For the NLO accuracy construction, we absorb the identity terms back into the $z$-integral as a $\delta$-function term, which writes for $\sigma^\gamma$ as:
    \begin{equation}
        \Ical_{2-L,q}[f] \equiv \int_x^1 x f(z) \delta(z-x) \ud z + \Ical_{2-L,q}[f],
    \end{equation}
    which is the same as if we had used the $C_2$ and $C_L$ coefficient functions given in Sec.~\ref{sec:nlo-nc}.
    We also define the shorthands:
    \begin{align}
        \Ccal_{2-L,g}[f] & \coloneqq \qty( \Ccal_{2,g}[f] - \frac{y^2}{Y_+} \Ccal_{L,g}[f]),
        \\
        \Ical^{L+H}_{r,q,Q}[f] & \coloneqq \Ical^{L+H}_{2,q,Q}[f] - \frac{y^2}{Y_+}\Ical^{L+H}_{L,q,Q}[f],
    \end{align}
    and analogously for the gluon contribution $\mathcal{H}_{i,g}$ just with the replacement of $q \to g$.
    The notation and integral operators of the NLO forward operator were defined in Sec.~\ref{sec:forward-problem-nlo} and the matrix-valued integral equation framing was discussed in Sec.~\ref{sec:tensor-ip-nlo}.

\section{Smoothness of the parton distribution functions} \label{app:pdf-smooth}

    We stated our belief in the smoothness---or at least high regularity---of the PDFs as a ``postulate'' in Sec.~\ref{sec:intro}. However, this might well be a mathematically provable property of the PDFs. Taking the definition given in Ref.~\cite{Collins:1987pm:collinear-factorization} for the PDF $f_q$ of a quark flavor $q$:
    \begin{equation}
        f_{q,h}(x,Q) \coloneqq \frac{1}{4\pi} \int_\mathbb{R} e^{-i x p^+ y^-} \langle p | \bar \psi_q(0,y^-,0^T) \gamma^+ P \psi_q(0) | p \rangle_R \ud y^-,
    \end{equation}
    where the superscript $+$ and $-$ are the plus- and minus-components in light-cone coordinates, $\psi_q$ is the wavefunction of the quark in the hadron $h$, $P$ is a path-ordered exponential of the gluon field, and the subscript $R$ denotes that the quantity is defined after renormalization~\cite{Collins:1987pm:collinear-factorization}. Corresponding definitions hold for antiquarks and gluons. We conjecture the following:
    \begin{conjecture}
        Parton distribution functions $f_{q,h}(x,Q): [0,1] \times \mathbb{R}_+ \mapsto \mathbb{R}$ are smooth, or at least quite regular, functions of $x$, and perhaps also to some degree in $Q$.
    \end{conjecture}
    Sufficient regularity of the PDFs as functions of $x$, like smoothness or at least continuous differentiability, is required from the presumed solutions by the method constructed in Sec.~\ref{sec:discretization}. Assuming smoothness in $x$ is standard practice in all global analyses of the PDFs, which work with parametrizations like Eq.~\ref{eq:fit-ansatz}. Non-negativity has also been assumed to be satisfied, and recent progress has been made on verifying it to hold~\cite{Candido:2023ujx:positivity-of-pdf}. On the other hand the DGLAP evolution equations clearly describe the PDFs as solutions of a system of PDEs as functions of the scale $\mu^2$, often set to $Q^2$. However, at a fixed order of perturbation theory and a use of a \gls{flavor number scheme}, the PDF results of global analyses exhibit discontinuities in $Q^2$ at the boundaries of the \gls{flavor number scheme} matching\footnote{Global analyses of the PDFs enforce the continuity at the observable level of the problem, since the measured data appear without discontinuities. The fact that the pQCD theory has to be applied at a fixed truncated order of the expansion, forces one to create a discontinuity at the boundary of matching the \gls{flavor number scheme} regimes, which might suggest that the appearance of the discontinuity is in some sense artificial, and perhaps not an unavoidable fundamental feature of QCD and the PDFs.}. Whether this type of a discontinuity of the PDFs would appear in an ``all orders consistent theory of QCD global analysis'' is not quite clear. This conjecture is left for future work.

\section{Unpolarized two-loop Wilson coefficients for inclusive heavy flavor production in DIS}
\label{app:heavy-flavor-wilson-coeffs}
    \newcommand{\HA}{{\rm H}}

    In this section we recollect the heavy flavor production Wilson coefficient $H_{i,q}^{(2)}$ from Ref.~\cite{Blumlein:2019qze:Hq-heavy-wilson}. We need the harmonic polylogarithms~\cite{Remiddi:1999ew} defined recursively by
    \begin{equation}
    \HA_{b,\vec{a}}(z) \coloneqq \int_0^z dy f_b(y) \HA_{\vec{a}}(y),~~\HA_\emptyset = 1,~~b, a_i \in \{-1,0,1\}.
    \end{equation}
    We will also need the following integration kernel function elements known as letters:
    \begin{equation}\label{eq:HPL1}
    f_0(z) = \frac{1}{z},~~~~
    f_1(z) = \frac{1}{1-z},~~~~
    f_{-1}(z) = \frac{1}{1+z},
    \end{equation}
    and
    \begin{align}
        f_{w_1}(t) &= \frac{1}{1 - k t},
        \\
        f_{w_2}(t) &= \frac{1}{1 + k t},
        \\
        f_{w_3}(t) &= \frac{1}{\beta + t},
        \\
        f_{w_4}(t) &= \frac{1}{\beta - t},
        \\
        f_{w_5}(t) &= \frac{1}{k - z - ( 1 - z ) k t },
        \\
        f_{w_6}(t) &= \frac{1}{k + z - ( 1 - z ) k t },
        \\
        f_{w_7}(t) &= \frac{1}{k - z + ( 1 - z ) k t },
        \\
        f_{w_8}(t) &= \frac{1}{k + z + ( 1 - z ) k t },
        \\
        f_{w_9}(t) &= \frac{t}{k^2 \left(1 - t^2 \left(1 - z^2\right)\right)-z^2},
        \\
        f_{w_{10}}(t) &= \frac{1}{t \sqrt{1-t^2} \sqrt{1-k^2 t^2}},
        \\
        f_{w_{11}}(t) &= \frac{t}{\sqrt{1-t^2} \sqrt{1-k^2 t^2}},
        \\
        f_{w_{12}}(t) &= \frac{t}{\sqrt{1-t^2} \sqrt{1-k^2 t^2} \left(k^2 \left(1 - t^2 \left(1 - 
        z^2\right)\right)-z^2\right)} ,
    \end{align}
    where $\beta \equiv \beta(z, m^2, Q^2) \coloneqq \sqrt{1 - \frac{4 m^2}{Q^2}\frac{z}{1-z}}$, and $k \equiv k(z, \beta) \coloneqq \frac{\sqrt{z}}{\sqrt{1-(1-z)\beta^2}}$.
    With this machinery the heavy flavor Wilson coefficient functions are~\cite{Blumlein:2019qze:Hq-heavy-wilson}:
    \begin{align}
    H_{L,q}^{(2),\text{PS}} &= 
    \cf T_F \Biggl\{-\frac{8 P_1}{3 z} \biggl\{
      k \biggl[ \HA_{w_{1}}^2 - \HA_{w_{2}}^2
    + (1-z) \bigl( \HA_{w_{5},w_{1}} 
          + \HA_{w_{6},w_{2}}
          - \HA_{w_{7},w_{2}}
    \nonumber \\ &
          - \HA_{w_{8},w_{1}}
          - \HA_{w_{5}} \HA_{w_{1}}
          + \HA_{w_{8}} \HA_{w_{1}}
          - \HA_{w_{6}} \HA_{w_{2}}
          + \HA_{w_{7}} \HA_{w_{2}}
    \bigr) \biggr]
    + 2 \bigl(  \HA_{w_{1},w_{4}}
          + \HA_{w_{2},w_{4}}
          + \HA_{w_{3},w_{1}}
    \nonumber \\ &
          + \HA_{w_{3},w_{2}}
    \bigr)
    -\bigl(
              2 \HA_{w_{3}}        
            - 6 \ln(k)
            + \ln\big(1-k^2\big)
            - \ln(k^2 - z^2)
            + 2 \ln \big(k^2-z\big)
    \bigr) \bigl[ \HA_{w_{1}} 
    \nonumber \\ &
    + \HA_{w_{2}} \bigr]
    \biggr\}
    -\frac{16 (1-z) \beta P_2}{3 z} \ln (k^2 - z^2 )
    -\frac{16 (1-z) \beta  P_3}{9 k^2 z}
    +\frac{8 (1-k^2) (1-z) P_4}{3 k^4 z} 
    \biggl[
            \HA_{w_{5},0}
    \nonumber \\ &
          - \HA_{w_{6},0}
          + \HA_{w_{7},0}
          - \HA_{w_{8},0}
          - \bigl(
            \HA_{w_{5}}
          - \HA_{w_{6}}
          + \HA_{w_{7}}
          - \HA_{w_{8}}
          \bigr) \HA_{0}
    \biggr]
    + \frac{16 (1-k^2) P_4}{3 k^4 z} \bigl(
            \HA_{w_{1}}
    \nonumber \\ &
          + \HA_{w_{2}}
    \bigr) \HA_0
    +\frac{32  P_5}{3 k^2} \bigl( \HA_{-1} \HA_1 - 2 \HA_{-1,1} \bigr)
    +\frac{32 P_6}{3 k^4 z} \bigl( \HA_{w_{1},0} + \HA_{w_{2},0} \bigr)
    +\frac{16 P_7}{3 k^4} \bigl( \HA_1 \HA_{w_{1}} 
    \nonumber \\ &
    - \HA_{-1} \HA_{w_{2}} \bigr)
    +\frac{16 P_8}{3 k^4} \bigl( \HA_1 \HA_{w_{2}} - \HA_{-1} \HA_{w_{1}} \bigr)
    -\frac{64 P_9}{3 k^2 z \beta } \HA_{w_{3}}
    -\frac{16 (1-k^2) (1-z^2) P_{10}}{3 k^2} 
    \biggl[
          \HA_{w_{9},1} 
    \nonumber \\ &
            + \HA_{w_{9},-1}
          - (1-z) k \bigl(
                  \HA_{w_{9},w_{5}}
                + \HA_{w_{9},w_{6}}
                + \HA_{w_{9},w_{7}}
                + \HA_{w_{9},w_{8}}
          \bigr)
    \biggr]
    -\frac{16 P_{11}}{3 k^2} \bigl( \HA_1^2 - \HA_{-1}^2 \bigr)
    \nonumber \\ &
    - \frac{(1-z) P_{12}}{3 z^{3/2} k^3} \biggl[
            \HA_{w_{10},w_{5}}
          - \HA_{w_{10},w_{6}}
          + \HA_{w_{10},w_{7}}
          - \HA_{w_{10},w_{8}}
          - k \bigl(
                  \HA_{w_{5},w_{11}}
                + \HA_{w_{6},w_{11}}
                + \HA_{w_{7},w_{11}}
    \nonumber \\ &
                + \HA_{w_{8},w_{11}}
          \bigr)
          + k \bigl(
                  \HA_{w_{5}}
                + \HA_{w_{6}}
                + \HA_{w_{7}}
                + \HA_{w_{8}}
          \bigr) \HA_{w_{11}}
          -\frac{2}{1-z} \bigl( \HA_{w_{10},w_{1}} + \HA_{w_{10},w_{2}} \bigr)
    \biggr]
    \nonumber \\ &
    +\frac{4 (1+k) (1-z) P_{13}}{3 k^4} \bigl(
            \HA_{w_{6},-1}
          - \HA_{w_{8},1}
          + \HA_{w_{8}} \HA_1
          - \HA_{w_{6}} \HA_{-1}
    \bigr)
    \nonumber \\ &
    +\frac{4 (1-k) (1-z) P_{14}}{3 k^4} \bigl(
            \HA_{w_{5},-1}
          - \HA_{w_{7},1}
          + \HA_{w_{7}} \HA_1
          - \HA_{w_{5}} \HA_{-1}
    \bigr)
    +\frac{8 P_{15}}{3 k^4 z} \bigl( \HA_{w_{1},1} - \HA_{w_{2},-1} \bigr)
    \nonumber \\ &
    -\frac{4 (1-z) P_{16}}{3 k^4} \bigl(
            \HA_{w_{6},1}
          - \HA_{w_{8},-1}
          - \HA_{w_{6}} \HA_1
          + \HA_{w_{8}} \HA_{-1}
    \bigr)
    -\frac{4 (1-z) P_{17}}{3 k^4} \bigl(
            \HA_{w_{5},1}
          - \HA_{w_{7},-1}
    \nonumber \\ &
          - \HA_{w_{5}} \HA_1
          + \HA_{w_{7}} \HA_{-1}
    \bigr)
    -\frac{2 (1-k^2) P_{18}}{3 \sqrt{z} k^3}
    \biggl[
            \HA_{w_{12},1}
          + \HA_{w_{12},-1}
          + (1-z) k \bigl(
                  \HA_{w_{5},w_{12}}
                + \HA_{w_{6},w_{12}}
    \nonumber \\ &
                + \HA_{w_{7},w_{12}}
                + \HA_{w_{8},w_{12}}
          \bigr)
          - (1-z) k \bigl(
                  \HA_{w_{5}}
                + \HA_{w_{6}}
                + \HA_{w_{7}}
                + \HA_{w_{8}}
          \bigr) \HA_{w_{12}}
    \biggr]
    -\frac{8 P_{19}}{3 k^4 z} \bigl( \HA_{w_{1},-1} 
    \nonumber \\ &
    - \HA_{w_{2},1} \bigr)
    +\frac{2 P_{20}}{9 k^2 z (1-k \beta)} \HA_{w_{1}}
    -\frac{2 P_{21}}{9 k^2 z (1+k \beta)} \HA_{w_{2}}
    +\frac{(1-z) P_{22}}{3 k^3 z (k (z - 2) + z ) (1-k \beta)} \HA_{w_{5}}
    \nonumber \\ &
    +\frac{2 P_{23}}{9 k^4 z \big(k^2 (z-2)^2-z^2\big)} \HA_1
    -\frac{2 P_{24}}{9 k^4 z \big(k^2 (z-2)^2-z^2\big)} \HA_{-1}
    \nonumber \\ &
    -\frac{(1-z) P_{25}}{3 k^3 z (k (z-2)-z) (1+k \beta)} \HA_{w_{6}}
    +\frac{(1-z) P_{26}}{3 k^3 z (k (z-2)+z) (1+k \beta)} \HA_{w_{7}}
    \nonumber \\ &
    +\frac{(1-z) P_{27}}{3 k^3 z (k (z-2)-z) (1-k \beta)} \HA_{w_{8}}
    -32 (1-z)^2 z ( \ln (z) + \ln (1-z) ) \bigl( 2 \beta - \HA_1 - \HA_{-1} \bigr)
    \nonumber \\ &
    - 64 z \big(3-z+\frac{z}{k^2}\big) \ln (k) \bigl( \HA_1 + \HA_{-1} \bigr)
    + \frac{16 (-1+z) \beta }{3 z} \big(3-k^2-4 z-4 z^2\big) \bigl( 6 \ln (k) 
    \nonumber \\ &
    - \ln \big(1-k^2\big) - 2 \ln \big(k^2-z\big) - 2 \HA_0 \bigr)
    -\frac{64 z \big(k^2 (z-3)-z\big)}{3 k^2} \biggl[
            \HA_1 \HA_0
          + \HA_{-1,0}
          - \HA_{0,1}
    \nonumber \\ &
          - \HA_{1,w_{4}}
          - \HA_{-1,w_{4}}
          - \HA_{w_{3},1}
          - \HA_{w_{3},-1}
          + \biggl(
                \frac{1}{2} \ln \big(1-k^2\big)
                + \ln \big(k^2-z\big)
                + \HA_{w_{3}}
          \biggr) 
    \nonumber \\ & \times \bigl( \HA_1 + \HA_{-1} \bigr)            
    \biggr]
    -\frac{32 z}{3 k^2} \big(z+k^2 \big(6-7 z+3 z^2\big)\big) \ln (k^2-z^2) \bigl( \HA_1 + \HA_{-1} \bigr)\Biggr\}
    \nonumber \\ &
    +\frac{1}{2} P_{gq}^{(0)} \otimes \bar{h}_{L,g}^{(1)} \ln \left( \frac{Q^2}{\mu_F^2} \right)
    - P_{gq}^{(0)} \otimes \bar{b}_{L,g}^{(1)}  ~,
    \label{eq:heavy-wilson-HLq}
    \end{align}
    \begin{align}
    H_{1,q}^{(2),\text{PS}} &= 
    \cf T_F \Biggl\{-\frac{4 (1-z) P_{28}}{k^2} 
    \bigl(
            \HA_{w_{6},-1}
          - \HA_{w_{8},1}
          + \HA_1 \HA_{w_{8}}
          - \HA_{-1} \HA_{w_{6}}
    \bigr)
    \nonumber \\ &
    -\frac{8 P_{29}}{3 k^3} \bigl( \HA_1 \HA_{w_{1}} - \HA_{-1} \HA_{w_{2}} \bigr)
    -\frac{8 P_{30}}{3 k^3} \HA_1 \HA_{w_{2}} 
    +\frac{8 \big(k^2-z\big) P_{30}}{3 k^5 (1-z) \beta ^2} \HA_{w_{1}} \HA_{-1}
    \nonumber \\ &
    +\frac{4 (1-z) P_{31}}{k^2} 
    \bigl(  
            \HA_{w_{5},-1}
          - \HA_{w_{7},1}
          + \HA_1 \HA_{w_{7}}
          - \HA_{-1} \HA_{w_{5}}
    \bigr)
    +\frac{8 P_{32}}{3 z} 
    \biggl[
          k \bigl( \HA_{w_{1}}^2 - \HA_{w_{2}}^2 \bigr)
    \nonumber \\ &
          + 2 \bigl( \HA_{w_{1},w_{4}} + \HA_{w_{2},w_{4}} + \HA_{w_{3},w_{1}} + \HA_{w_{3},w_{2}} \bigr)
          + \bigl( \HA_{w_{1}} + \HA_{w_{2}} \bigr) \bigl[ 6 \ln (k) + \ln (k^2-z^2) \bigr]
    \nonumber \\ &
          + k (1-z) \bigl(
                  \HA_{w_{5},w_{1}}
                + \HA_{w_{6},w_{2}}
                - \HA_{w_{7},w_{2}}
                - \HA_{w_{8},w_{1}}
                - \HA_{w_{1}} \HA_{w_{5}}
                - \HA_{w_{2}} \HA_{w_{6}}
                + \HA_{w_{2}} \HA_{w_{7}}
    \nonumber \\ &
                + \HA_{w_{1}} \HA_{w_{8}}
          \bigr)
          -\bigl( \HA_{w_{1}} + \HA_{w_{2}} \bigl) \bigl[ \ln \big(1-k^2\big) + 2 \ln \big(k^2-z)
            + 2 \HA_{w_{3}} \bigr]
    \biggr]
    \nonumber \\ &
    +\frac{16 (1-z) \beta  P_{33}}{9 k^2 z}
    +\frac{32 P_{34}}{3 k^4}
    \biggl[
            \HA_{0,1}
          - \HA_{-1,0}
          - \HA_0 \HA_1
          + \HA_{1,w_{4}}
          + \HA_{w_{3},1}
          + \HA_{w_{3},-1}
          + \HA_{-1,w_{4}}
    \nonumber \\ &
          - \bigl( \HA_1 + \HA_{-1} \bigr) \bigl( \frac{1}{2} \ln \big(1-k^2\big) + \ln \big(k^2-z\big) + \HA_{w_{3}} \bigr)
    \biggr]
    -\frac{32 (1-z^2) P_{35}}{3 k^2}
    \biggl[
            \HA_{w_{9},1}
    \nonumber \\ &
          + \HA_{w_{9},-1}
          - (1-z) k \bigl( \HA_{w_{9},w_{5}} + \HA_{w_{9},w_{6}} + \HA_{w_{9},w_{7}} + \HA_{w_{9},w_{8}} \bigr)
    \biggr]
    +\frac{4 (1-z) P_{36}}{3 k^3} \bigl(  \HA_{w_{5},1} 
    \nonumber \\ &
    - \HA_{w_{7},-1} - \HA_1 \HA_{w_{5}} + \HA_{-1} \HA_{w_{7}} \bigr)
    +\frac{4 (1-z) P_{37}}{3 k^3} \bigl( \HA_{w_{6},1} - \HA_{w_{8},-1} - \HA_1 \HA_{w_{6}} + \HA_{-1} \HA_{w_{8}} \bigr)
    \nonumber \\ &
    +\frac{16 P_{38}}{3 k^4} \bigl( \HA_{-1} \HA_1 - 2 \HA_{-1,1} \bigr)
    -\frac{16 (1-z) \beta P_{39}}{3 k^2 z} \ln (k^2-z^2)
    -\frac{8 P_{40}}{3 k^3 z} \bigl( \HA_{w_{1},1} - \HA_{w_{2},-1} \bigr)
    \nonumber \\ &
    -\frac{8 P_{41}}{3 k^3 z} \HA_{w_{2},1}
    -\frac{16 (1-z) \beta P_{42}}{3 k^2 z}
    \biggl[
            \ln \big(1-k^2\big)
          + 2 \ln \big(k^2-z\big)
          - 6 \ln (k)
          + 2 \HA_0
    \nonumber \\ &
          + 4 \HA_{w_{3}}
    \biggr]
    -\frac{16 P_{43}}{3 k^2 z} \bigl( \HA_{w_{1},0} + \HA_{w_{2},0} \bigr)
    -\frac{8 P_{44}}{3 k^4} \bigl( \HA_1^2 - \HA_{-1}^2 \bigr)
    +\frac{16  P_{45}}{3 k^2 z} \bigl( \HA_{w_{1}} + \HA_{w_{2}} \bigr) \HA_0
    \nonumber \\ &
    +\frac{8 (1-z) P_{45}}{3 k^2 z}
    \biggl[
            \HA_{w_{5},0}
          - \HA_{w_{6},0}
          + \HA_{w_{7},0}
          - \HA_{w_{8},0}
          - \bigl(
                  \HA_{w_{5}}
                - \HA_{w_{6}}
                + \HA_{w_{7}}
                - \HA_{w_{8}}
          \bigr) \HA_0
    \biggr]
    \nonumber \\ &
    +\frac{4 P_{46}}{3 z^{3/2} k^3}
    \biggl[
            2 \HA_{w_{10},w_{1}}
          + 2 \HA_{w_{10},w_{2}}
          - (1-z) \biggl(
                  \HA_{w_{10},w_{5}}
                - \HA_{w_{10},w_{6}}
                + \HA_{w_{10},w_{7}}
                - \HA_{w_{10},w_{8}}
    \nonumber \\ &
                - k \bigl(
                        \HA_{w_{5},w_{11}}
                      + \HA_{w_{6},w_{11}}
                      + \HA_{w_{7},w_{11}}
                      + \HA_{w_{8},w_{11}}
                \bigr)
                + k \bigl(
                        \HA_{w_{5}}
                      + \HA_{w_{6}}
                      + \HA_{w_{7}}
                      + \HA_{w_{8}}
                \bigr) \HA_{w_{11}}
          \biggr)
    \nonumber \\ &
          + 2 k ( 1 - k^2 ) z (1 - z) \biggl(
                  \HA_{w_{5},w_{12}}
                + \HA_{w_{6},w_{12}}
                + \HA_{w_{7},w_{12}}
                + \HA_{w_{8},w_{12}}
                - \bigl(
                        \HA_{w_{5}}
                      + \HA_{w_{6}}
                      + \HA_{w_{7}}
    \nonumber \\ &
                      + \HA_{w_{8}}
                \bigr) \HA_{w_{12}}
          \biggr)
          + 2 (1-k^2) z \bigl( \HA_{w_{12},1} + \HA_{w_{12},-1} \bigr)
    \biggr]
    +\frac{8 P_{47}}{9 k^2 z (1+k \beta)} \HA_{w_{2}}
    \nonumber \\ &
    -\frac{8 P_{48}}{9 k^2 z (1-k \beta)} \HA_{w_{1}}
    -\frac{4 (1-z)^2 P_{49}}{3 k^3 z (k (z-2)-z)} \HA_{w_{6}}
    -\frac{4 (1-z)^2 P_{50}}{3 k^3 z (k (z-2)+z)} \HA_{w_{5}}
    \nonumber \\ &
    -\frac{4 (1-z)^2 P_{51}}{3 k^3 z (k (z-2)+z)} \HA_{w_{7}}
    -\frac{4 (1-z)^2 P_{52}}{3 k^3 z (k (z-2)-z)} \HA_{w_{8}}
    -\frac{8 P_{55}}{3 k^5 (1-z) z \beta ^2} \HA_{w_{1},-1}
    \nonumber \\ &
    -\frac{8 P_{53}}{9 k^4 z (1+\beta ) \big(k^2 (z-2)^2-z^2\big)} \HA_1
    +\frac{8 P_{54}}{9 k^4 z (1-\beta ) \big(k^2 (z-2)^2-z^2\big)} \HA_{-1}
    \nonumber \\ &
    -\biggl[
             \frac{16 \big(1+k^2\big)\big(1-3 k^2\big) z^2}{3 k^4} \ln (k^2-z^2)
            +16 (1-z) \bigl( \ln (1-z) + \ln (z) \bigr)
    \nonumber \\ &
            +32 \biggl(3(1-z)+\frac{ \big(1+k^2\big)\big(1-3 k^2\big) z^2}{k^4}\biggr) \ln (k)
    \biggr] \bigl( \HA_1 + \HA_{-1} \bigr)
    \nonumber \\ &
    - 8 \frac{2k^2 +\big( 3 k^2 - 1 \big) z}{k^2}
    \biggl[
          4  \HA_{0,1,1}
          + 4 \HA_{0,-1,1}
          - 20 \HA_{1,1,1}
          - 4 \HA_{1,1,w_{4}}
          - 4 \HA_{1,-1,w_{4}}
    \nonumber \\ &
          + 4 \HA_{w_{3},1,1}
          - 4 \HA_{w_{3},1,-1}
          + 4 \HA_{w_{3},-1,1}
          - 4 \HA_{w_{3},-1,-1}
          - 4 \HA_{-1,1,0}
          - 16 \HA_{-1,1,1}
          + 4 \HA_{-1,1,w_{4}}
    \nonumber \\ &
          - 4 \HA_{-1,-1,0}
          - 16 \HA_{-1,-1,1}
          + 4 \HA_{-1,-1,w_{4}}
          - 20 \HA_{-1,-1,-1}
    +2 \bigl(
              \HA_1^2
             - 2 \HA_{-1,1}
    \bigr) \HA_0
    \nonumber \\ &
    +2\bigl(       
            -4 \HA_{-1,1}
            + \HA_1^2
            - \HA_{-1}^2
            +2 \HA_1 \HA_{-1}
    \bigr) \HA_{w_{3}}
    +\bigl(
             4 \HA_{-1,1}
            -5 \HA_{-1}^2
            +5 \HA_{1}^2
            -4 \HA_{0,1}
    \nonumber \\ &
            -4 \HA_{0,-1}
            -4 \HA_{w_{3},1}
            -4 \HA_{w_{3},-1}
    \bigr) \HA_1
    +\bigl(
             4 \HA_0 \HA_1
            -  \HA_1^2
            +4 \HA_{w_{3},1}
            +4 \HA_{w_{3},-1}
            +12 \HA_{-1,1}
    \nonumber \\ &
            +5 \HA_{-1}^2
    \bigr) \HA_{-1}
          - \bigl[ \ln \big(1-k^2\big) - \ln (k^2-z^2) + 2 \ln \big(k^2-z\big) - 6 \ln (k) \bigr]
    \nonumber \\ & \times
          \bigl( 4 \HA_{-1,1} + \HA_{-1}^2 - \HA_1^2 - 2 \HA_{-1} \HA_1 \bigr)
    \biggr] 
    -\frac{16 (1-z) \big(z-k^2 (2+3 z)\big)}{k} 
    \biggl[
            \HA_{1,w_{4},w_{5}}
    \nonumber \\ &
          + \HA_{1,w_{4},w_{6}}
          + \HA_{1,w_{4},w_{7}}
          + \HA_{1,w_{4},w_{8}}
          - \HA_{w_{5},1,1}
          + \HA_{w_{5},1,-1}
          - \HA_{w_{5},w_{3},1}
          + \HA_{w_{5},w_{3},-1}
    \nonumber \\ &
          - \HA_{w_{6},1,1}
          + \HA_{w_{6},1,-1}
          - \HA_{w_{6},w_{3},1}
          + \HA_{w_{6},w_{3},-1}
          - \HA_{w_{7},w_{3},1}
          + \HA_{w_{7},w_{3},-1}
          + \HA_{w_{7},-1,1}
    \nonumber \\ &
          - \HA_{w_{7},-1,-1}
          - \HA_{w_{8},w_{3},1}
          + \HA_{w_{8},w_{3},-1}
          + \HA_{w_{8},-1,1}
          - \HA_{w_{8},-1,-1}
          - \HA_{-1,w_{4},w_{5}}
          - \HA_{-1,w_{4},w_{6}}
    \nonumber \\ &
          - \HA_{-1,w_{4},w_{7}}
          - \HA_{-1,w_{4},w_{8}}
          + k \bigl(
                  \HA_{w_{2},w_{4},w_{5}}
                + \HA_{w_{2},w_{4},w_{6}}
                + \HA_{w_{2},w_{4},w_{7}}
                + \HA_{w_{2},w_{4},w_{8}}
    \nonumber \\ &
                - \HA_{w_{1},w_{4},w_{5}}
                - \HA_{w_{1},w_{4},w_{6}}
                - \HA_{w_{1},w_{4},w_{7}}
                - \HA_{w_{1},w_{4},w_{8}}
                + \HA_{w_{5},1,w_{1}}
                - \HA_{w_{5},1,w_{2}}
                + \HA_{w_{5},w_{3},w_{1}}
    \nonumber \\ &
                - \HA_{w_{5},w_{3},w_{2}}
                + \HA_{w_{6},1,w_{1}}
                - \HA_{w_{6},1,w_{2}}
                + \HA_{w_{6},w_{3},w_{1}}
                - \HA_{w_{6},w_{3},w_{2}}
                + \HA_{w_{7},w_{3},w_{1}}
                - \HA_{w_{7},w_{3},w_{2}}
    \nonumber \\ &
                - \HA_{w_{7},-1,w_{1}}
                + \HA_{w_{7},-1,w_{2}}
                + \HA_{w_{8},w_{3},w_{1}}
                - \HA_{w_{8},w_{3},w_{2}}
                - \HA_{w_{8},-1,w_{1}}
                + \HA_{w_{8},-1,w_{2}}
          \bigr)
    \nonumber \\ &
          + \bigl\{
                  \HA_{w_{3},1}
                - \HA_{w_{3},-1} 
                + \HA_{-1,1}
                + k \bigl[
                        \HA_{w_{1},1}
                      - \HA_{w_{2},1}
                      - \HA_{w_{3},w_{1}}
                      + \HA_{w_{3},w_{2}}
                \bigr]
          \bigr\} \bigl( \HA_{w_{5}} + \HA_{w_{6}} \bigr)
    \nonumber \\ &
          + \bigl\{
                  \HA_{w_{3},1}
                - \HA_{w_{3},-1}
                - \HA_{-1,1}
                - \HA_{-1,-1}
                + k \bigl[
                        \HA_{w_{2},-1}
                      - \HA_{w_{1},-1}
                      - \HA_{w_{3},w_{1}}
                      + \HA_{w_{3},w_{2}}
                \bigr]
          \bigr\}
    \nonumber \\ & \times
          \bigl( \HA_{w_{7}} + \HA_{w_{8}} \bigr)
          + \bigl(
                  \HA_{w_{5},1}
                + \HA_{w_{5},w_{3}}
                + \HA_{w_{6},1}
                + \HA_{w_{6},w_{3}}
                + \HA_{w_{7},w_{3}}
                - \HA_{w_{7},-1}
                + \HA_{w_{8},w_{3}}
    \nonumber \\ &
                - \HA_{w_{8},-1}
                - \bigl[    
                        \HA_{w_{5}}
                      + \HA_{w_{6}}
                      + \HA_{w_{7}}
                      + \HA_{w_{8}}
                \bigr] \HA_{w_{3}}
          \bigr) \bigl( \HA_1 - \HA_{-1} \bigr)
          - k \bigl(
                  \HA_{w_{5},1}
                + \HA_{w_{5},w_{3}}
    \nonumber \\ &
                + \HA_{w_{6},1}
                + \HA_{w_{6},w_{3}}
                + \HA_{w_{7},w_{3}}
                - \HA_{w_{7},-1}
                + \HA_{w_{8},w_{3}}
                - \HA_{w_{8},-1}
                - \bigl[    
                        \HA_{w_{5}}
                      + \HA_{w_{6}}
                      + \HA_{w_{7}}
    \nonumber \\ &
                      + \HA_{w_{8}}
                \bigr] \HA_{w_{3}}
          \bigr) \bigl( \HA_{w_{1}} - \HA_{w_{2}} \bigr)
          + \bigl( \HA_{w_{7}} + \HA_{w_{8}} \bigr) \HA_1 \HA_{-1}
          - \frac{1}{2} \bigl( \HA_{w_{5}} + \HA_{w_{6}} \bigr) \HA_1^2
    \biggr]
    \nonumber \\ &
    + 16 \big(z-k^2 (2+3 z)\big) \bigl[ \HA_{w_{1},1} + \HA_{w_{1},-1} - \HA_{w_{2},1} - \HA_{w_{2},-1} \bigr] \bigl( \HA_{w_{1}} - \HA_{w_{2}} \bigr)
    \nonumber \\ &
    + \frac{ 32 ( k^2 (2+3 z) - z ) }{k} 
    \biggl[
            \HA_{w_{1},1,0}
          + \HA_{w_{1},1,1}
          - \HA_{w_{1},1,w_{4}}
          - \HA_{w_{1},1,-1}
          + \HA_{w_{1},-1,0}
          + \HA_{w_{1},-1,1}
    \nonumber \\ &
          - \HA_{w_{1},-1,w_{4}}
          - \HA_{w_{1},-1,-1}
          - \HA_{w_{2},1,0}
          - \HA_{w_{2},1,1}
          + \HA_{w_{2},1,w_{4}}
          + \HA_{w_{2},1,-1}
          - \HA_{w_{2},-1,0}
    \nonumber \\ &
          - \HA_{w_{2},-1,1}
          + \HA_{w_{2},-1,w_{4}}
          + \HA_{w_{2},-1,-1}
          + \HA_{w_{3},1,w_{1}}
          - \HA_{w_{3},1,w_{2}}
          + \HA_{w_{3},-1,w_{1}}
          - \HA_{w_{3},-1,w_{2}}
    \nonumber \\ &
          + \frac{1}{2} \bigl[
                  \HA_{w_{1},1}
                + \HA_{w_{1},-1}
                - \HA_{w_{2},1}
                - \HA_{w_{2},-1}
          \bigr] \bigl( 2 \HA_{w_{3}} + \HA_1 - \HA_{-1} \bigr)
          + \frac{1}{4} \bigl[ 
                  \HA_1^2 
                - 4 \HA_{w_{3},-1} 
    \nonumber \\ &
                - 4 \HA_{w_{3},1} 
                - 4 \HA_{-1,1} 
                - \HA_{-1}^2  
                + 2 \HA_{-1} \HA_1 
          \bigr] \bigl( \HA_{w_{1}} - \HA_{w_{2}}  \bigr)
          + \frac{1}{2} \bigl[
                  \HA_{w_{2},-1}
                - \HA_{w_{1},1}
                - \HA_{w_{1},-1}
    \nonumber \\ &
                + \HA_{w_{2},1}
          \bigr] \bigl( 
                  6 \ln (k) 
                - \ln \big(1-k^2\big) 
                + \ln (k^2-z^2) 
                - 2 \ln \big(k^2-z\big) 
          \bigr)
    \biggr]
    \nonumber \\ &
    +32 (1-z) \beta  \bigl( \ln (1-z) + \ln (z) \bigr)\Biggr\}
    +\frac{1}{2} P_{gq}^{(0)} \otimes \bar{h}_{1,g}^{(1)} \ln \left( \frac{Q^2}{\mu_F^2} \right) 
    - P_{gq}^{(0)} \otimes \bar{b}_{1,g}^{(1)}  ~,
    \label{eq:heavy-wilson-H1q}
\end{align}
    where the polynomials are:
\begin{align*}
    P_1={}&k^4+k^2 (2-6 z)-12 z^2+6 z-3, 
    \\ 
    P_2={}&-k^2+12 z^3-16 z^2-4 z+3, 
    \\
    P_3={}&8 k^4+k^2 \left(-25 z^2-28 z+12\right)+9 z^2, 
    \\
    P_4={}&k^6+k^4 \left(3-6 z^2\right)-4 z^4, 
    \\
    P_5={}&k^2 \left(z^2-3 z-1\right)-z^2-3 z+1, 
    \\
    P_6={}&k^8+k^6 \left(-3 z^2-3 z+2\right)-3 k^4 \left(z^2-z+1\right)-2 k^2 z^4+2 z^4, 
    \\
    P_7={}&3 k^6 (z-1)-2 k^5 z \left(3 z^2-7 z+6\right)+k^4 (3-9 z)-2 k^3 z^2+2 k^2 z^3-2 z^3, 
    \\
    P_8={}&3 k^6 (z-1)+2 k^5 z \left(3 z^2-7 z+6\right)+k^4 (3-9 z)+2 k^3 z^2+2 k^2 z^3-2 z^3, 
    \\
    P_9={}&k^4+k^2 \left(4 z^2+3 z-3\right)+z \left(-4 z^2-4 z+3\right), 
    \\
    P_{10}={}&k^2 \left(5 z^2-2\right)+3 z^2, 
    \\
    P_{11}={}&k^2 \left(5 z^2-15 z+1\right)-5 z^2+3 z-1, 
    \\
    P_{12}={}&k^4 \left(-80 z^3+35 z^2+30 z-9\right)+2 k^2 z \left(19 z^2-10 z-9\right)+3 z^2 \left(5 z^2+2 z+1\right), 
    \\
    P_{13}={}&6 k^5 (z-1)+k^4 \left(-4 z^3+21 z^2-30 z+8\right)+k^3 \left(4 z^3-21 z^2+12 z-2\right)+3 k^2 z^2\\ &
    +k z^2 (4 z-3)-4 z^3, 
    \\
    P_{14}={}&6 k^5 (z-1)+k^4 \left(4 z^3-21 z^2+30 z-8\right)+k^3 \left(4 z^3-21 z^2+12 z-2\right)-3 k^2 z^2\\ &
    +k z^2 (4 z-3)+4 z^3, 
    \\
    P_{15}={}&3 k^8-6 k^6 \left(z^2+2 z-1\right)+k^5 z \left(12 z^3-25 z^2+6\right)-3 k^4 \left(6 z^2-4 z+3\right)-2 k^3 z \bigl(z^2
    \\ &
    -6 z+3\bigr)-4 k^2 z^4+3 k z^3+4 z^4, 
    \\
    P_{16}={}&6 k^6 (z-1)+k^5 \left(20 z^3-35 z^2+24 z+2\right)+k^4 (6-18 z)+2 k^3 \left(2 z^3-5 z^2+6 z-1\right)
    \\ &
    +4 k^2 z^3-3 k z^2-4 z^3, 
    \\
    P_{17}={}&-6 k^6 (z-1)+k^5 \left(20 z^3-35 z^2+24 z+2\right)+6 k^4 (3 z-1)
    \\ &
    +2 k^3 \left(2 z^3-5 z^2+6 z-1\right)-4 k^2 z^3-3 k z^2+4 z^3, 
    \\
    P_{18}={}&k^4 \left(80 z^3-35 z^2-30 z+9\right)+2 k^2 z \left(-19 z^2+10 z+9\right)-3 z^2 \left(5 z^2+2 z+1\right), 
    \\
    P_{19}={}&3 k^8-6 k^6 \left(z^2+2 z-1\right)+k^5 \left(-12 z^4+25 z^3-6 z\right)-3 k^4 \left(6 z^2-4 z+3\right)
    \\ &
    +2 k^3 z \left(z^2-6 z+3\right)-4 k^2 z^4-3 k z^3+4 z^4, 
    \\
    P_{20}={}&16 \beta  k^7-40 k^6+8 \beta  k^5 \left(18 z^2+3 z-5\right)+8 k^4 \left(36 z^3-66 z^2-15 z+17\right)
    \\ &
    +3 \beta  k^3 \left(192 z^4-344 z^3+69 z^2+82 z-31\right)-3 k^2 \left(192 z^4-248 z^3-59 z^2+50 z-7\right)
    \\ &
    +3 \beta  k z \left(25 z^2-6 z-3\right)+3 z \left(-25 z^2+6 z+3\right), 
    \\
    P_{21}={}&16 \beta  k^7+40 k^6+8 \beta  k^5 \left(18 z^2+3 z-5\right)-8 k^4 \left(36 z^3-66 z^2-15 z+17\right)
    \\ &
    +3 \beta  k^3 \left(192 z^4-344 z^3+69 z^2+82 z-31\right)+3 k^2 \left(192 z^4-248 z^3-59 z^2+50 z-7\right)
    \\ &
    +3 \beta  k z \left(25 z^2-6 z-3\right)+3 z \left(25 z^2-6 z-3\right), 
    \\
    P_{22}={}&8 k^8 (z-2) (\beta  (z-1)+1)-8 k^7 \left(-2 \beta +\beta  z^3+(1-8 \beta ) z^2+(9 \beta -4) z+2\right)
    \\ &
    +k^6 \left(-66 \beta +(68 \beta -96) z^4+(328-186 \beta ) z^3+(17 \beta -288) z^2+(167 \beta -24) z+48\right)
    \\ &
    +k^5 \bigl(-30 \beta -192 \beta  z^5+4 (207 \beta -41) z^4+(314-935 \beta ) z^3+3 (47 \beta +5) z^2+
    \\ &
    (188 \beta -199) z+66\bigr)
    +k^4 \bigl(-192 (\beta -1) z^5+4 (94 \beta -183) z^4-15 (9 \beta -41) z^3
    \\ &
    +(83-52 \beta ) z^2+(3 \beta -100) z-18\bigr)
    +k^3 z \bigl(-6 \beta +192 z^4+7 (\beta -40) z^3+(7-18 \beta ) z^2
    \\ &
    +(17 \beta +20) z+21\bigr)
    +k^2 (z-1) z \left((4 \beta -7) z^2+(3 \beta +11) z-6\right)
    \\ &
    -k (z-1) z^2 ((3 \beta +4) z+3)+3 (z-1) z^3, 
    \\
    P_{23}={}&
    72 k^8 (z-2)^2 (\beta  (z-1)+1)
    +k^6 \bigl(108 (8 \beta -7)+8 (36 \beta +29) z^5-2 (576 \beta +539) z^4
    \\ &
    +(576 \beta +1807) z^3+3 (768 \beta -563) z^2-1440 (2 \beta -1) z\bigr)
    +k^4 z \bigl(-16 (18 \beta +17) z^4
    \\ &
    +208 z^3+(504 \beta +95) z^2-3 (72 \beta +145) z+360\bigr)
    +k^2 z^2 \bigl(43 z^3+99 z^2-150 z+36\bigr)
    \\ &
    -3 z^4 (z+3), 
    \\
    P_{24}={}&72 k^8 (z-2)^2 (\beta  (z-1)-1)
    +k^6 \bigl(108 (8 \beta +7)+8 (36 \beta -29) z^5-2 (576 \beta -539) z^4
    \\ &
    +(576 \beta -1807) z^3+3 (768 \beta +563) z^2-1440 (2 \beta +1) z\bigr)
    -k^4 z \bigl(16 (18 \beta -17) z^4
    \\ &
    +208 z^3+(95-504 \beta ) z^2+3 (72 \beta -145) z+360\bigr)
    -k^2 z^2 \bigl(43 z^3+99 z^2-150 z+36\bigr)
    \\ &
    +3 z^4 (z+3), 
    \\
    P_{25}={}&8 k^8 (z-2) (\beta  (z-1)+1)
    +8 k^7 \bigl(-2 \beta +\beta  z^3+(1-8 \beta ) z^2+(9 \beta -4) z+2\bigr)
    \\ &
    +k^6 \bigl(-66 \beta +(68 \beta -96) z^4+(328-186 \beta ) z^3+(17 \beta -288) z^2+(167 \beta -24) z+48\bigr)
    \\ &
    +k^5 \bigl(30 \beta +192 \beta  z^5+(164-828 \beta ) z^4+(935 \beta -314) z^3-3 (47 \beta +5) z^2
    \\ &
    +(199-188 \beta ) z-66\bigr)
    +k^4 \bigl(-192 (\beta -1) z^5+4 (94 \beta -183) z^4-15 (9 \beta -41) z^3
    \\ &
    +(83-52 \beta ) z^2+(3 \beta -100) z-18\bigr)
    -k^3 z \bigl(-6 \beta +192 z^4+7 (\beta -40) z^3
    \\ &
    +(7-18 \beta ) z^2+(17 \beta +20) z+21\bigr)
    +k^2 (z-1) z \bigl((4 \beta -7) z^2+(3 \beta +11) z-6\bigr)
    \\ &
    +k (z-1) z^2 ((3 \beta +4) z+3)+3 (z-1) z^3, 
    \\
    P_{26}={}&-8 k^8 (z-2) (\beta  (z-1)-1)
    +8 k^7 \bigl(-2 (\beta +1)+\beta  z^3-(8 \beta +1) z^2+(9 \beta +4) z\bigr)
    \\ &
    -k^6 \bigl(-6 (11 \beta +8)+(68 \beta +96) z^4-2 (93 \beta +164) z^3+(17 \beta +288) z^2+(167 \beta +24) z\bigr)
    \\ &
    +k^5 \bigl(30 \beta +192 \beta  z^5-4 (207 \beta +41) z^4+(935 \beta +314) z^3-3 (47 \beta -5) z^2
    \\ &
    -(188 \beta +199) z+66\bigr)
    +k^4 \bigl(192 (\beta +1) z^5-4 (94 \beta +183) z^4+15 (9 \beta +41) z^3
    \\ &
    +(52 \beta +83) z^2-(3 \beta +100) z-18\bigr)
    +k^3 z \bigl(6 \beta +192 z^4-7 (\beta +40) z^3
    \\ &
    +(18 \beta +7) z^2+(20-17 \beta ) z+21\bigr)
    -k^2 (z-1) z \bigl((4 \beta +7) z^2+(3 \beta -11) z+6\bigr)
    \\ &
    +k (z-1) z^2 ((3 \beta -4) z-3)+3 (z-1) z^3, 
    \\
    P_{27}={}&8 k^8 (z-2) (\beta  (z-1)-1)
    +8 k^7 \bigl(-2 (\beta +1)+\beta  z^3-(8 \beta +1) z^2+(9 \beta +4) z\bigr)
    \\ &
    +k^6 \bigl(-6 (11 \beta +8)+(68 \beta +96) z^4-2 (93 \beta +164) z^3+(17 \beta +288) z^2+(167 \beta +24) z\bigr)
    \\ &
    +k^5 \bigl(30 \beta +192 \beta  z^5-4 (207 \beta +41) z^4+(935 \beta +314) z^3-3 (47 \beta -5) z^2
    \\ &
    -(188 \beta +199) z+66\bigr)
    +k^4 \bigl(-192 (\beta +1) z^5+4 (94 \beta +183) z^4-15 (9 \beta +41) z^3
    \\ &
    -(52 \beta +83) z^2+(3 \beta +100) z+18\bigr)
    +k^3 z \bigl(6 \beta +192 z^4-7 (\beta +40) z^3
    \\ &
    +(18 \beta +7) z^2+(20-17 \beta ) z+21\bigr)
    +k^2 (z-1) z \bigl((4 \beta +7) z^2+(3 \beta -11) z+6\bigr)
    \\ &
    +k (z-1) z^2 ((3 \beta -4) z-3)-3 (z-1) z^3
    \\
    P_{28}={}&3 k^4 (z-2)+k^3 (20-14 z)+6 k^2 (z+1)+2 k z-z,
    \\
    P_{29}={}&9 k^5 (z-2)-6 k^4 z^2+18 k^3 (z+1)-4 k^2 z^2-3 k z+2 z^2,
    \\
    P_{30}={}&9 k^5 (z-2)+6 k^4 z^2+18 k^3 (z+1)+4 k^2 z^2-3 k z-2 z^2,
    \\
    P_{31}={}&3 k^4 (z-2)+2 k^3 (7 z-10)+6 k^2 (z+1)-2 k z-z,
    \\
    P_{32}={}&3 k^4-2 k^2 (9 z+2)+18 z-7,
    \\
    P_{33}={}&30 k^4+k^2 \bigl(-60 z^2+63 z+28\bigr)+16 z^2,
    \\
    P_{34}={}&3 k^4 \bigl(z^2+z-1\bigr)+2 k^2 z^2-z^2,
    \\
    P_{35}={}&3 k^4 \bigl(z^2+3\bigr)+k^2 \bigl(2 z^2+3\bigr)-z^2,
    \\
    P_{36}={}&-9 k^5 (z-2)+6 k^4 \bigl(2 z^2-7 z+10\bigr)-18 k^3 (z+1)+2 k^2 z (4 z+3)+3 k z-4 z^2,
    \\
    P_{37}={}&9 k^5 (z-2)+6 k^4 \bigl(2 z^2-7 z+10\bigr)+18 k^3 (z+1)+2 k^2 z (4 z+3)-3 k z-4 z^2,
    \\
    P_{38}={}&3 k^4 (z-8) z+k^2 \bigl(2 z^2+9 z-3\bigr)-z^2,
    \\
    P_{39}={}&3 k^4-k^2 \bigl(6 z^2+7\bigr)+2 z^2,
    \\
    P_{40}={}&9 k^7-3 k^5 \bigl(3 z^2+12 z+4\bigr)-6 k^4 z^2 (2 z+11)-3 k^3 \bigl(6 z^2-12 z+7\bigr)
    \\ &
    -2 k^2 z \bigl(4 z^2-9 z+6\bigr)+3 k z^2+4 z^3,
    \\
    P_{41}={}&9 k^7-3 k^5 \bigl(3 z^2+12 z+4\bigr)+6 k^4 z^2 (2 z+11)-3 k^3 \bigl(6 z^2-12 z+7\bigr)
    \\ &
    +2 k^2 z \bigl(4 z^2-9 z+6\bigr)+3 k z^2-4 z^3,
    \\
    P_{42}={}&-3 k^4+k^2 \bigl(6 z^2+6 z+7\bigr)-2 z^2,
    \\
    P_{43}={}&6 k^6-k^4 \bigl(9 z^2+18 z+8\bigr)-2 k^2 \bigl(9 z^2-9 z+7\bigr)+3 z^2,
    \\
    P_{44}={}&3 k^4 \bigl(5 z^2+14 z-6\bigr)+k^2 \bigl(10 z^2-9 z+3\bigr)-5 z^2,
    \\
    P_{45}={}&3 k^6-k^4 \bigl(9 z^2+4\bigr)-k^2 \bigl(18 z^2+7\bigr)+3 z^2,
    \\
    P_{46}={}&3 k^4 \bigl(6 z^3+9 z^2-z+2\bigr)+k^2 z \bigl(3 z^2+8 z+9\bigr)-z^2 (3 z+1),
    \\
    P_{47}={}&6 \beta  k^7+24 k^6+2 \beta  k^5 \bigl(27 z^2+27 z+28\bigr)+2 k^4 \bigl(9 z^2+27 z-2\bigr)
    \\ &
    -\beta  k^3 \bigl(36 z^3+27 z^2-93 z+52\bigr)+k^2 \bigl(-36 z^3+21 z^2+93 z-10\bigr)
    \\ &
    +3 \beta  k z \bigl(4 z^2+z-1\bigr)+3 z \bigl(4 z^2-3 z-1\bigr),
    \\
    P_{48}={}&6 \beta  k^7-24 k^6+2 \beta  k^5 \bigl(27 z^2+27 z+28\bigr)-2 k^4 \bigl(9 z^2+27 z-2\bigr)
    \\ &
    -\beta  k^3 \bigl(36 z^3+27 z^2-93 z+52\bigr)+k^2 \bigl(36 z^3-21 z^2-93 z+10\bigr)
    \\ &
    +3 \beta  k z \bigl(4 z^2+z-1\bigr)+3 z \bigl(-4 z^2+3 z+1\bigr),
    \\
    P_{49}={}&-6 (\beta -1) k^7 (z-2)+6 k^6 z (\beta +z-6)+k^5 \bigl(-28 \beta +3 (4 \beta -3) z^3-3 (8 \beta -5) z^2
    \\ &
    +2 (7 \beta -22) z+40\bigr)+k^4 \bigl((9-12 \beta ) z^3-8 z^2+(30-14 \beta ) z+12\bigr)
    \\ &
    +2 k^3 z \bigl(-2 \beta  z^2+(4 \beta +2) z+7\bigr)+2 k^2 z \bigl(2 \beta  z^2+z-1\bigr)+k (z-3) z^2-z^3,
    \\
    P_{50}={}&-6 (\beta -1) k^7 (z-2)-6 k^6 z (\beta +z-6)+k^5 \bigl(-28 \beta +3 (4 \beta -3) z^3-3 (8 \beta -5) z^2
    \\ &
    +2 (7 \beta -22) z+40\bigr)+k^4 \bigl(3 (4 \beta -3) z^3+8 z^2+2 (7 \beta -15) z-12\bigr)
    \\ &
    +2 k^3 z \bigl(-2 \beta  z^2+(4 \beta +2) z+7\bigr)-2 k^2 z \bigl(2 \beta  z^2+z-1\bigr)+k (z-3) z^2+z^3,
    \\
    P_{51}={}&6 (\beta +1) k^7 (z-2)-6 k^6 z (-\beta +z-6)+k^5 \bigl(28 \beta -3 (4 \beta +3) z^3+3 (8 \beta +5) z^2
    \\ &
    -2 (7 \beta +22) z+40\bigr)-k^4 \bigl(3 (4 \beta +3) z^3-8 z^2+2 (7 \beta +15) z+12\bigr)
    \\ &
    +2 k^3 z \bigl(2 \beta  z^2+(2-4 \beta ) z+7\bigr)+2 k^2 z \bigl(2 \beta  z^2-z+1\bigr)+k (z-3) z^2+z^3,
    \\
    P_{52}={}&6 (\beta +1) k^7 (z-2)+6 k^6 z (-\beta +z-6)+k^5 \bigl(28 \beta -3 (4 \beta +3) z^3+3 (8 \beta +5) z^2
    \\ &
    -2 (7 \beta +22) z+40\bigr)+k^4 \bigl(3 (4 \beta +3) z^3-8 z^2+2 (7 \beta +15) z+12\bigr)
    \\ &
    +2 k^3 z \bigl(2 \beta  z^2+(2-4 \beta ) z+7\bigr)-2 k^2 z \bigl(2 \beta  z^2-z+1\bigr)+k (z-3) z^2-z^3,
    \\
    P_{53}={}&54 \beta  k^8 (z-2)^2 z-3 k^6 \bigl(-24 (\beta +1)+(\beta -35) z^5+(5 \beta +113) z^4-(47 \beta +125) z^3
    \\ &
    +6 (15 \beta +31) z^2-240 z\bigr)+k^4 z \bigl(72 (3 \beta -4)+(59 \beta -193) z^4+(187-173 \beta ) z^3
    \\ &
    +2 (82 \beta -143) z^2-6 (17 \beta +5) z\bigr)-k^2 z^2 \bigl(12 (\beta +1)+3 (23 \beta -37) z^3
    \\ &
    +(11-25 \beta ) z^2+(103 \beta -167) z\bigr)+z^4 (3 \beta +13 \beta  z-23 z+3),
    \\
    P_{54}={}&54 \beta  k^8 (z-2)^2 z-3 k^6 \bigl(-24 (\beta -1)+(\beta +35) z^5+(5 \beta -113) z^4+(125-47 \beta ) z^3
    \\ &
    +6 (15 \beta -31) z^2+240 z\bigr)+k^4 z \bigl(72 (3 \beta +4)+(59 \beta +193) z^4-(173 \beta +187) z^3
    \\ &
    +2 (82 \beta +143) z^2-6 (17 \beta -5) z\bigr)-k^2 z^2 \bigl(12 (\beta -1)+3 (23 \beta +37) z^3
    \\ &
    -(25 \beta +11) z^2+(103 \beta +167) z\bigr)+z^4 (3 \beta +13 \beta  z+23 z-3),
    \\
    P_{55}={}&9 \beta ^2 k^9 (z-1)+k^7 \bigl(12 \beta ^2+\bigl(9-54 \beta ^2\bigr) z^2+6 \bigl(7 \beta ^2-3\bigr) z\bigr)+6 k^6 z^2 \bigl(-11 \beta ^2
    \\ &
    +3 \beta ^2 z^2+8 \beta ^2 z+z\bigr)+k^5 \bigl(21 \beta ^2-9 z^3+18 \bigl(3 \beta ^2+2\bigr) z^2+\bigl(18-75 \beta ^2\bigr) z\bigr)
    \\ &
    +2 k^4 z \bigl(-6 \beta ^2+\bigl(6 \beta ^2-3\bigr) z^3+\bigl(2-15 \beta ^2\bigr) z^2+15 \beta ^2 z\bigr)
    \\ &
    -3 k^3 z^2 (6 z+7)-2 k^2 z^3 \bigl(-3 \beta ^2+\bigl(3 \beta ^2+2\bigr) z+1\bigr)+3 k z^3+2 z^4.
    \end{align*}
    The additional terms that appear are, using the variable $\kappa \coloneqq \frac{m^2}{Q^2}$, for $F_L^{\gamma,Q\bar Q}$~\cite{Blumlein:2019qze:Hq-heavy-wilson}:
    \begin{align}
    P_{gq}^{(0)} \otimes h_{L,g}^{(1)} &= \cf T_F \left\{ 64\beta (1-z) \frac{1 + 6\kappa - (8\kappa + 2)z - (8\kappa + 2)z^2}{3z(1 + 4\kappa)} - \frac{64}{3} z (3 + 4\kappa z) \ln \left( \frac{1 - \beta}{1 + \beta} \right) \right. \nonumber \\
    &\quad \left. + \frac{64}{3} \frac{4\kappa (1 + 3\kappa) - 6\kappa (1 + 4\kappa)z + 3(1 + 4\kappa)^2 z^2}{z(1 + 4\kappa)^{3/2}} \ln \left( \frac{\sqrt{1 + 4\kappa} - \beta}{\sqrt{1 + 4\kappa} + \beta} \right) \right\} ,
    \end{align}
    \begin{align}
        P_{gq}^{(0)} \otimes \bar{b}_{L,g}^{(1)} ={} &
        \cf T_F \Biggl\{-\frac{32 (1-z) \big( 3 - 4 z - 6 z^2 \big) \beta }{3 z}
        +\frac{8}{3} z (3 + 4 z \kappa ) \ln^2 \left( \frac{1-\beta }{1+\beta } \right)
        \nonumber \\ &
        - \frac{64}{3} z (3+4 z \kappa ) \bigl[ {\Li}_2\big(\frac{1-\beta }{2}\big) - {\Li}_2(1-\beta ) - {\Li}_2(-\beta ) \bigr]
        \nonumber \\ &
        - \frac{8}{3 z (1+4 \kappa )^{5/2} } 
                \biggl[
                2 \kappa ^2 (1+\kappa )
                -3 z \kappa ^2 (1+4 \kappa )
                +3 z^2 (1 + 4 \kappa )^2 
                \big(
                        \kappa 
                        +\sqrt{1+4 \kappa }
                \big)
        \nonumber \\ &
                +4 z^3 \kappa  (1+4 \kappa )^{5/2}
        \biggr] \ln^2(1-z)
        -\frac{8 \kappa R_3 }{3 z (1+4 \kappa )^{5/2}} 
        \biggl[
        \ln^2 \left( \frac{\sqrt{1+4 \kappa }-1}{\sqrt{1+4 \kappa }+1} \right)
        \nonumber \\ &
        + \ln^2 \left( \frac{ \sqrt{1+4 \kappa } - \beta }{ \sqrt{1+4 \kappa } + \beta } \right)
        - 4 \ln \big( \kappa \big) \ln \left( \frac{ \sqrt{1+4 \kappa } - 1 }{ \sqrt{1+4 \kappa } + 1} \right)
        - 8 {\Li}_2 \left( \frac{1}{1-\sqrt{1+4 \kappa }} \right)
        \nonumber \\ &
        + 8 {\Li}_2 \left( \frac{1}{1+\sqrt{1+4 \kappa }} \right)
        + 8 {\Li}_2 \left( \frac{ \sqrt{1+4 \kappa } - 1 }{ \sqrt{1+4 \kappa } + 1 } \right)
        - 8 \ln(2) \ln \left( \frac{ \sqrt{1+4 \kappa } -1 }{ \sqrt{1+4 \kappa } + 1} \right)
        \nonumber \\ &
        + 8 {\Li}_2 \left( \frac{\beta - \sqrt{1+4 \kappa }}{\beta +\sqrt{1+4 \kappa }} \right)
        - 8 {\Li}_2 \left( \frac{ \big( \sqrt{1+4 \kappa } - 1 \big) \big( \sqrt{1+4 \kappa } - \beta \big)}{\big(1+\sqrt{1+4 \kappa }\big)\big(\beta +\sqrt{1+4 \kappa } \big)} \right)
        \nonumber \\ & 
        - 2 \ln(1-z)  \ln \left( \frac{\sqrt{1+4 \kappa } - 1}{\sqrt{1+4 \kappa } + 1} \right)
        \biggr]
        + \frac{64}{3} z (3+4 z \kappa ) \ln (\beta ) \ln (2)
        \nonumber \\ &
        +\frac{16 R_7}{3 z (1+4 \kappa )^{5/2} } 
        \ln \left(\frac{\sqrt{1+4 \kappa }-1}{\sqrt{1+4 \kappa }+1}\right) 
        \ln \left(\frac{\sqrt{1+4 \kappa }-\beta}{\sqrt{1+4 \kappa }+\beta}\right) 
        \nonumber \\ &
        +\frac{32 R_5}{3 z (1+4 \kappa )^{5/2}} \biggl[
          {\Li}_2 \left( \frac{ \sqrt{1+4 \kappa } - \beta }{ \sqrt{1+4 \kappa } + 1} \right)
        + {\Li}_2 \left( \frac{ \sqrt{1+4 \kappa } - 1 }{ \sqrt{1+4 \kappa } + \beta } \right)
        \biggr]
        \nonumber \\ &
        -\frac{32 R_4}{3 z (1+4 \kappa )^{3/2} } 
        \ln \left(\frac{ \sqrt{1+4 \kappa } - \beta }{ \sqrt{1+4 \kappa } + \beta }\right)
        - \frac{32 R_2}{3 z (1+4 \kappa )^{3/2} } \biggl[
          2 {\Li}_2 \left (- \frac{\beta }{\sqrt{1+4 \kappa }} \right)
        \nonumber \\ &
        - 2 {\Li}_2 \left( \frac{\beta }{\sqrt{1+4 \kappa }} \right)
        +   {\Li}_2 \left( \frac{ \sqrt{1+4 \kappa } - 1}{ \sqrt{1+4 \kappa } - \beta } \right)
        +   {\Li}_2 \left( \frac{ \sqrt{1+4 \kappa } + \beta }{ \sqrt{1+4 \kappa} + 1} \right)
        \nonumber \\ &
        - 2 \ln(\beta) \ln \left(\frac{\sqrt{1+4 \kappa } -\beta }{\sqrt{1+4 \kappa } + \beta} \right)
        \biggr]
        +\frac{32}{3 z (1+4 \kappa )^{5/2} } \biggl[
                 6 \kappa ^2 (1+\kappa )
                -9 z \kappa ^2 (1+4 \kappa )
        \nonumber \\ &
                +3 z^2 (1 + 4 \kappa )^2 \big(3 \kappa -\sqrt{1+4 \kappa }\big)
                -4 z^3 \kappa  (1+4 \kappa )^{5/2}
        \biggr] \zeta(2)
        + \frac{32 \beta  R_1}{3 z (1+4 \kappa )} \ln (1-z)
        \nonumber \\ &
        +\frac{16 R_6}{3 z (1+4 \kappa )^{5/2} } 
        \ln (1-z) \ln \left( \frac{ \sqrt{1+4 \kappa } - \beta }{ \sqrt{1+4 \kappa } + \beta } \right)               
        - \frac{16}{3} z (3+4 z \kappa ) \biggl[        
                \ln \left(\frac{1-\beta }{1+\beta }\right)
        \nonumber \\ &
                - \ln (z)
                + 2 \ln (\beta )
                - \ln (\kappa )
        \biggr] \ln (1-z)
        -\frac{32 \beta  R_1}{3 z (1+4 \kappa )} \ln (z)
        \nonumber \\ &
        + \frac{16}{3} z (3+4 z \kappa) \biggl[
                \ln \left(\frac{1-\beta }{1+\beta }\right)
                + 2 \ln (\beta )
                - \ln (\kappa )
        \biggr] \ln (z)
        -\frac{8}{3} z (3+4 z \kappa) \ln^2(z)
        \nonumber \\ &
        + \frac{64 \beta  R_1}{3 z (1 + 4 \kappa)} \ln (\beta )
        - \frac{32}{3} z (3+4 z \kappa) \biggl[
                \ln \left(\frac{1-\beta }{1+\beta }\right)        
                - \ln (\kappa )
        \biggr] \ln (\beta )
        \nonumber \\ &
        -\biggl[
                \frac{32}{3} \left( 3 - 6 z - 4 z^2 \kappa - \frac{1 + 6 \kappa }{z( 1 + 4 \kappa )} \right)
                +\frac{16}{3} z (3+4 z \kappa) \ln (\kappa )
        \biggr] \ln \left( \frac{1-\beta }{1+\beta } \right)
        \nonumber \\ &
        -\frac{8}{3} z (3+4 z \kappa) \ln^2(\kappa )\Biggr\}
                ~,
    \end{align}
    where the polynomials that appear are~\cite{Blumlein:2019qze:Hq-heavy-wilson}:
    \begin{align*}
    R_1 &= 6\kappa + (8\kappa + 2)z^3 - (14\kappa + 3)z + 1 \ , \\
    R_2 &= 4\kappa(1 + 3\kappa) + 3(1 + 4\kappa)^2 z^2 - 6\kappa(1 + 4\kappa)z \ , \\
    R_3 &= 2\kappa(1 + \kappa) + 3(1 + 4\kappa)^2 z^2 - 3\kappa(1 + 4\kappa)z \ , \\
    R_4 &= 24\kappa^2 + 12\kappa - 3(1 + 4\kappa)^2 z + 6(1 + 4\kappa)^2 z^2 + 1 \ , \\
    R_5 &= 4\kappa \left(11\kappa^2 + 6\kappa + 1\right) - 6\kappa \left(12\kappa^2 + 7\kappa + 1\right) z + 3(1 + 2\kappa)(1 + 4\kappa)^2 z^2 \ , \\
    R_6 &= 2\kappa \left(23\kappa^2 + 13\kappa + 2\right) - 3\kappa \left(28\kappa^2 + 15\kappa + 2\right) z + 3(1 + 3\kappa)(1 + 4\kappa)^2 z^2 \ , \\
    R_7 &= 2\kappa \left(25\kappa^2 + 15\kappa + 2\right) - 3\kappa \left(36\kappa^2 + 17\kappa + 2\right) z + 3(1 + 5\kappa)(1 + 4\kappa)^2 z^2 \ .
    \end{align*}
    For $F_1$ the remaining terms are~\cite{Blumlein:2019qze:Hq-heavy-wilson}:
    \begin{align}
    P_{gq}^{(0)} \otimes h_{1,g}^{(1)} &= \cf T_F \left\{ (1 + z - 2z\kappa) \left[ -32 \ln^2 \left( \frac{1 - \beta}{1 + \beta} \right) - 64 \text{Li}_2 \left( \frac{1 - \beta}{2} \right) + 64 \text{Li}_2 \left( \frac{1 + \beta}{2} \right) \right.\right. \nonumber \\
    &\quad - 64 \text{Li}_2 \left( \frac{\beta + 1}{1 - \sqrt{1 + 4\kappa}} \right) + 64 \text{Li}_2 \left( \frac{\beta - 1}{\sqrt{1 + 4\kappa} - 1} \right) + 64 \text{Li}_2 \left( \frac{1 - \beta}{1 + \sqrt{1 + 4\kappa}} \right) \nonumber \\
    &\quad - 64 \text{Li}_2 \left( \frac{1 + \beta}{1 + \sqrt{1 + 4\kappa}} \right) + \left( -64 \ln(1 + \beta) - 128 \ln\left(1 + \sqrt{1 + 4\kappa}\right) \right. \nonumber \\
    &\quad + 128 \ln\left(\beta + \sqrt{1 + 4\kappa}\right) - 64 \ln\left( \frac{\sqrt{1 + 4\kappa} - 1}{\sqrt{1 + 4\kappa} + 1} \right) + 64 \ln\left( \frac{\sqrt{1 + 4\kappa} - \beta}{\sqrt{1 + 4\kappa} + \beta} \right) \nonumber \\
    &\quad \left.\left. + 64 \ln(2) \right) \ln\left( \frac{1 - \beta}{1 + \beta} \right) \right] - \frac{64(1 - z)\beta}{3z(1 + 4\kappa)} \Big( 3z(1 + 4\kappa) + 2z^2(1 - 2\kappa)(1 + 4\kappa) \nonumber \\
    &\quad + 2(1 + 7\kappa) \Big) - \frac{32}{3} \left( 3 - 3z - 4z^2(1 - 2\kappa)(1 + 2\kappa) \right) \ln\left( \frac{1 - \beta}{1 + \beta} \right) \nonumber \\
    &\quad \left. - \frac{128}{3z(1 + 4\kappa)^{3/2}} \left( 1 + 9(1 - z)\kappa + 2(7 - 18z)\kappa^2 \right) \ln\left( \frac{\sqrt{1 + 4\kappa} - \beta}{\sqrt{1 + 4\kappa} + \beta} \right) \right\} ,
    \end{align}
    \begin{align}
        P_{gq}^{(0)} \otimes \bar{b}_{1,g}^{(1)} ={}&
        \cf T_F \Biggl\{\frac{2 (1+k)^3 R_8}{3 k^4 z}
        \biggl[
                 k \HA_{w_{1}}
                -k \HA_{w_{2}}
                + \ln (1-k^2)
                - \ln (1-z)
        \biggr] \HA_0
        +\frac{32 R_9}{3 z} \bigl( \HA_{w_{1}} 
        \nonumber \\ &
        + \HA_{w_{2}} \bigr)
        -\frac{R_{10}}{6 k^2 z} \ln (1-k) \HA_{w_{2}}
        +\frac{R_{11}}{6 k^2 z} \bigl[ \ln (1-k) \HA_{w_{1}} + \ln (1+k) \HA_{w_{2}} \bigr]
        +\frac{8  R_{12}}{3 z}  
        \nonumber \\ &
        \times        \biggl[\HA_{w_{1},-1} 
              - \HA_{w_{2},1} 
              + \HA_{w_{2},-1} 
              + 2 \ln(k) \bigl( 
                      \HA_{w_{1}} 
                    + \HA_{w_{2}} 
              \bigr) 
        \biggr]
        +\frac{96 k z (1+z)}{3 z} \bigl( \HA_{w_{1},-1} - \HA_{w_{2},1} 
        \nonumber \\ &
        - \HA_{w_{2},-1} \bigr)
        +\frac{-16 R_{12}}{3 z} \biggl( \HA_{w_{1},0} + \HA_{w_{2},0} + \frac{1}{2} \HA_{w_{1},1} \biggr)
        +\frac{96 k z (1+z)}{3 z} \HA_{w_{1},1}
        \nonumber \\ &
        -\frac{\big(1-3 k^2\big) R_{13}}{6 k^3 z} \biggl[
                \ln ^2(1-k) 
              - \ln ^2(1+k)
              - \ln (1-z) \bigl\{ \ln (1-k) - \ln (1+k) \bigr\}
        \biggr]
        \nonumber \\ &
        +\frac{R_{14}}{6 k^2 z} \ln (1+k) \HA_{w_{1}}
        +\frac{16 R_{15}}{3 k^4} \HA_1 \HA_{-1}
        + \frac{16 R_{16}}{3 k^4} \biggl[
                2 \HA_{0,1}
              - 2 \HA_{-1,0}
              - 2 \HA_1 \HA_0
        \nonumber \\ &
              - \bigl[ \ln (1-k^2) - 2 \ln (k) \bigr] \bigl( \HA_1 + \HA_{-1} \bigr)
        \biggr]
        +\frac{16 (1-z) \beta  R_{17}}{3 k^2 z}
        -\frac{8 R_{18}}{3 k^4 z} \ln (2) \biggl[
                \ln (1-z)
        \nonumber \\ &
              - \ln (1-k^2)
              - k \bigl( \HA_{w_{1}} - \HA_{w_{2}} \bigr)
        \biggr]
        +\frac{16 R_{19}}{3 k^4 z (1-\beta )} \big(z-k^2 (1-(1-z) \beta)\big) \biggl[ \ln (1-k^2) 
        \nonumber \\ &
         - 2 \ln (k) \biggr]
        -\frac{32 (1-z) \beta  R_{20}}{3 k^2 z} \HA_0
        -\frac{8 R_{21}}{3 k^4 z} \HA_1
        +\frac{8 R_{22}}{3 k^4 z} \HA_{-1}
        -\frac{8}{3} \biggl[3+9 z
        \nonumber \\ &
        -\frac{\big(1+k^2\big)\big(1-3 k^2\big) z^2}{k^4}\biggr] \bigl( \HA_1^2 - \HA_{-1}^2 \bigr)
        +\frac{32}{3} \biggl[9+3 z+\frac{\big(1+k^2\big)\big(1-3 k^2\big) z^2}{k^4}\biggr] \HA_{-1,1}
        \nonumber \\ &
        +\big(\frac{16 z}{k}-16 k (2+3 z)\big) \biggl[
              - 2 \HA_{w_{1},1,0}
              -   \HA_{w_{1},1,1}
              +   \HA_{w_{1},1,-1}
              - 2 \HA_{w_{1},-1,0}
              -   \HA_{w_{1},-1,1}
        \nonumber \\ &
              +   \HA_{w_{1},-1,-1}
              + 2 \HA_{w_{2},1,0}
              +   \HA_{w_{2},1,1}
              -   \HA_{w_{2},1,-1}
              + 2 \HA_{w_{2},-1,0}
              +   \HA_{w_{2},-1,1}
        \nonumber \\ &
              -   \HA_{w_{2},-1,-1}
              - \bigl( \zeta(2) - \ln ^2(2) \bigr) \bigl( \HA_{w_{1}} - \HA_{w_{2}} \bigr)
              - \bigl( \HA_{w_{1},1} 
                    + \HA_{w_{1},-1} 
                    - \HA_{w_{2},1} 
                    - \HA_{w_{2},-1} 
              \bigr) 
        \nonumber \\ & \times
        \bigl\{ \ln (1-k^2) - 2 \ln(k) \bigr\}
        \biggr]
        + \big(2+\big(3-\frac{1}{k^2}\big) z\big)
        \biggl[ 
              -\frac{8}{3} \bigl( \HA_{-1}^3 + \HA_1^3 \bigr)
              - 32 \HA_{-1,1} \HA_{-1}
        \nonumber \\ &
              + 32 \HA_{-1,0,1}
              + 64 \HA_{-1,1,0}
              + 64 \HA_{-1,1,1}
              + 32 \HA_{-1,-1,0}
              + 64 \HA_{-1,-1,1}
              - 32 \HA_{0,1,1}
        \nonumber \\ &
              + 16 \bigl[ \ln (1-z) - \ln( 1 - k^2 ) \bigr]  \bigl( \ln ^2(2) - \zeta(2) \bigr)
              + 8 \bigl( \HA_{-1} - 2 \HA_0 \bigr) \HA_1^2
              + 8 \bigl( \HA_{-1}^2 
        \nonumber \\ &
        + 4 \HA_{0,1} - 4 \HA_{-1,0} - 4 \HA_{-1,1} \bigr)  \HA_1
              - 8 \bigl[ \ln(1-k^2) - 2 \ln(k) \bigr] 
              \bigl\{
                            2 \HA_{-1}\HA_1
                          - 4 \HA_{-1,1}
        \nonumber \\ &
                          + \HA_1^2
                          - \HA_{-1}^2
              \bigr\}            
        \biggr]\Biggr\},
    \end{align}
    using the polynomials~\cite{Blumlein:2019qze:Hq-heavy-wilson}:
    \begin{align*}
        R_{8} &= 99k^6 - 297k^5 + 270k^4 - 18k^3 - 77k^2 + 39k - 8 , \\
        R_{9} &= k^4 + k^2(3z + 2) + 6z - 3 , \\
        R_{10} &= 9k^8 + 48k^6(3z - 2) + k^4(214 - 552z) + 48k^2(9z - 5) - 24z + 17 , \\
        R_{11} &= 9k^8 + 48k^6(3z - 4) + 6k^4(4z + 57) - 16k^2(9z + 1) - 24z + 17 , \\
        R_{12} &= 3k^4 - 2k^2(9z + 2) + 18z - 7 , \\
        R_{13} &= 3k^6 + k^4(48z - 47) + k^2(77 - 72z) + 24z - 17 , \\
        R_{14} &= -9k^8 - 48k^6(3z - 2) + k^4(552z - 214) - 48k^2(9z - 5) + 24z - 17 , \\
        R_{15} &= 3k^4(z^2 - z - 3) + 2k^2 z^2 - z^2 , \\
        R_{16} &= 3k^4(z^2 + z - 1) + 2k^2 z^2 - z^2 , \\
        R_{17} &= 2k^4 + k^2(2z^2 + 9z + 12) - 2z^2 , \\
        R_{18} &= 9k^4 z(z + 3) + 2k^2(3z^2 - 9z + 5) - 3z^2 + 3z - 2 , \\
        R_{19} &= 3k^4 - k^2(6z^2 + 6z + 7) + 2z^2 , \\
        R_{20} &= -3k^4 + k^2(6z^2 + 6z + 7) - 2z^2 , \\
        R_{21} &= 6k^6(\beta(z - 1) + 1) + k^4\left(14(\beta - 1) - 2(6\beta - 5)z^3 + 3z^2 - 2(\beta - 15)z\right) \nonumber \\
        &\quad + k^2 z^2(-4\beta + 4(\beta - 1)z + 3) + 2z^3 , \\
        R_{22} &= 6k^6(\beta(z - 1) - 1) - k^4\left(-14(\beta + 1) + 2(6\beta + 5)z^3 + 3z^2 + 2(\beta + 15)z\right) \nonumber \\
        &\quad + k^2 z^2(-4\beta + 4(\beta + 1)z - 3) - 2z^3 .
    \end{align*}

    With the above, Eqs.~\eqref{eq:heavy-wilson-HLq} and~\eqref{eq:heavy-wilson-H1q}, the Wilson coefficient for the $F_2^{\gamma,Q\bar Q}$ contribution is~\cite{Blumlein:2019qze:Hq-heavy-wilson}
    \begin{align}
    \label{eq:heavy-wilson-H2q}
            H_{2,q}^{(2),\text{PS}} = \frac{1}{2} \left( H_{1,q}^{(2),\text{PS}} + 3 H_{L,q}^{(2),\text{PS}} \right).
    \end{align}



\section{Unpolarized two-loop Wilson coefficients for inclusive electromagnetic DIS}

\label{app:nnlo}

    The quark coefficient functions in electromagnetic DIS are composed of two distinct contributions starting at order $\as^2$~\cite{Moch:1999eb,Vermaseren:2005qc}:
    \begin{equation}
        x^{-1} F_i^\gamma = C_{i,\text{ns}} \otimes q_{\text{ns}} + \langle e^2 \rangle \left( C_{i,q} \otimes q_{\text{s}} + C_{i,g} \otimes g \right),
    \end{equation}
    where the quark coefficient $C_{i,q}$ can be decomposed into the preceding non-singlet (ns) term $C_{i,\text{ns}}$ and an additional pure singlet (ps) contribution:
    \begin{equation}
        C_{i,q}(z) \equiv C_{i,\mathrm{ns}}(z) + C_{i,\mathrm{ps}}(z).
    \end{equation}
    The quark PDF states refer the singlet flavor state in Eq.~\eqref{eq:pdf-basis:singlet}, and non-singlet in Eqs.~\eqref{eq:pdf-basis:non-singlet-v}, \eqref{eq:pdf-basis:non-singlet-pm}, respectively. This implies that starting at order $\as^2$ the choice of flavor basis of the PDFs has a concrete effect on the specific form of the coefficient functions. That is of course not a problem as long as every part of the theory is written consistently in the same basis. This new feature arises only now at NNLO as this is the first order where the pure singlet contribution is non-zero~\cite{Vermaseren:2005qc}, i.e. it is an order $\as^2$ effect.

    The results are expressed in terms of harmonic polylogarithms $\HA_{\vec{m}}$~\cite{Remiddi:1999ew}, where $\vec{m}$ is an integer multi-index, which use the polylogarithms:
    \begin{align}
       \Li_2(x) &= - \int_0^x \frac{dz}{z}\ln(1-z)\, , 
       \\
        \Li_3(x) &= \int_0^x \frac{dz}{z}\Li_2(z)\, .
    \label{Li2} 
    \end{align}
    Ref.~\cite{Moch:1999eb,Vermaseren:2005qc} use the harmonic polylogarithms:
    \begin{align}
            \HA_{0}(z) &= \ln{z} \, ,          \\ 
            \HA_{1}(z) &= \int_0^z \frac{dz'}{1-z'} = - \ln(1-z) \, ,\\ 
            \HA_{-1}(z) &= \int_0^z \frac{dz'}{1+z'} = \ln(1+z) \, , \\
            \HA_{-1,-1}(z)  &=
                 \frac{1}{2} \ln^2(1+z)
              \, ,\\
            \HA_{-1,0}(z)  &=  
                 \ln{z} \ln(1+z) + \Li_2(-z) 
              \, ,\\
            \HA_{-1,1}(z)  &=
               - \frac{1}{2} \zeta(2)
               + \frac{1}{2} \ln^2{2}
               - \ln(1+z) \ln{2}
               + \Li_2\!\! \left(\frac{1\!+\!z}{2}\right)
              \, ,\\
           \HA_{0,-1}(z)  &=
               - \Li_2(-z)
              \, ,\\
           \HA_{0,0}(z)  &=  
                 \frac{1}{2}\ln^2{z}
              \, ,\\
           \HA_{0,1}(z)  &=  
                 \Li_2(z) 
              \ , \\
           \HA_{1,-1}(z)  &= 
                 \frac{1}{2} \zeta(2)
               - \frac{1}{2} \ln^2{2}
               - \ln(1-z) \ln(1+z)
               + \ln(1+z) \ln{2}
               - \Li_2\!\! \left(\frac{1\!+\!z}{2}\right)
              \, ,\\
           \HA_{1,0}(z)  &=  
               - \ln{z} \ln(1-z) 
               - \Li_2(z) 
              \, ,\\
           \HA_{1,1}(z)  &=  
                 \frac{1}{2} \ln^2(1-z) 
              \, ,\\
            \HA_{0,0,0}(z) &=
            \frac{1}{6} \ln^3{z}
            \, ,\\
            \HA_{1,0,0}(z) &=
            - \frac{1}{2} \ln^2z \ln(1\!-\!z)
            - \ln{z} \Li_2(z)
            + \Li_3(z)
            \, ,\\
            \HA_{1,1,0}(z) &=
            \zeta(3)
            + \ln(1\!-\!z) \zeta(2)
            - \Li_3(1\!-\!z)
            \, ,\\
            \HA_{1,1,1}(z) &=
            - \frac{1}{6} \ln^3(1\!-\!z)
            \, ,\\
            \HA_{-1,0,0}(z) &=
            \frac{1}{2} \ln^2{z} \ln(1\!+\!z)
            + \ln{z} \Li_2( - z)
            - \Li_3( - z)
            \, ,\\
            \HA_{-1,-1,0}(z) &=
            \zeta(3)
            - \ln(1\!+\!z) \zeta(2)
            + \frac{1}{6} \ln^3(1\!+\!z)
            - \Li_3\!\! \left(\frac{1}{1\!+\!z}\right).
    \end{align}
    We also used the identities $\HA_{2} \equiv \HA_{0,1} \equiv \Li_2$ and $\HA_{3} \equiv \HA_{0,0,1} \equiv \Li_3$~\cite{Remiddi:1999ew} when comparing the results of Refs.~\cite{Moch:1999eb,Vermaseren:2005qc}.
    The polynomials that appear are~\cite{Moch:1999eb,Vermaseren:2005qc}:
        \begin{align}
        p_{\rm{qq}}(z) &=
                    \frac{2}{1- z}
                  - 1
                  - z\, , \\
        p_{\rm{qg}}(z) &=  
                  1
                  - 2 z
                  + 2 z^2\, , \\
        p_{\rm{gq}}(z) &=   
                    \frac{2}{z}
                  - 2
                  + z, \\
        p_{\rm{gg}}(z) &=
                    \frac{1}{1- z}
                  + \frac{1}{z}
                  - 2
                  + z
                  - z^2\, .
        \end{align}
    The order-$\as^2$ coefficient functions are expressed in terms of the above harmonic polylogarithms~\cite{Moch:1999eb,Vermaseren:2005qc}:
        \begin{align}
        c^{(2)}_{2,\rm ns}(z) &= \cf \left( \cf - \frac{C_{\mathrm{A}}}{2} \right) \Biggl(
            \frac{8}{5} \left( 9 p_{qg}(-z) (1 - z) + p_{gq}(-z) (1 - z^{-1}) + (37 + 17z) \right) \HA_{-1,0}(z) \nonumber \\
            &\quad + 4 p_{qq}(-z) \left( 7\zeta(3) + 6 \HA_{-2,0}(z) + 4 \HA_{-1,2}(z) - 8 \HA_{-1,-1,0}(z) + 10 \HA_{-1,0,0}(z) \right. \nonumber \\
            &\quad \left. - 3 \HA_{0,0,0}(z) - 8 \HA_{-1}(z) \zeta(2) - 2 \HA_{0}(z) + 2 \HA_{0}(z) \zeta(2) - 2 \Li_3(z) \right) \nonumber \\
            &\quad - \frac{72}{5} p_{qg}(z) \left( (1 + z) (\zeta(2) - \HA_{0,0}(z)) - 1 - \HA_{0}(z) \right) + \frac{8}{5} p_{gq}(z) (1 - \HA_{0}(z)) \nonumber \\
            &\quad + 8 (1 - 5z) \left( \HA_{1,0,0}(z) - \HA_{1}(z) \zeta(2) \right) - 8 (1 + 5z) \left( 2 \HA_{-1,-1,0}(z) - \HA_{-1,0,0}(z) + \HA_{-1}(z) \zeta(2) \right)
        \Biggr) \nonumber \\
        &\quad + \cf n_f \Biggl(
            \frac{1}{54} p_{qq}(z) \left( 247 - 144 \zeta(2) + 180 \HA_{0,0}(z) + 72 \HA_{1,0}(z) + 72 \HA_{1,1}(z) \right. \nonumber \\
            &\quad \left. + 342 \HA_{0}(z) + 174 \HA_{1}(z) + 144 \Li_2(z) \right) - \frac{1}{3} (7 + 19z) \HA_{0}(z) - \frac{1}{3} (1 + 13z) \HA_{1}(z) \nonumber \\
            &\quad - \frac{1}{18} (23 + 243z) + \delta(1 - z) \left( \frac{457}{36} + \frac{4}{3} \zeta(3) + \frac{38}{3} \zeta(2) \right)
        \Biggr) \nonumber \\
        &\quad + \cf^2 \Biggl(
            16 \HA_{-2,0}(z) + \frac{1}{5} (33 + 37z) \HA_{0,0}(z) + \frac{1}{4} p_{qq}(z) \left( 51 + 128 \zeta(3) + 48 \zeta(2) \right. \nonumber \\
            &\quad \left. + 96 \HA_{-2,0}(z) - 12 \HA_{0,0}(z) - 72 \HA_{1,0}(z) - 72 \HA_{1,1}(z) - 96 \HA_{1,2}(z) - 96 \HA_{2,0}(z) \right. \nonumber \\
            &\quad \left. - 112 \HA_{2,1}(z) - 32 \HA_{0,0,0}(z) - 48 \HA_{1,0,0}(z) - 128 \HA_{1,1,0}(z) - 96 \HA_{1,1,1}(z) \right. \nonumber \\
            &\quad \left. + 122 \HA_{0}(z) + 96 \HA_{0}(z) \zeta(2) + 54 \HA_{1}(z) + 32 \HA_{1}(z) \zeta(2) - 48 \Li_2(z) - 96 \Li_3(z) \right) \nonumber \\
            &\quad - \frac{1}{2} (43 + 63z) \HA_{0}(z) + \frac{1}{2} (59 - 109z) \HA_{1}(z) - 4 (1 - 19z) \zeta(3) \nonumber \\
            &\quad + 2 (1 + z) \left( 7 \HA_{1,0}(z) + 2 \HA_{2,0}(z) + 2 \HA_{2,1}(z) + 5 \HA_{0,0,0}(z) - 4 \HA_{0}(z) \zeta(2) + 4 \Li_3(z) \right) \nonumber \\
            &\quad + 2 (5 + 9z) \left( \HA_{1,1}(z) + 2 \Li_2(z) \right) - \frac{4}{5} (7 + 13z) \zeta(2) - \frac{1}{4} (93 + 209z) \nonumber \\
            &\quad + \delta(1 - z) \left( \frac{331}{8} - 78 \zeta(3) + 69 \zeta(2) + 6 \zeta(2)^2 \right)
        \Biggr) \nonumber \\
        &\quad + C_{\mathrm{A}} \cf \Biggl(
            -\frac{1}{108} p_{qq}(z) \left( 3155 - 216 \zeta(3) - 1584 \zeta(2) + 1296 \HA_{-2,0}(z) + 1980 \HA_{0,0}(z) \right. \nonumber \\
            &\quad \left. + 792 \HA_{1,0}(z) + 792 \HA_{1,1}(z) + 432 \HA_{1,2}(z) + 648 \HA_{0,0,0}(z) + 864 \HA_{1,0,0}(z) \right. \nonumber \\
            &\quad \left. - 432 \HA_{1,1,0}(z) + 4302 \HA_{0}(z) - 432 \HA_{0}(z) \zeta(2) + 2202 \HA_{1}(z) - 1296 \HA_{1}(z) \zeta(2) \right. \nonumber \\
            &\quad \left. + 1584 \Li_2(z) + 432 \Li_3(z) \right) + \frac{1}{6} (71 + 323z) \HA_{0}(z) - \frac{17}{6} (5 - 19z) \HA_{1}(z) \nonumber \\
            &\quad - \frac{4}{5} (9 + 16z) (\zeta(2) - \HA_{0,0}(z)) + \frac{1}{36} (139 + 3159z) - 8 (5 \zeta(3) z + \HA_{-2,0}(z)) \nonumber \\
            &\quad - \delta(1 - z) \left( \frac{5465}{72} - \frac{140}{3} \zeta(3) + \frac{251}{3} \zeta(2) - \frac{71}{5} \zeta(2)^2 \right)
        \Biggr) \, ,
        \end{align}
        \begin{align}
        c^{(2)}_{2,\rm ps}(z) &= \cf n_f \Biggl(
            \frac{8}{3} \left( -p_{qg}(-z) + p_{gq}(-z) \right) \HA_{-1,0}(z) \nonumber \\
            &\quad + \frac{8}{27} p_{qg}(z) \left( 28 + 27 \zeta(2) - 36 \HA_{0,0}(z) - 9 \HA_{1,0}(z) - 9 \HA_{1,1}(z) \right. \nonumber \\
            &\quad \left. - 24 \HA_{0}(z) + 6 \HA_{1}(z) - 27 \Li_2(z) \right) \nonumber \\
            &\quad + \frac{4}{27} p_{gq}(z) \left( 43 - 18 \zeta(2) + 18 \HA_{1,0}(z) + 18 \HA_{1,1}(z) - 39 \HA_{1}(z) \right) \nonumber \\
            &\quad + \frac{8}{9} (71 - 49z) \HA_{0}(z) + \frac{4}{3} (16 - 13z) \HA_{1}(z) + 8 (1 - 2z) \Li_2(z) \nonumber \\
            &\quad + 12 (1 - z) \left( \HA_{1,0}(z) + \HA_{1,1}(z) \right) \nonumber \\
            &\quad - \frac{2}{3} (1 + z) \left( 12 \zeta(3) + 12 \HA_{-1,0}(z) - 13 \HA_{0,0}(z) - 12 \HA_{2,0}(z) - 12 \HA_{2,1}(z) \right. \nonumber \\
            &\quad \left. - 30 \HA_{0,0,0}(z) + 24 \HA_{0}(z) \zeta(2) - 24 \Li_3(z) \right) \nonumber \\
            &\quad + \frac{22}{3} (3 - 5z) - \frac{8}{3} (5 - z) \zeta(2)
        \Biggr) \, ,
        \end{align}
        \begin{align}
        c^{(2)}_{2,\rm g}(z) &= \cf n_f \Biggl(
            \frac{4}{15} \left( p_{gq}(-z) (1 - z^{-1}) + (217 + 117z) \right) \HA_{-1,0}(z) \nonumber \\
            &\quad + \frac{1}{15} (639 - 1004z) \HA_{0,0}(z) - \frac{8}{5} p_{qg}(-z) \left( 6 (1 - z) \HA_{-1,0}(z) \right. \nonumber \\
            &\quad \left.- 5 (2 \HA_{-2,0}(z) - 2 \HA_{-1,-1,0}(z) + \HA_{-1,0,0}(z) - \HA_{-1}(z) \zeta(2)) \right) \nonumber \\
            &\quad + \frac{2}{5} p_{qg}(z) \left( 6 (19 + 4z) (\zeta(2) - \HA_{0,0}(z)) - (9 - 90 \zeta(3) + 90 \HA_{1,0}(z) \right. \nonumber \\
            &\quad \left. + 90 \HA_{1,1}(z) + 60 \HA_{1,2}(z) + 40 \HA_{2,0}(z) + 50 \HA_{2,1}(z) + 50 \HA_{0,0,0}(z) \right. \nonumber \\
            &\quad \left. + 30 \HA_{1,0,0}(z) + 40 \HA_{1,1,0}(z) + 50 \HA_{1,1,1}(z) + 54 \HA_{0}(z) - 60 \HA_{0}(z) \zeta(2) \right. \nonumber \\
            &\quad \left. + 30 \HA_{1}(z) - 40 \HA_{1}(z) \zeta(2) + 90 \Li_2(z) + 60 \Li_3(z)) \right) \nonumber \\
            &\quad + \frac{4}{15} p_{gq}(z) (1 - \HA_{0}(z)) + \frac{1}{3} (16 - 61z) \HA_{0}(z) - 2 (1 - 8z) \HA_{1}(z) \nonumber \\
            &\quad + 4 (5 - 4z) \Li_2(z) - 4 (1 - 18z) \zeta(3) + 2 (1 - 2z) \left( 8 \HA_{-2,0}(z) + 2 \HA_{2,0}(z) \right. \nonumber \\
            &\quad \left. + 2 \HA_{2,1}(z) + 5 \HA_{0,0,0}(z) + 4 \HA_{1,0,0}(z) - 4 \HA_{0}(z) \zeta(2) - 4 \HA_{1}(z) \zeta(2) + 4 \Li_3(z) \right) \nonumber \\
            &\quad - 8 (1 + 2z) \left( 2 \HA_{-1,-1,0}(z) - \HA_{-1,0,0}(z) + \HA_{-1}(z) \zeta(2) \right) \nonumber \\
            &\quad + 2 (5 + 4z) \left( \HA_{1,0}(z) + \HA_{1,1}(z) \right) - \frac{4}{15} (111 - 176z) \zeta(2) - \frac{1}{3} (117 - 121z)
        \Biggr) \nonumber \\
        &\quad + C_{\mathrm{A}} n_f \Biggl(
            \frac{8}{3} \left( p_{gq}(-z) - 3 (4 + 3z) \right) \HA_{-1,0}(z) + \frac{4}{3} \HA_{0,0}(z) (47 + 35z) \nonumber \\
            &\quad + 2 (23 - 6z) \HA_{1,0}(z) + 6 (7 - 2z) \HA_{1,1}(z) + 4 (5 + 14z) \HA_{0,0,0}(z) \nonumber \\
            &\quad + \frac{4}{3} p_{qg}(-z) \left( 6 \HA_{-2,0}(z) + 10 \HA_{-1,0}(z) + 6 \HA_{-1,2}(z) \right. \nonumber \\
            &\quad \left. + 9 \HA_{-1,0,0}(z) - 6 \HA_{-1}(z) \zeta(2) \right) \nonumber \\
            &\quad - \frac{1}{54} p_{qg}(z) \left( 4493 - 648 \zeta(3) - 3996 \zeta(2) + 3492 \HA_{0,0}(z) + 2412 \HA_{1,0}(z) \right. \nonumber \\
            &\quad \left. + 2196 \HA_{1,1}(z) + 216 \HA_{1,2}(z) + 432 \HA_{2,0}(z) + 432 \HA_{2,1}(z) \right. \nonumber \\
            &\quad \left. + 432 \HA_{1,0,0}(z) + 648 \HA_{1,1,0}(z) + 216 \HA_{1,1,1}(z) + 6270 \HA_{0}(z) \right. \nonumber \\
            &\quad \left. - 432 \HA_{0}(z) \zeta(2) + 4710 \HA_{1}(z) - 216 \HA_{1}(z) \zeta(2) + 3996 \Li_2(z) + 432 \Li_3(z) \right) \nonumber \\
            &\quad + \frac{4}{27} p_{gq}(z) \left( 43 - 18 \zeta(2) + 18 \HA_{1,0}(z) + 18 \HA_{1,1}(z) - 39 \HA_{1}(z) \right) \nonumber \\
            &\quad + \frac{1}{9} (1567 - 338z) \HA_{0}(z) + \frac{1}{3} (289 - 52z) \HA_{1}(z) + 2 (33 - 2z) \Li_2(z) \nonumber \\
            &\quad - 4 (1 - 2z) \left( \HA_{1,0,0}(z) - \HA_{1}(z) \zeta(2) \right) \nonumber \\
            &\quad - 4 (1 + 2z) \left( 2 \HA_{-2,0}(z) - 2 \HA_{-1,-1,0}(z) + \HA_{-1,0,0}(z) - \HA_{-1}(z) \zeta(2) \right) \nonumber \\
            &\quad - 8 (1 + 3z) \left( \zeta(3) + 2 \HA_{0}(z) \zeta(2) - 2 \Li_3(z) \right) + 8 (1 + 4z) \left( \HA_{2,0}(z) + \HA_{2,1}(z) \right) \nonumber \\
            &\quad + \frac{7}{6} (105 - 46z) - \frac{2}{3} (107 - 10z) \zeta(2)
        \Biggr) \, ,
        \end{align}
        where $C_A = N_c = 3$ is a QCD constant.
        The coefficients for the longitudinal structure functions are~\cite{SanchezGuillen:1990iq,Zijlstra:1991qc,Moch:1999eb,Vermaseren:2005qc}:
        \begin{align}
        c^{(2)}_{L,\rm ns}(z) &= \cf \left( \cf - \frac{C_{\mathrm{A}}}{2} \right) \Biggl(
            \frac{16}{5} \left( 3 p_{qg}(-z) (1 - z) + 2 p_{gq}(-z) (1 - z^{-1}) + (19 + 9z) \right) \HA_{-1,0}(z) \nonumber \\
            &\quad - \frac{48}{5} p_{qg}(z) \left( (1 + z) (\zeta(2) - \HA_{0,0}(z)) - (1 + \HA_{0}(z))+ \frac{32}{5} p_{gq}(z) (1 - \HA_{0}(z)) \right)
        \Biggr) \nonumber \\
        &\quad + \cf n_f \Biggl(
            \frac{4}{9} (6 - 25z) - \frac{8}{3} z \left( 2 \HA_{0}(z) + \HA_{1}(z) \right)
        \Biggr) \nonumber \\
        &\quad + \cf^2 \Biggl(
            - 4 (2 + 7z) \HA_{1}(z) - \frac{8}{5} (3 + 2z) \left( 2 \HA_{0,0}(z) + 5 \HA_{0}(z)+ 2 (6 - 31z)  \right) \nonumber \\
            &\quad + \frac{8}{5} (6 - z) \zeta(2) + 8z \Bigl( 8 \zeta(3) + 2 \HA_{1,0}(z) + 2 \HA_{1,1}(z) - 8 \HA_{-1,-1,0}(z)  \nonumber \\
            &\quad + 4 \HA_{-1,0,0}(z) - 4 \HA_{1,0,0}(z) - 4 \HA_{-1}(z) \zeta(2) + 4 \HA_{1}(z) \zeta(2) + 3 \Li_2(z) \Bigr)
        \Biggr) \nonumber \\
        &\quad + C_{\mathrm{A}} \cf \Biggl(
            \frac{8}{3} (3 + 14z) \HA_{0}(z) - \frac{8}{5} (3 + 7z) (\zeta(2) - \HA_{0,0}(z)) \nonumber \\
            &\quad - \frac{2}{9} (66 - 317z) - \frac{4}{3} z \Bigl( 24 \zeta(3) - 24 \HA_{-1,-1,0}(z) + 12 \HA_{-1,0,0}(z) \nonumber \\
            &\quad - 12 \HA_{1,0,0}(z) - 12 \HA_{-1}(z) \zeta(2) - 23 \HA_{1}(z) + 12 \HA_{1}(z) \zeta(2) \Bigr)
        \Biggr) \, ,
        \end{align}

        \begin{align}
        c^{(2)}_{L,g}&(z) ={}
        \cf n_f \Biggl(
          \frac{16}{15} \left( - 6 p_{qg}(-z) (1 - z) + p_{gq}(-z) (1 - z^{-1}) + (7 - 3z) \right) \HA_{-1,0}(z) 
          \nonumber \\
          & + \frac{32}{15} (3 - 13z) \HA_{0,0}(z) 
          + \frac{8}{5} p_{qg}(z) \left( 4(1 + z)(\zeta(2) - \HA_{0,0}(z)) + (21 + 6 \HA_{0}(z) \right.
          \nonumber \\ 
          & \left. + 10 \HA_{1}(z)) \right)
          + \frac{16}{15} p_{gq}(z) (1 - \HA_{0}(z)) 
          - \frac{8}{3} (7 + 8z) \HA_{0}(z) 
          - 8 (3 - z) \HA_{1}(z) 
          \nonumber \\
          & - 16 z \Li_2(z)  - \frac{16}{15} (6 - 11z) \zeta(2) 
          - \frac{8}{3} (15 - 2z)
        \Biggr) 
        \nonumber \\
        & + C_\text{A} n_f \Biggl(
          16 p_{qg}(-z) \HA_{-1,0}(z) \nonumber \\
          & - \frac{8}{9} p_{qg}(z) \Big( 53 - 18\zeta(2) + 18 \HA_{1,0}(z) + 18 \HA_{1,1}(z) + 117 \HA_{0}(z) + 87 \HA_{1}(z) + 18 \Li_2(z) \Big) \nonumber \\
          & - \frac{8}{9} p_{gq}(z) (1 + 3 \HA_{1}(z)) 
          + 40 (3 - 2z) \HA_{0}(z) 
          + 8 (11 - z) \HA_{1}(z) 
          + 16 (1 + 4z) \Li_2(z) \nonumber \\
          & - 16 (1 + 2z) \zeta(2) 
          + \frac{8}{3} (19 - z) 
          - 16 \left( \HA_{-1,0}(z) - 6z \HA_{0,0}(z) - \HA_{1,0}(z) - \HA_{1,1}(z) \right)
        \Biggr) \, .
        \end{align}

\newpage

\section{High-energy physics nomenclature}
    \label{sec:hep-nomenclature}
    For the convenience of the mathematically versed reader, we provide a glossary of nomenclature that can hopefully be at least a little helpful, along with Refs.~\cite{ParticleDataGroup:2024cfk,Ellis:1996mzs:qcd-collider-phys,Sterman:1995:handbook-qcd,Griffiths:1987tj,Peskin:0201503972}. Our intention is to give at least a basic description of all the basic concepts, quantities, and variables that go to define the mathematical expressions of the inverse problem to make this inverse problem approach to the \gls{global analysis} of \glspl{parton distribution function} more accessible.

    \subsection*{Key elements of the mathematical structure of the inverse problem}

    \begin{itemize}
        \item \Gls{parton distribution function} (PDF). Smooth, possible non-negative, functions to be inferred from data. Multiple distinct PDFs ($N=13$) contribute to measurements, and none can be perfectly isolated experimentally, which leads to a coupled inference problem.
        \item \Gls{world data} is the collective name for data composed of measurements of physically distinct experiments that fall into some broader common class, such as scattering experiments on free protons. Since the physical detail of each measurement is different, the full problem has to simultaneously solve different kinds of integral equations for the PDFs.
        \item \Gls{reduced cross section} and \gls{structure function} measurements are the primary experimental data to solve the PDFs from. Mathematically they exhibit as integral equations of the PDFs.
        \item DGLAP evolution equation (see Sec.~\ref{sec:tensor-ip-lo}) is a partial differential equation of the PDFs, and the inferred PDFs should at least approximately satisfy the DGLAP equation.
    \end{itemize}

    \printnoidxglossaries

\end{document}

%% file: defs.tex
\newcommand{\lo}{{\textnormal{LO}}}
\newcommand{\nlo}{{\textnormal{NLO}}}

\newcommand{\ud}{\, \mathrm{d}}

\newcommand{\cf}{C_\mathrm{F}}

\newcommand{\gev}{\ \textrm{GeV}}

\newcommand{\as}{\alpha_{\mathrm{s}}}
\newcommand{\aem}{\alpha_{\mathrm{em}}}

\newcounter{diag}
\newcounter{subdiag}[diag]


%% file: glossary.tex
\usepackage{glossaries}

\newglossaryentry{anti}{
    name={anti-},
    description={The ``bar'' notation in this work refers the anti-particle $\bar q$ of a given particle $q$, mainly the \glspl{quark}. Quarks and leptons have anti-particles that are otherwise identical to their counterparts, aside from the equal but opposite electrical charge}}

\newglossaryentry{ansatz}{
    name={ansatz},
    description={}}

\newglossaryentry{bjorkenx}{
    name={Bjorken-$x$},
    description={$\xbj$, or often just $x$, is a dimensionless kinematic variable of high-energy particle scattering defined as $x \coloneqq \frac{Q^2}{2 q \cdot P}$. $x \in (0,1)$, relate to fraction of the proton longitudinal momentum}}

\newglossaryentry{boson}{
    name={boson},
    description={By definition bosons are quantum mechanical particles that have an integer spin. This work is concerned with the elementary bosons of the standard model, which act as carriers of the elementary forces: the photon $\gamma$ mediates the electro-magnetic force, gluon $g$ mediates the strong nuclear force, and the $W^+, W^-$, and $Z$ bosons mediate the weak nuclear force. An example of a composite boson would be the Helium-4 nucleus, which has an integer spin since it is composed of two protons and two neutrons, both of which have half-integer spins}}

\newglossaryentry{deeply inelastic scattering}{
    name={deeply inelastic scattering},
    description={(DIS) High-energy collision of a lepton probe and a hadronic target, typically a proton. Deeply refers to the high resolution into the structure of the target, and inelastic implies that the target is broken apart in the collision}}

\newglossaryentry{GeV}{
    name={GeV},
    description={is the abbreviation for gigaelectronvolt, which is a unit of energy used in high-energy physics. Giga- is the SI prefix for a billion $(10^9)$ electronvolts (eV), which is the base unit of energy; $1 \, \mathrm{eV} \approx 1.602 \times 10^{-19}$ joules (J). However, in the natural units used in high-energy physics the speed of light is set to unity $c\equiv1$, which means that the unit of mass becomes also expressible in $\mathrm{GeV}$ via an abuse of notation: $ 1 \, \mathrm{kg} \approx 5.6 \times10^{26} \frac{\gev}{c^2} \equiv 5.6 \times10^{26} \gev$, which means that particle masses are often expressed in $\frac{\mathrm{eV}}{c^2} \equiv \mathrm{eV}$ as it is a more natural scale of reference for elementary particles, than the regular SI units}}

\newglossaryentry{global analysis}{
    name={global analysis},
    description={Historical terminology for the inference problem of leveraging all possible experimental data to solve the parton distribution functions of the proton, and to test the parton model itself in the process. Practically this has meant that model parameters have been constrained or fit against data from proton--proton, electron--proton, and various other high-energy particle collider experiments}}

\newglossaryentry{electromagnetic}{
    name={electromagnetic},
    description={force is one of the four elementary forces of the standard model. In quantum field theory, the theoretical language of the standard model, the electro-magnetic force is mediated by the virtual photon, which is a gauge \gls{boson} denoted by $\gamma^*$}}

\newglossaryentry{explicit inverse problem}{
    name={explicit inverse problem},
    description={Inverse or inference problem where it is mathematically clear to see how to solve an unknown function or quantity from experimental data. Contrast this with \textit{implicit inverse problem}}}

\newglossaryentry{fine structure constant}{
    name={fine structure constant},
    description={is the name for the quantified strength of the electro-magnetic interaction, which is often denoted by $\alpha$, or by $\aem$ to clarify its distinction from the strong coupling $\as$. Naively it can be thought of as a physical constant of nature with the approximate value $\aem \approx \frac{1}{137}$. However, technically quantum field theoretic effects will adjust the size of $\aem$ depending on the momentum scale of the interaction, completely analogously to the running of the strong coupling. Experimentally it has been determined to eleven significant digits: $\aem= 1/137.035 999 084(21)$~\cite{ParticleDataGroup:2024cfk}}}

\newglossaryentry{flavor}{
    name={flavor},
    description={Quarks, the color charged elementary particles of the Standard Model, come in six distinct kinds, known as flavors. They are called: up, down, strange, charm, bottom, and top}}

\newglossaryentry{forward operator}{
    name={forward operator},
    description={is a mathematical operation in the formalism of inverse problems that maps an unknown quantity to the space of data that is attained in a measurement. For example, in computed tomography the forward operator is the X-ray transform that maps the interior distribution of the subject to the space of sinograms in which the measurements of a CT scan are defined. In this work we define a forward operator that maps the unknown quantity to multiple distinct measurement spaces simultaneously}}

\newglossaryentry{fractional charge}{
    name={fractional charge},
    description={The electric charges of elementary particles are quantified in comparison to the charge of the electron, which is defined to have the electric charge $-1e$. The quarks are special since they are the only known elementary particles that have relative charges of $\pm \frac{1}{3}$ or $\pm \frac{2}{3}$}}

\newglossaryentry{gluon}{
    name={gluon},
    description={Elementary particle of the Standard Model and the force carrier of the strong nuclear force that binds together the quarks that form the proton, and that also binds the protons and neutrons together within the positively charged atomic nucleus}}

\newglossaryentry{hadron}{
    name={hadron},
    description={Hadron $H$ is a composite particle which is composed of quarks that are bound together by the strong nuclear force. Quarks carry color charge, and yet all hadrons are color neutral. Important examples of hadrons are the proton and neutron, which belong to the so-called conventional hadrons over a hundred of which are known. Exotic hadrons are particles composed of four or more quarks, of which $\sim 80$ states have been experimentally discovered~\cite{ParticleDataGroup:2024cfk}}}

\newglossaryentry{hadronic tensor}{
    name={hadronic tensor},
    description={Hadronic tensor $W_{\mu\nu}$ is a mathematical object that encodes the structure of the proton in four-momentum space. The indices $\mu, \nu$ are the indices of the four momenta of the in an out going parton at an interaction vertex \cite{Peskin:0201503972}. Interactions with the \glspl{parton} in the proton are theoretically described by contractions of a probe tensor such as that of a lepton with the hadronic tensor, which enables the prediction of interaction probabilities}}

\newglossaryentry{heavy quark}{
    name={heavy quark},
    description={
    There are six quarks in the standard model and they all have different masses. Up, down, and strange are light enough to be (almost) always approximated as massless in high-energy scattering, whereas charm, bottom, and top have masses that command attention in the theoretical picture, and so are collectively known as heavy quarks. Basically, whenever the mass of a quark is taken into account, the quark is called heavy, and when it's ``light'' it is taken to be massless}}

\newglossaryentry{inclusive}{
    name={inclusive},
    description={is used synonymously with ``total'' or ``inclusive of all'', whereas exclusive is used in the sense of ``only'' or ``strictly limited to''}}

\newglossaryentry{inelasticity}{
    name={inelasticity},
    description={Define $y$. Anything helpful to explain it?}}

\newglossaryentry{interaction}{
    name={interaction},
    description={The standard model, and quantum field theories generally, describe interactions of particles as points in spacetime where quantities like energy and momentum are instantaneously exchanged between the interacting particles}}

\newglossaryentry{implicit inverse problem}{
    name={implicit inverse problem},
    description={Inference problem where it is non-trivial to determine what the mathematical structure is to solve the unknown quantity of interest without fitting model parameters by solving the forward problem}}

\newglossaryentry{leading order}{
    name={leading order},
    description={Perturbative QCD breaks physical theory calculations into an expansion series in the strength of the \gls{strong coupling} $\as$, whose fixed order terms can be calculated. The term leading order refers to the $0$th order term of the expansion, i.e. $\as^0$ typically, but this can vary as some sub-dominant physical processes can be overall be suppressed by a power of $\as$, meaning that such processes can be considered as leading order at the order $\as^1$}}

\newglossaryentry{lepton}{
    name={lepton},
    description={
    Leptons are a class of elementary particle in the standard model. The most familiar of them is the electron $(e^-)$, whose anti-particle is known as the positron $(e^+)$. In addition to them, there are the muon $\mu^\pm$ and tau $\tau^\pm$ leptons, and for each main type of lepton, also known as ``generation'', there are neutrinos and anti-neutrinos: $\nu_e, \bar \nu_e, \nu_\mu, \bar \nu_\mu, \nu_\tau, \bar \nu_\tau$.
    Leptons, especially electron and positron, are important for this work as they are clean probes to study the structure of \glspl{hadron} in scattering experiments}}

\newglossaryentry{flavor number scheme}{
    name={flavor number scheme},
    description={Theoretical framework to describe QCD physics over broad regimes of data where the momentum scale of the scattering affects whether a quark can be approximated as massless or not. This is significant because only in the massless limit are the quarks and gluons described by \glspl{parton distribution function}}}

\newglossaryentry{next-to-leading order}{
    name={next-to-leading order},
    description={(NLO) refers to the order of the \gls{perturbative QCD} expansion in $\as$ beyond the first dominant order. Each successive order of correction in the perturbative expansion in the \gls{strong coupling} $\as$, at least initially, should produce more accurate prescription of the physical phenomenon. For the global analysis inverse problem for the PDFs, the state-of-the-art is at NNNLO (N$^3$LO), and the early orders of corrections have been of substantial importance, i.e. fairly large in magnitude, which is why any approach of global analysis of the PDFs needs to be able to handle higher order corrections systematically in the perturbative expansion}}

\newglossaryentry{non-perturbative}{
    name={non-perturbative},
    description={A quantity or problem that cannot be solved with conventional means within \gls{perturbative QCD}. The most important non-perturbative quantity for this work are the parton distribution functions, or at least the functional initial condition of their $Q^2$ evolution, which is what is conventionally constrained in the inverse problem of \gls{global analysis}}}

\newglossaryentry{nuclear modification}{
    name={nuclear modification},
    description={effects are non-trivial changes to the PDFs of a proton or neutron when they are bound within an atomic nucleus}}

\newglossaryentry{observable}{
    name={observable},
    description={is a quantity that is physically measurable, even in principle. In classical mechanics they are real valued quantities such as distance or energy, whereas in quantum mechanics they are defined by linear operators that satisfy specific transformation laws that theoretically describes how the measurements of different observers are precisely related to one another. A non-trivial example of a quantum mechanical observable is the Wigner quasiprobability distribution, which has been measured for photon~\cite{Lvovsky:2009:photon-wigner-tomo} and phonon states~\cite{Chu2018-wigner-tomo}. Their complement, non-observable quantities such as the parametrization of a light-like geodesic, an individual quantum wavefunction including its phase, or the electromagnetic vector potential, cannot be measured under any circumstance because they have associated mathematical ambiguities such as gauge transformations, which allows a degree of mathematical freedom in the choice of the quantity, which is not reflected in measurements in the real world}}

\newglossaryentry{parton}{
    name={parton},
    description={is the collective name for the \glspl{quark} and \glspl{gluon} that comprise \glspl{hadron} such as protons and neutrons}}

\newglossaryentry{parton distribution function}{
     name={parton distribution function},
     description={(PDF) are smooth functions $f_a(x)$ that quantify the structure of the proton or other hadron for \gls{flavor} $a$. For a complete picture the PDFs must be known for all the \glspl{flavor}. Their prediction from first principles, i.e. computation in the standard model, is a very hard task, so the global analysis approach was developed to constrain them from experimental data. The goal of this paper is to develop a mathematical methodology for their solution from data without fitting model parameters. Conventionally the PDFs have been assumed to be non-negative, however recent theory work has put this presumption on a more solid footing~\cite{Candido:2023ujx:positivity-of-pdf}}}

\newglossaryentry{parton distribution tensor}{
     name={parton distribution tensor},
     description={(PDT) is a name coined in this work for the function valued vector $\pdtbf$ defined in Eq.~\eqref{eq:pdt-definition}, which is composed of the canonical \glspl{parton distribution function} to give a mathematical meaning for the postulate that the parton distribution functions are meaningful only as a whole}}

\newglossaryentry{perturbative QCD}{
    name={perturbative QCD},
    description={The full theory of \gls{quantum chromodynamics} is notoriously challenging to work with without simplifying assumptions. One of the most successful approaches is that of perturbative QCD, in which calculations are broken down in an expansion in $\as$, the \gls{strong coupling}, which requires $\as \ll 1$ to successfully develop a power series in $\as$. However, as the name of the \gls{strong coupling} implies, this does not always hold, so the framework of perturbative QCD is unfortunately not universally applicable. In this work we assume that perturbative QCD is applicable, which in practice constrains what data this approach is applicable to}}

\newglossaryentry{phenomenological}{
    name={phenomenological},
    description={Phenomenology is an approach of theoretical physics where approximations or simplifications are made in the accuracy of the physics or mathematical methods to make a problem more tractable. QCD is a notoriously difficult quantum field theory to work with, and some aspects require so-called phenomenological methods to be able to reach a quantified comparison between the physical theoretical calculations and predictions, and experimental data}}

\newglossaryentry{quantum chromodynamics}{
    name={quantum chromodynamics},
    description={(QCD) is the part of the Standard Model that describes the strong nuclear force and the elementary particles affected by it, quarks and gluons}}

\newglossaryentry{quark}{
    name={quark},
    description={Elementary particle of the Standard Model that carries color charge. Protons and neutrons are made of up and down quarks. There are six flavors of quarks in total: up, down, strange, charm, bottom, and top}}

\newglossaryentry{reduced cross section}{
    name={reduced cross section},
    description={Smooth scalar function $\sigma$ of one or more scalar variables. Cross sections are related to the probability of a scattering event, and they are one of the fundamental measurable quantities in particle scattering experiments. Typically denoted with $\sigma$, and additional sub- and superscripts are used to specify what is the exact physical quantity in question. ``Reduced'' refers to the convention that a group of separable natural constants and other multipliers are divided out of the data, which simplifies the definition of the observable quantity from a practical perspective. This is perhaps analogous to reporting measurements of distance in relation to some relative unit of distance, instead of reporting the experimental data in absolute units of measurement such as meters}}

\newglossaryentry{renormalization}{
    name={renormalization},
    description={is a mathematical framework to treat, and ideally understand, infinities that arise in quantum field theoretic calculations in a manner that enables the theory to precisely describe the physical world, which does not exhibit features implied by said non-physical infinities}}

\newglossaryentry{running coupling}{
     name={running coupling},
     description={
     The \gls{strong coupling} $\as(\mu^2)$ becomes weaker at higher momentum scales $\mu^2 \gg 1 \gev^2$, and this decay of the interaction strength is known as running of the coupling. This weakening of the strong force is what enables the perturbation theory approach to QCD at high momentum}}

\newglossaryentry{sea quark}{
    name={sea quark},
    description={Hadrons contain in addition to \glspl{valence quark} so-called sea quarks, which are a quantum mechanical background contribution to the structure. They are described in the standard model as pairs of a quark and an antiquark that exist momentarily as a quantum mechanical fluctuation of the structure of the parent hadron}}

\newglossaryentry{sinogram}{
    name={sinogram},
    description={}}

\newglossaryentry{standard model}{
    name={standard model},
    description={}}

\newglossaryentry{strong coupling}{
     name={strong coupling},
     description={is the name for the magnitude of the strong nuclear force as described by the standard model. See also \gls{running coupling}. Denoted by $\as \equiv \as(\mu^2)$, which is typically a smooth or piece-wise smooth scalar function of one variable $\mu^2$, which is often fixed at $\mu^2 = Q^2$. It has been determined experimentally~\cite{CMS:2024trs:running-as,Deur:2016tte:running-as}, and is a critical quantity in QCD}}

\newglossaryentry{structure function}{
    name={structure function},
    description={Smooth functions $F_i(x,Q^2)$ that quantify the structure of a hadron such as the proton. From the perspective of this work they are an important stepping stone between experimental data and the detailed structure of the proton described by the \gls{parton distribution function}}}

\newglossaryentry{valence quark}{
    name={valence quark},
    description={Valence quarks are the primary internal constituents of hadrons: the proton is composed of two up quarks and a down quark. Technically, valence quarks are defined to be the quarks that determine the quantum numbers of a hadron, which are externally observable invariants. Hadrons also contain \glspl{sea quark}}}

\newglossaryentry{virtuality}{
    name={virtuality},
    description={is a scalar quantity denoted by $Q^2$. By physical definition, it is the magnitude of the four-momentum of a photon that is off-shell, i.e. a photon that is a temporary excitation in a quantum field that mediates electro-magnetic interaction between charged particles. Virtuality is the Fourier conjugate of spatial resolution in deep inelastic scattering, which means that high-virtuality photons need to be produced in scattering to achieve the shortest scale resolution in to the target hadron}}

\newglossaryentry{weak nuclear}{
    name={weak nuclear},
    description={force is one of the four elementary forces of the standard model, and the one that is responsible for example of radioactive beta decay of nuclei. The part of standard model that theoretically describes the weak nuclear interaction is known as electro-weak theory. Weak nuclear force is mediated by the $W^+$, $W^-$, and $Z$ \glspl{boson}, and from the perspective of this work---and global analysis generally---weak nuclear interaction is crucial since it enables theoretically distinct sensitivity to the flavor structure of hadrons than is allowed by only considering electro-magnetic interaction}}

\newglossaryentry{world data}{
    name={world data},
    description={High-energy physics terminology for all the globally and historically collected particle collider data. This means that in the \gls{global analysis} inverse problem one considers experimental data from all relevant experiments that have been performed, the collective data from all of which is known as \textit{world data}}}

%% file: refs.bib
@book{Griffiths:1987tj,
  address = {New York, USA},
  author = {Griffiths, D.},
  publisher = {John Wiley \& Sons},
  title = {Introduction to Elementary Particles},
  year = 1987
}

@book{Peskin:0201503972,
  asin = {0201503972},
  author = {Peskin, Michael E. and Schroeder, Daniel V.},
  description = {An Introduction to Quantum Field Theory (Frontiers in Physics): Michael E. Peskin, Daniel V. Schroeder},
  dewey = {530.143},
  ean = {9780201503975},
  isbn = {0201503972},
  publisher = {Perseus Books},
  title = {An Introduction to Quantum Field Theory (Frontiers in Physics)},
  year = 2008
}

@book{Collins:1984renormalization,
  title={Renormalization: an introduction to renormalization, the renormalization group and the operator-product expansion},
  author={Collins, John C},
  year={1984},
  publisher={Cambridge university press}
}

@article{tHooft:1972tcz,
    author = "'t Hooft, Gerard and Veltman, M. J. G.",
    title = "{Regularization and Renormalization of Gauge Fields}",
    doi = "10.1016/0550-3213(72)90279-9",
    journal = "Nucl. Phys. B",
    volume = "44",
    pages = "189--213",
    year = "1972"
}

@article{tHooft:1973mfk,
    author = "'t Hooft, Gerard",
    title = "{Dimensional regularization and the renormalization group}",
    doi = "10.1016/0550-3213(73)90376-3",
    journal = "Nucl. Phys. B",
    volume = "61",
    pages = "455--468",
    year = "1973"
}

@article{Weinberg:1973xwm,
    author = "Weinberg, Steven",
    title = "{New approach to the renormalization group}",
    doi = "10.1103/PhysRevD.8.3497",
    journal = "Phys. Rev. D",
    volume = "8",
    pages = "3497--3509",
    year = "1973"
}

@article{Deur:2016tte:running-as,
    author = "Deur, Alexandre and Brodsky, Stanley J. and de Teramond, Guy F.",
    title = "{The QCD Running Coupling}",
    eprint = "1604.08082",
    archivePrefix = "arXiv",
    primaryClass = "hep-ph",
    reportNumber = "JLAB-PHY-16-2199, SLAC-PUB-16448, DOE/OR/23177-3645, DOE-OR-23177-3645",
    doi = "10.1016/j.ppnp.2016.04.003",
    journal = "Nucl. Phys.",
    volume = "90",
    pages = "1",
    year = "2016"
}

@article{CMS:2024trs:running-as,
    author = "Chekhovsky, Vladimir and others",
    collaboration = "CMS",
    title = "{Determination of the strong coupling and its running from measurements of inclusive jet production}",
    eprint = "2412.16665",
    archivePrefix = "arXiv",
    primaryClass = "hep-ex",
    reportNumber = "CMS-SMP-24-007, CERN-EP-2024-327",
    doi = "10.1016/j.physletb.2025.139651",
    journal = "Phys. Lett. B",
    volume = "868",
    pages = "139651",
    year = "2025"
}

@article{Altarelli:1977zs,
    author = "Altarelli, Guido and Parisi, G.",
    title = "{Asymptotic Freedom in Parton Language}",
    reportNumber = "LPTENS-77-6",
    doi = "10.1016/0550-3213(77)90384-4",
    journal = "Nucl. Phys. B",
    volume = "126",
    pages = "298--318",
    year = "1977"
}

@article{Dokshitzer:1977sg,
    author = "Dokshitzer, Yuri L.",
    title = "{Calculation of the Structure Functions for Deep Inelastic Scattering and e+ e- Annihilation by Perturbation Theory in Quantum Chromodynamics.}",
    journal = "Sov. Phys. JETP",
    volume = "46",
    pages = "641--653",
    year = "1977"
}

@article{Lipatov:1974qm,
    author = "Lipatov, L. N.",
    title = "{The parton model and perturbation theory}",
    journal = "Yad. Fiz.",
    volume = "20",
    pages = "181--198",
    year = "1974"
}

@article{Gribov:1972ri,
    author = "Gribov, V. N. and Lipatov, L. N.",
    title = "{Deep inelastic e p scattering in perturbation theory}",
    reportNumber = "IPTI-381-71",
    journal = "Sov. J. Nucl. Phys.",
    volume = "15",
    pages = "438--450",
    year = "1972"
}

@article{Martin:2008cn:dglap,
    author = "Martin, Alan D.",
    editor = "Fiore, R. and Papa, A. and Royon, C.",
    title = "{Proton structure, Partons, QCD, DGLAP and beyond}",
    eprint = "0802.0161",
    archivePrefix = "arXiv",
    primaryClass = "hep-ph",
    reportNumber = "IPPP-08-03, DCPT-08-06",
    journal = "Acta Phys. Polon. B",
    volume = "39",
    pages = "2025--2062",
    year = "2008"
}

@article{Moch:2004pa:dglap-general-nnnlo,
    author = "Moch, S. and Vermaseren, J. A. M. and Vogt, A.",
    title = "{The Three loop splitting functions in QCD: The Nonsinglet case}",
    eprint = "hep-ph/0403192",
    archivePrefix = "arXiv",
    reportNumber = "DESY-04-047, SFB-CPP-04-09, NIKHEF-04-001",
    doi = "10.1016/j.nuclphysb.2004.03.030",
    journal = "Nucl. Phys. B",
    volume = "688",
    pages = "101--134",
    year = "2004"
}

@article{Vogt:2004mw:dglap-general-nnnlo,
    author = "Vogt, A. and Moch, S. and Vermaseren, J. A. M.",
    title = "{The Three-loop splitting functions in QCD: The Singlet case}",
    eprint = "hep-ph/0404111",
    archivePrefix = "arXiv",
    reportNumber = "NIKHEF-04-004, DESY-04-060, SFB-CPP-04-12",
    doi = "10.1016/j.nuclphysb.2004.04.024",
    journal = "Nucl. Phys. B",
    volume = "691",
    pages = "129--181",
    year = "2004"
}

@article{Candido:2022tld:EKO,
    author = "Candido, Alessandro and Hekhorn, Felix and Magni, Giacomo",
    title = "{EKO: evolution kernel operators}",
    eprint = "2202.02338",
    archivePrefix = "arXiv",
    primaryClass = "hep-ph",
    reportNumber = "TIF-UNIMI-2022-2, Nikhef-2022-003",
    doi = "10.1140/epjc/s10052-022-10878-w",
    journal = "Eur. Phys. J. C",
    volume = "82",
    number = "10",
    pages = "976",
    year = "2022"
}

@article{Candido:2023ujx:positivity-of-pdf,
    author = "Candido, Alessandro and Forte, Stefano and Giani, Tommaso and Hekhorn, Felix",
    title = "{On the positivity of $\overline{\textrm{MS}}$ parton distributions}",
    eprint = "2308.00025",
    archivePrefix = "arXiv",
    primaryClass = "hep-ph",
    reportNumber = "TIF-UNIMI-2023-21",
    doi = "10.1140/epjc/s10052-024-12681-1",
    journal = "Eur. Phys. J. C",
    volume = "84",
    number = "3",
    pages = "335",
    year = "2024"
}

@article{Lappi:2023lmi:phys-pdf-basis,
    author = {Lappi, Tuomas and M{\"a}ntysaari, Heikki and Paukkunen, Hannu and Tevio, Mirja},
    title = "{Evolution of structure functions in momentum space}",
    eprint = "2304.06998",
    archivePrefix = "arXiv",
    primaryClass = "hep-ph",
    doi = "10.1140/epjc/s10052-023-12365-2",
    journal = "Eur. Phys. J. C",
    volume = "84",
    number = "1",
    pages = "84",
    year = "2024",
    note = "[Erratum: Eur.Phys.J.C 85, 88 (2025)]"
}

@article{Lappi:2024dvv:phys-pdf-basis,
    author = {Lappi, Tuomas and M{\"a}ntysaari, Heikki and Paukkunen, Hannu and Tevio, Mirja},
    title = "{Next-to-leading order evolution of structure functions without PDFs}",
    eprint = "2412.09589",
    archivePrefix = "arXiv",
    primaryClass = "hep-ph",
    doi = "10.1140/epjc/s10052-025-14134-9",
    journal = "Eur. Phys. J. C",
    volume = "85",
    number = "4",
    pages = "429",
    year = "2025"
}

@article{Buckley:2014ana:LHAPDF,
    author = {Buckley, Andy and Ferrando, James and Lloyd, Stephen and Nordstr{\"o}m, Karl and Page, Ben and R{\"u}fenacht, Martin and Sch{\"o}nherr, Marek and Watt, Graeme},
    title = "{LHAPDF6: parton density access in the LHC precision era}",
    eprint = "1412.7420",
    archivePrefix = "arXiv",
    primaryClass = "hep-ph",
    reportNumber = "GLAS-PPE-2014-05, MCNET-14-29, IPPP-14-111, DCPT-14-222",
    doi = "10.1140/epjc/s10052-015-3318-8",
    journal = "Eur. Phys. J. C",
    volume = "75",
    pages = "132",
    year = "2015"
}

@article{Rosenbluth:1950yq,
    author = "Rosenbluth, M. N.",
    title = "{High Energy Elastic Scattering of Electrons on Protons}",
    doi = "10.1103/PhysRev.79.615",
    journal = "Phys. Rev.",
    volume = "79",
    pages = "615--619",
    year = "1950"
}

@article{Ernst:1960zza,
    author = "Ernst, F. J. and Sachs, R. G. and Wali, K. C.",
    title = "{Electromagnetic form factors of the nucleon}",
    doi = "10.1103/PhysRev.119.1105",
    journal = "Phys. Rev.",
    volume = "119",
    pages = "1105--1114",
    year = "1960"
}

@article{Sachs:1962zzc,
    author = "Sachs, R. G.",
    title = "{High-Energy Behavior of Nucleon Electromagnetic Form Factors}",
    doi = "10.1103/PhysRev.126.2256",
    journal = "Phys. Rev.",
    volume = "126",
    pages = "2256--2260",
    year = "1962"
}

@article{Sterman:1995:handbook-qcd,
  title = {Handbook of perturbative QCD},
  author = {Sterman, George and Smith, John and Collins, John C. and Whitmore, James and Brock, Raymond and Huston, Joey and Pumplin, Jon and Tung, Wu-Ki and Weerts, Hendrik and Yuan, Chien-Peng and Kuhlmann, Stephen and Mishra, Sanjib and Morf\'{\i}n, Jorge G. and Olness, Fredrick and Owens, Joseph and Qiu, Jianwei and Soper, Davison E.},
  journal = {Rev. Mod. Phys.},
  volume = {67},
  issue = {1},
  pages = {157--248},
  numpages = {0},
  year = {1995},
  month = {Jan},
  publisher = {American Physical Society},
  doi = {10.1103/RevModPhys.67.157},
  url = {https://link.aps.org/doi/10.1103/RevModPhys.67.157}
}

@article{Gao:2017yyd:global-analysis-modern-rev,
    author = "Gao, Jun and Harland-Lang, Lucian and Rojo, Juan",
    title = "{The Structure of the Proton in the LHC Precision Era}",
    eprint = "1709.04922",
    archivePrefix = "arXiv",
    primaryClass = "hep-ph",
    doi = "10.1016/j.physrep.2018.03.002",
    journal = "Phys. Rept.",
    volume = "742",
    pages = "1--121",
    year = "2018"
}

@article{Ethier:2020way:global-analysis-modern-rev,
    author = "Ethier, Jacob J. and Nocera, Emanuele R.",
    title = "{Parton Distributions in Nucleons and Nuclei}",
    eprint = "2001.07722",
    archivePrefix = "arXiv",
    primaryClass = "hep-ph",
    reportNumber = "Nikhef/2020-003",
    doi = "10.1146/annurev-nucl-011720-042725",
    journal = "Ann. Rev. Nucl. Part. Sci.",
    volume = "70",
    pages = "43--76",
    year = "2020"
}

@article{Amoroso:2022eow:global-analysis-modern-rev,
    author = "Amoroso, S. and others",
    title = "{Snowmass 2021 Whitepaper: Proton Structure at the Precision Frontier}",
    eprint = "2203.13923",
    archivePrefix = "arXiv",
    primaryClass = "hep-ph",
    reportNumber = "Edinburgh 2022/08, FERMILAB-PUB-22-222-QIS-SCD-T, MPP-2022-32,
  SLAC-PUB-17652, SMU-HEP-22-02, TIF-UNIMI-2022-6",
    doi = "10.5506/APhysPolB.53.12-A1",
    journal = "Acta Phys. Polon. B",
    volume = "53",
    number = "12",
    pages = "12-A1",
    year = "2022"
}

@article{Huston:2023ofk:global-analysis-modern-rev,
    author = "Huston, Joey and Rabbertz, Klaus and Zanderighi, Giulia",
    title = "{Quantum Chromodynamics}",
    eprint = "2312.14015",
    archivePrefix = "arXiv",
    primaryClass = "hep-ph",
    month = "12",
    year = "2023"
}

@article{PDF4LHCWorkingGroup:2022cjn:pdf-comparison,
    author = "Ball, Richard D. and others",
    collaboration = "PDF4LHC Working Group",
    title = "{The PDF4LHC21 combination of global PDF fits for the LHC Run III}",
    eprint = "2203.05506",
    archivePrefix = "arXiv",
    primaryClass = "hep-ph",
    reportNumber = "Edinburgh 2021/31, FERMILAB-PUB-22-121-QIS-SCD-T, MSUHEP-22-010, SMU-HEP-22-01, Nikhef 2021-033",
    doi = "10.1088/1361-6471/ac7216",
    journal = "J. Phys. G",
    volume = "49",
    number = "8",
    pages = "080501",
    year = "2022"
}

@article{ParticleDataGroup:2024cfk,
    author = "Navas, S. and others",
    collaboration = "Particle Data Group",
    title = "{Review of particle physics}",
    doi = "10.1103/PhysRevD.110.030001",
    journal = "Phys. Rev. D",
    volume = "110",
    number = "3",
    pages = "030001",
    year = "2024"
}

@article{Gluck:1980cp,
    author = "Gluck, M. and Hoffmann, E. and Reya, E.",
    title = "{Scaling Violations and the Gluon Distribution of the Nucleon}",
    reportNumber = "DO-TH 80/13",
    doi = "10.1007/BF01547675",
    journal = "Z. Phys. C",
    volume = "13",
    pages = "119",
    year = "1982"
}

@article{Duke:1983gd,
    author = "Duke, D. W. and Owens, J. F.",
    title = "{Q**2 Dependent Parametrizations of Parton Distribution Functions}",
    reportNumber = "FSU-HEP-831115",
    doi = "10.1103/PhysRevD.30.49",
    journal = "Phys. Rev. D",
    volume = "30",
    pages = "49--54",
    year = "1984"
}

@article{Eichten:1984eu,
    author = "Eichten, E. and Hinchliffe, I. and Lane, Kenneth D. and Quigg, C.",
    title = "{Super Collider Physics}",
    reportNumber = "FERMILAB-PUB-84-017-T, LBL-16875, DOE-ER-01545-345",
    doi = "10.1103/RevModPhys.56.579",
    journal = "Rev. Mod. Phys.",
    volume = "56",
    pages = "579--707",
    year = "1984",
    note = "[Addendum: Rev.Mod.Phys. 58, 1065--1073 (1986)]"
}

@article{Ball:2016spl,
    author = "Ball, Richard D. and Nocera, Emanuele R. and Rojo, Juan",
    title = "{The asymptotic behaviour of parton distributions at small and large $x$}",
    eprint = "1604.00024",
    archivePrefix = "arXiv",
    primaryClass = "hep-ph",
    doi = "10.1140/epjc/s10052-016-4240-4",
    journal = "Eur. Phys. J. C",
    volume = "76",
    number = "7",
    pages = "383",
    year = "2016"
}

@article{Carrazza:2021yrg:pdf-parametrization,
    author = "Carrazza, Stefano and Cruz-Martinez, Juan M. and Stegeman, Roy",
    title = "{A data-based parametrization of parton distribution functions}",
    eprint = "2111.02954",
    archivePrefix = "arXiv",
    primaryClass = "hep-ph",
    reportNumber = "TIF-UNIMI-2021-18",
    doi = "10.1140/epjc/s10052-022-10136-z",
    journal = "Eur. Phys. J. C",
    volume = "82",
    number = "2",
    pages = "163",
    year = "2022"
}

@article{Regge:1959mz,
    author = "Regge, T.",
    title = "{Introduction to complex orbital momenta}",
    doi = "10.1007/BF02728177",
    journal = "Nuovo Cim.",
    volume = "14",
    pages = "951",
    year = "1959"
}

@book{Collins:1977jy:intro-to-regge,
    author = "Collins, P. D. B.",
    title = "{An Introduction to Regge Theory and High Energy Physics}",
    doi = "10.1017/9781009403269",
    isbn = "978-1-009-40326-9, 978-1-009-40329-0, 978-1-009-40328-3, 978-0-521-11035-8",
    publisher = "Cambridge University Press",
    year = "1977"
}

@article{Brodsky:1973kr,
    author = "Brodsky, Stanley J. and Farrar, Glennys R.",
    title = "{Scaling Laws at Large Transverse Momentum}",
    reportNumber = "SLAC-PUB-1290",
    doi = "10.1103/PhysRevLett.31.1153",
    journal = "Phys. Rev. Lett.",
    volume = "31",
    pages = "1153--1156",
    year = "1973"
}

@article{Hou:2019efy,
    author = "Hou, Tie-Jiun and others",
    title = "{New CTEQ global analysis of quantum chromodynamics with high-precision data from the LHC}",
    eprint = "1912.10053",
    archivePrefix = "arXiv",
    primaryClass = "hep-ph",
    reportNumber = "MSUHEP-19-025, PITT-PACC-1911, SMU-HEP-19-03",
    doi = "10.1103/PhysRevD.103.014013",
    journal = "Phys. Rev. D",
    volume = "103",
    number = "1",
    pages = "014013",
    year = "2021"
}

@article{NNPDF:2021njg,
    author = "Ball, Richard D. and others",
    collaboration = "NNPDF",
    title = "{The path to proton structure at 1{\%} accuracy}",
    eprint = "2109.02653",
    archivePrefix = "arXiv",
    primaryClass = "hep-ph",
    reportNumber = "Edinburgh 2021/12, Nikhef-2021-013, TIF-UNIMI-2021-11",
    doi = "10.1140/epjc/s10052-022-10328-7",
    journal = "Eur. Phys. J. C",
    volume = "82",
    number = "5",
    pages = "428",
    year = "2022"
}

@article{Bailey:2020ooq-MSHT,
    author = "Bailey, S. and Cridge, T. and Harland-Lang, L. A. and Martin, A. D. and Thorne, R. S.",
    title = "{Parton distributions from LHC, HERA, Tevatron and fixed target data: MSHT20 PDFs}",
    eprint = "2012.04684",
    archivePrefix = "arXiv",
    primaryClass = "hep-ph",
    reportNumber = "IPPP/20/58",
    doi = "10.1140/epjc/s10052-021-09057-0",
    journal = "Eur. Phys. J. C",
    volume = "81",
    number = "4",
    pages = "341",
    year = "2021"
}

@article{Harland-Lang:2014zoa,
    author = "Harland-Lang, L. A. and Martin, A. D. and Motylinski, P. and Thorne, R. S.",
    title = "{Parton distributions in the LHC era: MMHT 2014 PDFs}",
    eprint = "1412.3989",
    archivePrefix = "arXiv",
    primaryClass = "hep-ph",
    reportNumber = "LCTS-2014-47, IPPP-14-97, DCPT-14-194",
    doi = "10.1140/epjc/s10052-015-3397-6",
    journal = "Eur. Phys. J. C",
    volume = "75",
    number = "5",
    pages = "204",
    year = "2015"
}

@article{Martin:2009iq,
    author = "Martin, A. D. and Stirling, W. J. and Thorne, R. S. and Watt, G.",
    title = "{Parton distributions for the LHC}",
    eprint = "0901.0002",
    archivePrefix = "arXiv",
    primaryClass = "hep-ph",
    reportNumber = "IPPP-08-95, DCPT-08-190, CAVENDISH-HEP-08-16",
    doi = "10.1140/epjc/s10052-009-1072-5",
    journal = "Eur. Phys. J. C",
    volume = "63",
    pages = "189--285",
    year = "2009"
}

@article{EuropeanMuon:1983wih,
    author = "Aubert, J. J. and others",
    collaboration = "European Muon",
    title = "{The ratio of the nucleon structure functions $F2_n$ for iron and deuterium}",
    reportNumber = "CERN-EP/83-14",
    doi = "10.1016/0370-2693(83)90437-9",
    journal = "Phys. Lett. B",
    volume = "123",
    pages = "275--278",
    year = "1983"
}

@article{Eskola:2016oht,
    author = "Eskola, Kari J. and Paakkinen, Petja and Paukkunen, Hannu and Salgado, Carlos A.",
    title = "{EPPS16: Nuclear parton distributions with LHC data}",
    eprint = "1612.05741",
    archivePrefix = "arXiv",
    primaryClass = "hep-ph",
    doi = "10.1140/epjc/s10052-017-4725-9",
    journal = "Eur. Phys. J. C",
    volume = "77",
    number = "3",
    pages = "163",
    year = "2017"
}

@article{Eskola:2021nhw,
    author = "Eskola, Kari J. and Paakkinen, Petja and Paukkunen, Hannu and Salgado, Carlos A.",
    title = "{EPPS21: a global QCD analysis of nuclear PDFs}",
    eprint = "2112.12462",
    archivePrefix = "arXiv",
    primaryClass = "hep-ph",
    doi = "10.1140/epjc/s10052-022-10359-0",
    journal = "Eur. Phys. J. C",
    volume = "82",
    number = "5",
    pages = "413",
    year = "2022"
}

@article{Paakkinen:2025pcq,
    author = "Paakkinen, Petja",
    title = "{Hard probes and nuclear PDFs}",
    eprint = "2504.21610",
    archivePrefix = "arXiv",
    primaryClass = "hep-ph",
    doi = "10.1051/epjconf/202533901011",
    journal = "EPJ Web Conf.",
    volume = "339",
    pages = "01011",
    year = "2025"
}

@article{Accardi:2012qut:EIC,
    author = "Accardi, A. and others",
    editor = "Deshpande, A. and Meziani, Z. E. and Qiu, J. W.",
    title = "{Electron Ion Collider: The Next QCD Frontier}: {Understanding the glue that binds us all}",
    eprint = "1212.1701",
    archivePrefix = "arXiv",
    primaryClass = "nucl-ex",
    reportNumber = "BNL-98815-2012-JA, JLAB-PHY-12-1652",
    doi = "10.1140/epja/i2016-16268-9",
    journal = "Eur. Phys. J. A",
    volume = "52",
    number = "9",
    pages = "268",
    year = "2016"
}

@article{AbdulKhalek:2021gbh:EIC,
    author = "Abdul Khalek, R. and others",
    title = "{Science Requirements and Detector Concepts for the Electron-Ion Collider}: {EIC Yellow Report}",
    eprint = "2103.05419",
    archivePrefix = "arXiv",
    primaryClass = "physics.ins-det",
    reportNumber = "BNL-220990-2021-FORE, JLAB-PHY-21-3198, LA-UR-21-20953",
    doi = "10.1016/j.nuclphysa.2022.122447",
    journal = "Nucl. Phys. A",
    volume = "1026",
    pages = "122447",
    year = "2022"
}

@article{Anderle:2021wcy:EicC,
    author = "Anderle, Daniele P. and others",
    title = "{Electron-ion collider in China}",
    eprint = "2102.09222",
    archivePrefix = "arXiv",
    primaryClass = "nucl-ex",
    reportNumber = "Frontiers of Physics, Volume 16 Issue (6):64701, 2021",
    doi = "10.1007/s11467-021-1062-0",
    journal = "Front. Phys. (Beijing)",
    volume = "16",
    number = "6",
    pages = "64701",
    year = "2021"
}

@article{LHeCStudyGroup:2012zhm,
    author = "Abelleira Fernandez, J. L. and others",
    collaboration = "LHeC Study Group",
    title = "{A Large Hadron Electron Collider at CERN: Report on the Physics and Design Concepts for Machine and Detector}",
    eprint = "1206.2913",
    archivePrefix = "arXiv",
    primaryClass = "physics.acc-ph",
    reportNumber = "SLAC-R-999, CERN-OPEN-2012-015, LHEC-NOTE-2012-001-GEN",
    doi = "10.1088/0954-3899/39/7/075001",
    journal = "J. Phys. G",
    volume = "39",
    pages = "075001",
    year = "2012"
}

@article{LHeC:2020van,
    author = "Agostini, P. and others",
    collaboration = "LHeC, FCC-he Study Group",
    title = "{The Large Hadron{\textendash}Electron Collider at the HL-LHC}",
    eprint = "2007.14491",
    archivePrefix = "arXiv",
    primaryClass = "hep-ex",
    reportNumber = "CERN-ACC-Note-2020-0002, JLAB-ACP-20-3180",
    doi = "10.1088/1361-6471/abf3ba",
    journal = "J. Phys. G",
    volume = "48",
    number = "11",
    pages = "110501",
    year = "2021"
}

@article{H1:2008rkk:fl,
    author = "Aaron, F. D. and others",
    collaboration = "H1",
    title = "{Measurement of the Proton Structure Function F(L)(x, Q**2) at Low x}",
    eprint = "0805.2809",
    archivePrefix = "arXiv",
    primaryClass = "hep-ex",
    reportNumber = "DESY-08-053",
    doi = "10.1016/j.physletb.2008.05.070",
    journal = "Phys. Lett. B",
    volume = "665",
    pages = "139--146",
    year = "2008"
}

@article{ZEUS:2009nwk:fl,
    author = "Chekanov, S. and others",
    collaboration = "ZEUS",
    title = "{Measurement of the Longitudinal Proton Structure Function at HERA}",
    eprint = "0904.1092",
    archivePrefix = "arXiv",
    primaryClass = "hep-ex",
    reportNumber = "DESY-09-046",
    doi = "10.1016/j.physletb.2009.10.050",
    journal = "Phys. Lett. B",
    volume = "682",
    pages = "8--22",
    year = "2009"
}

@article{H1:2010fzx:fl,
    author = "Aaron, F. D. and others",
    collaboration = "H1",
    title = "{Measurement of the Inclusive e{{\textbackslash}pm}p Scattering Cross Section at High Inelasticity y and of the Structure Function $F_L$}",
    eprint = "1012.4355",
    archivePrefix = "arXiv",
    primaryClass = "hep-ex",
    reportNumber = "DESY-10-228",
    doi = "10.1140/epjc/s10052-011-1579-4",
    journal = "Eur. Phys. J. C",
    volume = "71",
    pages = "1579",
    year = "2011"
}

@article{NewMuon:1996fwh,
    author = "Arneodo, M. and others",
    collaboration = "New Muon",
    title = "{Measurement of the proton and deuteron structure functions, F2(p) and F2(d), and of the ratio sigma-L / sigma-T}",
    eprint = "hep-ph/9610231",
    archivePrefix = "arXiv",
    doi = "10.1016/S0550-3213(96)00538-X",
    journal = "Nucl. Phys. B",
    volume = "483",
    pages = "3--43",
    year = "1997"
}

@article{Tvaskis:2016uxm:jlabfl,
    author = "Tvaskis, V. and others",
    title = "{Measurements of the separated longitudinal structure function $F_L$ from hydrogen and deuterium targets at low $Q^2$}",
    eprint = "1606.02614",
    archivePrefix = "arXiv",
    primaryClass = "nucl-ex",
    reportNumber = "JLAB-PHY-16-2295",
    doi = "10.1103/PhysRevC.97.045204",
    journal = "Phys. Rev. C",
    volume = "97",
    number = "4",
    pages = "045204",
    year = "2018"
}

@article{JeffersonLabHallCE94-110:2004nsn;jlabfl,
    author = "Liang, Y. and others",
    collaboration = "Jefferson Lab Hall C E94-110",
    title = "{Measurement of R={\ensuremath{\sigma}}L/{\ensuremath{\sigma}}T and the separated longitudinal and transverse structure functions in the nucleon-resonance region}",
    eprint = "nucl-ex/0410027",
    archivePrefix = "arXiv",
    reportNumber = "JLAB-PHY-04-45",
    doi = "10.1103/PhysRevC.105.065205",
    journal = "Phys. Rev. C",
    volume = "105",
    number = "6",
    pages = "065205",
    year = "2022"
}

@article{ZEUS:2002pdo:F3,
    author = "Chekanov, S. and others",
    collaboration = "ZEUS",
    title = "{Measurement of high Q**2 e- p neutral current cross-sections at HERA and the extraction of xF(3)}",
    eprint = "hep-ex/0208040",
    archivePrefix = "arXiv",
    reportNumber = "DESY-02-113",
    doi = "10.1140/epjc/s2003-01163-y",
    journal = "Eur. Phys. J. C",
    volume = "28",
    pages = "175--201",
    year = "2003"
}

@article{NuTeV:2005wsg,
    author = "Tzanov, M. and others",
    collaboration = "NuTeV",
    title = "{Precise measurement of neutrino and anti-neutrino differential cross sections}",
    eprint = "hep-ex/0509010",
    archivePrefix = "arXiv",
    reportNumber = "FERMILAB-PUB-05-453-E",
    doi = "10.1103/PhysRevD.74.012008",
    journal = "Phys. Rev. D",
    volume = "74",
    pages = "012008",
    year = "2006"
}

@misc{NuTeV:2005:hepdata.11120,
    author = "{NuTeV Collaboration}",
    title = "{Precise measurement of neutrino and anti-neutrino differential cross sections.}",
    howpublished = "{HEPData (collection)}",
    year = 2005,
    note = "\url{https://doi.org/10.17182/hepdata.11120}"
}

@article{CHORUS:2005cpn,
    author = "Onengut, G. and others",
    collaboration = "CHORUS",
    title = "{Measurement of nucleon structure functions in neutrino scattering}",
    reportNumber = "CERN-PH-EP-2005-048",
    doi = "10.1016/j.physletb.2005.10.062",
    journal = "Phys. Lett. B",
    volume = "632",
    pages = "65--75",
    year = "2006"
}

@misc{CHORUS:2005:hepdata.6187,
    author = "{CHORUS Collaboration}",
    title = "{Measurement of nucleon structure functions in neutrino scattering.}",
    howpublished = "{HEPData (collection)}",
    year = 2006,
    note = "\url{https://doi.org/10.17182/hepdata.6187}"
}

@article{ArgoNeuT:2011bms,
    author = "Anderson, C. and others",
    collaboration = "ArgoNeuT",
    title = "{First Measurements of Inclusive Muon Neutrino Charged Current Differential Cross Sections on Argon}",
    eprint = "1111.0103",
    archivePrefix = "arXiv",
    primaryClass = "hep-ex",
    reportNumber = "FERMILAB-PUB-11-591-PPD",
    doi = "10.1103/PhysRevLett.108.161802",
    journal = "Phys. Rev. Lett.",
    volume = "108",
    pages = "161802",
    year = "2012"
}

@article{ArgoNeuT:2014rlj,
    author = "Acciarri, R. and others",
    collaboration = "ArgoNeuT",
    title = "{Measurements of Inclusive Muon Neutrino and Antineutrino Charged Current Differential Cross Sections on Argon in the NuMI Antineutrino Beam}",
    eprint = "1404.4809",
    archivePrefix = "arXiv",
    primaryClass = "hep-ex",
    reportNumber = "FERMILAB-PUB-14-104-E",
    doi = "10.1103/PhysRevD.89.112003",
    journal = "Phys. Rev. D",
    volume = "89",
    number = "11",
    pages = "112003",
    year = "2014"
}

@article{MicroBooNE:2019nio,
    author = "Abratenko, P. and others",
    collaboration = "MicroBooNE",
    title = "{First Measurement of Inclusive Muon Neutrino Charged Current Differential Cross Sections on Argon at $E_\nu\sim$0.8 GeV with the MicroBooNE Detector}",
    eprint = "1905.09694",
    archivePrefix = "arXiv",
    primaryClass = "hep-ex",
    reportNumber = "FERMILAB-PUB-19-235-ND",
    doi = "10.1103/PhysRevLett.123.131801",
    journal = "Phys. Rev. Lett.",
    volume = "123",
    number = "13",
    pages = "131801",
    year = "2019"
}

@article{MicroBooNE:2023foc,
    author = "Abratenko, P. and others",
    collaboration = "MicroBooNE",
    title = "{Measurement of three-dimensional inclusive muon-neutrino charged-current cross sections on argon with the MicroBooNE detector}",
    eprint = "2307.06413",
    archivePrefix = "arXiv",
    primaryClass = "hep-ex",
    reportNumber = "FERMILAB-PUB-23-368-ND",
    doi = "10.1016/j.physletb.2025.139939",
    journal = "Phys. Lett. B",
    volume = "870",
    pages = "139939",
    year = "2025"
}

@article{MicroBooNE:2018xad,
    author = "Adams, C. and others",
    collaboration = "MicroBooNE",
    title = "{Comparison of $\nu_\mu$-Ar multiplicity distributions observed by MicroBooNE to GENIE model predictions}",
    eprint = "1805.06887",
    archivePrefix = "arXiv",
    primaryClass = "hep-ex",
    reportNumber = "FERMILAB-PUB-18-077-ND",
    doi = "10.1140/epjc/s10052-019-6742-3",
    journal = "Eur. Phys. J. C",
    volume = "79",
    number = "3",
    pages = "248",
    year = "2019"
}

@article{NINJA:2020gbg:water,
    author = "Hiramoto, A. and others",
    collaboration = "NINJA",
    title = "{First measurement of $\overline{\nu}_{\mu}$ and $\nu_{\mu}$ charged-current inclusive interactions on water using a nuclear emulsion detector}",
    eprint = "2008.03895",
    archivePrefix = "arXiv",
    primaryClass = "hep-ex",
    doi = "10.1103/PhysRevD.102.072006",
    journal = "Phys. Rev. D",
    volume = "102",
    number = "7",
    pages = "072006",
    year = "2020"
}

@article{T2K:2019dgm:water,
    author = "Abe, K. and others",
    collaboration = "T2K",
    title = "{Measurement of the $\nu_{\mu}$ charged-current cross sections on water, hydrocarbon, iron, and their ratios with the T2K on-axis detectors}",
    eprint = "1904.09611",
    archivePrefix = "arXiv",
    primaryClass = "hep-ex",
    doi = "10.1093/ptep/ptz070",
    journal = "PTEP",
    volume = "2019",
    number = "9",
    pages = "093C02",
    year = "2019"
}

@article{T2K:2013nor:carbon,
    author = "Abe, K. and others",
    collaboration = "T2K",
    title = "{Measurement of the inclusive $\nu_\mu$ charged current cross section on carbon in the near detector of the T2K experiment}",
    eprint = "1302.4908",
    archivePrefix = "arXiv",
    primaryClass = "hep-ex",
    doi = "10.1103/PhysRevD.87.092003",
    journal = "Phys. Rev. D",
    volume = "87",
    number = "9",
    pages = "092003",
    year = "2013"
}

@article{T2K:2018lnf:carbon,
    author = "Abe, K. and others",
    collaboration = "T2K",
    title = "{Measurement of inclusive double-differential $\nu_\mu$ charged-current cross section with improved acceptance in the T2K off-axis near detector}",
    eprint = "1801.05148",
    archivePrefix = "arXiv",
    primaryClass = "hep-ex",
    doi = "10.1103/PhysRevD.98.012004",
    journal = "Phys. Rev. D",
    volume = "98",
    pages = "012004",
    year = "2018"
}

@article{WA25:1983qub:Adler-exp,
    author = "Allasia, D. and others",
    collaboration = "WA25",
    title = "{Measurement of the Neutron and Proton Structure Functions From Neutrino and Anti-neutrinos Scattering in Deuterium}",
    reportNumber = "PRINT-83-0840",
    doi = "10.1016/0370-2693(84)90488-X",
    journal = "Phys. Lett. B",
    volume = "135",
    pages = "231",
    year = "1984"
}

@article{NewMuon:1991hlj,
    author = "Amaudruz, P. and others",
    collaboration = "New Muon",
    title = "{The Gottfried sum from the ratio F2(n) / F2(p)}",
    reportNumber = "CERN-PPE-91-05",
    doi = "10.1103/PhysRevLett.66.2712",
    journal = "Phys. Rev. Lett.",
    volume = "66",
    pages = "2712--2715",
    year = "1991"
}

@article{NewMuon:1996uwk,
    author = "Arneodo, M. and others",
    collaboration = "New Muon",
    title = "{Accurate measurement of F2(d) / F2(p) and R**d - R**p}",
    eprint = "hep-ex/9611022",
    archivePrefix = "arXiv",
    doi = "10.1016/S0550-3213(96)00673-6",
    journal = "Nucl. Phys. B",
    volume = "487",
    pages = "3--26",
    year = "1997"
}

@article{Leung:1992yx:GLS,
    author = "Leung, W. C. and others",
    title = "{A Measurement of the Gross-Llewellyn-Smith Sum Rule from the CCFR $xF_3$ Structure Function}",
    reportNumber = "NEVIS-1460, FERMILAB-PUB-92-205",
    doi = "10.1016/0370-2693(93)91386-2",
    journal = "Phys. Lett. B",
    volume = "317",
    pages = "655--659",
    year = "1993"
}

@inproceedings{NuTeVCCFR:1995wdi:GLS,
    author = "Harris, Deborah A. and others",
    collaboration = "NuTeV CCFR",
    title = "{A Measurement of $\alpha^- s (Q^{2)}$ from the Gross-Llewellyn-Smith sum rule}",
    booktitle = "{30th Rencontres de Moriond: QCD and High-energy Hadronic Interactions}",
    eprint = "hep-ex/9506010",
    archivePrefix = "arXiv",
    reportNumber = "FERMILAB-CONF-95-144",
    pages = "247--250",
    year = "1995"
}

@article{Kim:1998kia:GLS,
    author = "Kim, J. H. and others",
    title = "{A Measurement of $\alpha_s (Q^2$) from the Gross-Llewellyn Smith Sum Rule}",
    eprint = "hep-ex/9808015",
    archivePrefix = "arXiv",
    reportNumber = "FERMILAB-PUB-98-0474-PPD",
    doi = "10.1103/PhysRevLett.81.3595",
    journal = "Phys. Rev. Lett.",
    volume = "81",
    pages = "3595--3598",
    year = "1998"
}

@article{Adler:1965ty,
    author = "Adler, Stephen L.",
    title = "{Sum rules giving tests of local current commutation relations in high-energy neutrino reactions}",
    doi = "10.1103/PhysRev.143.1144",
    journal = "Phys. Rev.",
    volume = "143",
    pages = "1144--1155",
    year = "1966"
}

@article{Gottfried:1967kk,
    author = "Gottfried, Kurt",
    title = "{Sum rule for high-energy electron - proton scattering}",
    doi = "10.1103/PhysRevLett.18.1174",
    journal = "Phys. Rev. Lett.",
    volume = "18",
    pages = "1174",
    year = "1967"
}

@article{Gross:1969jf,
    author = "Gross, David J. and Llewellyn Smith, Chris H.",
    title = "{High-energy neutrino - nucleon scattering, current algebra and partons}",
    doi = "10.1016/0550-3213(69)90213-2",
    journal = "Nucl. Phys. B",
    volume = "14",
    pages = "337--347",
    year = "1969"
}

@book{Ellis:1996mzs:qcd-collider-phys,
    author = "Ellis, R. Keith and Stirling, W. James and Webber, B. R.",
    title = "{QCD and collider physics}",
    doi = "10.1017/CBO9780511628788",
    isbn = "978-0-511-82328-2, 978-0-521-54589-1",
    publisher = "Cambridge University Press",
    volume = "8",
    month = "2",
    year = "2011"
}

@article{Collins:1987pm:collinear-factorization,
    author = "Collins, John C. and Soper, Davison E.",
    title = "{The Theorems of Perturbative QCD}",
    reportNumber = "OITS-350",
    doi = "10.1146/annurev.ns.37.120187.002123",
    journal = "Ann. Rev. Nucl. Part. Sci.",
    volume = "37",
    pages = "383--409",
    year = "1987"
}

@article{Collins:1989gx:collinear-factorization-long,
    author = "Collins, John C. and Soper, Davison E. and Sterman, George F.",
    title = "{Factorization of Hard Processes in QCD}",
    eprint = "hep-ph/0409313",
    archivePrefix = "arXiv",
    reportNumber = "ITP-SB-89-31",
    doi = "10.1142/9789814503266_0001",
    journal = "Adv. Ser. Direct. High Energy Phys.",
    volume = "5",
    pages = "1--91",
    year = "1989"
}

@article{Braun:2022gzl:rev-twist,
    author = "Braun, Vladimir M.",
    title = "{Higher Twists}",
    eprint = "2212.02887",
    archivePrefix = "arXiv",
    primaryClass = "hep-ph",
    doi = "10.1051/epjconf/202227401012",
    journal = "EPJ Web Conf.",
    volume = "274",
    pages = "01012",
    year = "2022"
}

@article{Schienbein:2007gr:target-mass-corrections,
    author = "Schienbein, Ingo and others",
    title = "{A Review of Target Mass Corrections}",
    eprint = "0709.1775",
    archivePrefix = "arXiv",
    primaryClass = "hep-ph",
    reportNumber = "JLAB-THY-07-718",
    doi = "10.1088/0954-3899/35/5/053101",
    journal = "J. Phys. G",
    volume = "35",
    pages = "053101",
    year = "2008"
}

@article{Blumlein:2012bf:dis-review,
    author = "Blumlein, Johannes",
    title = "{The Theory of Deeply Inelastic Scattering}",
    eprint = "1208.6087",
    archivePrefix = "arXiv",
    primaryClass = "hep-ph",
    reportNumber = "DESY-12-096, DO-TH-12-19, SFB-CPP-12-38, LPN-12-056",
    doi = "10.1016/j.ppnp.2012.09.006",
    journal = "Prog. Part. Nucl. Phys.",
    volume = "69",
    pages = "28--84",
    year = "2013"
}

@article{Ma:1974cj:dis-scalar-higgs,
    author = "Ma, Ernest",
    title = "{Deep-Inelastic Lepton-Hadron Scattering: The Higgs Scalar Contribution}",
    reportNumber = "Print-74-1022 (UC,SANTA BARBARA)",
    doi = "10.1103/PhysRevD.10.2298",
    journal = "Phys. Rev. D",
    volume = "10",
    pages = "2298",
    year = "1974"
}

@article{Soar:2009yh:dis-scalar-higgs,
    author = "Soar, G. and Moch, S. and Vermaseren, J. A. M. and Vogt, A.",
    title = "{On Higgs-exchange DIS, physical evolution kernels and fourth-order splitting functions at large x}",
    eprint = "0912.0369",
    archivePrefix = "arXiv",
    primaryClass = "hep-ph",
    reportNumber = "LTH-857, DESY-09-211, SFB-CPP-09-119, NIKHEF-09-031",
    doi = "10.1016/j.nuclphysb.2010.02.003",
    journal = "Nucl. Phys. B",
    volume = "832",
    pages = "152--227",
    year = "2010"
}

@article{Ellis:1975ap,
    author = "Ellis, John R. and Gaillard, Mary K. and Nanopoulos, Dimitri V.",
    title = "{A Phenomenological Profile of the Higgs Boson}",
    reportNumber = "CERN-TH-2093",
    doi = "10.1016/0550-3213(76)90382-5",
    journal = "Nucl. Phys. B",
    volume = "106",
    pages = "292",
    year = "1976"
}

@article{ATLAS:2012yve:Higgs,
    author = "Aad, Georges and others",
    collaboration = "ATLAS",
    title = "{Observation of a new particle in the search for the Standard Model Higgs boson with the ATLAS detector at the LHC}",
    eprint = "1207.7214",
    archivePrefix = "arXiv",
    primaryClass = "hep-ex",
    reportNumber = "CERN-PH-EP-2012-218",
    doi = "10.1016/j.physletb.2012.08.020",
    journal = "Phys. Lett. B",
    volume = "716",
    pages = "1--29",
    year = "2012"
}

@article{Bjorken:1969ja,
    author = "Bjorken, J. D. and Paschos, Emmanuel A.",
    title = "{Inelastic Electron Proton and gamma Proton Scattering, and the Structure of the Nucleon}",
    reportNumber = "SLAC-PUB-0572",
    doi = "10.1103/PhysRev.185.1975",
    journal = "Phys. Rev.",
    volume = "185",
    pages = "1975--1982",
    year = "1969"
}

@article{Altarelli:1978id,
    author = "Altarelli, Guido and Ellis, R. Keith and Martinelli, G.",
    title = "{Leptoproduction and Drell-Yan Processes Beyond the Leading Approximation in Chromodynamics}",
    reportNumber = "MIT-CTP-723",
    doi = "10.1016/0550-3213(78)90067-6",
    journal = "Nucl. Phys. B",
    volume = "143",
    pages = "521",
    year = "1978",
    note = "[Erratum: Nucl.Phys.B 146, 544 (1978)]"
}

@article{Furmanski:1981cw,
    author = "Furmanski, W. and Petronzio, R.",
    title = "{Lepton - Hadron Processes Beyond Leading Order in Quantum Chromodynamics}",
    reportNumber = "CERN-TH-3046",
    doi = "10.1007/BF01578280",
    journal = "Z. Phys. C",
    volume = "11",
    pages = "293",
    year = "1982"
}

@article{Altarelli:1978FL,
title = {Transverse momentum of jets in electroproduction from quantum chromodynamics},
journal = {Physics Letters B},
volume = {76},
number = {1},
pages = {89-94},
year = {1978},
issn = {0370-2693},
doi = {https://doi.org/10.1016/0370-2693(78)90109-0},
author = {G. Altarelli and G. Martinelli},
}

@article{Zijlstra:1992qd,
    author = "Zijlstra, E. B. and van Neerven, W. L.",
    title = "{Order alpha-s**2 QCD corrections to the deep inelastic proton structure functions F2 and F(L)}",
    reportNumber = "INLO-PUB-1-92",
    doi = "10.1016/0550-3213(92)90087-R",
    journal = "Nucl. Phys. B",
    volume = "383",
    pages = "525--574",
    year = "1992"
}

@article{vanNeerven:1991nn,
    author = "van Neerven, W. L. and Zijlstra, E. B.",
    title = "{Order alpha-s**2 contributions to the deep inelastic Wilson coefficient}",
    reportNumber = "INLO-PUB-3-91",
    doi = "10.1016/0370-2693(91)91024-P",
    journal = "Phys. Lett. B",
    volume = "272",
    pages = "127--133",
    year = "1991"
}

@article{Zijlstra:1991qc,
    author = "Zijlstra, E. B. and van Neerven, W. L.",
    title = "{Contribution of the second order gluonic Wilson coefficient to the deep inelastic structure function}",
    reportNumber = "INLO-PUB-6-91",
    doi = "10.1016/0370-2693(91)90301-6",
    journal = "Phys. Lett. B",
    volume = "273",
    pages = "476--482",
    year = "1991"
}

@article{SanchezGuillen:1990iq,
    author = "Sanchez Guillen, J. and Miramontes, J. and Miramontes, M. and Parente, G. and Sampayo, O. A.",
    title = "{Next-to-leading order analysis of the deep inelastic R = sigma-L / sigma-total}",
    reportNumber = "MAD-PH-576, US-FT-10-90",
    doi = "10.1016/0550-3213(91)90340-4",
    journal = "Nucl. Phys. B",
    volume = "353",
    pages = "337--345",
    year = "1991"
}

@article{Moch:1999eb,
    author = "Moch, S. and Vermaseren, J. A. M.",
    title = "{Deep inelastic structure functions at two loops}",
    eprint = "hep-ph/9912355",
    archivePrefix = "arXiv",
    reportNumber = "NIKHEF-99-030",
    doi = "10.1016/S0550-3213(00)00045-6",
    journal = "Nucl. Phys. B",
    volume = "573",
    pages = "853--907",
    year = "2000"
}

@article{Moch:2004xu,
    author = "Moch, S. and Vermaseren, J. A. M. and Vogt, A.",
    title = "{The Longitudinal structure function at the third order}",
    eprint = "hep-ph/0411112",
    archivePrefix = "arXiv",
    reportNumber = "SFB-CPP-04-65, DCPT-04-130, IPPP-04-65, NIKHEF-04-013, DESY-04-210",
    doi = "10.1016/j.physletb.2004.11.063",
    journal = "Phys. Lett. B",
    volume = "606",
    pages = "123--129",
    year = "2005"
}

@article{Vermaseren:2005qc,
    author = "Vermaseren, J. A. M. and Vogt, A. and Moch, S.",
    title = "{The Third-order QCD corrections to deep-inelastic scattering by photon exchange}",
    eprint = "hep-ph/0504242",
    archivePrefix = "arXiv",
    reportNumber = "NIKHEF-05-006, DCPT-05-28, IPPP-05-14, DESY-05-063, SFB-CPP-05-13",
    doi = "10.1016/j.nuclphysb.2005.06.020",
    journal = "Nucl. Phys. B",
    volume = "724",
    pages = "3--182",
    year = "2005"
}

@article{Blumlein:2022gpp:nnnlo-f2-f3,
    author = {Bl{\"u}mlein, J. and Marquard, P. and Schneider, C. and Sch{\"o}nwald, K.},
    title = "{The massless three-loop Wilson coefficients for the deep-inelastic structure functions F$_{2}$, F$_{L}$, xF$_{3}$ and g$_{1}$}",
    eprint = "2208.14325",
    archivePrefix = "arXiv",
    primaryClass = "hep-ph",
    reportNumber = "DESY 22-123, DO-TH 22/20, TTP 22-057, RISC Report Series 22-12, SAGEX-22-30, RISC Report Series 22-12,
  SAGEX-22-30",
    doi = "10.1007/JHEP11(2022)156",
    journal = "JHEP",
    volume = "11",
    pages = "156",
    year = "2022"
}

@article{Ablinger:2026xza:nnnlo-vfns,
    author = {Ablinger, J. and Behring, A. and Bl{\"u}mlein, J. and De Freitas, A. and von Manteuffel, A. and Schneider, C. and Sch{\"o}nwald, K.},
    title = "{The variable flavor number scheme to three-loop order}",
    eprint = "2607.05235",
    archivePrefix = "arXiv",
    primaryClass = "hep-ph",
    reportNumber = "DESY 26--064, RISC Report number 26-06, CERN-TH-2026-113, MPP-2026-89",
    journal = "PoS",
    volume = "LL2026",
    pages = "025",
    year = "2026"
}

@article{Moch:2007rq:cc-nnlo,
    author = "Moch, S. and Rogal, M. and Vogt, A.",
    title = "{Differences between charged-current coefficient functions}",
    eprint = "0708.3731",
    archivePrefix = "arXiv",
    primaryClass = "hep-ph",
    reportNumber = "DESY-07-048, SFT-CPP-07-13, LTH-756",
    doi = "10.1016/j.nuclphysb.2007.09.022",
    journal = "Nucl. Phys. B",
    volume = "790",
    pages = "317--335",
    year = "2008"
}

@article{Moch:2008fj:cc-nnlo,
    author = "Moch, S. and Vermaseren, J. A. M. and Vogt, A.",
    title = "{Third-order QCD corrections to the charged-current structure function F(3)}",
    eprint = "0812.4168",
    archivePrefix = "arXiv",
    primaryClass = "hep-ph",
    reportNumber = "DESY-08-197, SFB-CPP-08-105, NIKHEF-08-032, LTH-815",
    doi = "10.1016/j.nuclphysb.2009.01.001",
    journal = "Nucl. Phys. B",
    volume = "813",
    pages = "220--258",
    year = "2009"
}

@article{Moch:2007gx:cc-nnlo-neutrino,
    author = "Moch, S. and Rogal, M.",
    title = "{Charged current deep-inelastic scattering at three loops}",
    eprint = "0704.1740",
    archivePrefix = "arXiv",
    primaryClass = "hep-ph",
    reportNumber = "DESY-07-002, SFB-CPP-07-01",
    doi = "10.1016/j.nuclphysb.2007.05.008",
    journal = "Nucl. Phys. B",
    volume = "782",
    pages = "51--78",
    year = "2007"
}

@article{Davies:2016ruz:cc-nnlo-neutrino,
    author = "Davies, J. and Vogt, A. and Moch, S. and Vermaseren, J. A. M.",
    title = "{Non-singlet coefficient functions for charged-current deep-inelastic scattering to the third order in QCD}",
    eprint = "1606.08907",
    archivePrefix = "arXiv",
    primaryClass = "hep-ph",
    reportNumber = "LTH-1089, DESY-16-110, NIKHEF-2016-030, LTH 1089, DESY 16-110, Nikhef 2016-030",
    doi = "10.22323/1.265.0059",
    journal = "PoS",
    volume = "DIS2016",
    pages = "059",
    year = "2016"
}

@article{Ingelman:1987zv,
    author = "Ingelman, G. and Ruckl, R.",
    title = "{Determination of Quark Distributions in $e p$ Collisions}",
    reportNumber = "DESY-87-140",
    doi = "10.1016/0370-2693(88)91158-6",
    journal = "Phys. Lett. B",
    volume = "201",
    pages = "369--374",
    year = "1988"
}

@article{Bardeen:1978yd:nc,
    author = "Bardeen, William A. and Buras, A. J. and Duke, D. W. and Muta, T.",
    title = "{Deep Inelastic Scattering Beyond the Leading Order in Asymptotically Free Gauge Theories}",
    reportNumber = "FERMILAB-PUB-78-042-T",
    doi = "10.1103/PhysRevD.18.3998",
    journal = "Phys. Rev. D",
    volume = "18",
    pages = "3998",
    year = "1978"
}

@article{Cooper-Sarkar:1997pqx,
    author = "Cooper-Sarkar, Amanda M. and Devenish, R. C. E. and De Roeck, A.",
    title = "{Structure functions of the nucleon and their interpretation}",
    eprint = "hep-ph/9712301",
    archivePrefix = "arXiv",
    reportNumber = "OUNP-97-10, DESY-97-226",
    doi = "10.1142/S0217751X98001670",
    journal = "Int. J. Mod. Phys. A",
    volume = "13",
    pages = "3385--3586",
    year = "1998"
}

@article{Risse:2025smp:heavy-cc,
    author = "Risse, Peter and Bertone, Valerio and Je{\v{z}}o, Tomas and Kova{\v{r}}{\'\i}k, Karol and Kusina, Aleksander and Olness, Fredrick and Schienbein, Ingo",
    title = "{Heavy quark mass effects in charged-current deep-inelastic scattering at approximate NNLO in the Aivazis-Collins-Olness-Tung scheme}",
    eprint = "2504.13317",
    archivePrefix = "arXiv",
    primaryClass = "hep-ph",
    reportNumber = "MS-TP-25-07, SMU-PHY-25-01, JLAB-THY-25-4279, IFJPAN-IV-2025-9",
    doi = "10.1103/hmlr-p7zb",
    journal = "Phys. Rev. D",
    volume = "112",
    number = "11",
    pages = "114004",
    year = "2025"
}

@article{Gao:2021fle:heavy-cc,
    author = "Gao, Jun and Hobbs, T. J. and Nadolsky, P. M. and Sun, ChuanLe and Yuan, C. -P.",
    title = "{General heavy-flavor mass scheme for charged-current DIS at NNLO and beyond}",
    eprint = "2107.00460",
    archivePrefix = "arXiv",
    primaryClass = "hep-ph",
    reportNumber = "SMU-HEP-21-07, MSUHEP-21-012, FERMILAB-PUB-21-842-SCD",
    doi = "10.1103/PhysRevD.105.L011503",
    journal = "Phys. Rev. D",
    volume = "105",
    number = "1",
    pages = "L011503",
    year = "2022"
}

@article{Remiddi:1999ew,
    author = "Remiddi, E. and Vermaseren, J. A. M.",
    title = "{Harmonic polylogarithms}",
    eprint = "hep-ph/9905237",
    archivePrefix = "arXiv",
    reportNumber = "NIKHEF-99-005, TTP-99-08",
    doi = "10.1142/S0217751X00000367",
    journal = "Int. J. Mod. Phys. A",
    volume = "15",
    pages = "725--754",
    year = "2000"
}

@article{Blumlein:2016xcy:Hq-heavy-wilson-nonsinglet,
    author = {Bl{\"u}mlein, Johannes and Falcioni, Giulio and De Freitas, Abilio},
    title = "{The Complete $O(\alpha_s^2)$ Non-Singlet Heavy Flavor Corrections to the Structure Functions $g_{1,2}^{ep}(x,Q^2)$, $F_{1,2,L}^{ep}(x,Q^2)$, $F_{1,2,3}^{\nu(\bar{\nu})}(x,Q^2)$ and the Associated Sum Rules}",
    eprint = "1605.05541",
    archivePrefix = "arXiv",
    primaryClass = "hep-ph",
    reportNumber = "DESY-15-171, DO-TH-15-14",
    doi = "10.1016/j.nuclphysb.2016.06.018",
    journal = "Nucl. Phys. B",
    volume = "910",
    pages = "568--617",
    year = "2016"
}

@article{Blumlein:2019qze:Hq-heavy-wilson,
    author = {Bl{\"u}mlein, J. and De Freitas, A. and Raab, C. G. and Sch{\"o}nwald, K.},
    title = "{The unpolarized two-loop massive pure singlet Wilson coefficients for deep-inelastic scattering}",
    eprint = "1903.06155",
    archivePrefix = "arXiv",
    primaryClass = "hep-ph",
    reportNumber = "DESY-19-038, DO-TH-18/25",
    doi = "10.1016/j.nuclphysb.2019.114659",
    journal = "Nucl. Phys. B",
    volume = "945",
    pages = "114659",
    year = "2019"
}

@article{Forte:2010ta:FONLL,
    author = "Forte, Stefano and Laenen, Eric and Nason, Paolo and Rojo, Juan",
    title = "{Heavy quarks in deep-inelastic scattering}",
    eprint = "1001.2312",
    archivePrefix = "arXiv",
    primaryClass = "hep-ph",
    reportNumber = "IFUM-949-FT, NIKHEF-2010-001, ITP-UU-10-03, ITFA-2010-01",
    doi = "10.1016/j.nuclphysb.2010.03.014",
    journal = "Nucl. Phys. B",
    volume = "834",
    pages = "116--162",
    year = "2010"
}

@article{Kramer:2000hn:s-acot,
    author = {Kr{\"a}mer, Michael and Olness, Fredrick I. and Soper, Davison E.},
    title = "{Treatment of heavy quarks in deeply inelastic scattering}",
    eprint = "hep-ph/0003035",
    archivePrefix = "arXiv",
    reportNumber = "EDINBURGH-2000-02",
    doi = "10.1103/PhysRevD.62.096007",
    journal = "Phys. Rev. D",
    volume = "62",
    pages = "096007",
    year = "2000"
}

@article{Tung:2001mv:acot-chi,
    author = "Tung, Wu-Ki and Kretzer, Stefan and Schmidt, Carl",
    editor = "Grindhammer, Guenter and Kniehl, B. A. and Kramer, G. and Ochs, W.",
    title = "{Open heavy flavor production in QCD: Conceptual framework and implementation issues}",
    eprint = "hep-ph/0110247",
    archivePrefix = "arXiv",
    doi = "10.1088/0954-3899/28/5/321",
    journal = "J. Phys. G",
    volume = "28",
    pages = "983--996",
    year = "2002"
}

@article{Kretzer:2003it:acot-chi,
    author = "Kretzer, S. and Lai, H. L. and Olness, F. I. and Tung, W. K.",
    title = "{CTEQ6 parton distributions with heavy quark mass effects}",
    eprint = "hep-ph/0307022",
    archivePrefix = "arXiv",
    reportNumber = "MSU-HEP-030101, BNL-NT-03-2, RBRC-325",
    doi = "10.1103/PhysRevD.69.114005",
    journal = "Phys. Rev. D",
    volume = "69",
    pages = "114005",
    year = "2004"
}

@article{Thorne:1997ga:mscheme-tr,
    author = "Thorne, R. S. and Roberts, R. G.",
    title = "{An Ordered analysis of heavy flavor production in deep inelastic scattering}",
    eprint = "hep-ph/9709442",
    archivePrefix = "arXiv",
    reportNumber = "RAL-TR-97-049",
    doi = "10.1103/PhysRevD.57.6871",
    journal = "Phys. Rev. D",
    volume = "57",
    pages = "6871--6898",
    year = "1998"
}

@article{Thorne:1997uu:mscheme-tr,
    author = "Thorne, R. S. and Roberts, R. G.",
    title = "{A Practical procedure for evolving heavy flavor structure functions}",
    eprint = "hep-ph/9711223",
    archivePrefix = "arXiv",
    reportNumber = "RAL-TR-97-061",
    doi = "10.1016/S0370-2693(97)01580-3",
    journal = "Phys. Lett. B",
    volume = "421",
    pages = "303--311",
    year = "1998"
}

@article{Thorne:2006qt:mscheme-tr,
    author = "Thorne, R. S.",
    title = "{A Variable-flavor number scheme for NNLO}",
    eprint = "hep-ph/0601245",
    archivePrefix = "arXiv",
    reportNumber = "CAVENDISH-HEP-2006-01",
    doi = "10.1103/PhysRevD.73.054019",
    journal = "Phys. Rev. D",
    volume = "73",
    pages = "054019",
    year = "2006"
}

@Inbook{Klein2012,
author="Klein, Sebastian",
title="Deeply Inelastic Scattering",
bookTitle="Charm Production in Deep Inelastic Scattering: Mellin Moments of Heavy Flavor Contributions to F2(x,Q^2) at NNLO",
year="2012",
publisher="Springer Berlin Heidelberg",
address="Berlin, Heidelberg",
pages="17--46",
isbn="978-3-642-23286-2",
doi="10.1007/978-3-642-23286-2_2",
url="https://doi.org/10.1007/978-3-642-23286-2_2"
}

@article{Witten:1975bh,
    author = "Witten, Edward",
    title = "{Heavy Quark Contributions to Deep Inelastic Scattering}",
    reportNumber = "PRINT-75-1059 (PRINCETON)",
    doi = "10.1016/0550-3213(76)90111-5",
    journal = "Nucl. Phys. B",
    volume = "104",
    pages = "445--476",
    year = "1976"
}

@article{Babcock:1977fi,
    author = "Babcock, John and Sivers, Dennis W. and Wolfram, Stephen",
    title = "{QCD Estimates for Heavy Particle Production}",
    reportNumber = "ANL-HEP-PR-77-68",
    doi = "10.1103/PhysRevD.18.162",
    journal = "Phys. Rev. D",
    volume = "18",
    pages = "162",
    year = "1978"
}

@article{Shifman:1977yb,
    author = "Shifman, Mikhail A. and Vainshtein, A. I. and Zakharov, Valentin I.",
    title = "{Remarks on Charm Electroproduction in QCD}",
    reportNumber = "ITEP-19-1977",
    doi = "10.1016/0550-3213(78)90020-2",
    journal = "Nucl. Phys. B",
    volume = "136",
    pages = "157",
    year = "1978"
}

@article{Leveille:1978px,
    author = "Leveille, J. P. and Weiler, Thomas J.",
    title = "{Characteristics of Heavy Quark Leptoproduction in QCD}",
    reportNumber = "LTH 43, ICTP/77-78/25a",
    doi = "10.1016/0550-3213(79)90420-6",
    journal = "Nucl. Phys. B",
    volume = "147",
    pages = "147--173",
    year = "1979"
}

@article{Laenen:1992zk-heavyq-nlo,
    author = "Laenen, Eric and Riemersma, S. and Smith, J. and van Neerven, W. L.",
    title = "{Complete O (alpha-s) corrections to heavy flavor structure functions in electroproduction}",
    reportNumber = "ITP-SB-92-09",
    doi = "10.1016/0550-3213(93)90201-Y",
    journal = "Nucl. Phys. B",
    volume = "392",
    pages = "162--228",
    year = "1993"
}

@article{Riemersma:1994hv-heavyq-nlo,
    author = "Riemersma, S. and Smith, J. and van Neerven, W. L.",
    title = "{Rates for inclusive deep inelastic electroproduction of charm quarks at HERA}",
    eprint = "hep-ph/9411431",
    archivePrefix = "arXiv",
    reportNumber = "SMU-HEP-94-25, ITP-SB-94-59, INLO-PUB-16-94",
    doi = "10.1016/0370-2693(95)00036-K",
    journal = "Phys. Lett. B",
    volume = "347",
    pages = "143--151",
    year = "1995"
}

@article{Buza:1995ie:orignal-heavy-nlo,
    author = "Buza, M. and Matiounine, Y. and Smith, J. and Migneron, R. and van Neerven, W. L.",
    title = "{Heavy quark coefficient functions at asymptotic values Q**2 {\ensuremath{>}}{\ensuremath{>}} m**2}",
    eprint = "hep-ph/9601302",
    archivePrefix = "arXiv",
    reportNumber = "NIKHEF-95-070, ITP-SB-95-59, INLO-PUB-22-95",
    doi = "10.1016/0550-3213(96)00228-3",
    journal = "Nucl. Phys. B",
    volume = "472",
    pages = "611--658",
    year = "1996"
}

@article{Buza:1996wv-heavyq-nlo,
    author = "Buza, M. and Matiounine, Y. and Smith, J. and van Neerven, W. L.",
    title = "{Charm electroproduction viewed in the variable flavor number scheme versus fixed order perturbation theory}",
    eprint = "hep-ph/9612398",
    archivePrefix = "arXiv",
    reportNumber = "NIKHEF-96-027, ITP-SB-96-66, DESY-96-258, INLO-PUB-22-96",
    doi = "10.1007/BF01245820",
    journal = "Eur. Phys. J. C",
    volume = "1",
    pages = "301--320",
    year = "1998"
}

@article{Jiang:2025frv:topq:eic,
    author = "Jiang, Xu-Hui and Liu, Yiming and Yan, Bin",
    title = "{Probing top-quark electroweak couplings indirectly at the Electron-Ion Collider}",
    eprint = "2507.21477",
    archivePrefix = "arXiv",
    primaryClass = "hep-ph",
    reportNumber = "CPTNP-2025-022",
    doi = "10.1103/w9wl-cjzq",
    journal = "Phys. Rev. D",
    volume = "112",
    number = "11",
    pages = "L111303",
    year = "2025"
}

@article{Formaggio:2012cpf:neutrino-dis-review,
    author = "Formaggio, J. A. and Zeller, G. P.",
    title = "{From eV to EeV: Neutrino Cross Sections Across Energy Scales}",
    eprint = "1305.7513",
    archivePrefix = "arXiv",
    primaryClass = "hep-ex",
    reportNumber = "FERMILAB-PUB-12-785-E",
    doi = "10.1103/RevModPhys.84.1307",
    journal = "Rev. Mod. Phys.",
    volume = "84",
    pages = "1307--1341",
    year = "2012"
}

@article{Bjorken:1969in:neutrinoDIS,
    author = "Bjorken, J. D. and Paschos, Emmanuel A.",
    title = "{High-Energy Inelastic Neutrino-Nucleon Interactions}",
    reportNumber = "SLAC-PUB-0678",
    doi = "10.1103/PhysRevD.1.3151",
    journal = "Phys. Rev. D",
    volume = "1",
    pages = "3151--3160",
    year = "1970"
}

@article{LlewellynSmith:1971uhs:neutrinoDIS,
    author = "Llewellyn Smith, C. H.",
    title = "{Neutrino Reactions at Accelerator Energies}",
    reportNumber = "SLAC-PUB-0958",
    doi = "10.1016/0370-1573(72)90010-5",
    journal = "Phys. Rept.",
    volume = "3",
    pages = "261--379",
    year = "1972"
}

@article{Zijlstra:1992kj,
    author = "Zijlstra, E. B. and van Neerven, W. L.",
    title = "{Order alpha-s**2 correction to the structure function F3 (x, Q**2) in deep inelastic neutrino - hadron scattering}",
    reportNumber = "INLO-PUB-12-92",
    doi = "10.1016/0370-2693(92)91277-G",
    journal = "Phys. Lett. B",
    volume = "297",
    pages = "377--384",
    year = "1992"
}

@article{Heinz:2000bk:qgp-plasma,
    author = "Heinz, Ulrich W. and Jacob, Maurice",
    title = "{Evidence for a new state of matter: An Assessment of the results from the CERN lead beam program}",
    eprint = "nucl-th/0002042",
    archivePrefix = "arXiv",
    month = "1",
    year = "2000"
}

@article{CMS:2016gox:dark,
    author = "Khachatryan, Vardan and others",
    collaboration = "CMS",
    title = "{Search for dark matter particles in proton-proton collisions at $ \sqrt{s}=8 $ TeV using the razor variables}",
    eprint = "1603.08914",
    archivePrefix = "arXiv",
    primaryClass = "hep-ex",
    reportNumber = "CMS-EXO-14-004, CERN-EP-2016-025",
    doi = "10.1007/JHEP12(2016)088",
    journal = "JHEP",
    volume = "12",
    pages = "088",
    year = "2016"
}

@article{Alekhin:2015byh:dark,
    author = "Alekhin, Sergey and others",
    title = "{A facility to Search for Hidden Particles at the CERN SPS: the SHiP physics case}",
    eprint = "1504.04855",
    archivePrefix = "arXiv",
    primaryClass = "hep-ph",
    reportNumber = "CERN-SPSC-2015-017, SPSC-P-350-ADD-1",
    doi = "10.1088/0034-4885/79/12/124201",
    journal = "Rept. Prog. Phys.",
    volume = "79",
    number = "12",
    pages = "124201",
    year = "2016"
}

@inproceedings{Hong:2017avi:dark,
    author = "Hong, Tae Min",
    title = "{Dark matter searches at the LHC}",
    booktitle = "{5th Large Hadron Collider Physics Conference}",
    eprint = "1709.02304",
    archivePrefix = "arXiv",
    primaryClass = "hep-ex",
    reportNumber = "ATL-PHYS-PROC-2017-100",
    month = "9",
    year = "2017"
}

@misc{oma4:invdip-github,
    author = {H{\"a}nninen, Henri and Kykk{\"a}nen, Antti and Schl{\"u}ter, Hj{\o}rdis},
    title = "{Reconstruction of the Dipole Amplitude in the Dipole Picture as a mathematical Inverse Problem (software)}",
    archivePrefix = "Github",
    URL = {https://github.com/hhannine/inversedipole},
    year = "2025"
}

@misc{oma-talk-2025:invdip-qcd-seminar,
    author = {H{\"a}nninen, Henri},
    title = "{Turning the dipole picture of deep inelastic scattering into a definition for an integral transform (talk, slides avail.)}",
    note = "{Centre of Excellence in Quark Matter seminar series}",
    URL = {https://indico.global/event/16327/},
    year = "2025"
}

@article{Armesto:2022mxy:pdf-saturation,
    author = {Armesto, Nestor and Lappi, Tuomas and M{\"a}ntysaari, Heikki and Paukkunen, Hannu and Tevio, Mirja},
    title = "{Signatures of gluon saturation from structure-function measurements}",
    eprint = "2203.05846",
    archivePrefix = "arXiv",
    primaryClass = "hep-ph",
    doi = "10.1103/PhysRevD.105.114017",
    journal = "Phys. Rev. D",
    volume = "105",
    number = "11",
    pages = "114017",
    year = "2022"
}

@phdthesis{Hekhorn:2019nlf-felix-phd,
    author = "Hekhorn, Felix",
    title = "{Next-to-Leading Order QCD Corrections to Heavy-Flavour Production in Neutral Current DIS}",
    eprint = "1910.01536",
    archivePrefix = "arXiv",
    primaryClass = "hep-ph",
    doi = "10.15496/publikation-34811",
    school = "Tubingen U., Math. Inst.",
    year = "2019"
}

@phdthesis{Davies:2016bwb:cc-dis-davies-phd,
    author = "Davies, Joshua",
    title = "{NNNLO and All-Order Corrections to Splitting and Coefficient Functions in Deep-Inelastic Scattering}",
    doi = "10.17638/03003745",
    school = "U. Liverpool (main)",
    year = "2016"
}

@article{Hekhorn:2018ywm:phd-ref-article1,
    author = "Hekhorn, Felix and Stratmann, Marco",
    title = "{Next-to-Leading Order QCD Corrections to Inclusive Heavy-Flavor Production in Polarized Deep-Inelastic Scattering}",
    eprint = "1805.09026",
    archivePrefix = "arXiv",
    primaryClass = "hep-ph",
    doi = "10.1103/PhysRevD.98.014018",
    journal = "Phys. Rev. D",
    volume = "98",
    number = "1",
    pages = "014018",
    year = "2018"
}

@misc{Hekhorn:heavy-quark-yadism-leproHQ-zenodo,
    author = "Barontini, A. and Candido, A. and Hekhorn, F. and Magni, G. and Laurenti, N. and Rabemananjara, T. R. and Schwan, C. and Stegeman, R.",
    title = "{NNPDF/yadism [Computer software]. Zenodo.}",
    url = "https://doi.org/10.5281/zenodo.21036392",
    doi = "10.5281/zenodo.21036392",
    year = "2026"
}

@article{Barontini:2024xgu:flavor-number-scheme-names-and-intro,
    author = "Barontini, Andrea and Candido, Alessandro and Hekhorn, Felix and Magni, Giacomo and Stegeman, Roy",
    title = "{An FONLL prescription with coexisting flavor number PDFs}",
    eprint = "2408.07383",
    archivePrefix = "arXiv",
    primaryClass = "hep-ph",
    reportNumber = "Nikhef-2024-014, Edinburgh 2024/5, TIF-UNIMI-2024-13",
    doi = "10.1007/JHEP10(2024)004",
    journal = "JHEP",
    volume = "10",
    pages = "004",
    year = "2024"
}

@article{Barontini:2023vmr:pineline-fktable-discretization,
    author = "Barontini, Andrea and Candido, Alessandro and Cruz-Martinez, Juan M. and Hekhorn, Felix and Schwan, Christopher",
    title = "{Pineline: Industrialization of high-energy theory predictions}",
    eprint = "2302.12124",
    archivePrefix = "arXiv",
    primaryClass = "hep-ph",
    reportNumber = "TIF-UNIMI-2023-4, CERN-TH-2023-021",
    doi = "10.1016/j.cpc.2023.109061",
    journal = "Comput. Phys. Commun.",
    volume = "297",
    pages = "109061",
    year = "2024"
}

@article{DelDebbio:2021whr:fktable,
    author = "Del Debbio, Luigi and Giani, Tommaso and Wilson, Michael",
    title = "{Bayesian approach to inverse problems: an application to NNPDF closure testing}",
    eprint = "2111.05787",
    archivePrefix = "arXiv",
    primaryClass = "hep-ph",
    doi = "10.1140/epjc/s10052-022-10297-x",
    journal = "Eur. Phys. J. C",
    volume = "82",
    number = "4",
    pages = "330",
    year = "2022"
}

@article{Candido:2024rkr:yadism,
    author = "Candido, Alessandro and Hekhorn, Felix and Magni, Giacomo and Rabemananjara, Tanjona R. and Stegeman, Roy",
    title = "{Yadism: yet another deep-inelastic scattering module}",
    eprint = "2401.15187",
    archivePrefix = "arXiv",
    primaryClass = "hep-ph",
    reportNumber = "CERN-TH-2024-015, Edinburgh 2024/4, Nikhef-2024-002",
    doi = "10.1140/epjc/s10052-024-12972-7",
    journal = "Eur. Phys. J. C",
    volume = "84",
    number = "7",
    pages = "698",
    year = "2024"
}

@misc{yadism:manual:online,
    author = "Candido, Alessandro and Hekhorn, Felix and Magni, Giacomo and Rabemananjara, Tanjona R. and Stegeman, Roy",
    key = "Isospin",
    title = "Yadism: yet another deep-inelastic scattering module",
    url = "https://yadism.readthedocs.io/en/latest/theory/misc.html#isospin"
}

@article{10.1093/ptep/ptac097,
    author = "{Particle Data Group}",
    title = "{Review of Particle Physics}",
    journal = {Progress of Theoretical and Experimental Physics},
    volume = {2022},
    number = {8},
    pages = {083C01},
    year = {2022},
    month = {08},
    issn = {2050-3911},
    doi = {10.1093/ptep/ptac097},
    url = {https://doi.org/10.1093/ptep/ptac097},
}

@article{Gelis:2010nm,
    author = "Gelis, Francois and Iancu, Edmond and Jalilian-Marian, Jamal and Venugopalan, Raju",
    title = "{The Color Glass Condensate}",
    eprint = "1002.0333",
    archivePrefix = "arXiv",
    primaryClass = "hep-ph",
    doi = "10.1146/annurev.nucl.010909.083629",
    journal = "Ann. Rev. Nucl. Part. Sci.",
    volume = "60",
    pages = "463--489",
    year = "2010"
}

@article{Abramowicz:2015mha:HERAIInewcombined,
      author         = "Abramowicz, H. and others",
      title          = "{Combination of measurements of inclusive deep inelastic
                        ${e^{\pm }p}$ scattering cross sections and QCD analysis
                        of HERA data}",
      collaboration  = "H1, ZEUS",
      journal        = "Eur. Phys. J.",
      volume         = "C75",
      year           = "2015",
      number         = "12",
      pages          = "580",
      doi            = "10.1140/epjc/s10052-015-3710-4",
      eprint         = "1506.06042",
      archivePrefix  = "arXiv",
      primaryClass   = "hep-ex",
      reportNumber   = "DESY-15-039",
      SLACcitation   = "%%CITATION = ARXIV:1506.06042;%%"
}

@misc{Abramowicz:2015mha:HERAIInewcombined:hepdata.68951,
    author = "{H1Collaboration} and {ZEUS Collaboration}",
    title = "{Combination of Measurements of Inclusive Deep Inelastic $e^{\pm}p$ Scattering Cross Sections and QCD Analysis of HERA Data}",
    howpublished = "{HEPData (collection)}",
    year = 2015,
    note = "\url{https://doi.org/10.17182/hepdata.68951}"
}

@article{Abramowicz:2018flt:HERAII-charm-bottom-combined,
      author         = "Abramowicz, H. and others",
      title          = "{Combination and QCD analysis of charm and beauty
                        production cross-section measurements in deep inelastic
                        $ep$ scattering at HERA}",
      collaboration  = "H1, ZEUS",
      journal        = "Eur. Phys. J.",
      volume         = "C78",
      year           = "2018",
      number         = "6",
      pages          = "473",
      doi            = "10.1140/epjc/s10052-018-5848-3",
      primaryClass   = "hep-ex",
      reportNumber   = "DESY 18-037, DESY-18-037",
      SLACcitation   = "%%CITATION = ARXIV:1804.01019;%%"
}

@article{Riemann1859,
  author    = {Bernhard Riemann},
  title     = {Ueber die Anzahl der Primzahlen unter einer gegebenen Gr{\"o}sse},
  journal   = {Monatsberichte der Berliner Akademie},
  year      = {1859},
  month     = {November},
  pages     = {671--680}
}

@book{hochstadt1989integraleq,
  title={Integral Equations},
  author={Hochstadt, H.},
  isbn={9780471504047},
  lccn={73004230},
  series={Wiley Classics Library},
  url={https://books.google.fi/books?id=gUbLEAAAQBAJ},
  year={1989},
  publisher={Wiley}
}

@book{Polyanin2008handbook:integral-equations,
  title={Handbook of Integral Equations: Second Edition},
  author={Polyanin, A. D. and Manzhirov, A.V.},
  isbn={9780203881057},
  series={Handbooks of mathematical equations},
  url={https://books.google.fi/books?id=M4JjK9OvVqkC},
  year={2008},
  publisher={CRC Press}
}

@book{Moslehian:2023:matrix-operator-equations-applications-integral-equations,
   title =     {Matrix and Operator Equations and Applications (Mathematics Online First Collections)},
   author =    {Mohammad Sal Moslehian},
   publisher = {Springer},
   isbn =      {303125385X; 9783031253850},
   year =      {2023},
   url =       {libgen.li/file.php?md5=657e67bc5f2b82ba75f148178506e67b}}

@article{Wiener1931:wiener-hopf-original,
  author    = {Wiener, Norbert and Hopf, Eberhard},
  title     = {{\"U}ber eine Klasse singul{\"a}rer Integralgleichungen},
  journal   = {Sitzungsberichte der Preussischen Akademie der Wissenschaften, Physikalisch-Mathematische Klasse},
  year      = {1931},
  pages     = {696--706}
}

@book{Hopf1934:wiener-hopf-original,
  author    = {Hopf, Eberhard},
  title     = {Mathematical Problems of Radiative Equilibrium},
  series    = {Cambridge Tracts in Mathematics and Mathematical Physics},
  number    = {31},
  publisher = {Cambridge University Press},
  address   = {Cambridge},
  year      = {1934}
}

@article{Lawrie2007:wiener-hopf-review,
  author   = {Lawrie, J. B. and Abrahams, I. D.},
  title    = {A brief historical perspective of the {Wiener}--{Hopf} technique},
  journal  = {Journal of Engineering Mathematics},
  year     = {2007},
  volume   = {59},
  number   = {4},
  pages    = {351--358},
  doi      = {10.1007/s10665-007-9195-x},
  url      = {https://doi.org/10.1007/s10665-007-9195-x}
}

@article{GohbergKrein1958:wiener-hopf-systems,
  author    = {Gohberg, I. C. and Kre{\u{\i}}n, M. G.},
  title     = {Systems of integral equations on a half line with kernels depending on the difference of arguments},
  journal   = {Uspehi Mat. Nauk},
  volume    = {13},
  year      = {1958},
  number    = {2 (80)},
  pages     = {3--72},
  note      = {English transl., Amer. Math. Soc. Transl. (2) 14 (1960), 217--287. MR 21 \#1506; 22 \#3954}
}

@article{Geleg:1966:wiener-hopf-nonint-kernel,
  author  = {Gelig, A. Kh.},
  title   = {The absolute stability of nonlinear control systems with distributed parameters in critical cases},
  journal = {Automat. Remote Control},
  volume  = {27},
  page = {5},
  number = {4},
  year    = {1966}
}

@article{Kisil:2021:wiener-hopf-methods,
    author = {Kisil, Anastasia V. and Abrahams, I. David and Mishuris, Gennady and Rogosin, Sergei V.},
    title = {The Wiener–Hopf technique, its generalizations and applications: constructive and approximate methods},
    journal = {Proceedings of the Royal Society A: Mathematical, Physical and Engineering Sciences},
    volume = {477},
    number = {2254},
    pages = {20210533},
    year = {2021},
    month = {10},
    issn = {1364-5021},
    doi = {10.1098/rspa.2021.0533},
    url = {https://doi.org/10.1098/rspa.2021.0533},
}

@incollection{Corduneanu:1973:wiener-hopf,
title = {4 Wiener-Hopf Equations},
author = {Constantin Corduneanu},
series = {Mathematics in Science and Engineering},
publisher = {Elsevier},
volume = {104},
pages = {143-175},
year = {1973},
booktitle = {Integral Equations and Stability of Feedback Systems},
issn = {0076-5392},
doi = {https://doi.org/10.1016/S0076-5392(08)61735-X},
url = {https://www.sciencedirect.com/science/article/pii/S007653920861735X}
}

@article{Volterra1,
    author = {Volterra, Vito},
    title = "{Sulle inversione degli integrali definiti, Nota I}",
    journal = {Atti R. Accad. Sci. Torino},
    volume={31},
    pages={311--323},
    year = {1896}
}

@article{Volterra2,
    author = {Volterra, Vito},
    title = "{Sulle inversione degli integrali definiti, Nota II}",
    journal = {Atti R. Accad. Sci. Torino},
    volume={31},
    pages={400--408},
    year = {1896}
}

@article{Volterra3,
    author = {Volterra, Vito},
    title = "{Sulle inversione degli integrali definiti}",
    journal = {Rend. R. Accad. Lincei},
    volume={5},
    pages={177--l85},
    year = {1896}
}

@article{Volterra4,
    author = {Volterra, Vito},
    title = {Sulle inversione degli integrali multipli},
    journal = {Rend. R. Accad. Lincei},
    volume={5},
    pages={289--300},
    year = {1896}
}

@article{Volterra5,
    author = {Volterra, Vito},
    title = {Sopra alcune questioni di inversione di integrali definiti},
    journal = {Ann. Mat. Pura Appl.},
    volume={25},
    pages={139--178},
    year = {1897}
}

@book{brunner2017volterra,
  title={Volterra integral equations: an introduction to theory and applications},
  author={Brunner, Hermann},
  volume={30},
  year={2017},
  publisher={Cambridge University Press}
}

@article{JohnFritz:1938:xraytransform,
author = {Fritz John},
title = {{The ultrahyperbolic differential equation with four independent variables}},
volume = {4},
journal = {Duke Mathematical Journal},
number = {2},
publisher = {Duke University Press},
pages = {300 -- 322},
year = {1938},
doi = {10.1215/S0012-7094-38-00423-5},
URL = {https://doi.org/10.1215/S0012-7094-38-00423-5}
}

@book{Natterer:2001:mathematicalmethods-img-rec,
author = {Natterer, Frank and Wübbeling, Frank},
title = {Mathematical Methods in Image Reconstruction},
publisher = {Society for Industrial and Applied Mathematics},
year = {2001},
doi = {10.1137/1.9780898718324},
address = {},
edition   = {},
URL = {https://epubs.siam.org/doi/abs/10.1137/1.9780898718324},
eprint = {https://epubs.siam.org/doi/pdf/10.1137/1.9780898718324}
}

@book{Hansen:2021ct,
author = {Hansen, Per Christian and Jørgensen, Jakob and Lionheart, William R. B.},
title = {Computed Tomography: Algorithms, Insight, and Just Enough Theory},
publisher = {Society for Industrial and Applied Mathematics},
year = {2021},
doi = {10.1137/1.9781611976670},
address = {Philadelphia, PA},
edition   = {},
URL = {https://epubs.siam.org/doi/abs/10.1137/1.9781611976670},
eprint = {https://epubs.siam.org/doi/pdf/10.1137/1.9781611976670}
}

@misc{clason2021regularizationinverseproblems,
      title={Regularization of Inverse Problems}, 
      author={Christian Clason},
      year={2021},
      eprint={2001.00617},
      archivePrefix={arXiv},
      primaryClass={math.FA}, 
}

@book{Mueller:2012:eit,
author = {Mueller, Jennifer L. and Siltanen, Samuli},
title = {Linear and Nonlinear Inverse Problems with Practical Applications},
publisher = {Society for Industrial and Applied Mathematics},
year = {2012},
doi = {10.1137/1.9781611972344},
address = {Philadelphia, PA},
edition   = {},
URL = {https://epubs.siam.org/doi/abs/10.1137/1.9781611972344},
}

@article{Goyal:2019:imagedenoising,
author = {Goyal, Bhawna and Dogra, Ayush and Agrawal, Sunil and Sohi, B and Sharma, Apoorav},
year = {2019},
month = {09},
pages = {},
title = {Image Denoising Review: From classical to state-of-the-art approaches},
volume = {55},
journal = {Information Fusion},
doi = {10.1016/j.inffus.2019.09.003}
}

@article{Nystrom:1930-nystrommethod-quadrature,
author = {E. J. Nystr{\"o}m},
title = {{Über Die Praktische Auflösung von Integralgleichungen mit Anwendungen auf Randwertaufgaben}},
volume = {54},
journal = {Acta Mathematica},
publisher = {Institut Mittag-Leffler},
pages = {185 -- 204},
year = {1930},
doi = {10.1007/BF02547521},
URL = {https://doi.org/10.1007/BF02547521}
}

@book{kress1989linear-inteq-methods-quadrature,
  title={Linear Integral Equations},
  author={Kress, R.},
  isbn={9783540506164},
  lccn={89011289},
  series={Applied mathematical sciences},
  url={https://books.google.fi/books?id=BsU0rgEACAAJ},
  year={1989},
  publisher={World Publishing Company}
}

@book{hansen1998rank:stacked-form,
   title={Rank-deficient and discrete ill-posed problems: numerical aspects of linear inversion},
   author={Hansen, Per Christian},
   year={1998},
   publisher={SIAM}
}

@article{Tikhonov:1943,
   title={On the stability of inverse problems (transl.)},
   volume={39},
   journal={Doklady Akademii Nauk SSSR},
   author={Tikhonov, Andrey Nikolayevich},
   year={1943},
   pages={195–198} }

@article{Tikhonov:1963,
   title={Solution of incorrectly formulated problems and the regularization method (transl.)},
   volume={151},
   journal={Doklady Akademii Nauk SSSR},
   author={Tikhonov, Andrey Nikolayevich},
   year={1963},
   pages={501–504} }

@book{Tikhonov:1977,
   title={Solution of Ill-posed Problems},
   publisher={Washington: Winston \& Sons},
   author={Tikhonov, Andrey Nikolayevich and Arsenin, V. Y.},
   year={1977},
   isbn={0-470-99124-0} }

@article{Phillips:1962,
    author = {Phillips, David L.},
    title = {A Technique for the Numerical Solution of Certain Integral Equations of the First Kind},
    year = {1962},
    publisher = {Association for Computing Machinery},
    address = {New York, NY, USA},
    volume = {9},
    number = {1},
    issn = {0004-5411},
    url = {https://doi.org/10.1145/321105.321114},
    doi = {10.1145/321105.321114},
    journal = {J. ACM},
    month = 1,
    pages = {84–97},
    numpages = {14}}

@article{Hoerl:1962,
   title={Application of Ridge Analysis to Regression Problems},
   volume={58},
   journal={Chemical Engineering Progress},
   author={Hoerl, Arthur E.},
   year={1962},
   pages={54–59} }

@article{Kaczmarz:1937,
   title={Angenäherte Auflösung von Systemen linearer Gleichungen},
   volume={35},
   journal={Bulletin International de l'Académie Polonaise des Sciences et des Lettres},
   author={Kaczmarz, Stefan},
   year={1937},
   pages={355–357} }

@article{Gordon:1970:ART,
    title = {Algebraic Reconstruction Techniques (ART) for three-dimensional electron microscopy and X-ray photography},
    journal = {Journal of Theoretical Biology},
    volume = {29},
    number = {3},
    pages = {471-481},
    year = {1970},
    issn = {0022-5193},
    doi = {https://doi.org/10.1016/0022-5193(70)90109-8},
    author = {Richard Gordon and Robert Bender and Gabor T. Herman},
    }

@article{Cimmino:1938,
   title={Calcolo approssimato per le soluzioni dei sistemi di equazioni lineari},
   volume={XVI, Series II, Anno IX 1},
   journal={La Ricerca Scientifica},
   author={Cimmino, G.},
   year={1938},
   pages={326–333} }

@article{hanke1992regularization:large-ip,
    title={Regularization with differential operators: an iterative approach},
    author={Hanke, Martin},
    journal={Numerical functional analysis and optimization},
    volume={13},
    number={5-6},
    pages={523--540},
    year={1992},
    publisher={Taylor \& Francis}
}

@book{leugering2014trends:pde-opti,
  title={Trends in PDE Constrained Optimization},
  author={Leugering, G. and Benner, P. and Engell, S. and Griewank, A. and Harbrecht, H. and Hinze, M. and Rannacher, R. and Ulbrich, S.},
  isbn={9783319050836},
  series={International Series of Numerical Mathematics},
  year={2014},
  publisher={Springer International Publishing}
}

@book{Biegler-pde-opti,
author = {Biegler, Lorenz and Ghattas, Omar and Heinkenschloss, Matthias and Keyes, David and Waanders, Bart},
year = {2007},
month = {01},
isbn={978-0-89871-621-4},
publisher={Society for Industrial and Applied Mathematics},
series={Computational Science \& Engineering},
title = {Real-time PDE-constrained Optimization},
doi = {https://doi.org/10.1137/1.9780898718935}
}

@article{Mang_2018:pde-constr-medical,
   title={PDE-constrained optimization in medical image analysis},
   volume={19},
   ISSN={1573-2924},
   url={http://dx.doi.org/10.1007/s11081-018-9390-9},
   DOI={10.1007/s11081-018-9390-9},
   number={3},
   journal={Optimization and Engineering},
   publisher={Springer Science and Business Media LLC},
   author={Mang, Andreas and Gholami, Amir and Davatzikos, Christos and Biros, George},
   year={2018},
   month=6,
   pages={765–812} }

@article{Hanninen:2025iuv,
    author = {H{\"a}nninen, Henri and Kykk{\"a}nen, Antti and Schl{\"u}ter, Hj{\o}rdis},
    title = "{Reconstruction of the dipole amplitude in the dipole picture as a mathematical inverse problem}",
    eprint = "2509.05005",
    archivePrefix = "arXiv",
    primaryClass = "hep-ph",
    doi = "10.1103/7fhf-7fp4",
    journal = "Phys. Rev. D",
    volume = "112",
    number = "9",
    pages = "094026",
    year = "2025"
}

@article{Camarena:2019rmj:cosmic-distance,
    author = "Camarena, David and Marra, Valerio",
    title = "{A new method to build the (inverse) distance ladder}",
    eprint = "1910.14125",
    archivePrefix = "arXiv",
    primaryClass = "astro-ph.CO",
    doi = "10.1093/mnras/staa770",
    journal = "Mon. Not. Roy. Astron. Soc.",
    volume = "495",
    number = "3",
    pages = "2630--2644",
    year = "2020"
}

@article{Metz:2008ib:pdf-lorentz-invariance-violation,
    author = "Metz, A. and Schweitzer, P. and Teckentrup, T.",
    title = "{Lorentz invariance relations between parton distributions and the Wandzura-Wilczek approximation}",
    eprint = "0810.5212",
    archivePrefix = "arXiv",
    primaryClass = "hep-ph",
    doi = "10.1016/j.physletb.2009.08.052",
    journal = "Phys. Lett. B",
    volume = "680",
    pages = "141--147",
    year = "2009"
}

@inbook{Lorce:2025aqp:pdfs-and-generalizations-wigner,
    author = "Lorc{\'e}, C{\'e}dric and Metz, Andreas and Pasquini, Barbara and Schweitzer, Peter",
    title = "{Parton Distribution Functions and their Generalizations}",
    eprint = "2507.12664",
    archivePrefix = "arXiv",
    primaryClass = "hep-ph",
    doi = "10.1016/B978-0-443-26598-3.00088-2",
    month = "7",
    pages = "167-218",
    publisher={Elsevier},
    year = "2025"
}

@article{Ji:2003ak:pdf-wigner,
    author = "Ji, Xiang-dong",
    title = "{Viewing the proton through 'color' filters}",
    eprint = "hep-ph/0304037",
    archivePrefix = "arXiv",
    doi = "10.1103/PhysRevLett.91.062001",
    journal = "Phys. Rev. Lett.",
    volume = "91",
    pages = "062001",
    year = "2003"
}

@article{Belitsky:2003nz:pdf-wigner,
    author = "Belitsky, Andrei V. and Ji, Xiang-dong and Yuan, Feng",
    title = "{Quark imaging in the proton via quantum phase space distributions}",
    eprint = "hep-ph/0307383",
    archivePrefix = "arXiv",
    doi = "10.1103/PhysRevD.69.074014",
    journal = "Phys. Rev. D",
    volume = "69",
    pages = "074014",
    year = "2004"
}

@article{Hillery:Wigner-1984,
title = {Distribution functions in physics: Fundamentals},
journal = {Physics Reports},
volume = {106},
number = {3},
pages = {121-167},
year = {1984},
issn = {0370-1573},
doi = {https://doi.org/10.1016/0370-1573(84)90160-1},
author = {M. Hillery and R.F. O'Connell and M.O. Scully and E.P. Wigner},
}

@article{Wigner:1932,
  title = {On the Quantum Correction For Thermodynamic Equilibrium},
  volume = {40},
  ISSN = {0031-899X},
  url = {http://dx.doi.org/10.1103/PhysRev.40.749},
  DOI = {10.1103/physrev.40.749},
  number = {5},
  journal = {Physical Review},
  publisher = {American Physical Society (APS)},
  author = {Wigner,  E.},
  year = {1932},
  month = 6,
  pages = {749–759}
}

@article{Brodsky:1980pb:Hoyer-intrinsic-charm,
    author = "Brodsky, S. J. and Hoyer, P. and Peterson, C. and Sakai, N.",
    title = "{The Intrinsic Charm of the Proton}",
    reportNumber = "NORDITA-80-18",
    doi = "10.1016/0370-2693(80)90364-0",
    journal = "Phys. Lett. B",
    volume = "93",
    pages = "451--455",
    year = "1980"
}

@article{Ball:2015dpa:intrinsic-charm-dis,
    author = "Ball, Richard D. and Bonvini, Marco and Rottoli, Luca",
    title = "{Charm in Deep-Inelastic Scattering}",
    eprint = "1510.02491",
    archivePrefix = "arXiv",
    primaryClass = "hep-ph",
    reportNumber = "EDINBURGH-2015-06, CERN-PH-TH-2015-118, OUTP-15-25P",
    doi = "10.1007/JHEP11(2015)122",
    journal = "JHEP",
    volume = "11",
    pages = "122",
    year = "2015"
}

@article{NNPDF:2022qks,
    author = "Ball, Richard D. and Candido, Alessandro and Cruz-Martinez, Juan and Forte, Stefano and Giani, Tommaso and Hekhorn, Felix and Kudashkin, Kirill and Magni, Giacomo and Rojo, Juan",
    collaboration = "NNPDF",
    title = "{Evidence for intrinsic charm quarks in the proton}",
    eprint = "2208.08372",
    archivePrefix = "arXiv",
    primaryClass = "hep-ph",
    reportNumber = "Nikhef 2021-032, Edinburgh 2021/28, TIF-UNIMI-2021-21",
    doi = "10.1038/s41586-022-04998-2",
    journal = "Nature",
    volume = "608",
    number = "7923",
    pages = "483--487",
    year = "2022"
}

@article{NNPDF:2024dpb:mhou,
    author = "Ball, Richard D. and others",
    collaboration = "NNPDF",
    title = "{Determination of the theory uncertainties from missing higher orders on NNLO parton distributions with percent accuracy}",
    eprint = "2401.10319",
    archivePrefix = "arXiv",
    primaryClass = "hep-ph",
    reportNumber = "TIF-UNIMI-2023-22, Edinburgh 2023/34, CERN-TH-2024-009",
    doi = "10.1140/epjc/s10052-024-12772-z",
    journal = "Eur. Phys. J. C",
    volume = "84",
    number = "5",
    pages = "517",
    year = "2024"
}

@article{Chu2018-wigner-tomo,
  title     = "Creation and control of multi-phonon Fock states in a bulk
               acoustic-wave resonator",
  author    = "Chu, Yiwen and Kharel, Prashanta and Yoon, Taekwan and Frunzio,
               Luigi and Rakich, Peter T and Schoelkopf, Robert J",
  journal   = "Nature",
  publisher = "Springer Science and Business Media LLC",
  volume    =  {563},
  number    =  {7733},
  pages     = "666--670",
  month     =  {11},
  year      =  {2018}
}

@article{Lvovsky:2009:photon-wigner-tomo,
  title = {Continuous-variable optical quantum-state tomography},
  author = {Lvovsky, A. I. and Raymer, M. G.},
  journal = {Rev. Mod. Phys.},
  volume = {81},
  issue = {1},
  pages = {299--332},
  numpages = {0},
  year = {2009},
  month = {Mar},
  publisher = {American Physical Society},
  doi = {10.1103/RevModPhys.81.299},
  url = {https://link.aps.org/doi/10.1103/RevModPhys.81.299}
}

@article{Fusaoka:1998vc,
    author = "Fusaoka, Hideo and Koide, Yoshio",
    title = "{Updated estimate of running quark masses}",
    eprint = "hep-ph/9712201",
    archivePrefix = "arXiv",
    reportNumber = "AMU-97-02, US-97-07",
    doi = "10.1103/PhysRevD.57.3986",
    journal = "Phys. Rev. D",
    volume = "57",
    pages = "3986--4001",
    year = "1998"
}

@article{Gell-Mann:1964ewy,
    author = "Gell-Mann, Murray",
    title = "{A Schematic Model of Baryons and Mesons}",
    doi = "10.1016/S0031-9163(64)92001-3",
    journal = "Phys. Lett.",
    volume = "8",
    pages = "214--215",
    year = "1964"
}

@techreport{Zweig:1964ruk,
      author        = "Zweig, G",
      title         = "{An SU$_3$ model for strong interaction symmetry and its
                       breaking}",
      institution   = "CERN",
      reportNumber  = "CERN-TH-401",
      address       = "Geneva",
      year          = "1964",
      url           = "https://cds.cern.ch/record/352337",
      doi           = "10.17181/CERN-TH-401",
}

@inbook{Zweig:1964jf,
    author = "Zweig, G.",
    title = "{An SU(3) model for strong interaction symmetry and its breaking. Version 2}",
    booktitle = "{DEVELOPMENTS IN THE QUARK THEORY OF HADRONS. VOL. 1. 1964 - 1978}",
    reportNumber = "CERN-TH-412, NP-14146, PRINT-64-170",
    publisher = "Hadronic Press",
    doi = "10.17181/CERN-TH-412",
    pages = "22--101",
    month = "2",
    year = "1964"
}

@article{Gell-Mann:1961omu,
    author = "Gell-Mann, Murray",
    title = "{The Eightfold Way: A Theory of strong interaction symmetry}",
    reportNumber = "CTSL-20, TID-12608",
    doi = "10.2172/4008239",
    month = "3",
    year = "1961"
}

@article{Neeman:1961jhl,
    author = "Ne'eman, Yuval",
    editor = "Ruffini, R. and Verbin, Y.",
    title = "{Derivation of strong interactions from a gauge invariance}",
    doi = "10.1016/0029-5582(61)90134-1",
    journal = "Nucl. Phys.",
    volume = "26",
    pages = "222--229",
    year = "1961"
}

@article{Bjorken:1964gz,
    author = "Bjorken, J. D. and Glashow, S. L.",
    title = "{Elementary Particles and SU(4)}",
    doi = "10.1016/0031-9163(64)90433-0",
    journal = "Phys. Lett.",
    volume = "11",
    pages = "255--257",
    year = "1964"
}
